# ASPECTS OF HIGHER-SPIN THEORY
# WITH FERMIONS

This thesis was defended behind closed doors on April 16th, and publicly on April 18th 2014, in front of the following jury at U.L.B., in Brussels:

President: Prof. Michel TYTGAT

Secretary: Prof. Glenn BARNICH

Advisor: Prof. Marc HENNEAUX

Prof. Nicolas BOULANGER,
*Université de Mons, Belgique*
Prof. Matthias GABERDIEL,
*Eidgenössische Technische Hochschule Zürich, Schweiz*
Prof. Augusto SAGNOTTI,
*Scuola Normale Superiore di Pisa, Italia*
Dr. Ricardo TRONCOSO,
*Centro de Estudios Científicos de Valdivia, Chile*

# ASPECTS OF HIGHER-SPIN THEORY

# WITH FERMIONS

Gustavo Lucena Gómez

glucenag@ulb.ac.be

Service de Physique Théorique et Mathématique

The work presented in this thesis was carried out at the *Université Libre de Bruxelles* (U.L.B.), Department of Physics, in the *Physique Théorique et Mathématique* Service. It was supported by the *Fonds pour la Formation à la Recherche dans l'Industrie et dans l'Agriculture* (F.R.I.A.), of which the author is a Research Fellow (*aspirant*); and by the International Solvay Institutes (I.S.I.), as well as partially by IISN-Belgium (convention 4.4514.08), by the *Communauté Francaise de Belgique* through the ARC program and by the ERC 'SyDuGraM' Advanced Grant.

Printed in April, 2014 by *Presses universitaires de Bruxelles ASBL* — Avenue Franklin Roosevelt 50, CP 149, B-1050, Belgique

Typeset in LaTeX with the memoir class.

# Contents

*Fir d'Ann, mat Léift*

# Abstract

The present thesis is divided into three parts. In Part I we address a problem within Higher-Spin Gauge Theory in dimension three: namely, that of computing the asymptotic symmetry algebra of supersymmetric models, describing an infinite spectrum of integer and half-integer higher-spin fields. In Part II we investigate higher-spin theories in dimension four or greater, where we classify the consistent cross interactions between free gauge fermions of arbitrary spin and a photon or a graviton. A third part supplements the bulk of the manuscript with technical appendices.

Part I is concerned with the Higher-Spin Theory extending the anti-de Sitter orthosymplectic Supergravity in three dimensions. After recalling the construction of the latter we exhibit the structure of the former, and then explain how to generalize the boundary conditions for Supergravity to the higher-spin case. Following the usual procedure, we compute the form of the residual gauge parameter and then identify the Poisson-bracket algebra governing the asymptotic dynamics. It is found to be a nonlinear, supersymmetric algebra of the $\mathcal{W}_\infty$ type with same central charge as pure Gravity in the Virasoro sector, which is a subalgebra thereof. The simply supersymmetric case is treated explicitly whereas the details of the extended cases are relegated to the appendices.

Part II deals with the interaction problem for gauge fermions coupled to Electromagnetism and Gravity in flat spacetime of arbitrary dimension. First we recall the so-called BRST-Antifield techniques, which reformulate the deformation problem as a cohomological one, recasting the familiar Noether procedure for finding out interactions in a mathematically systematic way. We then use these methods to classify and obtain expressions for the gauge-invariant cubic couplings between a symmetric tensor-spinor and a spin-1 and spin-2 gauge field. With no input from previous works, we find the complete list of interaction terms with minimal assumptions and in particular shed light on the quartic obstructions to full consistency.

# Credits

The original contributions of this thesis are based on the following papers:

- ♭ **M. Henneaux**, **G. Lucena Gómez**, **J. Park**, **S.-J. Rey**, "Super-W(infinity) Asymptotic Symmetry of Higher-Spin $AdS_3$ Supergravity", *JHEP* 1206 (2012), p. 037, arXiv:1203.5152

- ♮ **M. Henneaux**, **G. Lucena Gómez**, **R. Rahman**, "Higher-Spin Fermionic Gauge Fields and Their Electromagnetic Coupling", *JHEP* 1208 (2012), p. 093, arXiv:1206.1048

- ♯ **M. Henneaux**, **G. Lucena Gómez**, **R. Rahman**, "Gravitational Interactions of Higher-Spin Fermions", *JHEP* 1401 (2014), p. 087, arXiv:1310.5152

Along the way, we have also contributed to the proceedings of the eighth *Modave Summer School in Mathematical Physics* with the following lecture notes, which contain review material used in the text:

- ♪ **G. Lucena Gómez**, "Higher-Spin Theory - Part II: enter dimension three", *Proceedings of Science* ModaveVIII (2012), p. 003, arXiv:1307.3200

# Acknowledgments

The work presented in this thesis as well as its writing would have been difficult, if not impossible to carry out without the help, support and companionship I have been surrounded by along these four years of doctorate. Accordingly, hereafter I attempt at thanking all the persons I would like to hold partly responsible for these achievements.

Foremost among all the people I am indebted to is my PhD advisor, Marc HENNEAUX. Working with him, benefiting from his insight and knowledge as well as from his counsel has been the greatest privilege, and a much pleasant and exciting one too. The time will come, I hope, when I am able to show the full extent of my gratitude to the bright teacher I have been lucky enough to take lessons from.

Also, for quite some part of my doctorate years, I have been under the impression of having a second advisor: M. Rakibur RAHMAN, and going on with these acknowledgments would be impossible without first mentioning the mentoring, the guiding and the example of dedication he has granted me with. He is also the reason why I now know so many jokes, and for that if not for the rest I owe him my deepest thankfulness.

Before continuing with these acknowledgments any further, I very much want to express my recognition to all the members of the *Service de Physique Théorique et Mathématique*, which it has been a pleasure to be part of these four years. First of all, I owe my office mate Pierre–Henry LAMBERT a great deal of gratitude for his companionship. For four years of good time in a noisy box: thanks a lot! Next, it is urgent for me to thank the other PhD students of my generation: Micha M. MOSKOVIC and Diego REDIGOLO, to whom I owe more than they think. Their friendship, intellectual generosity and, for the former, help in computer matters is impossible to forget. Then, among the older students my special thanks go to Antonin ROVAI. If not for him and his ability to convey his fascination for the field I might have never started my Physics undergraduate. Also, for many a good moment, on the football pitch and outside of it, my warmest thanks go to Eduardo CONDE PENA and Ignacio CORTESE 'Big Nacho' MOMBELLI as well as to all the other participants of *Más Fútbol*. Finally, these four years of work would have been technically impossible to carry out without our splendid team of secretaries: Marie–France, Fabienne, Delphine, Dominique and Isabelle ! Asking for their ever-friendly help has been my pleasure.

Another, recent member of the Service is Andrea Campoleoni, whom I cannot afford to not thank. His availability to discuss physics, his many remarks and advices as well as his proofreading of part of this manuscript have been crucial. As for people abroad, equally essential discussions have been much appreciated with Evgeny Skvortsov, who is also responsible for part of the proofreading, and with Massimo Taronna as well as with Xavier Bekaert. Debating higher-spin issues with them has been a repeated pleasure ! Moreover, my gratitude also goes to many persons with whom I have had fruitful exchanges of ideas, such as Mikhail Vasiliev, Euihun Joung, Paul Dempster, Karapet Mkrtchyan, Dario Francia, Pan Kessel, Stefan Fredenhagen, Slava Didenko, Per Sundell, Dimitri Ponomarev, Laura Donnay and many others.

Furthermore, I feel compelled to underline two persons who have shaped the path which has taken me to the beginning of this PhD program. I feel a special gratitude towards my high-school Physics teacher Jean–Luc Ottinger, who first of all woke up in me the fascination for the field. Then, during my first year as a Physics student, I much abused the time and patience of Nicolas Bougard, without the help of whom I probably would not have made it so well through the Differential Calculus final. During those undergraduate years, it was also a pleasure to share time and stupid ideas with my comrades. Martin, Micha and most specially Florian: those were the days !

I also want to warmly thank the members of the jury of my PhD viva, let alone my advisor: Glenn Barnich, Matthias Gaberdiel, Augusto Sagnotti, Ricardo Troncoso, Michel Tytgat and especially Nicolas Boulanger, for his support and for many enlightening discussions.

Last but not least, I owe my family: my sister, for never failing to believe in me, my ever-encouraging mother and most especially my father, for passing on to me his interest in science and in work well done.

There is, in fact, one more person. For waking me up more than a thousand mornings, and for lifting my life to a higher level of happiness, I owe her more than she would ever admit. My love and gratitude for her have no boundaries.

# General Presentation

The last few years have witnessed a great renewal of interest in the field of Higher-Spin Gauge Theory. For example, the earliest investigations of the now rich and fruitful framework of so-called Minimal Model Holography dates back only to 2010, when two collaborations independently computed the asymptotic symmetry algebra of some $AdS_3$ Higher-Spin Gravity models [1, 2], which were found to be of the $\mathcal{W}$-type previously investigated in the late eighties and early nineties by many authors [3, 4]. Following these ideas, a few months later M. Gaberdiel and R. Gopakumar proposed a conformal dual for those higher-spin gravities: the so-called Minimal Models, which were known to enjoy $\mathcal{W}$-symmetry [5] — Minimal Model Holography was born, and has been under intensive study ever since [6]. And for good reason: the higher-dimensional equivalent of Minimal Model Holography, relating Vasiliev Theory in $AdS_4$ to Conformal Vector Models in dimension three, was already being addressed before, but the level of computational intricacy on the bulk side was somewhat back-setting. By working, instead, with the much simpler $AdS_3$ version of Vasiliev Theory, and also taking advantage of the control granted by the boundary $\mathcal{W}$-symmetry, one hopes to explore the promising field of Higher-Spin Holography in a more tractable setup.

It was, however, the so-called Vector Model Holography [7, 8] — the higher-dimensional parent of Minimal Model Holography — which actually triggered the immense interest in Higher-Spin Holography, and to some extent in Higher-Spin Theory in general. More precisely, it was the first non-trivial checks of the correspondence between a version of Vasiliev Theory in $AdS_4$ and a Conformal Vector Model for the $O(N)$ group which really sparked the whole field in 2009, thanks to impressive computations by S. Giombi and X. Yin [9]. In fact, at that time the proposal of applying the ideas of Holography to higher spins had been around for quite some time, and a fruitful version of the correspondence in that context was conjectured by various authors, starting with the insightful contributions made during 2000–2002 [10–14]. However, the intricacy inherent to the formulation of Vasiliev Theory had to wait for the authors in question to successfully tackle it. The interest in such a version of Holography is easily understood: it held inside the promise of being a different and not-too-trivial framework to test the ideas put forward by J. M. Maldacena and developed also by others toward the end of the past century [15–17], and which were first applied to more complicated bulk theories such as Type IIB String Theory compactified on $AdS_5 \times S^5$. Indeed, it is very tempting to be able to explore the holographic principle without having to deal with the full-fledged String Theory [18]. Moreover, one further hopes to investigate Holography without the aid of Supersymmetry, since non-supersymmetric versions of Vasiliev Theory exist, perhaps thus unveiling the precise role played by Supersymmetry in the context of the correspondence.

Holography is, however, just one among many interesting aspects of Higher-Spin Theory. For example, the study of the Lagrangian formulation of Vasiliev Theory, already initiated by Fradkin and Vasiliev in the vierbein formalism [19, 20], has been studied more recently with other techniques, such as the so-called ambient-space formalism, with which the investigations have partially returned to a metric-like form, for example in [21]. Moreover, the cubic couplings in Minkowski space have also been the subject of a renewed attention, and a systematic study initiated in 2006 [22–24] has been followed by many contributions, aiming at understanding precisely the reason for the no-go theorems in flat spacetime, among other things. Other interesting works include the first precise calculations relating String Theory and Higher-Spin Theory [25], where flat-space couplings have been obtained from the tensionless limit of open, bosonic String Theory in non-critical Minkowski spacetime. Another recent area of investigation is that of higher-spin black holes [26, 27], the understanding of which should shed light on the nature of higher-spin gauge symmetries, also called generalized diffeomorphisms. As for the 'coupling-classification' program at the level of the Lagrangian, the perturbative approach mentioned above shall certainly provide further insight into the structure of the interactions. At any rate, in the AdS case it may bring us closer to an action principle for Vasiliev Theory, and in the flat setup it could lead to a better understanding of its relation with String Theory. It is the author's belief that gaining control over Higher-Spin Theory should be a prime concern Mathematical Physics in the years to come.

Higher-Spin Theory is, however, far from being a novelty, and indeed it has a rather long history. In fact, the very meaning of the word 'higher' in the term Higher-Spin has changed over the past century. One has in mind, of course, the Rarita–Schwinger theory, developed for the description of fields of spin $\frac{3}{2}$ or higher in 1941 [28]. In particular, the spin-$\frac{3}{2}$ field already exhibits some of the features characterizing higher-spin fields, as for example the obstruction to minimal electromagnetic coupling of massless fields in flat space. Nowadays, given the consistent incorporation of the Rarita–Schwinger field in standard theories such as Supergravity, it is no longer regarded as a higher-spin field in the standard lore. However, when it was first introduced no such thing as Supersymmetry had been invented yet, and the concept of a field of spin $\frac{3}{2}$ or higher was truly regarded as exotic. Interestingly, what W. Rarita and J. Schwinger did was really to reformulate the 1939 Fierz–Pauli theory of particles of arbitrary half-integer spin [29, 30] in terms of the usual symmetric tensor-spinors we are now so used to. Anyhow, the very first steps into the direction of describing particles of spin higher than a half were taken even before, first of all by E. Majorana in 1932 [31], in Italian, and subsequently by P. A. M. Dirac in his paper of 1936 [32], where he himself states that *the present paper will have no immediate physical application.* [...] *Further, the underlying theory is of considerable mathematical interest.*

As explained above, the tensor-spinor equations describing higher-spin gauge fields were determined a long time ago, at the beginning of the forties. However, the Lagrangian then remained a problem for more than thirty years, and we had to wait until 1978, when J. Fang and C. Fronsdal finally cleared out the issue of the Lagrangian formalism applied to gauge fields of arbitrary spin [33, 34]. In fact, the stage for their work was set by L. P. S. Singh and C. R. Hagen in 1974 [35, 36], when they constructed Lagrangians for the massive fields. Also, one should recall that the latter investigations were themselves made possible by the previous breakthrough of E. P. Wigner and V. Bargmann, who found positivity of the energy in Field Theory to be equivalent to the requirement that one-particle states carry irreducible and unitary representations of the Poincaré group [37, 38]. Those Lagrangians obtained in the seventies were either free or coupled to external electromagnetic fields, and such works then triggered the search for higher-spin interactions, which began on four-dimensional Minkowski space with the work of A. Bengtsson, I. Bengtsson and L. Brink [39] on spin-3 couplings in the Light-Cone gauge and the covariant approach used by F. Berends, G. Burgers and H. van Dam [40, 41]. These works gave the remarkable result that spin-3 gauge fields could have consistent cubic self-couplings, which opened the door for the more advanced investigations to come. The consistency to all orders of a theory involving such cubic couplings was discussed only marginally, and at the same time many no-go theorems were proved that forbid, under some assumptions, the existence of a fully consistent theory in flat spacetime [42–44], thereby complementing the earlier prediction by S. Weinberg that low-energy exchanges of particles with spin higher than two are incompatible with the physical requirement that spurious polarizations do not contribute to the S-Matrix [45]. However, as flat-space Higher-Spin Theory was being found to be more and more constrained, the efforts of M. Vasiliev began with the reformulation of free higher-spin dynamics along the lines of the vierbein formulation of Gravity, first in flat spacetime in 1980 [46], and then on constant-curvature spaces in 1987 [47].

This formalism was then used by the same author, together with E. S. Fradkin, to build AdS interactions in a striking fashion: the interactions terms were seen to come with inverse powers of the cosmological constant [19]! Although strange at first glance, this feature nonetheless provided an explanation for the difficulty in constructing higher-spin interactions in Minkowski spaces. In AdS spacetimes, the no-go theorems did not apply, and by the early nineties the basics of Vasiliev Theory were laid out. This theory describes infinitely many gauge fields of increasing spin coupled to a massless spin-2 tensor in a four-dimensional anti-de Sitter background [48–50]. However, despite such an amazing breakthrough the enthusiasm regarding these theories remained rather discrete, much like what happened with the Rarita–Schwinger Theory. Amusingly, in full analogy with the latter, Vasiliev Theory only provides the equations of motion, and not the associated action. Perhaps the latter feature, together with the complexity pertaining to the so-called unfolded formulation in which Vasiliev Theory is cast, contributed to the lack of attention during the nineties. As mentioned earlier, it was partly after the turn of the century, with the tools of AdS/CFT at hand, that the skepticism regarding Vasiliev's work was overcome and that the role Higher-Spin Theory might play began to be appreciated.

Nowadays the physics community has largely overcome the prudence regarding Higher-Spin Gauge Theory, and indeed a great deal of information has been gathered — see [51–53] for recent reviews. However, despite the enormous progress, one cannot help the feeling that much is still to be uncovered, and indeed many promising research directions are open. The most pressing and intriguing issue is perhaps that of the relation between Higher-Spin Gauge Theory and String Theory, the investigation of which started with [54]. As is well known, String Theory contains infinite towers of higher-spin states in its spectrum, which instead are massive. A bridge between the two is yet to be discovered, but we point out, however, the recent and interesting investigation in that direction carried out in [55]. Then, with the motivation of String Theory in mind, one is inevitably led to supersymmetric higher-spin theories and their properties, and it is important per se to gain as much insight as possible about supersymmetric setups. Such is the standpoint of this thesis, in which we use different approaches to have different supersymmetric higher-spin theories under better control. The manuscript is divided into three parts: in Part I we tackle supersymmetric Higher-Spin Theory in dimension three, while in Part II we study higher-dimensional setups. Finally, a third part contains technical appendices, referred to in the main text.

In Part I we study a supersymmetric three-dimensional Higher-Spin Gravity model with infinite spectrum of gauge fields. Just like pure Gravity, higher-spin gauge fields are also topological in dimension three. This is reflected in the fact that one can cast the action in the form of a Chern–Simons term [56], in full analogy with what one can do for Einstein–Hilbert Gravity and Supergravity [57, 58]. Because they propagate no local degrees of freedom, one is led to study the boundary dual theory — at least when one considers $AdS_3$ as a background, which is our primary interest. One should thus study first the asymptotic symmetry algebra, which should govern the asymptotic boundary dynamics. Such is precisely the aim of the first part, namely, to consider the supersymmetric version of the model studied in [1] and explore its asymptotic symmetries, thereby unveiling a nonlinear, supersymmetric version of the previously-found $\mathcal{W}_\infty$-algebra. The first chapter contains pedagogical material, where we recall the construction of higher-spin theories in dimension three via the Chern–Simons formulation [59], which is followed by a chapter containing the explicit computation of the asymptotic symmetries that preserve generalized Henneaux–Maoz–Schwimmer boundary conditions for the gauge connection one-form [60]. In the last and third chapter we then comment on our results [61] and on related topics. An 'invitation' to the first half of this thesis is found at the beginning of Part I.

In Part II we address higher-spin theories in generic dimension greater than three, by considering gauge fermions of arbitrary spin together with a photon or a graviton. Unlike in three dimensions, where the vanishing of the Weyl tensor allows for minimal coupling of higher-spin fields with Gravity, in dimension four and greater minimal coupling is inconsistent in flat space. However, higher-derivative couplings may still exist, and a careful and systematic study of the cubic vertices is thus needed. To do so, we use the powerful machinery of the so-called BRST–BV formalism: it conveniently recasts the interaction problem into a deformation one, which can then be formulated in precise cohomological terms. In this instance we are interested in flat spacetime propagation of our gauge tensor-spinors, which again is partly motivated by the potential relation with String Theory. Also, despite the many studies on higher-spin couplings in dimension greater or equal to four, one may notice a 'gap' concerning fermions, with [25] and [62] among the exceptions. With the assumptions of Poincaré invariance, locality and parity invariance, and with no input from previous works, we thus find out and classify all the consistent couplings of a spin-$s$ gauge tensor-spinor of the symmetric type and a photon, which is carried out in the second chapter [63]. The third chapter then reproduces the analogous computations in the case of gravitational coupling, to find noticeably more complicated yet neat and appealing results [64]. The last and fourth chapter discusses the implications of our results and their connection with previous works, where among other things we note that our findings are in complete agreement with the restrictions imposed on the allowed vertices by the works of Metsaev [62]. The reader will also find an 'invitation' to this second half of the manuscript at the beginning of Part II.

# Part I

# Dimension 3

> *It was in old days, with our learned men,*
> *an interesting and oft-investigated question,*
> *'What is the origin of light?'*
> *and the solution of it has been repeatedly attempted,*
> *with no other result than to crowd our lunatic asylums*
> *with the would-be solvers. [...]*
> *I - alas; I alone in Flatland - know now only too well*
> *the true solution of this mysterious problem;*
> *but my knowledge cannot be made intelligible*
> *to a single one of my countrymen;*
> *and I am mocked at - I, the sole possessor of the truths of Space*
> *and of the theory of the introduction of Light*
> *from the world of three Dimensions -*
> *as if I were the maddest of the mad!*

From the book **Flatland: A Romance of Many Dimensions** [65], by Edwin ABBOTT ABBOTT, under the pseudonym of 'A Square'.

# Invitation

Over the past few decades, the study of Gravity in dimension 3 has proved most fruitful (see [66] for a review), one of the main reasons being the 1986 discovery of the Brown–Henneaux central charge in its asymptotic symmetry algebra [67]. Let us recall that three-dimensional Gravity is topological, which can be seen as the key feature making it less complicated than its higher-dimensional versions. Moreover, the reformulation of it as a Chern–Simons gauge theory in the late nineties [57, 58] made its study easier and even more appealing. This is also one of the prime reasons for the study of three-dimensional Higher-Spin Gauge Theory: it is still topological and hence simpler than the corresponding higher-dimensional setups [68], which is reflected in that the Chern–Simons formulation still holds [56]. The insights provided by the study of three-dimensional Gravity, the necessity of understanding higher-spin theories better and the importance of exploring the AdS/CFT correspondence in new setups are all good reasons to study Higher-Spin Theory in dimension 3, which is the topic of the first part of this thesis.

Foremost among the many interesting aspects of three-dimensional Higher-Spin Theory is perhaps that of the holographic correspondence, the so-called Minimal Model Holography (see [6] for a review), relating the bulk theory to the so-called $\mathcal{W}$-minimal models, which stand among the best-understood interacting conformal field theories. The subject is still relatively young (see General Presentation above) but it has already witnessed a rather high degree of attention, and many advanced features have now been uncovered which go beyond the matching of the classical symmetries, which are described by so-called $\mathcal{W}$-algebras. For example, the spectrum of the boundary minimal models, the agreement at the level of three-point functions [69–72], some aspects of higher-spin black holes [73–75], partition functions [76–78], the quantization of the symmetries [79, 80] and other features are nowadays better understood [6]. Nonetheless, the fact that the boundary dynamics are governed by a $\mathcal{W}$-type algebra is remarkable in itself. Indeed, those algebras had been studied before, and were seen to appear in many other areas of physics [4]. Therefore, on top of the usual, holography-driven motivations for studying the asymptotic symmetry algebra of three-dimensional higher-spin models, the presence of $\mathcal{W}$-algebras at spatial infinity thereof further calls for a systematic and thorough investigation in this direction.

In the case of higher spins, the corresponding supersymmetric setups have been less intensively addressed. With the latter motivations spelled out, however, it seems important to have access to supersymmetric versions of Minimal Model Holography [81–91], and in particular to the asymptotic symmetries of higher-spin supergravities in $D = 3$. Moreover, the possibility of relating the bulk theories to string embeddings further nurtures the need for supersymmetric versions of the bosonic investigations (see e.g. [61]). The aim of this first part is, precisely, that of exhibiting the asymptotic symmetries of some higher-spin supergravities [61], which we do in Chapter 2, after recalling the procedure for computing asymptotic symmetries in the Chern–Simons formalism for the case of pure Gravity. More precisely, we deal explicitly with the $osp(1, 2|\mathbb{R})$-based models involving an infinite tower of gauge fields and then comment on the extended higher-spin supergravities. The basics of three-dimensional Higher-Spin Theory as well as their construction from the Chern–Simons standpoint are recapitulated in Chapter 1.

We find the boundary dynamics to be governed by some supersymmetric $\mathcal{W}_\infty$-algebra, which we do by means of employing the oscillator realization of our bulk higher-spin gauge algebra, recalled first. In fact, the latter $\mathcal{W}$-algebra is found to be nonlinear, in agreement with previous results which we generalize [1]. Nevertheless, we shall prove that the isometry algebra of the bulk vacuum solution is a subalgebra thereof, up to nonlinearities. Other subalgebras are investigated, and we also give a closed form for the first nonlinear commutation relations in the simply supersymmetric case. The generalization of the results to the extended case is then sketched, and in Chapter 3 we discuss our findings and comment on various related topics, touching upon bulk symmetries and subalgebras, $\lambda$-deformations, previous approaches to $\mathcal{W}_\infty$ algebras, etc. In particular, the role of nonlinearities and Supersymmetry is emphasized, as well as the comparison with previous approaches dealing with similar algebraic structures [4].

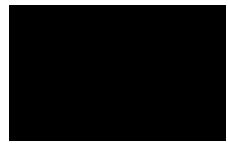

CHAPTER 1

# Higher Spins in Dimension 3

In Chapter 2 we compute the asymptotic symmetry algebra of higher-spin supergravities in three dimensions, taking ample advantage of the Chern–Simons reformulation of the latter as a gauge theory for a connection one-form valued in the superalgebra of the vacuum isometries. In the present chapter we thus recall such a formulation, and for the sake of pedagogy we start with the simplest case of pure Gravity in Section 1.1, for which we recall the frame formalism and then switch to the gauge picture [57]. Higher spins are only introduced in Section 1.2, where the same logic is followed, although at the free level. Interactions are then introduced in the Chern–Simons context, by means of finite- or infinite-dimensional higher-spin algebras [56]. In fact, our interest is in (supersymmetric) higher-spin theories describing infinitely many gauge fields of arbitrarily-high spin, and accordingly we deal with infinite-dimensional algebras.

The following material is meant to be somewhat pedagogical, and the reader familiar with three-dimensional Gravity in its Chern–Simons form shall find it harmless to skip Section 1.1. Similarly, the reader familiar with three-dimensional Higher-Spin Theory may move to Chapter 2 without much loss of valuable information. As mentioned in the above invitation, our focus is on Supersymmetry, and in Chapter 2 it is the supersymmetric higher-spin models that we are interested in. However, in the sequel of this introductory chapter we shall mostly approach the material leaving fermions out for the sake of conciseness, only quoting the corresponding supersymmetric versions at the end of each section.

## 1.1 Gravity as a Gauge Theory in Dimension 3

What shall be done in Section 1.2 for higher-spin fields is here reviewed in the context of Gravity and Supergravity. In Subsection 1.1.1 we shall recall first, in generic dimension, the formulation of Gravity in terms of the vielbein and the spin-connection and will then specialize to three-dimensions, which Subsection 1.1.2 will start with to then make contact with the gauge formulation of Gravity as a Chern–Simons theory. After this rather pedestrian introduction without supersymmetry we quote the final results for non-extended as well as for extended AdS$_3$ Supergravity in Subsection 1.1.3, namely, we display the rewriting of the latter as a Chern–Simons action term for some gauge superalgebra and give the definition of the superconnection in terms of the Supergravity fields.

Let us stress that, in the case of (super-) gravity, all of the aforementioned reformulations will be carried out at the non-linear level, whereas for higher spins we will only introduce interactions once the Chern–Simons formulation is at hand. Note, however, that it is in principle possible to introduce interactions at the level of the metric-like formulation for higher spins, but it is perhaps less clean than doing it in the gauge picture [92].

### 1.1.1 The Frame Formulation of Gravity

This subsection relies, among others, on [93], which we recommend to the reader unfamiliar with the subject, for only a pragmatic review is provided in the present introduction. Many other references on this subject are available, among which we shall highlight the mathematically-oriented one [94] as well as the rather earlier one [95]. Note that only the three-dimensional version of the present subsection will be of use for us but, as it is not much effort to do so, for the sake of completeness we shall start in general dimension $D$ and will only particularize to $D=3$ at the end of the subsection.

#### The Vielbein

Pure Gravity [96] is described by the Einstein–Hilbert action in $D$ spacetime dimensions with cosmological constant $\Lambda$ (here for $c = 1$, which we assume throughout this presentation unless otherwise specified):

$$S_{\text{EH}} \equiv S_{\text{EH}}[g] \equiv \frac{1}{16\pi \text{G}} \int_{\mathcal{M}_D} (R - 2\Lambda)\sqrt{-g}\, \text{d}^D x, \tag{1.1}$$

where G is the $D$-dimensional Newton constant, $g$ is the determinant of the metric $g_{\mu\nu}$, $R$ is the Ricci scalar and $\mathcal{M}_D$ is the spacetime manifold.

The equations of motion one derives from the above action read

$$G_{\mu\nu} + \Lambda g_{\mu\nu} = 0, \qquad (1.2)$$

where $G_{\mu\nu} \equiv R_{\mu\nu} - \frac{1}{2}g_{\mu\nu}R$ is the Einstein tensor and $R \equiv g^{\mu\nu}R_{\mu\nu}$ is the Ricci scalar, the contraction of the Ricci tensor with the inverse metric $g^{\mu\nu}$. The usual rewriting of the above equations without involving the Ricci scalar is then

$$R_{\mu\nu} = \tfrac{2}{D-2} g_{\mu\nu}\Lambda. \qquad (1.3)$$

Let us now introduce the so-called vielbein by the relation

$$g_{\mu\nu} \equiv e_\mu^a e_\nu^b \eta_{ab}, \qquad (1.4)$$

with our conventions for the signature of the Minkowski metric $\eta_{ab}$ being $(-+\cdots+)$. The Latin indices are usually referred to as 'frame' indices. The relation is, however, invariant under the so-called local Lorentz transformations (LLTs) of the vielbein

$$e_\mu^a(x) \to \Lambda^a{}_b(x) e_\mu^b(x), \qquad (1.5)$$

with the matrix $\Lambda(x) \in \mathrm{SO}(D-1,1)$ (the Lorentz group) at all spacetime points $x$. Now, the vielbein is a $D \times D$ matrix at each spacetime point, of which we can eliminate as many components as the dimension of the Lie algebra $\mathrm{so}(D-1,1)$ (at each spacetime point), that is, $D(D-1)/2$, which leaves us with $D(D+1)/2$ independent components: the number of independent components of the $D$-dimensional metric. Thinking of (1.4) as a mere change of variables for Gravity, the transformation law (1.5) simply originates in that the change of variables is not one-to-one and some redundancy is introduced (which we just saw can be 'gauged away' using LLTs).

Our vielbein is a hybrid object; it has both a spacetime and a frame index. We already displayed its transformation rules with respect to its frame index (LLTs), which resulted from a redundancy in our change of variables. With respect to its spacetime index, the tensor nature of the metric forces it to transform as a covector under the diffeomorphism group.

The spacetime indices are always raised and lowered using the metric $g_{\mu\nu}$ and its inverse, but the metric governing the frame indices is always the Minkowski one $\eta_{ab}$. This is actually related to a conceptually important fact: the so-called tangent frame defined by the vielbein is orthogonal at any spacetime point, as the following relations illustrate:

$$e_\mu^a e_b^\mu = \delta_b^a, \quad e_\mu^a e_a^\nu = \delta_\mu^\nu, \qquad (1.6)$$

where the Latin indices have been raised and lowered with $\eta_{ab}$. The vielbein $e^\mu_a$, with both indices swapped, is called the inverse vielbein (and it is indeed so if we think of it as a matrix). Also note the useful relations

$$\eta_{ab} = g_{\mu\nu} e^\mu_a e^\nu_b, \quad \sqrt{-g} = e \equiv \det(e^a_\mu). \tag{1.7}$$

We thus see that, at every spacetime point, the vielbein is providing us with some frame in which the metric looks flat (the tangent frame), and those vielbeins should be really thought of as being the matrices implementing a change of basis. For a more in-depth understanding of the geometrical interpretation of this object we refer to [94].

**The Spin-Connection**

Working in the tangent frame will force us to consider various objects having frame indices, such as the vielbein, that we already encountered, and we want to be able to derive such objects. For simplicity, let us first focus on a frame vector, that is, an object having only one frame index (upstairs), $V^a$. Proceeding in an analogous way to what is done in metric-like Gravity, we require the derivative of our tangent frame vector to be a tangent frame tensor, which uniquely leads us to introducing the so-called spin-connection $\omega^{ab}_\mu$; a hybrid object having the following transformation rules under local Lorentz transformations (avoiding to spell out spacetime indices, which remain unaffected by such transformations):

$$\omega'^a{}_b(x) = (\Lambda^{-1})^a{}_c(x) d\Lambda^c{}_b(x) + (\Lambda^{-1})^a{}_c(x) \omega^c{}_d(x) \Lambda^d{}_b(x), \tag{1.8}$$

which ensure that the tangent frame covariant derivative,

$$D_\mu V^a = \partial_\mu V^a + \omega^a{}_{b\mu} V^b, \tag{1.9}$$

is an 'LLT tensor' (transforms covariantly under LLTs). The resemblance with standard Gravity is manifest, as the above transformation law for the spin-connection bears much resemblance with that of the Christoffel symbols (spacetime connection) of Gravity in its metric formulation. However, let us again insist on that, in the above transformation rule, spacetime indices are not affected, and hence nor are the coordinates.

Some comments are now in order. First, the above derivation rules — for both spacetime and tangent frame indices — of course contain more terms when one derives higher-order tensors (see [93]). Second, both rules are to be combined when deriving hybrid objects (such as the vielbein or the spin-connection). That 'full' derivation will be noted $\hat{\nabla}$. This will help us distinguish all three types of derivation rules one could use to derive a hybrid object: one could derive it with respect to its spacetime indices ($\nabla$), with respect to its tangent frame indices ($D$), or with respect to both ($\hat{\nabla}$). As an example, we give

$$\hat{\nabla}_\mu e^a_\nu = \partial_\mu e^a_\nu + \omega^a{}_{b\mu} e^b_\nu - \Gamma^\rho_{\mu\nu} e^a_\rho, \tag{1.10a}$$

$$De^a = \mathrm{d}e^a + \omega^a{}_b \wedge e^b, \tag{1.10b}$$

where the $\Gamma$'s are the familiar Christoffel symbols for Gravity and $\wedge$ is the wedge product on spacetime indices. In fact, the last expression above can be seen to be the torsion of our connection, while the first one is the so-called 'vielbein postulate' [93] (a rewriting of the the relation between the Christoffel symbols and the spin-connection). We now want to solve for the spin-connection, that is, to impose conditions such that we can uniquely find some $\omega = \omega[e]$. In the metric formalism, one imposes metric-compatibility as well as zero-torsion ($T^\mu \equiv \nabla \mathrm{d}x^\mu$) for the spacetime connection, which uniquely leads to the well-known Christoffel expression for $\Gamma^\mu_{\nu\rho}$ in terms of the metric and its derivatives. A similar thing happens in the tangent frame. Indeed, imposing metric-compatibility as well as zero torsion (setting to zero the last expression above) one uniquely finds

$$\omega^{ab}_\mu[e] = 2e^{\nu[a} \partial_{[\mu} e^{b]}_{\nu]} - e^{\nu[a} e^{b]\sigma} e_{\mu c} \partial_\nu e^c_\sigma. \tag{1.11}$$

Let us also display the condition of metric compatibility, which in spacetime means $\nabla_\mu g_{\nu\rho} = 0$ and translates to the tangent frame as $D\eta = 0$. Note, however, that the latter is simply equivalent to $\omega^{ab}_\mu = -\omega^{ba}_\mu$.

**Remark** : it is interesting to notice that, in spacetime, it is the zero-torsion condition which is equivalent to the symmetry of the connection (in its two lower indices), whereas in the frame it is the metric-compatibility condition which implies antisymmetry in the two Latin indices.

Although we are not giving all the details of how to arrive at (1.11) (see [93]), the key point here is that one *can* solve for $\omega = \omega[e]$ and, furthermore, the conditions uniquely leading to the solution are precisely the 'tangent frame translation' of the conditions one usually imposes in standard gravity. This analogy between the frame picture and the usual metric formulation can actually be pushed further, which we do in the sequel. In fact, as we have stressed above, the vielbeins can be seen as a change of basis. Quantities such as the torsion (or the curvature, defined below) can be defined intrinsically, and one can then write them in a frame basis or in a coordinate one — those will be two expressions for the same object. For more information and, in particular, for intrinsic definitions of geometrical quantities we refer to [97].

**The Curvature**

Recalling the known expression for the Riemann tensor $R[\Gamma]$, one is naturally led to consider the following object, called the curvature, and which shall be proved to be the rewriting of the Riemann tensor in a mixed basis (see below):
$$R^{ab}[\omega] \equiv d\omega^{ab} + \omega^a{}_c \wedge \omega^{cb}. \tag{1.12}$$

It is a conceptually important point that we are also led to such an expression if we simply notice that the spin-connection transforms under LLTs just like a standard Yang–Mills gauge potential, and indeed, the above expression is precisely the standard Yang–Mills field strength for $\omega$. It is therefore a hybrid object we are dealing with, and we further note that it has two spacetime indices and two tangent-frame ones, and that it is a tensor with respect to both types of indices (under diffeomorphisms and LLTs respectively). This 'dual nature' goes even further, for the above tensor is blessed with two Bianchi-like identities:

$$R[\omega]^{ab} \wedge e_b = R_{(\mu\nu\rho)_{\text{CYCL.}}}{}^a = 0, \tag{1.13a}$$
$$dR[\omega]^{ab} + \omega^a{}_c \wedge R[\omega]^{cb} - R[\omega]^{ac} \wedge \omega_c{}^b = D_{(\mu} R_{\nu\rho)_{\text{CYCL.}}}{}^{ab} = 0, \tag{1.13b}$$

the first one being 'purely gravitational' (with no analogue in Yang–Mills Theory), and the second one being the standard gauge-theory identity simply stemming from the definition of the field strength. These Bianchi identities are heavily reminiscent of the ones endowing the Riemann curvature tensor. Actually, it is one of the most important basic results of the frame formulation of gravity that the following relation holds:

$$R[\Gamma]^\rho{}_{\mu\nu\sigma} = R[\omega]_{\mu\nu ab} e^{a\rho} e^b_\sigma. \tag{1.14}$$

Note that, knowing the relations linking the spacetime connection and the spin-connection to the metric and vielbein respectively, a direct check of such a relation seems doable but incredibly tedious. However, there exists a trick, which is the mere evaluation of $[\nabla_\mu, \nabla_\nu]e^\rho_a = 0$ and which leads to the result more easily. In fact, as aforementioned, the above equation simply expresses the change of basis for two of the indices of the Riemann tensor !

Another analogy between both formulations of the curvature is their transformation under a variation of $\Gamma$ and $\omega$ respectively. Those read

$$\delta R[\Gamma]^\rho_{\mu\nu\sigma} = \nabla_\mu(\delta\Gamma^\rho_{\nu\sigma}) - \nabla_\nu(\delta\Gamma^\rho_{\mu\sigma}), \tag{1.15a}$$

$$\delta R[\omega]_{\mu\nu ab} = D_\mu(\delta\omega_{\nu ab}) - D_\nu(\delta\omega_{\mu ab}), \tag{1.15b}$$

the latter of which will be useful in the sequel, when going back from the so-called first-order formalism to the 1.5-order formalism. Finally, let us point out that, as it should be, $R[\Gamma]$ and $R[\omega]$ both appear when one considers the commutator of two covariant derivatives ($D$ and $\nabla$ respectively), which is really the object characterizing the curvature of spacetime, that is, how much it fails to be flat — and we again refer to [93]. Thus, just as the spin-connection is a rewriting of the familiar spacetime connection in a mixed basis (also called 'non-holonomic'), so is the above curvature $R[\omega]$ simply an expression for the Riemann tensor with some indices changed to the frame basis provided by the vielbeins.

**The Action**

We are finally ready to rewrite the Einstein–Hilbert action (1.1) in terms of the vielbein. Indeed, the relation (1.14) leaves us only with the problem of rewriting the infinitesimal spacetime volume element, but, fortunately, the following relations are easily derived:

$$\begin{aligned} e\,\mathrm{d}^D x = e^0 \wedge \cdots \wedge e^{D-1} &= \frac{1}{D!}\epsilon_{a_1\ldots a_D} e^{a_1} \wedge \cdots \wedge e^{a_D} \\ &= \frac{e}{D!}\epsilon_{\mu_1\ldots\mu_D}\mathrm{d}x^{\mu_1} \wedge \cdots \wedge \mathrm{d}x^{\mu_D}, \end{aligned} \tag{1.16}$$

our conventions being $\epsilon_{0\ldots D-1} \equiv 1$. Indeed, plugging the above rewriting of the infinitesimal volume element $\mathrm{d}V$ as well as relation (1.14) in the action (1.1), and further using

$$e^{a_1} \wedge \cdots \wedge e^{a_p} \wedge e^{b_1} \wedge \cdots \wedge e^{b_q} = -\epsilon^{a_1\ldots a_p b_1\ldots b_q}\mathrm{d}V, \tag{1.17}$$

we finally find (for the $\Lambda = 0$ case),

$$\begin{aligned} S_{\text{EH}}[g[e]] &= \frac{1}{(D-2)!16\pi\mathrm{G}}\int_{\mathcal{M}_D}\epsilon_{abc_1\ldots c_{D-2}}e^{c_1} \wedge \cdots \wedge e^{c_{D-2}} \wedge R[\omega[e]]^{ab} \\ &\equiv S_{\text{SO}}[e,\omega[e]] \equiv S_{\text{SO}}[e], \end{aligned} \tag{1.18}$$

which in three dimensions reads (now including the obvious contribution from the cosmological constant)

$$S_{\text{SO}}[e] = \frac{1}{16\pi G} \int_{\mathcal{M}_3} \epsilon_{abc}\, e^a \wedge \left(R^{bc}[\omega] + \tfrac{1}{3}\Lambda e^b \wedge e^c\right), \qquad (1.19)$$

and where 'SO' stands for 'second-order' formalism, meaning that the spin-connection is thought of as depending on the vielbein, so that the action really depends only on the latter, which obeys a second-order differential equation. As the Einstein equations are also second-order equations in the metric, this terminology is also sometimes used to refer to the standard metric-like (Einstein–Hilbert) formalism. For the three-dimensional epsilon symbol we also use the convention $\epsilon_{123} \equiv 1$. Note that the dependence of the spin-connection on the vielbein is usually dropped when writing the action, just as we did for the last expression above. It is rather clear that, because the actions are equivalent, finding the equations of motion for the vielbein that the above action yields and going back to the metric formulation one should find the Einstein equations. This is a nice exercise that we shall not comment on more.

A natural thing to do now is to consider the same action, but forgetting about the relation between the spin-connection and the vielbein, that is, both objects are treated independently. Then, the variational principle yields equations for $\omega$ as well, in addition to those for the vielbein. As it turns out, these 'equations of motion' for $\omega$ are precisely the zero-torsion condition. Therefore, if we further demand that $\omega$ be antisymmetric in its two frame indices (which is equivalent to requiring it to be a Lorentz connection), its equation of motion allows us to solve for it, obtaining the expression (1.11). This way of thinking about the frame formulation of the action is called the 'first-order' formalism, because now the spin-connection is to be thought of as an auxiliary field (for which we can solve), and before integrating it out the vielbein thus obeys a first-order differential equation (which gives back Einstein equations if we plug in it the functional dependence of the spin-connection on the vielbein). To stress that it depends on $e$ and $\omega$ independently and to distinguish it from the second-order action (1.19), the first order action will be noted $S_{\text{FO}}[e,\omega]$, but it looks exactly like (1.19) except for the fact that $\omega$ is no longer to be understood *off-shell* as a function of the vielbein and its derivatives.

Note that, in order to find the equations of motion for the spin-connection, one first uses the aforegiven formula (1.15b). Then, combining the vielbein postulate (1.10a) with the integration formula

$$\int d^D x \sqrt{-g}\, \nabla_\mu V^\mu = \int d^D x\, \partial_\mu(\sqrt{-g}\, V^\mu) + \int d^D x \sqrt{-g}\, T^\nu_{\nu\mu} V^\mu, \quad (1.20)$$

one obtains that some combination of the torsion is equal to zero; an equation that one has to act on with vielbein fields in order to get the zero-torsion condition.

**Particularizing to Dimension 3**

From now on we shall work in three dimensions, where we can perform the standard 'dual' rewriting

$$R[\omega]_a \equiv \tfrac{1}{2}\epsilon_{abc} R[\omega]^{bc} \quad \Leftrightarrow \quad R[\omega]^{ab} = -\epsilon^{abc} R[\omega]_c, \quad (1.21)$$

and do the same for $\omega^a$ itself, thus obtaining

$$R[\omega]_a = d\omega_a - \tfrac{1}{2}\epsilon_{abc}\, \omega^b \wedge \omega^c. \quad (1.22)$$

The action (1.19) at $\Lambda = 0$ can then be rewritten as

$$S_{\text{FO}}[e,\omega] = \frac{2}{16\pi G} \int_{\mathcal{M}_3} e^a \wedge R[\omega]_a, \quad (1.23)$$

which we recall can only be written down in three dimensions. Note that we have moved to the first-order formalism, for it is the one we shall start from in order to pass to the Chern–Simons formulation.

Before moving on to the next subsection, let us display the linearized equations of motion corresponding to the first-order formalism. The reason why we only need display the linearized equations of motion is that, in Section 1.2, we shall start from a linearized higher-spin theory in order to try to build its non-linear completion. The linearized higher-spin equations of motion will thus be expressed in the frame formalism and, in order to have something to compare them to, we give hereafter the linearized equations of motion in the frame formalism for the $s = 2$ case. In linearizing the vielbein, we have adopted the notation $e \to \bar{e} + v$, where $\bar{e}$ is the background dreibein associated with the background metric (that of three-dimensional anti-de Sitter spacetime for example) via the usual formula (1.4). As for the spin-connection, we have used $\omega \to \bar{\omega} + \omega$, where $\bar{\omega}$ is some background, related to $\bar{e}$ via the usual zero-torsion condition

(the background is on-shell), that is, via equation (1.11). We find these excitations to satisfy

$$d\omega^a + \epsilon^{abc}\bar{\omega}_b \wedge \omega_c - \Lambda\epsilon^{abc}\bar{e}_b \wedge v_c = 0, \tag{1.24a}$$

$$dv^a + \epsilon^{abc}\bar{\omega}_b \wedge v_c + \epsilon^{abc}\bar{e}_b \wedge \omega_c = 0, \tag{1.24b}$$

which can be rewritten as

$$D\omega^a - \Lambda\epsilon^{abc}\bar{e}_b \wedge v_c = 0, \tag{1.25a}$$

$$Dv^a + \epsilon^{abc}\bar{e}_b \wedge \omega_c = 0, \tag{1.25b}$$

where the first one above is the linearized zero-torsion condition for the metric-compatible spin-connection $\omega^a$ and the second one is simply the linearized equation of motion for the dreibein. One may check that the above equations are invariant under linearized diffeomorphisms as well as infinitesimal local Lorentz transformations, as they should be.

### 1.1.2 Gravity as a Chern–Simons Theory in Dimension 3

As has been argued in the previous subsection, three-dimensional pure Gravity can be rewritten in the so-called first-order formalism, with the vielbein and the spin-connection being independent variables (the latter being an auxiliary field). Starting from the latter formulation, we shall now discuss the Achúcarro–Townsend–Witten result [57, 58] in which yet another formulation of Gravity is found (in three-dimensions), namely that of a gauge theory described by a Chern–Simons action with a connection one-form $A_\mu$ taking values in the Lie algebra of isometries of the vacuum solution. The vacuum solution being either Minkowski, anti-de Sitter or de Sitter, the relevant Lie algebras underlying our yet-to-be-formulated gauge description of $D = 3$ Gravity will be respectively iso(2, 1), so(2, 2) or so(3, 1).

The way in which we shall proceed is backwards, that is, we will start from the gauge theory we claim to be equivalent to three-dimensional Gravity and will then show it to be so. Much like in the previous subsection, we try to be as pedagogical as possible but will remain rather pragmatic in spirit, referring the reader to [57] for a more detailed and very enlightening discussion.

**Intuitive Arguments**

The frame formulation of Gravity has made many physicists try to combine the vielbein and the spin-connection into some iso($D-1, 1$)-valued one-form gauge field. Indeed, the vielbein (resp. the spin-connection) looks like an appealing candidate for the role of the coefficient of the gauge connection $A$ corresponding to the translation generators (resp. Lorentz generators) of iso$(D-1, 1)$. This is so because, as mentioned in the previous subsection, when formulated in terms of the vielbein and spin-connection Gravity already has some of the taste of a Yang–Mills-like gauge theory. However, there is an easy intuitive reason why this is likely to fail in dimension four [57] (or at the very least be somewhat unnatural). Indeed, looking back at (1.18) for $D = 4$ we see that it is of the schematic form (for $\Lambda = 0$)

$$S \sim \int_{\mathcal{M}_4} e \wedge e \wedge (\mathrm{d}\omega + \omega \wedge \omega), \tag{1.26}$$

so that the corresponding Yang–Mills-like action should look somewhat like

$$S \sim \int_{\mathcal{M}_4} \mathrm{Tr}\left(A \wedge A \wedge (\mathrm{d}A + A \wedge A)\right), \tag{1.27}$$

which does not exist in gauge theory (it is not gauge invariant).[1] However, in $D = 3$, we have the well-known Chern–Simons action, which roughly looks like

$$S \sim \int_{\mathcal{M}_3} \mathrm{Tr}\left(A \wedge (\mathrm{d}A + A \wedge A)\right). \tag{1.28}$$

It is precisely what one feels like trying when looking at (1.19) for $\Lambda = 0$ in dimension three!

The other indication that the three-dimensional scenario is specially suited for establishing such a correspondence has to do with bilinear forms on the relevant Lie algebras. Indeed, if one is to build a Chern–Simons action (or any Yang–Mills-like action) for some Lie algebra, one should first make sure that there exists some invariant, non-degenerate, symmetric and bilinear form on it. In the $\Lambda = 0$ case, it turns out that iso$(D-1, 1)$ admits such a form only for $D = 3$. Let us note, however, that despite the aforementioned difficulties in formulating Gravity in dimension four and higher as a gauge theory for the Poincaré group, past investigations nevertheless arrived at interesting results [98–100]. In fact, these works later inspired Vasiliev and other authors in their formulation of higher-spin dynamics [19, 47, 101] (see General Presentation).

---

[1] The same direction of investigation was explored in AdS spaces too [98, 99].

**The Gauge Algebras**

Let us start by considering the gauge algebras that we will have to work with in the sequel, which will also allow us to fix the conventions thereof. From now on we stick to $D = 3$, for which the commutation relations of our three different gauge algebras can be packaged into

$$[J_a, J_b] = \epsilon_{abc} J^c, \qquad [J_a, P_b] = \epsilon_{abc} P^c, \qquad [P_a, P_b] = \lambda \epsilon_{abc} J^c, \quad (1.29)$$

where the Latin indices $a, b, c = 1, 2, 3$ are raised and lowered with the three-dimensional Minkowski metric[2] $\eta_{ab}$ and its inverse $\eta^{ab}$, which are also chosen to have signature $(-++)$. Note that we have used again the 'three-dimensional rewriting'

$$J_a \equiv \tfrac{1}{2} \epsilon_{abc} J^{bc} \quad \Leftrightarrow \quad J^{ab} \equiv -\epsilon^{abc} J_c, \quad (1.30)$$

where $J_{ab}$ are the usual Lorentz generators ($P_a$ are of course the translation ones). For $\lambda = 0$, $\lambda < 0$ and $\lambda > 0$, the above relations describe respectively iso(2, 1), so(2, 2) and so(3, 1).

As aforementioned, iso($D - 1, 1$) admits a non-degenerate and invariant (symmetric and real) bilinear form[3] only for $D = 3$, which is unique in the space of such forms.[4] It reads

$$(J_a, P_b) = \eta_{ab}, \qquad (J_a, J_b) = (P_a, P_b) = 0. \quad (1.31)$$

As for so($D - 1, 2$) and so($D, 1$), they admit a non-degenerate, invariant bilinear form for any $D$, the particularization of which to $D = 3$ reads

$$(J_a, J_b) = \eta_{ab}, \qquad (J_a, P_b) = 0, \qquad (P_a, P_b) = \lambda \eta_{ab}. \quad (1.32)$$

For $D \neq 3$ they are both simple and the higher-dimensional equivalent of the above form is therefore unique (up to normalization) and proportional to the Killing form. For $D = 3$, however, the AdS one becomes semi-simple and undergoes the splitting

$$\mathrm{so}(2, 2) \simeq \mathrm{sl}(2|\mathbb{R}) \oplus \mathrm{sl}(2|\mathbb{R}), \quad (1.33)$$

so that, in addition to the above form they also admit (1.31).

**Remark** : note that (1.32) is degenerate for $\lambda = 0$, which is the reason iso($D - 1, 1$) only admits a non-degenerate form for $D = 3$, (1.31), which is a specificity of the three-dimensional case, as can be seen by noting that it corresponds to the invariant $\epsilon_{abc}J^{ab}P^c$. The latter can only be constructed in three-dimensions, and it is thus the 'epsilon magic' which is really at work here.

---

[2] This is important since if one takes the indices to be euclidean the commutation relations would describe, e.g. for $\lambda = 0$, iso(3) instead of iso(2, 1).

[3] For a general treatment of Lie algebras we recommend for example [102].

[4] It is not the Killing form, which is degenerate because iso(2, 1) is not semi-simple.

In the $\Lambda \geq 0$ case we do not have any choice for the bilinear form to use, but for $\Lambda < 0$ we do have the freedom of choosing our bilinear form among the two above ones. However, it seems somewhat more natural to use also in that case the one which also endows the isometry algebra of three-dimensional Minkowski spacetime. This choice will not be further justified in the present work (except by the fact that it will lead to a Chern–Simons action which will indeed reproduce the Einstein–Hilbert one), and we refer the interested reader to [57] for more discussions on the subject.

As we shall be most interested in the AdS case, let us already point out that the splitting (1.33) of so(2,2) explicitly reads

$$[J_a^+, J_b^+] = \epsilon_{abc}J^{+c}, \qquad [J_a^-, J_b^-] = \epsilon_{abc}J^{-c}, \qquad [J_a^+, J_b^-] = 0, \quad (1.34)$$

where

$$J_a^\pm \equiv \tfrac{1}{2}\bigl(J_a \pm \tfrac{1}{\sqrt{\lambda}}P_a\bigr). \quad (1.35)$$

Before moving to the next subsection we also note that, when expressed in terms of the $J^\pm$ generators of so(2,2), the above form (1.31) reads

$$(J_a^+, J_b^+) = \tfrac{1}{2}\eta_{ab}, \qquad (J_a^-, J_b^-) = -\tfrac{1}{2}\eta_{ab}, \qquad (J_a^+, J_b^-) = 0, \quad (1.36)$$

which we recall corresponds to the $\Lambda < 0$ case.

### The Action

Now that the algebraic aspects have been dealt with, let us work out the equivalence at the level of the actions between some Chern–Simons term with a connection one-form $A_\mu$ living in one of the above three-dimensional isometry algebras and three-dimensional Einstein–Hilbert Gravity with corresponding cosmological constant.

We begin by proving the equivalence in the $\Lambda = 0$ case. The identification of the off-shell degrees of freedom is the following:

$$A_\mu \equiv e_\mu^a P_a + \omega_\mu^a J_a, \quad (1.37)$$

where the generators $J_a$, $P_a$ of iso(2,1) satisfy (1.29) at $\lambda = 0$. If we now plug this expansion into the Chern–Simons action term below and use

(1.31) for the scalar product (trace) a straightforward computation yields

$$\begin{aligned} S_{\text{CS}}[A] &\equiv \kappa \int_{\mathcal{M}_3} \text{Tr}\left(A \wedge \mathrm{d}A + \tfrac{2}{3} A \wedge A \wedge A\right) \\ &= \kappa \int_{\mathcal{M}_3} e^a \wedge R[\omega]_a \\ &\equiv \kappa 16\pi \text{G}\, S_{\text{FO}}[e,\omega], \end{aligned} \tag{1.38}$$

where we have used $\epsilon_{abc}\epsilon^{ade} = \delta^e_b \delta^d_c - \delta^d_b \delta^e_c$. We thus conclude that, upon using our identification (1.37), $S_{\text{CS}}[A] = S_{\text{FO}}[e,\omega]$ provided we set $\kappa = 1/16\pi \text{G}$, which is called the Chern–Simons level. We also note that the matching of the actions assumes that the vielbein is invertible. We shall not dwell on this interesting issue here, and refer to [57] for an interesting discussion.

Before moving to the next subsection, let us work out — for we shall need them — the equations of motion derived from the above Chern–Simons action. In terms of the gauge connection they read

$$F[A] \equiv \mathrm{d}A + A \wedge A = 0, \tag{1.39}$$

as is well known, and in terms of $e$ and $\omega$ we easily find the corresponding expressions:

$$D_\mu e^a_\nu - D_\nu e^a_\mu = 0, \tag{1.40a}$$

$$\partial_\mu \omega^a_\nu - \partial_\nu \omega^a_\mu + \epsilon^{abc} \omega_{b\mu} \omega_{c\nu} = 0, \tag{1.40b}$$

where we use the standard abuse of notation $D_\mu e^a_\nu \equiv (D_\mu e_\nu)|_{P_a}$, with $|_{P_a}$ meaning taking the component along the $P_a$ generators. Note that, as it should be, the above equations of motion do coincide, at the linearized level, with the $\Lambda = 0$ version of (1.24).

**The Gauge Transformations**

There is a last non-trivial check to do before one can safely claim the two theories to be equivalent; namely, we need verify the gauge transformations on both sides to be the same. Indeed, while both the frame formulation and the Chern–Simons actions are manifestly invariant under diffeomorphisms, in the first-order formulation we also have the local Lorentz transformations as gauge symmetries, whereas in the Chern–Simons picture we have instead the full iso(2, 1) gauge symmetries. As we shall demonstrate, the homogeneous part of the iso(2, 1) gauge symmetries are easily seen to correspond to the LLTs in the first-order formalism but, as for the infinitesimal gauge translations of iso(2, 1), one has to show that they are not extra gauge symmetries (which would be bad for our rewriting of the action would then eliminate degrees of freedom in some sense) but, rather, that they correspond to some combination of the symmetries of the first-order formalism action.

The gauge transformations in the Chern–Simons picture are parametrized by a zero-form gauge parameter taking values in the gauge algebra,
$$u \equiv \rho^a P_a + \tau^a J_a, \qquad (1.41)$$
with $\rho^a$ and $\tau^a$ being infinitesimal parameters, and the transformation law for the gauge connection (sitting in the adjoint representation of the gauge algebra) is $A \to A + \delta A$ with
$$\delta A_\mu = \partial_\mu u + [A_\mu, u]. \qquad (1.42)$$
Upon now plugging the expression for $u$ and the decomposition of $A$ in terms of the dreibein and spin-connection in the above equation we can read off the variations of $e$ and $\omega$, which are all the (infinitesimal) local symmetries of the action in the gauge (Chern–Simons) picture and they can be decomposed into those generated by $\rho^a$, and those generated by $\tau^a$. Moreover, as we already explained, the Chern–Simons term is also manifestly invariant under diffeomorphisms because it is written in terms of forms. Then, when dealing with the first-order action we also have the local Lorentz transformations, which act on $e$ and $\omega$ as in (1.5) and (1.8) respectively.

Now, as shown in Appendix C.1, the later LLTs are quite easily seen to be in one-to-one correspondence with the gauge transformations generated by the parameters $\tau^a$ on the Chern–Simons side. As for the infinitesimal gauge transformations of the gauge picture generated by the $\rho^a$'s the story is a little more subtle, and indeed at first sight one wonders what they could correspond to in the frame formulation. Actually, we will show that the gauge transformations generated by the $\rho^a$'s are in fact not extra gauge transformations but, rather, they are simply some combination of diffeomorphisms and LLTs up to so-called trivial gauge transformations (see Appendix C.1). As already stated, this is well, since the point was to check that there are no extra gauge symmetries. Let us also stress that this fact is truly a three-dimensional feature and does not happen in dimension four and greater. Actually, this is precisely what prevents one from writing Gravity in dimension four and greater as a gauge theory simply by gauging the isometry group of the vacuum and employing a gauge-connection valued therein. Thus, we might say that 'only in three dimensions is Gravity a true gauge theory', and even there, we see that its action is that of Chern–Simons, which is not of the Yang–Mills type that we are more used to in standard gauge theory. Further note that a Yang–Mills action term in three dimensions would propagate scalar degrees of freedom, unlike Gravity which propagates none in dimension three.

This achieves the proof of the equivalence for the $\lambda = 0 = \Lambda$ case. Three-dimensional gravity is thus a gauge theory for iso$(2,1)$, the Poincaré algebra in dimension 3 for zero cosmological constant.

**Remark** : as we just said in dimension four and greater the first-order action (1.23) is only invariant under the homogeneous Poincaré group so$(2,1)$, not under the whole of iso$(2,1)$ — that is, gauge translations do not leave the action invariant. The action (1.23) is of course invariant under coordinate reparametrization (diffeomorphisms), but those Lie derivatives do not correspond (in the first order formalism at least) to gauge translations. Only in three-dimensions does that happen, thus allowing to rewrite three-dimensional gravity as a gauge theory for the whole Poincaré group. More information can be found in [93, 95] and we do not further dwell on that point.

**Cosmological Constant and Chiral Copies**

As announced, what we shall be interested in is the case with negative cosmological constant. Following the same reasoning as in the $\lambda = 0$ case (with same identifications of $e$ and $\omega$ via (1.37) and same bilinear form on the algebra) the equivalence can again be proven between the frame formulation and the Chern–Simons one based on so$(2,2)$. As this calculation is really close to the one we have just performed we shall not go through it again and we just quote what is different in the Chern–Simons picture, that is, the gauge transformations now read

$$\delta e_\mu^a = \partial_\mu \rho^a + \epsilon^{abc} e_{b\mu} \tau_c + \epsilon^{abc} \omega_{b\mu} \rho_c, \tag{1.43a}$$

$$\delta \omega_\mu^a = \partial_\mu \tau^a + \epsilon^{abc} \omega_{b\mu} \tau_c + \lambda \epsilon^{abc} e_{b\mu} \rho_c, \tag{1.43b}$$

and the way in which they correspond to diffeomorphisms and LLTs is the same as before. Further note that the identification of the actions now requires one to set

$$\kappa = \frac{l}{16\pi G} \tag{1.44}$$

and
$$\lambda = \Lambda, \tag{1.45}$$
so that the parameter $\lambda$ appearing in (1.29) is indeed the cosmological constant $\Lambda$ and $l$ is the AdS radius defined in the usual way by $\Lambda \equiv -1/l^2$. Note that $\kappa$ is often parametrized as $k/4\pi$ in the literature, which implies $k = l/4G$ in the AdS case.

The splitting (1.33) of so(2,2) into two chiral copies of sl(2|$\mathbb{R}$) makes it is possible to rewrite the Chern–Simons action term for $\Gamma_\mu \in$ so(2,2) as the sum of two Chern–Simons actions, each of them having their connections $A$ and $\tilde{A}$ in the first and second chiral copy of sl(2|$\mathbb{R}$) respectively. In the sequel we shall only deal with the first chiral copy but we wanted to still call the connection thereof $A$, which is why we have changed notations at this point. Actually, the decomposition of $\Gamma$ in terms of $e$ and $\omega$ is quite helpful in formulating this splitting precisely, for we see that

$$\Gamma = e^a P_a + \omega^a J_a = \big(\omega^a + \frac{e^a}{l}\big)J_a^+ + \big(\omega^a - \frac{e^a}{l}\big)J_a^- \equiv A^a J_a^+ + \tilde{A}^a J_a^- \equiv A + \tilde{A}, \tag{1.46}$$

where the $J_a^\pm$'s are defined by (1.35). Now, taking into acount both the commutation relations (1.34) and the bilinear form (1.36) written in terms of $J_a^\pm$, we see that the Chern–Simons action term for so(2,2) can be split as follows

$$S_{\text{CS}}[\Gamma = A + \tilde{A}] = S_{\text{CS}}[A] + \tilde{S}_{\text{CS}}[\tilde{A}] \equiv S_{\text{CS}}[A, \tilde{A}], \tag{1.47}$$

with each chiral copy having prefactor $\kappa = l/16\pi G$. Note that for the splitting of the kinetic piece one needs only notice that $(J_a^\pm, J_b^\mp) = 0$, whereas also $[J_a^\pm, J_b^\mp] = 0$ is needed to prove the splitting of the interaction piece. Both chiral copies $S_{\text{CS}}[A]$ and $\tilde{S}_{\text{CS}}[\tilde{A}]$ are the same actions except for one difference, which is that the $J_a^+$'s and $J_a^-$'s are equipped with bilinear forms having opposite signs (1.36). Equivalently, if one prefers to have both chiral copies equipped with the same bilinear form, one can instead declare

$$A \equiv \big(\omega^a + \frac{e^a}{l}\big)T_a, \tag{1.48a}$$

$$\tilde{A} \equiv \big(\omega^a - \frac{e^a}{l}\big)T_a, \tag{1.48b}$$

with the $T_a$ generators of sl(2|$\mathbb{R}$) satisfying the same commutation relations and scalar products as the $J_a^+$ ones (we changed notations not to confuse the reader). The decomposition of the action then reads

$$S_{\text{CS}}[\Gamma = e/l + \omega] = S_{\text{CS}}[A] - S_{\text{CS}}[\tilde{A}] = S_{\text{CS}}[A, \tilde{A}]. \tag{1.49}$$

It is of course no longer true that $\Gamma = A + \tilde{A}$ (nor does it make sense to write so anymore), but this decomposition, in which the connections $A$ and $\tilde{A}$ both lie in the first chiral copy of sl(2|$\mathbb{R}$) (to put it that way) is handier as the two action functionals are the same now also when seen as functionals of the components $A^a$ and $\tilde{A}^a$ — they are truly the same action functionals now. This formulation, where the metrics on both copies of sl(2|$\mathbb{R}$) have the same signature, will be of much use in the sequel, where we shall only treat the first chiral copy for many of our purposes. This is also the formulation that is most often encountered in the literature.

Note that the equations of motion now read

$$F[A] = 0, \quad F[\tilde{A}] = 0, \tag{1.50}$$

which, when combined as $F[A] \pm F[\tilde{A}] = 0$ and subsequently linearized are seen to yield those written in (1.24) (in the linearized limit). Also note that the gauge transformations are also split now, but we shall not display them here for the sake of conciseness. For future reference, let us also display the constraints of the theory, called the Chern–Simons–Gauss constraints, which read

$$\mathcal{G}_a \equiv \frac{k}{4\pi}(T_a, T_b)\epsilon^{ij}F[A]^b_{ij} = \frac{k}{8\pi}\eta_{ab}\epsilon^{ij}F[A]^b_{ij} \approx 0, \tag{1.51}$$

where the indices $i$ and $j$ are spatial indices, and the '$\approx$' symbol here stands for weak equalities (on the constraint surface [103]). The other copy's constraints read similarly.

### 1.1.3 Supergravity

As mentioned in the introduction to the present chapter, our goal is to arrive at a Chern–Simons formulation for Higher-Spin Supergravity on AdS$_3$, which we do in Section 1.2. In this subsection we thus comment on how to rewrite the various Supergravity theories in terms of a superconnection valued in some gauge algebra governed by a Chern–Simons action term. As announced at the beginning of this section, given that the rewriting of Supergravity as a gauge theory follows the same lines as that of pure Gravity we shall only quote the final results hereafter.

A Chern–Simons action for a gauge connection valued in the superalgebra of isometries of the vacuum solution of some Supergravity theory reproduces the dynamics of the latter, provided one identifies the field components appropriately [58]. In Chapter 2, the emphasis will be on the most common (and arguably simplest) Supergravity, namely the non-extended one describing a graviton and a complex gravitino $\psi_\mu$, built upon an $\text{osp}(1,2|\mathbb{R}) \oplus \text{osp}(1,2|\mathbb{R})$-invariant vacuum state. The $\text{osp}(1,2|\mathbb{R})$ Lie superalgebra contains, in addition to the familiar $\text{sl}(2|\mathbb{R})$ sector, the odd generators $R^\pm$, which obey the commutation relations

$$\begin{aligned}
[H, E] &= 2E, & [H, F] &= -2F, & [E, F] &= H, \\
[H, R^+] &= R^+, & [E, R^+] &= 0, & [F, R^+] &= R^-, \\
[H, R^-] &= -R^-, & [E, R^-] &= R^+, & [F, R^-] &= 0, \\
\{R^+, R^+\} &= -2iE, & \{R^-, R^-\} &= 2iF, & \{R^+, R^-\} &= iH,
\end{aligned} \quad (1.52)$$

with the scalar product reading like

$$\text{STr}(\Gamma) \equiv \Gamma_{11} - \text{Tr}\left(\Gamma_{\text{sp}(2)}\right) = \Gamma_{11} - \Gamma_{22} - \Gamma_{33} = -\Gamma_{22} - \Gamma_{33}, \quad (1.53)$$
$$(\Gamma, \Gamma') \equiv \text{STr}(\Gamma\Gamma'). \quad (1.54)$$

More information on the superalgebras of extended and non-extended Supergravity can be found in Appendix B, where the conventions pertaining to the above commutation relations and scalar products are also given. The Chern–Simons action $S[\Gamma, \tilde{\Gamma}] = S_{\text{CS}}[\Gamma] - S_{\text{CS}}[\tilde{\Gamma}]$ for $\Gamma_\mu \in \text{osp}(1,2|\mathbb{R})$ then reproduces the Supergravity action [93] provided the correct identification is made in the fermionic sector [58]. In fact, because of the commutation relations and scalar product above, the action $S_{\text{CS}}[\Gamma]$ splits as

$$\begin{aligned}
S_{\text{cs}}[\Gamma] &\equiv \frac{k}{4\pi} \int_{\mathcal{M}_3} \text{STr}\left(\Gamma \wedge d\Gamma + \tfrac{2}{3}\Gamma \wedge \Gamma \wedge \Gamma\right) \\
&= \frac{k}{4\pi} \int_{\mathcal{M}_3} \left[\text{Tr}\left(A \wedge dA + \tfrac{2}{3} A \wedge A \wedge A\right) + i\text{Tr}\left(\bar{\Psi} \wedge d\Psi + \bar{\Psi} \wedge A \wedge \Psi\right)\right],
\end{aligned} \quad (1.55)$$

where $\Gamma = A + \Psi$, with $A \in \text{sl}(2|\mathbb{R})$. As in the pure-Gravity $\text{AdS}_3$ case, the coefficient $k$ is a dimensionless, real-valued coupling constant of the theory, and it is related to the three-dimensional Newton's constant G and the AdS radius of curvature $\ell$ through $k = \frac{\ell}{4G}$. The cosmological constant is $\Lambda \equiv -\frac{1}{\ell^2}$, and with $k$ real the action is real-valued. Also, one can again prove that the gauge transformations 'match', much in the spirit of the pure-Gravity case (see previous subsection). The same result holds for all the other Supergravity cases, with similar identifications of the field components, and the complete list of $\text{AdS}_3$ supergravities can be found in Appendix B. However, in the present work will shall be mainly concerned with the $\text{osp}(1,2|\mathbb{R})$ and $\text{osp}(N,2|\mathbb{R})$ supergravities as well as with the higher-spin extensions thereof, and comments on higher-spin theories stemming from extending other supergravities are made in Chapter 3. We note that, in the extended $\text{osp}(N, 2|\mathbb{R})$ case, the above action contains extra terms involving the internal-algebra components of $\Gamma_\mu \in \text{osp}(N, 2|\mathbb{R})$, which for $N = 1$ (resp. $N = 2$) is trivial (resp. abelian) [58]. Let us also recommend reference [60] for a detailed treatment of the above rewriting in the extended-Supergravity case, where it is also noted that one can cast the action in the Chern–Simons form for all cases listed in Appendix B and not just for $\text{osp}(N, 2|\mathbb{R})$.

A property worth highlighting is the following: we know that supergravity 'contains' gravity, that is, it describes a (non-propagating) graviton together with other, lower-spin fields. At the algebraic level, the corresponding statement is that $\text{sl}(2|\mathbb{R})$ is a subalgebra of $\text{osp}(N, 2|\mathbb{R})$, so that one is assured to be able to rewrite the $\Gamma_\mu \in \text{osp}(N, 2|\mathbb{R})$ Chern–Simons action as a $A_\mu \in \text{sl}(2|\mathbb{R})$ Chern–Simons term (Einstein–Hilbert) plus the fermion term, plus some interaction (cross) terms — provided the identification of the components of $\Gamma_\mu$ in the $\text{sl}(2|\mathbb{R})$ subsector with the dreibein and spin-connection is the same as that of $A_\mu$. Thus, in exactly the same way, a Chern–Simons term for a 'higher-spin algebra' containing $\text{sl}(2|\mathbb{R})$ (resp. $\text{osp}(N, 2|\mathbb{R})$) as a subalgebra and with correct identification of the components in the Gravity (resp. Supergravity) sector will be a good candidate for a Higher-Spin Gravity (resp. Supergravity) theory. However, first we should find an identification of the higher-spin components such that the higher-spin free kinematics are reproduced in the linearized limit of such an action term. The following section is concerned with precisely these matters.

## 1.2 Higher-Spin Gravity as a Gauge Theory in $D = 3$

What has been done for pure gravity in the previous section will now be carried out for higher-spin fields ($s > 2$). However, the formulation of interacting higher-spin fields (with themselves and with gravity) is notoriously tricky and plagued by many obstructions [52]. Nevertheless, as we shall now expand on a little, all the renowned difficulties in building interactions are evaded in dimension 3.

First of all, as noted in [42], in flat spacetime the generically non-vanishing Weyl tensor in $D \geq 4$ precludes the existence of so-called 'Hypergravity' (a spin-$\frac{5}{2}$ field minimally coupled to gravity; in some sense the first non-trivial gravitational higher-spin interaction). Moreover, although the higher-spin fields may have cubic multipoles with Gravity, a fully consistent theory is incompatible with locality of the Lagrangian, at least for a finite number of fields — in fact, Part II of the manuscript is concerned with precisely this problem. One can nevertheless have fully consistent theories on AdS backgrounds, as a long-term effort by Vasiliev ended up proving. However, the formulation thereof, which calls upon a generalized frame formalism, requires so-called 'extra fields' and 'extra gauge symmetry' — these are auxiliary fields and associated gauge symmetries one is forced to introduce in order to formulate higher-spin fields in the frame formalism in dimension four and greater [46, 47]. These facts essentially complicate the introduction of interactions, although as we know Vasiliev's equations do exist [50, 104, 105].

In $D = 3$ these two complications do not arise. Indeed, the Weyl tensor vanishes in three dimensions, which does allow for interaction terms involving the minimal coupling to Gravity. Also, as noted for example in the first sections of [2], the frame formulation of three-dimensional higher-spin fields does not require so-called 'extra fields' and 'extra gauge symmetry', so that the construction of interactions in the AdS case via the generalized frame formalism is also made easier. In full analogy with the previous section, we should therefore present the interacting three-dimensional theory of higher-spin fields and then translate it to the Chern–Simons form. However, as we shall explain, for interacting higher spins in three dimensions the Chern–Simons rewriting is not only a reformulation of an existing, interacting theory. It is actually a way to introduce interactions, which one could also introduce at the level of the frame formulation [106–108] (and to some extent at the level of the metric formalism [92]) but which are much more easily treated in the gauge picture.

As we shall describe in Subsection 1.2.1, one can formulate higher-spin fields in some analogue of the frame formalism for Gravity that we reviewed in Subsection 1.1.1 (at the free level) and then from there move on, in Subsection 1.2.2, to the Chern–Simons picture for higher spins — much in the spirit of what was done for pure gravity. The problem of introducing interactions will then be easily dealt with, as it will be equivalent to the purely algebraic problem of finding suitable higher-spin Lie algebras (see also the end of the previous section). The formulation one then arrives to, firstly introduced by Blencowe [56], namely a Chern–Simons gauge theory for some (finite or infinite-dimensional) gauge algebra containing $so(2,2)$ (in the AdS$_3$ case), is the basis for almost every study of three-dimensional Higher-Spin Gravity today, and is what the present section is devoted to reach and then generalize to include fermions. Just as in Section 1.1, we first develop the material without supersymmetry in quite a detailed way, and in Subsection 1.2.3 display the corresponding higher-spin supergravities, which are the ones we study the asymptotic symmetries of in Chapter 2.

### 1.2.1 The Frame Formulation of Free Higher Spins

As aforementioned, some analogue of the frame formalism for Gravity also exists for higher-spin fields, and we now present it. Let us stress that all of what we shall present in this subsection takes place at the linearized level. Although we shall be mainly interested in the AdS$_3$ case, we shall develop as much of the material as possible in arbitrary dimension and in a generic constant-curvature background spacetime. Part of the material exposed in this subsection is also reviewed in [53].

**The Metric Formalism**

Let us review the Fronsdal (or metric) formulation of (free) higher-spin fields [33]. The Fronsdal equations of motion for a spin-$s$ gauge field described by a rank-$s$ symmetric tensor $\varphi_{\mu_1\ldots\mu_s}$ and propagating in the Minkowski $D$-dimensional background are given by

$$F_{\mu_1\ldots\mu_s} \equiv \Box\varphi_{\mu_1\ldots\mu_s} - s\partial_{(\mu_1}\partial^\lambda\varphi_{\mu_2\ldots\mu_s)\lambda} + \tfrac{1}{2}s(s-1)\partial_{(\mu_1}\partial_{\mu_2}\varphi_{\mu_3\ldots\mu_s)\lambda}{}^\lambda = 0, \tag{1.56}$$

where $F$ is the so-called Fronsdal tensor, which should be thought of as a higher-spin equivalent of the linearized Ricci tensor (which it boils down to for $s = 2$). Our symmetrization parenthesis have weight one, so that e.g. $2A_{(ij)} \equiv A_{ij} + A_{ji}$. The above equations are invariant under the gauge transformations

$$\delta\varphi_{\mu_1\ldots\mu_s} = s\partial_{(\mu_1}\xi_{\mu_2\ldots\mu_s)}, \quad \text{with } \xi^\lambda{}_{\lambda\mu_3\ldots\mu_{s-1}} = 0. \tag{1.57}$$

One can verify that the above equations of motion are equivalent to the ones obtained from varying the action

$$S = \int \mathrm{d}^D x\, \varphi^{\mu_1\ldots\mu_s}\big(F_{\mu_1\ldots\mu_s} - \tfrac{1}{4}(s-1)s\eta_{(\mu_1\mu_2}F_{\mu_3\ldots\mu_s)\lambda}{}^\lambda\big), \tag{1.58}$$

where the expression in parenthesis is the higher-spin analogue of the linearized Einstein tensor. In fact, the gauge invariance of the above Lagrangian requires the double-trace constraint $\varphi^{\lambda\rho}{}_{\lambda\rho\mu_5\ldots\mu_s} = 0$ to hold. As one can check, this condition can be obtained by considering derivatives of the Fronsdal tensor $F$ (which is on-shell zero), so that on-shell it is automatically imposed, as for example at the level of the equations of motion. However, at the level of the action (off-shell) we need to impose this constraint separately, and one can verify that it is preserved by the above gauge symmetries. The above action is thus the higher-spin analogue of what we would obtain if we were to linearize the Einstein–Hilbert action (1.1) (in brackets we find the analogue of the linearized Einstein tensor), and the above equations of motion (1.56) are the higher-spin counterparts of the linearized version of (1.3) (at $\Lambda = 0$).

Let us then move to fields propagating on constant-curvature backgrounds. We are thus looking for equations that should now be invariant under
$$\delta\varphi_{\mu_1...\mu_s} = s\nabla_{(\mu_1}\xi_{\mu_2...\mu_s)}, \quad \text{with } \xi^\lambda{}_{\lambda\mu_3...\mu_{s-1}} = 0, \quad (1.59)$$
where $\nabla$ stands for the covariant derivative (see previous section) associated with the background metric $\bar{g}_{\mu\nu}$ (that we will choose to be anti-de Sitter later on). The equations of motion are now
$$\begin{aligned}\hat{F}_{\mu_1...\mu_s} &\equiv F_{\mu_1...\mu_s} + \Lambda\big([(s^2 + (D-6)s - 2(D-3)]\varphi_{\mu_1...\mu_s} \\ &+ s(s-1)\bar{g}_{(\mu_1\mu_2}\varphi_{\mu_3...\mu_s)\lambda}{}^\lambda\big) \\ &= 0,\end{aligned} \quad (1.60)$$

where $\hat{F}$ is the 'AdS Fronsdal tensor' and the Fronsdal tensor $F$ itself is now understood as in (1.56) but with all derivatives replaced by covariant derivatives with respect to the background metric. Again imposing the double trace constraint on our field the free Lagrangian is fixed by the requirement of gauge invariance and reads exactly as (1.58) but with $F$ replaced by $\hat{F}$. The analogies with (1.1) and (1.3) are again clear.

It is important to note that the above equations of motion and Lagrangians are fixed by the requirement of invariance under the corresponding gauge transformations. As explained e.g. in [109], the interactions are 'even more' constrained and although Vasiliev's equations do exist and are fully non-linear, they still lack a satisfactory corresponding action principle.[5]

**Remark** : in three spacetime dimensions, the usual notion of spin for massless particles (the helicity) reduces to a mere distinction between bosons and fermions, that is, the little group is trivial. Nonetheless, one may wish to consider the same four-dimensional free equations describing some tensor field but in dimension three. It is then easy to see that, apart from the scalar (dual to the spin 1) and the spin-$\frac{1}{2}$ field, no degrees of freedom can propagate [111]. In particular, higher-spins do not propagate any local degree of freedom in dimension three, and neither does a graviton or a Rarita–Schwinger field. However, one may still wish to call a fully symmetric rank-$s$ tensor satisfying the $D = 3$-projected Fronsdal equation a spin-$s$ field. That is, of course, what we mean by a higher spin in three dimensions. In [56], Blencowe obtained precisely that object (or, rather, its translation in terms of the generalized dreibein and spin-connection) by means of projecting directly the four-dimensional equations written in terms of the frame fields onto three dimensions.

---

[5] An interesting action principle was proposed in [110]. It is, however, non-standard in the sense that it is formulated in a higher-dimensional spacetime that includes twistorial directions. In particular, it does not reproduce Fronsdal's action upon linearization.

**The Vielbein and Spin-Connection**

Let us now try to formulate the above higher-spin free kinematics along the lines of the frame formulation of Gravity. However, in Subsection 1.1.1 the reformulation of Gravity in terms of the vielbein and spin-connection was carried out at the non-linear level, whereas here we only have a linear theory to start from (see comments at the end of the previous subsection) so that we shall remain at the linearized level. Therefore, instead of the relation (1.4), what we are trying to generalize to the higher-spin case, rather, is its linearized version

$$\varphi_{\mu\nu} = 2\bar{e}^a_{(\mu} v_{\nu)a}, \qquad (1.61)$$

which is simply derived by plugging $g_{\mu\nu} \equiv \bar{g}_{\mu\nu} + \varphi_{\mu\nu}$ in (1.4) and defining $\bar{e}$ to be the background vielbein, associated with $\bar{g}_{\mu\nu}$, and defined together with $v^a_\mu$ by $e \equiv \bar{e} + v$. The above relation is now invariant under

$$\delta v^a_\mu = \alpha^a{}_b \bar{e}^b_\mu, \qquad (1.62)$$

because $\alpha^a{}_b \in \mathrm{so}(D-1, 1)$ (remember that Latin indices are raised and lowered with $\eta_{ab}$).

The above change of variables is then generalizable to higher-spin fields. Indeed, let us introduce some generalized vielbein $e^{a_1\ldots a_{s-1}}_\mu$. Of course, we have no higher-spin analogue of the full metric at hand, so that the only thing we can do is declare this object to be its own excitation (that is, we assume that the background generalized vielbeins vanish[6]) and try to relate it to $\varphi_{\mu_1...\mu_s}$ in a sensible way that generalizes (1.61). This was done in the founding paper [46], resulting in the arbitrary-spin expression

$$\varphi_{\mu_1...\mu_s} \equiv s\bar{e}^{a_1}_{(\mu_1}\ldots\bar{e}^{a_{s-1}}_{\mu_{s-1}}e_{\mu_s)a_1...a_{s-1}}, \tag{1.63}$$

which is invariant under

$$\delta e^{a_1...a_{s-1}}_\mu = \bar{e}_{b\mu}\alpha^{b,a_1...a_{s-1}}, \tag{1.64}$$

for

$$\alpha^{(b,a_1...a_{s-1})} = 0. \tag{1.65}$$

---

[6] As the generalized (or higher-spin) vielbeins have no background values, we shall stick to the notation $e^{a_1\ldots a_{s-1}}_\mu$.

Note that, because Latin indices are raised and lowered with the Minkowski metric the last condition above indeed coincides, in the $s = 2$ case, with the matrix $(\alpha^a{}_b) \in so(D-1,1)$. Now, in the standard frame approach to higher spins the generalized vielbein is chosen to be an irreducible Lorentz tensor in its frame indices, that is, we choose it to be symmetric and traceless in those same indices, i.e. we impose the conditions

$$e^{a_1...a_{s-1}}_\mu = e^{(a_1...a_{s-1})}_\mu, \quad e_{\mu b}{}^{ba_1...a_{s-3}} = 0, \tag{1.66}$$

the latter of which ensures the double-trace constraint on the field $\varphi_{\mu_1...\mu_s}$. Now, with such a choice of generalized vielbeins, our generalized LLT parameter $\alpha$ will have to satisfy

$$\alpha^{b,a_1...a_{s-1}} = \alpha^{b,(a_1...a_{s-1})}, \quad \alpha^{b,a_1...a_{s-3}c}{}_c = 0, \tag{1.67}$$

which, together with (1.65) implies

$$\alpha^{b,(a_1...a_{s-1})}{}_b = 0. \tag{1.68}$$

Then, much like in Gravity, the vielbein is just a covector with respect to its spacetime index, and in the present formulation our generalized vielbein will have covariant transformation rules under the 'generalized diffeomorphisms' (1.59) such that its application to (1.63) reproduces (1.59). What we obtain is simply

$$\delta e^{a_1...a_{s-1}}_\mu = (s-1)\bar{e}^{(a_1}_{\nu_1}\ldots\bar{e}^{a_{s-1})}_{\nu_{s-1}}\nabla_\mu\xi^{\nu_1...\nu_{s-1}}. \tag{1.69}$$

Again proceeding along the lines of what is known for Gravity one introduces some generalized spin-connection $\omega_\mu^{a,b_1...b_{s-1}}$, satisfying the same conditions (1.65), (1.67) and (1.68) as the parameter $\alpha$:

$$\omega^{b,a_1...a_{s-1}}_\mu = \omega^{b,(a_1...a_{s-1})}_\mu, \quad \omega^{b,a_1...a_{s-3}c}_\mu{}_c = 0, \quad \omega^{(b,a_1...a_{s-1})}_\mu = 0, \tag{1.70}$$

which together imply
$$\omega_\mu^{b,(a_1...a_{s-2})}{}_b = 0. \tag{1.71}$$

Note that for $s = 2$ the condition $\omega_\mu^{(b,a_1...a_{s-1})} = 0 = \alpha^{(b,a_1...a_{s-1})}$ is implied by the antisymmetry in the two only Latin indices then carried by $\omega$ and $\alpha$.

The last comment we shall make in the present subsection is that, in dimension 3, a 'dual' rewriting analogous to (1.21) can also be performed in the arbitrary-spin case so that, ultimately, our first-order formalism will deal with the generalized dreibein $e_\mu^{a_1...a_{s-1}}$ and a generalized spin-connection $\omega_\mu^{a_1...a_{s-1}}$.

**The Action and the Equations of Motion**

In his pioneering work [46], Vasiliev identified a first-order action for the generalized vielbeins and spin-connections such that, when solving for the auxiliary field $\omega$ in terms of $e$ and further recalling the definition (1.63), one recovers an action functional coinciding with that of Fronsdal (1.58). For the sake of conciseness we only give here its four-dimensional spin-$s$ expression at $\Lambda = 0$, which reads

$$S = \int d^4x\, \epsilon^{\mu\nu\rho\sigma} \epsilon^{abc}{}_\sigma \omega_{\rho,b,a}{}^{i_1...i_{s-2}} \left( \partial_\mu e_{\nu,i_1...i_{s-2}c} - \tfrac{1}{2} \omega_{\mu,\nu,i_1...i_{s-2}c} \right). \tag{1.72}$$

Such an action, if we believe it to be equivalent to the Fronsdal one (which it is), will be invariant under generalized LLTs as well as generalized diffeomorphisms. However, as one can check, it is also invariant under an extra gauge transformation, acting only on the spin-connection [46]. That extra gauge parameter can be checked to vanish in the $s = 2$ case but, most importantly, in the arbitrary-spin case it also vanishes in $D = 3$! The reason why this is a key point is that one of the difficulties in formulating higher-spin theories stems from the fact that this extra gauge symmetry calls for so-called 'extra (gauge) fields' associated with it (much like we can think of the spin-connection as the gauge field associated with the LLT gauge symmetry), and one is actually led through an iterative procedure which introduces several such auxiliary fields. Dealing with them is a notorious source of inconveniences in the higher-spin context, and the fact that they are not needed in three dimensions can be thought of as being one of the reasons why the three-dimensional case is simpler to deal with.

Actually, reference [46] only deals with the four-dimensional Minkowski case, and one has to refer to [47] in order to get the corresponding (A)dS expression, and to [112] for the fermionic treatment (see also [101] for the generic-dimension case). As for the three-dimensional scenario, it was first treated in [56], where the usual frame expressions for free higher-spin fields were projected onto three-dimensional spacetimes and then completed to yield a fully interacting theory. Before giving its expression, note that we shall not display frame-index contraction explicitly when it is thought to be obvious (see below). On the $AdS_3$ spacetime background, that we are most interested in, the obtained spin-$s$ expression is thus (now in terms of the 'dualized' spin-connection):

$$S = \int e \wedge D\omega + \tfrac{1}{2}\epsilon^{abc}\bar{e}_a \wedge (\Lambda e_b \wedge e_c - \omega_b \wedge \omega_c), \qquad (1.73)$$

and the corresponding spin-$s$ equations of motion thus read[7]

$$D\omega^{a_1...a_{s-1}} - \Lambda\epsilon^{aba_1}\bar{e}_a \wedge e_b^{a_2...a_{s-1}} = 0, \qquad (1.74a)$$

$$De^{a_1...a_{s-1}} + \epsilon^{aba_1}\bar{e}_a \wedge \omega_b^{a_2...a_{s-1}} = 0, \qquad (1.74b)$$

and indeed one can verify that they enjoy no extra gauge invariance of any sort — only diffeomorphisms and local Lorentz transformations. Regarding our notation for index contraction, note for example the first term in the above action, where evidently contraction of all the indices of $e$ with all the indices of (the dualized) $\omega$ is implied, their index structure being the same. The same goes, for example, for both terms within the brackets in the action; we assume contraction of all indices except the ones that are displayed (and which are contracted with the epsilon tensor). Let us further stress that, since the spin-2 dreibeins are denoted respectively by $\bar{e}_a$ (background) and $v_a$ (excitation), there can be no confusion with some higher-spin dreibein of which we display only one frame index, as in the above action — recall that the higher-spin dreibeins and spin-connections are always assumed to have zero background values. Finally, let us point out that the background spin-connection for the spin-2 enters the action via the covariant derivative $D$.

Although we don't give the proof [2] that the above action is indeed equivalent to the Fronsdal one we point out the enlightening similarity of the above equations with the linearized equations (1.24); the structure is really the same, and all we have done is deal with the extra indices in the only possible way. Let us also make it clear that the apparent discrepancy one might seem to notice between the above action and (1.72) simply lies in the fact that (1.72) is given on a flat background, where $\bar{e}$ is the trivial matrix and $\bar{\omega}$ is zero.

---

[7] For $s = 2$ these equations of course boil down to those given in (1.24), only recalling that in that case the excitation is ascribed the letter $v_a$.

### 1.2.2 A Chern–Simons Action for Higher-Spin Gravity

The idea is now that, much in the spirit of what we did for pure Gravity, we shall rewrite the above action (1.73) for higher spins as a Chern–Simons term whose gauge connection one-form takes values in some Lie algebra, the coefficients of which shall be identified with the generalized dreibein and spin-connection. Once again we shall proceed backwards, that is, we shall display some Chern–Simons action together with some identification of the components of its connection and then show how our action (1.73) is reproduced therefrom (at the free level).

**Requirements at the Linearized Level**

From the previous section it should be obvious that the action we are now looking for is some Chern–Simons term for a gauge connection taking values in an algebra containing sl(2|$\mathbb{R}$). What is now to be investigated is what requirements are imposed on such an algebra by the matching with (1.73) at the linearized level. Note that the action we look for is the difference of two copies of the Chern–Simons action for independent combinations of the dreibein and spin-connection, like in (1.49). However, as we shall see, much of the discussion can be carried out considering only the first copy (at least the purely algebraic considerations).

Let the $T_a$'s be the spin-2 generators of (1.48). Now, as we have seen in the previous subsection, the higher-spin off-shell degrees of freedom[8] we need to accommodate for come in the form of the generalized dreibeins $e^{a_1...a_{s-1}}$ and spin-connections $\omega^{a_1...a_{s-1}}$, which are symmetric in their (frame) indices as well as traceless. The combination $e^{a_1...a_{s-1}} + \omega^{a_1...a_{s-1}}$ is therefore to be identified with the coefficient of some higher-spin generator $T_{a_1...a_{s-1}}$, that we may assume to be symmetric and traceless in its indices (and correspondingly for the other copy). As is easy to check, the number of independent spin-$s$ generators $T_{a_1...a_{s-1}}$ is precisely $2(s-1)+1$, that is, the dimension of a spin-$s$ (or, rather, $s-1$) representation of sl(2|$\mathbb{R}$). The nice thing about it is that, because of the isomorphism sl(2|$\mathbb{R}$) $\simeq$ so(2,1), the components of our Chern–Simons connection corresponding to the spin-$s$ field come in the right number to form an irreducible spin-$(s-1)$ representation of the three-dimensional Lorentz group, so(2,1). Actually, this is exactly what we shall assume, namely that the spin-$s$ generators behave as irreducible Lorentz tensors, which can be seen to translate to

$$[T_a, T_{a_1...a_{s-1}}] = \epsilon^c{}_{a(a_1} T_{a_2...a_{s-1})c}. \qquad (1.75)$$

---

[8] The adjective off-shell is used to stress again that in dimension three there are no on-shell degrees of freedom.

The higher-spin algebra we are looking for is therefore some algebra containing sl(2|$\mathbb{R}$) and, besides, the higher-spin generators $T_{a_1...a_{s-1}}$ up to some spin, sitting in irreducible representations of the Lorentz algebra according to the above formula. Note that the mismatch between the spin of some generators and the representation of so(2, 1) they sit in comes from the fact that, on top of the frame indices, the connection further carries a spacetime index. The generators $T_{a_1...a_{s-1}}$, that we have said to have spin-$s$, are also sometimes said to have conformal spin $s-1$.

Two important comments are now in order. Firstly, it should be noted that, whatever our higher-spin algebra is in the end, in order to make sense of the Chern–Simons term it should be equipped with an invariant and non-degenerate bilinear form. If the searched-for algebra is semi-simple, then we know that the Killing form, which always exists, is non-degenerate (Cartan's Criterion). Moreover, if the algebra is simple, the Killing form is unique in the space of invariant bilinear forms. Interestingly, one can check that the only possibility for a bilinear form is[9]

$$(\Gamma, \Gamma) = \sum_{s=1}^{N} c_s \Gamma^{a_1...a_{s-1}} \Gamma_{a_1...a_{s-1}}, \qquad (1.76)$$

where the coefficients $c_s$ are left undetermined by the requirement of invariance under the commutation relations we already have at hand, namely those of sl(2|$\mathbb{R}$) as well as those in (1.75). The commutation relations among the higher-spin generators can potentially fix (some of) those coefficients, but for semi-simple Lie algebras there is at least one form (the Killing form) which corresponds to all the above coefficients being non-zero.

**Remark** : in the following we study finite- and infinite-dimensional algebras. Let us note, then, that in the infinite-dimensional case the definition of being simple is less clear. At any rate, the point is really to be able to construct a bilinear form with the desired properties, which we shall do anyhow, even for infinite dimension (see below).

The second comment to be made is that, assuming all $c_s$'s to be non-zero, whatever algebra we find will do the job. Namely, if we write a Chern–Simons theory for a gauge connection living in some higher-spin algebra containing sl(2|$\mathbb{R}$) and whose higher-spin generators satisfy (1.75),

---

[9] Note that the unicity of such a form assumes, implicitly, that its formulation is covariant.

the linearization thereof shall yield precisely the action (1.73) (upon identification of the components along the lines of $A = e + \omega$, and correspondingly for the second copy). This last point is really the key one, so let us phrase it differently: *when one linearizes the Chern–Simons action with proper identification of the degrees of freedom as above, the commutator of higher-spin generators with themselves is not used*. Only the commutators of higher-spin generators with spin-2 ones, and of course those of spin-2 generators with themselves are used. The reason for this is simple and lies in the fact that the higher-spin dreibeins and spin-connections have been assumed to have zero background values, as is easy to note trying to do the exercise. Another nice feature is that the coefficients $c_s$ are not used either when linearizing the action; indeed, at the free level the Chern–Simons term for our higher-spin algebra splits into a sum of free actions for the different spins which are involved, with the corresponding $c_s$ coefficients in front, which therefore play no role in recovering the Fronsdal system.

**Remark** : to be precise, it is the absolute value of the $c_s$ coefficients which plays no role in recovering the Fronsdal system. The relative signs of the coefficients are of some importance. Indeed, if the relative sign for the spin-2 and spin-3 sector is minus then the kinetic terms of both those sectors will have opposite signs, which is non-standard. This means that our above statement about the fact that the 'higher-higher' commutators do not affect the linearized limit of the theory should be refined: those commutators may constrain the bilinear form, which in turn may yield non-standard relative signs for the kinetic terms (if it is not degenerate). The example of the two non-compact real forms of sl(3), treated below, illustrates this point well.

The conclusion is thus that any Lie algebra containing sl(2|$\mathbb{R}$) whose higher-spin generators are irreducible Lorentz tensors and whose invariant bilinear form is non-degenerate shall yield a Chern–Simons action (with proper identification of the degrees of freedom) which, at the linearized level, agrees with the aforegiven free higher-spin system [113]. The beauty of it is that we have reduced the quest for an interacting higher-spin theory in three dimensions to an algebraic problem: that of finding some Lie algebra satisfying the above requirements.

Two points now deserve a clear stating. The first is about simplicity and the second is about diversity, and we shall expand on them in the following. The 'simplicity' aspect is that something as common and easy to deal with as sl($n$|$\mathbb{R}$) fits into the above scheme. The 'diversity' aspect is that many other Lie algebras satisfy the requirements. We shall now

proceed to expanding on those two points. However, let us already point out that we shall primarily be interested in infinite-dimensional algebras, so that the following part on finite-dimensional algebras is included for the sake of generality and so that one can compare it to the treatment of infinite-dimensional ones, addressed afterward. Moreover, in Chapter 2 we are interested in supersymmetric Lie algebras (of infinite dimension), and in Subsection 1.2.3 we use the construction explained hereafter for bosonic, infinite-dimensional algebras in order to construct our supersymmetric theories.

**Finite-Dimensional Algebras**

To the reader unfamiliar with the subject it might now come as a (good) surprise that, as we just said, something as 'simple' as sl($n$) fits in this scheme [2]. Its usual presentation is the set of $n \times n$ traceless matrices (which is really the $n$-dimensional representation of it), but there exists another presentation. Indeed, consider the $n$-dimensional representation of sl($2|\mathbb{R}$) and define the higher-spin generators to be the symmetrized products of the corresponding number of spin-2 generators (in their $n$-dimensional representation) minus the corresponding trace projections. One can then prove that the resulting algebra is in fact sl($n$), where $n-1$ is the maximum number of spin-2 generators we allow ourselves to take products of. As an example we give the commutation relations of sl(3) in this way:

$$\begin{aligned}
[T_a, T_b] &= \epsilon_{abc} T^c, \\
[T_a, T_{bc}] &= \epsilon^m{}_{a(b} T_{c)m}, \\
[T_{ab}, T_{cd}] &= \sigma(\eta_{a(c}\epsilon_{d)bm} + \eta_{b(c}\epsilon_{d)am})T^m,
\end{aligned} \tag{1.77}$$

where the $T_a$'s are defined to be the sl($2|\mathbb{R}$) generators in their three-dimensional representation and the $T_{ab}$'s are defined as

$$T_{ab} \equiv T_{(a}T_{b)} - \tfrac{1}{3}\eta_{ab}T_c T_d \, \eta^{cd} = T_{ba}, \tag{1.78}$$

a definition implying not only that $\eta^{ab}T_{ab} = 0$ identically but also that the $T_{ab}$'s themselves are traceless matrices (as can be checked), so that we are indeed reproducing some Lie algebra of traceless matrices, as is sl(3).

Note the presence of the $\sigma$ parameter in the commutator of two spin-3 generators, which labels the real form which is chosen. In fact, its absolute value can be changed by rescaling the generators, but its sign cannot; $\sigma < 0$ corresponds to sl($3|\mathbb{R}$) while $\sigma > 0$ corresponds to su(1, 2), the other non-compact real form of sl(3). As we have already pointed out, the last commutator hereabove does not affect the linearized limit, except for the relative sign of the spin-2 and spin-3 kinetic terms, with $\sigma < 0$ yielding a non-standard minus sign. Apart from those considerations (see below), any real form is thus a priori acceptable. Also, as can be checked, the bilinear form (1.76) is in this case non-degenerate, because sl(3) is simple.

The above scheme of things actually extends to the arbitrary-$n$ case of sl($n$), of which any non-compact real form is suited (a priori) to describe an interacting theory of higher spins up to spin $n$. The most used form, however, is sl($n|\mathbb{R}$). The reason for this is partly that it is simple to handle, and partly that for other real forms some of the kinetic terms for different higher-spins would have opposite relative signs [102].

**Remark** : of course since no on-shell degrees of freedom are propagated by our three-dimensional action one might wonder how relevant is the requirement that different higher-spin kinetic terms have the same relative sign (which is usually required to preserve unitarity). However, other pathological features may be seen to show up when using those different real forms, such as non-unitarity of the associated boundary theory [114].

What about other Lie algebras ? As is well known, any non-compact simple algebra contains sl($2|\mathbb{R}$) as a subalgebra and, moreover, all semi-simple Lie algebras admit non-compact real forms. However, one might still wonder about the spectrum, that is, the requirement of containing, besides sl($2|\mathbb{R}$), higher-spin generators forming irreducible representations of the three-dimensional Lorentz group. Actually, this is *also* guaranteed ! The argument is the following: consider any Lie algebra containing sl($2|\mathbb{R}$) as well as other generators, that we collectively denote $T_A$. Assuming that our algebra is of finite dimension, the generators $T_A$ form a direct sum of finite-dimensional representations of sl($2|\mathbb{R}$). The reason for it is the following: all of the $T_A$'s, taken together, certainly form some (finite-dimensional) representation of sl($2|\mathbb{R}$) (which can be seen by considering the matrices corresponding to the sl($2|\mathbb{R}$)-generators in the adjoint representation). Then, either this representation is irreducible, in which case we are done, or it is not, in which case it will split in some direct sum of irreducible representations (because of Weyl's theorem stating that any finite-dimensional representation of a semi-simple Lie algebra is completely reducible [102]).

The outcome of this analysis is thus that *any* non-compact form of *any* simple Lie algebra beyond sl($2|\mathbb{R}$) is suited to describe some higher-spin theory via the Chern–Simons picture. Of course, and this is an important precision, some of them might actually contain higher-spin generators for only *some* spins beyond spin 2, that is, the spectrum might not be that of one irreducible representation of every spin up to some value. Furthermore, the spectrum might even contain spins below spin-2, and moreover the kinetic terms may in general enter the action with some relative signs (see below).

As a final comment before moving on to the infinite-dimensional higher-spin algebras we shall point out that, given some non-compact algebra, in general one may declare different sets of (three) generators to be the sl(2|$\mathbb{R}$) subalgebra describing pure gravity. Making such a choice is called choosing some 'embedding' of sl(2|$\mathbb{R}$) into the higher-spin algebra. Among all possible embeddings, there is a special one, called 'principal embedding', which has the property that all the other generators split into irreducible representations with multiplicity one.[10] Differently put, it means that the rest of the generators should organize as (1.75), once for each spin present in the spectrum. In the case of sl(3), for example, there is only one non-principal embedding, corresponding to the splitting of sl(3) as $\mathbf{8} = \mathbf{3} \oplus 2 \times \mathbf{2} \oplus \mathbf{1}$, whereas the principal embedding that we have presented in (1.77) corresponds to the splitting $\mathbf{8} = \mathbf{3} \oplus \mathbf{5}$ (the representations are denoted in boldface by their dimensions). Let us note that non-principally embedded sl(2|$\mathbb{R}$)'s have also been studied and seem somewhat more difficult to analyze. In particular, the properties of the corresponding boundary theory seem to present some subtleties — see e.g. [114, 116]. In the sequel, when we study infinite-dimensional higher-spin algebras, the gravitational sl(2|$\mathbb{R}$) subalgebra shall be principally embedded therein.

Note that all sl($n$) Lie algebras admit a non-compact form such that sl(2|$\mathbb{R}$) can be principally embedded thereof. Actually, some embeddings of sl(2|$\mathbb{R}$) may not be compatible with some non-compact real forms of whatever higher-spin algebra we use. For example, we point out that for the case of sl($n$) the principal embedding thereof is only compatible with the maximally non-compact real form, sl($n|\mathbb{R}$), as well as with su($\frac{n}{2}, \frac{n}{2}$) (or su($\frac{n-1}{2}, \frac{n+1}{2}$) if $n$ is odd). Let us also mention that the maximally non-compact real form is compatible with any embedding and, conversely, the so-called 'normal' embedding is compatible with any real form. Last of all we also point out that one switches non-compact real forms for the principal embedding by multiplying all odd-spin generators by a factor of $i$, and for more information on such algebraic aspects we refer to [102].

---

[10] An equivalent definition [115] is that the number of irreducible representations appearing in the spectrum is smaller than the rank of the algebra (which is $n$ for sl($n$)).

**Infinite-Dimensional Algebras**

In the previous paragraphs we have been concerned with finding some completion to the commutation relations of sl(2|$\mathbb{R}$) together with (1.75). However, explicitly or implicitly, so far we have confined ourselves to exploring finite-dimensional Lie algebras. In the following we address the question of infinite-dimensional higher-spin algebras, which are the type of algebras underlying the theories we study in Chapter 2. For pedagogical purposes we stick to bosonic algebras, and supersymmetric extensions thereof, which are our actual interest, are introduced in Subsection 1.2.3.

The idea is that, along the lines of the construction of the sl(3) higher-spin generators in terms of products of spin-2 ones (see above), we may very well consider the same construction without limiting the degree of the products thereof. In such a way one would generate an infinite tower of higher-spin generators in representations of sl(2|$\mathbb{R}$). Such a construction of an infinite-dimensional (associative) algebra is actually rather standard and bears the fancy name of *universal enveloping algebra*, and it is denoted by $\mathcal{U}(\text{sl}(2|\mathbb{R}))$. Moreover, the universal enveloping algebra is some abstract construction [117] in which we build the higher-spin generators as products of the original ones for some abstract associative product, without considering the latter to be in some representation.[11] This is why, before obtaining our infinite-dimensional higher-spin algebra out of such a construction, there is one last step we need to perform; namely, quotienting by some value of the sl(2|$\mathbb{R}$)-Casimir $C_2 \equiv T_a T_b \eta^{ab}$. The Lie algebra we are looking at is thus

$$\text{B}[\lambda] \equiv \text{hs}[\lambda] \oplus \mathbb{I} \equiv \frac{\mathcal{U}(\text{sl}(2|\mathbb{R}))}{\langle C_2 - \frac{1}{4}(\lambda^2 - 1)\mathbb{I}\rangle}, \qquad (1.79)$$

where hs[$\lambda$] is the standard infinite-dimensional higher-spin algebra in three dimensions, first introduced in [118] and then firstly explored by the authors of [119–124]. Note that in the above expression we have also removed the identity, which is strictly speaking included in the universal enveloping construction, but which forms an ideal we are not interested in. Also note that, at this point, as defined by the above equation hs[$\lambda$] is only an associative algebra, because so is $\mathcal{U}(\text{sl}(2|\mathbb{R}))$. However, in the sequel we shall equip it with the natural bracket (the antisymmetrization of the associative product) so to make it a Lie algebra, and we shall keep the same notation hs[$\lambda$], which indeed usually denotes the Lie structure.

Quotienting as in the above relation is precisely the equivalent of considering the original sl(2|$\mathbb{R}$) generators to be in some representation,

---

[11] The universal enveloping techniques can be applied to any Lie algebra [117].

in which $C_2$ thus has *some* value $\lambda$,[12] and then taking products thereof. The parameter $\lambda$ is usually referred to as the *deformation parameter*, for reasons that shall become clear in the following. Let us further note that, in a theory of infinitely many scalar-coupled higher-spin fields in dimension three, the deformation parameter is related to the mass of the scalar [125, 126].

Differently put, we also need to match the desired spectrum, namely that of the correct number of generators $(2s+1)$ at each spin-level $s$, hence the need for quotienting. Indeed, if one does not quotient, the algebra actually contains an infinite number of spin-$s$ generators for a given $s$. That is because, if one does not identify $C_2$ with some value, then the trace of a spin-$s$ generator will be something transforming as a spin-$(s-2)$ generator but independent of those built by taking products of $s-3$ spin-2 ones (and one can take further traces). By quotienting one precisely relates those two kinds of objects, and a non-degenerate spectrum is thus obtained. Yet another way to understand the need for quotienting is to note that, otherwise, the Casimir would generate an ideal and the scalar product thereon (to be defined below) would then be degenerate. Let us also refer to [127] for a treatment of the universal enveloping construction in $\text{AdS}_D$.

As is guaranteed by the universal enveloping technique, sl(2|$\mathbb{R}$) is a subalgebra of hs[$\lambda$], just as in the case of sl(3) described above. Moreover, the algebra hs[$\lambda$] contains higher-spin generators in irreducible representations of sl(2|$\mathbb{R}$): thanks to the quotienting by some value of the Casimir, there are $2(s-1)+1$ spin-$s$ generators for each $s = 2, 3, \ldots$ — the sl(2|$\mathbb{R}$) generators being understood as having spin 2, as before. In this way we thus manage to build suitable higher-spin algebras (up to the existence of appropriate bilinear forms thereon), and one might even think about universally enveloping other Lie algebras containing sl(2|$\mathbb{R}$), such as sl(3|$\mathbb{R}$). This would potentially yield higher-spin theories with different spectra. However, this approach has been largely ignored in the literature, with [128] among the exceptions (see also [127]). Higher-spin theories based on hs[$\lambda$] are the most commonly studied among the infinite-dimensional ones.

**Remark** : it is of course wrong that any Lie structure comes from some associative one, and therefore on top of the aforementioned generalizations one could formally wonder about higher-spin Lie algebras whose Lie bracket is not the commutator of some associative product. Although in dimension four and greater it has been shown that such situations cannot arise [129], in dimension three, where at any rate one seems to have much more freedom, no such result has been obtained.

---

[12] The $\lambda$ parameter is also sometimes denoted by $\mu$.

Besides infinite-dimensional subalgebras, one may of course wonder about the relation between hs[$\lambda$] and its finite-dimensional cousins. However, it is an important point that sl($n$) is *not* a subalgebra thereof for $n \geq 3$. In fact, even for general $\lambda$, there is no finite-dimensional subalgebra of hs[$\lambda$] apart from sl(2|$\mathbb{R}$) [122]. However, let us point out that for integer values $\lambda = N$, hs[$\lambda$] does decompose into the sum of sl($n$|$\mathbb{R}$) and an infinite-dimensional ideal one can then quotient by [122]. As we shall be working at $\lambda = \frac{1}{2}$ we do not dwell on that point any longer, and shall simply point out that such is the reason why hs[$\lambda$] can be sometimes thought of as the 'analytic continuation' of sl($n$|$\mathbb{R}$). Note that there are infinite-dimensional subalgebras, such as the well-known one consisting of only the even-spin generators (odd powers of the spin-2 ones), that one can restrict oneself to in a consistent fashion.

Let us now turn to describing a way of realizing hs[$\lambda$]. Indeed, our higher-spin algebra hs[$\lambda$] is compactly defined by the above universal enveloping expression, but the latter does not grant one with any convenient way of handling it. Of course, one can always work out the commutation relations among higher-spin generators from the above definition (along the lines of the sl(3|$\mathbb{R}$) case), but such an approach is far from handy. A far more convenient mean of treating our algebra is the so-called *oscillator realization*, which can be thought of as a refinement of the universal enveloping procedure (in the sense that it automatically imposes non-trivial conditions on the spectrum), and we now introduce it. The starting point is to notice that the spin-2 sector, sl(2|$\mathbb{R}$), can be realized in the following way. Consider a pair of commuting 'oscillators' $q_\alpha$, with $\alpha = 1, 2$, satisfying the following relation:

$$[q_\alpha, q_\beta]_\star \equiv 2i\epsilon_{\alpha\beta}, \qquad (1.80)$$

where $\epsilon_{\alpha\beta} = -\epsilon_{\beta\alpha}$ is the two-dimensional $\epsilon$-symbol with conventions $\epsilon_{12} \equiv 1 = \epsilon^{12}$, with which we raise and lower the spinor indices of the $q_\alpha$'s. The $\star$-symbol denotes the associative product which the above Lie bracket corresponds to, and one can also formulate the above definition in terms of that product:

$$q_\alpha \star q_\beta \equiv q_\alpha q_\beta + i\epsilon_{\alpha\beta}, \qquad (1.81)$$

which is called the $\star$-product [130, 131]. One then defines the quadratic combinations to be

$$T_{\alpha\beta} \equiv -\tfrac{i}{4} q_{(\alpha} \star q_{\beta)} = -\tfrac{i}{4} q_\alpha q_\beta, \qquad (1.82)$$

where only three of them are independent, because $T_{12} = T_{21}$. The key observation is now that the Lie algebra of the quadratic generators $T_{\alpha\beta}$ is precisely sl(2|$\mathbb{R}$). Indeed, with the above definitions one easily checks that

$$[T_{\alpha\beta}, T_{\mu\nu}]_\star = -\tfrac{1}{8}(\epsilon_{\nu\beta}T_{\mu\alpha} + \epsilon_{\mu\beta}T_{\nu\alpha} + \epsilon_{\nu\alpha}T_{\mu\beta} + \epsilon_{\mu\alpha}T_{\nu\beta}), \qquad (1.83)$$

which can be seen to reproduce the familiar sl(2|$\mathbb{R}$) commutation relations upon performing redefinitions (see Appendix B).

Evidently, such a realization does not make the handling of sl(2|$\mathbb{R}$) any simpler — quite the contrary — but it allows us to generalize it in the following, natural way. Let us no longer restrict our attention to quadratic combinations of our oscillators and allow instead for generators of arbitrary degree (higher than two) in the $q$'s. Having in mind the universal enveloping construction, we shall nonetheless restrict the degree of the generators to be even (and the identity component is not included). In this way one generates an infinite dimensional Lie algebra, and one can check that it corresponds to hs[$\lambda$] for some value of $\lambda$. Indeed, the higher-spin generators defined as symmetric products of the oscillators (of even degree) do correspond to taking symmetrized products of our spin-2 generators, so that the construction is really a reformulation of the universal enveloping technique. Moreover one can verify that the above higher-spin generators do form irreducible representations of the spin-2 sector. However, one may wonder where is the deformation parameter, $\lambda$, in such an oscillator construction. In fact, in the latter realization the quotient is automatically taken! The reason why this can happen is because we have specified more than the product (or commutation) relations among the $T_{\alpha\beta}$'s: we also know about how the $q_\alpha$ oscillators themselves commute to each other. To be fully convinced we should compute the Casimir $C_2$ in this formulation, the comparison of which with the universal enveloping construction tells us that the above oscillator realization corresponds to $\lambda = \tfrac{1}{2}$. Such a value of the deformation parameter is called the *undeformed case*, for reasons that shall me made clear hereafter. Thus, we can realize our higher-spin algebra at $\lambda = \tfrac{1}{2}$ as the algebra of linear combinations of our generators $T_{\alpha_1...\alpha_s}$ of even degree $s$ in the $q_\alpha$'s under the Lie bracket derived from (1.80).

Although we shall be primary interested in the so-called undeformed (supersymmetric) case, at this point we owe it to the reader to answer the following question: what about hs[$\lambda$] at $\lambda \neq \tfrac{1}{2}$? The answer is positive: there is a way to *deform* the oscillator relations (1.80) in such a way as to generate, upon considering generators of arbitrary (even) degree, hs[$\lambda$] at generic $\lambda$ [122]. The so-called *deformed* oscillator relations read

$$[q_\alpha, q_\beta]_\star \equiv 2i(1 + (2\lambda - 1)K)\epsilon_{\alpha\beta}, \tag{1.84}$$

where $K \equiv (-)^{N_q}$ is the so-called Klein operator and $N_q$ counts the number of $q$ oscillators to its right, so that $K$ behaves as $(-)^{N_q} q_\alpha = -q_\alpha (-)^{N_q}$. One can check that the quadratic sector still reproduces $sl(2|\mathbb{R})$ and is independent of $\lambda$, but the higher commutation relations will of course depend on the deformation parameter. One might ask, however, whether the hs[$\lambda$] algebras at different values of $\lambda$ are really different algebras or whether they are isomorphic, and it was shown that they differ [119, 120]. Evidently, at $\lambda = \frac{1}{2}$ one recovers the undeformed commutation relations of (1.80).

We have thus managed to realize a set of generators of the hs[$\lambda$] algebra as powers of our deformed oscillators $q_\alpha$ (we keep the same notation). However, it might be felt that the above approach is not making the handling of hs[$\lambda$] particularly simple. Indeed, the procedure to compute the $\star$-product (or $\star$-commutator) of two higher-spin generators involves successively making use of the formula (1.81) and identifying the produced generators. The structure of the $\star$-product of two higher-spin generators is clear but the details have to be worked out in quite a painful manner. As it turns out, there is a simpler way to deal with the oscillators, but which only works for the undeformed case: instead of taking our generators to be defined as symmetric $\star$-products of the oscillators we define them as simple products of the $q_\alpha$'s, and we further define the $\star$-product of any two polynomials $f$ and $g$ in the $q_\alpha$'s to be

$$(f \star g)(q'') \equiv \exp\left(i\,\epsilon_{\alpha\beta}\frac{\partial}{\partial q_\alpha}\frac{\partial}{\partial q'_\beta}\right) f(q)g(q')\big|_{q=q'=q''}, \tag{1.85}$$

where $f(q) \equiv f(q_1, q_2)$ and so on. In this way we have 'solved' for the $\star$-product, and one can check the above formula to imply the relations (1.80). Moreover, we define the Lie bracket defining our Lie algebra of $q$-polynomials to be

$$[f, g]_\star \equiv \frac{1}{2i}\left(f \star g - g \star f\right), \tag{1.86}$$

where the prefactor is a matter of conventions. It might seem as if the definition of the $\star$-product is now more complicated (it involves the exponential of differential operators), but it is really simpler, in the sense that we now have an explicit expression for it. Moreover, the above product law is in fact the so-called Moyal bracket, more familiar in the context of Quantum Mechanics. Again, one checks that the relations (1.80) and (1.81) are reproduced, so that the algebra is indeed the same. We further point out that in the following we shall often switch to the basis

$$X_{(p,q)} \equiv X_{\underbrace{1\ldots 1}_{p}\underbrace{2\ldots 2}_{q}} \equiv \frac{1}{p!q!}(q_1)^p(q_2)^q, \qquad \tfrac{p+q}{2} \in \mathbb{N}_0. \tag{1.87}$$

Our algebra is thus simply the space of polynomials of even degree (but zero) in the oscillators $q_\alpha$, equipped with the above Lie bracket. A downside of the latter standpoint is that it cannot be generalized to arbitrary $\lambda$, namely, an explicit expression such as (1.85) cannot be obtained in the deformed case.[13] To the best of our knowledge, this curious fact still lacks a deeper justification, if there is any. Anyhow, as we shall be primarily interested in the undeformed case we are much content with the above formulation.

We are almost ready to move to the next and last subsection of this chapter, where we address the supersymmetrization of this setup, but first let us discuss the reality conditions and bilinear forms on the above realization. First we discuss the reality conditions: so far the aforedefined algebras are complex, namely, the coefficients of the polynomials in the $q_\alpha$ variables are arbitrary in $\mathbb{C}$. Actually, since one starts with sl$(2|\mathbb{R})$ and then constructs its universal enveloping algebra, our hs$(1, 1)$ already 'comes in some real form', and it is actually the maximally non-compact one — the analogue of sl$(n|\mathbb{R})$. As it turns out, starting from there one may multiply all odd-spin generators (even powers of the spin-2 ones) by a factor of $i$ and obtain another real form, this time the analogue of su$(\frac{n}{2}, \frac{n}{2})$. Note, however, that it is the maximally non-compact real form which is usually referred to when speaking of hs$(1, 1)$, as it is the (only) one compatible with the universal enveloping technique if one assumes real coefficients.[14] Furthermore, one can actually prove that there are no other real forms thereof [122]. Let us also make clear that sl$(2|\mathbb{R})$ is, by construction, 'principally embedded' in hs$(1, 1)$.[15] The analogy with the finite-dimensional case is therefore complete (see above). Also note that in the oscillator realization presented above the maximally non-compact real form is obtained by restricting to linear combinations of the $X_{(p,q)}$ generators of (1.87) with real coefficients. The reason why this clashes with (1.82), where the sl(2|$\mathbb{R}$) generators have been defined with an '$i$' in front, is because of our normalization convention (1.86), which includes $i$.

---

[13] There is a way to achieve this but one needs use a different set of oscillators [128].
[14] Indeed, taking products of sl$(2, \mathbb{R})$ generators and considering linear combinations with real coefficients thereof does not leave any freedom and in such a way one always produces the maximally non-compact one.
[15] Although, as should be noted, there is no clear notion of principal embedding for infinite-dimensional algebras.

Last but not least we display the bilinear, symmetric, invariant and non-degenerate bilinear form one can define on hs(1, 1), namely

$$\text{STr}(f) \equiv 2f(0), \tag{1.88}$$

where the factor 2 is again a matter of convention. As can be checked, this definition coincides with the bilinear form of sl(2|$\mathbb{R}$) in the spin-2 subsector. More details about hs(1, 1) can be found in Appendix B.

### 1.2.3 Higher-Spin Supergravity

As aforementioned, in the following we are interested in infinite-dimensional supersymmetric models, and we thus look for supersymmetric extensions of hs(1, 1). For the sake of generality we might want to wonder about supersymmetric versions of hs[$\lambda$] but, however, as it turns out the deformed algebra hs[$\lambda$] at $\lambda \neq \frac{1}{2}$ cannot be supersymmetrized in any obvious way beyond $\mathcal{N} = 2$. Differently put, one may supersymmetrize the non-deformed version, hs(1, 1), thereby obtaining some infinite dimensional higher-spin superalgebra, but the latter does not admit a simple deformation anymore, except when the number of supersymmetries $\mathcal{N}$ is one or two. Indeed, one may realize osp($N$, 2|$\mathbb{R}$) in terms of oscillators, along the lines of the above realization of sl(2|$\mathbb{R}$): on top of the $q_\alpha$ oscillators, which are Grassmann even, we add the Grassmann-odd oscillators $\psi_i$ with $i = 1, \ldots, N$. The superalgebra osp($N$, 2|$\mathbb{R}$) is then obtained by considering the polynomials of total degree 2 in all the oscillators (and imposing suitable reality conditions on the coefficients) and redefining the $\star$-product as

$$(f \star g)(z'') \equiv \exp\left(i\,\epsilon_{\alpha\beta}\frac{\partial}{\partial q_\alpha}\frac{\partial}{\partial q'_\beta} + \delta_{ij}\frac{\overleftarrow{\partial}}{\partial \psi_i}\frac{\overrightarrow{\partial}}{\partial \psi'_j}\right)f(z)g(z')\big|_{z=z'=z''}. \tag{1.89}$$

In Appendix B we show that such a construction indeed reproduces the familiar osp($N$, 2|$\mathbb{R}$) commutation relations, and if one considers polynomials of all (even) degrees but zero with some reality conditions it yields the supersymmetric shs($N$, 2|$\mathbb{R}$) higher-spin algebra — the $N$-extended supersymmetric version of hs(1, 1). The obstruction regarding deformation now stems from the fact that, if one deforms the oscillators along the lines of (1.84), the osp($N$, 2|$\mathbb{R}$) commutation relations are no longer reproduced for $N \geq 3$, unlike what happens for sl(2|$\mathbb{R}$) (for $N = 1, 2$ the internal

algebra of $\text{osp}(N, 2|\mathbb{R})$ is abelian and the deformed oscillator realization still holds good).

One might think about forgetting the oscillator realization and simply considering the universal enveloping algebra of $\text{osp}(N, 2|\mathbb{R})$ further quotiented by some ideals of the latter. Indeed, there is no a priori obstruction to doing this, even for $N \geq 3$ and at arbitrary value of $\lambda$. However, as $N$ increases the superalgebra $\text{osp}(N, 2|\mathbb{R})$ possesses more than one Casimir, and other complications also arise [127]. In fact, a definition akin to (1.79) in the case of $\text{osp}(N, 2|\mathbb{R})$ would yield $\text{shs}(N, 2|\mathbb{R})$ (defined in terms of its oscillator realization) for $N = 1$ only, with $C_2$ replaced by the $\text{osp}(1, 2|\mathbb{R})$ quadratic Casimir [6]. Moreover, the latter statement holds for any value of the deformation parameter, so that even the undeformed versions $\text{shs}(N, 2|\mathbb{R})$, which exist at least in terms of the above oscillator realization, are difficult to reproduce from the universal enveloping standpoint.

**Remark** : there is an intuitive way of understanding the difficulty in bringing together the deformation and the supersymmetrization, which is the following. Let us consider Prokushkin–Vasiliev Theory, which couples the infinite tower of higher-spin gauge fields associated with $\text{hs}[\lambda]$ to a scalar field whose squared mass is $m^2 = -\frac{1}{4}(\lambda^2 - \frac{1}{4})$. In the deformed case $\lambda \neq \frac{1}{2}$ the scalar field is massive, and in a supersymmetric theory with more than two supersymmetries this would imply the presence of *massive* fields of spin $\frac{3}{2}$ or higher, much in contrast with the *gauge* theory Prokushkin–Vasiliev is supposed to be [125, 126].

As we shall be interested in the generic case of extended supersymmetric models with $\mathcal{N} \geq 1$, we shall simplify our task by dropping the deformation parameter, and consider the $\text{shs}(N, 2|\mathbb{R})$ superalgebras generated by the above oscillators. In full analogy with $\text{hs}(1, 1)$, they are realized as the space of polynomials of total even degree (but zero) in the oscillators $q_\alpha$ and $\psi_i$, equipped with the above $\star$-product, and with appropriate reality conditions imposed on the coefficients. More details about $\text{shs}(N, 2|\mathbb{R})$ are given in Appendix B, and we shall only point out here that the bilinear form is defined in exactly the same way as for $\text{hs}(1, 1)$ (see previous subsection).

In fact, in Chapter 2, when we compute asymptotic symmetries of these models, the case we shall explicitly treat is the non-extended supersymmetric case, that is, $\text{shs}(1, 2|\mathbb{R}) \equiv \text{shs}(1, 1)$ (the corresponding results are only sketched for the extended version and the details of the corresponding computations are relegated to Appendix C). As it turns out, the non-extended supersymmetric algebra $\text{shs}(1, 1)$ admits an alternative realization, which forgoes the Grassmann-odd oscillators and deals with the bosonic ones only, the $q_\alpha$'s. It will be the final word of this chapter to present such an alternative formulation, which is the one we make use of in Chapter 2.

The alternative, simpler realization of $shs(1,1)$ is achieved by considering the same construction as for $hs(1,1)$ but relaxing the condition that the degree of the polynomials should be even. We now allow for all degrees to appear in the spectrum, and work with the basis

$$X_{(p,q)} \equiv X_{\underbrace{1\ldots1}_{p}\underbrace{2\ldots2}_{q}} \equiv \frac{1}{p!q!}(q_1)^p(q_2)^q, \qquad p+q \in \mathbb{N}_0, \qquad (1.90)$$

where $q_1, q_2$ are the two commuting spinor variables used above. Let us stress again that, as the above definition states, we do not consider the (unique) generator with no indices (zero degree polynomials). The product is the one displayed in (1.85), and the commutator of (1.86) now becomes a graded bracket:

$$[f,g\}_\star \equiv \frac{1}{2i}\left(f \star g - (-)^{\pi_f \pi_g} g \star f\right), \qquad (1.91)$$

where $\pi_f$ is the parity of $f$ (assumed to have definite degree in the oscillators), defined to be 0 when $f$ is of even degree and 1 otherwise. The reality conditions on the even-degree polynomials are the same as in the $hs(1,1)$ case, which is again expressed as taking linear combination of the above basis with real coefficients. Let us note that, in the sequel, we shall often take our coefficients to satisfy Grassmann-parity conditions (in the 'physical' Grassmann algebra) according to their parity, and we shall thus use the standard Lie bracket 1.86, which of course boils down to the above superbracket when the Grassmann coefficients are pulled out. More information about these algebras and, in particular, about reality and parity conditions is given in Appendix B.3.

The generators (basis elements) with $n = p+q$ indices above are said to carry spin $n/2 + 1$ (conformal spin $n/2$). The generators with two and one indices are thus carrying spin 2 and $3/2$ (conformal spin 1 and $1/2$), which is indeed consistent with the fact that upon truncation to $p+q \leq 2$, $shs(1,2|\mathbb{R})$ is seen to reduce to $osp(1,2|\mathbb{R})$ (see below), which we know is (half of) the superalgebra describing type-$(1,1)$ three-dimensional Supergravity, which in turn is known to contain the graviton and the gravitino: fields of spin 2 and $3/2$ respectively (conformal spin 1 and $1/2$).

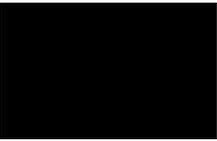

CHAPTER 2

# Asymptotic Symmetries

In the previous chapter we have introduced Higher-Spin Theory in dimension 3 or, rather, higher-spin theories in dimension 3. As these theories are all topological, a natural way of studying those living on AdS spacetimes is via the holographic principle, and a first step one should take therefore is the computation of the asymptotic symmetries of these models, which is what this chapter is devoted to. Let us argue that, because the full holographic correspondence between higher-spin theories in $AdS_3$ and two-dimensional conformal models is not only quite involved but also currently an ongoing topic of investigation [6], in the present manuscript we have chosen to focus on the asymptotic symmetries and shall touch upon the full holographic setup towards the end of the chapter.

We thus start by reviewing the asymptotic symmetry algebra of pure Gravity in $AdS_3$ and how one can obtain it in the Chern–Simons gauge picture, arriving at the famous Virasoro algebra with Brown–Henneaux central charge at the end of Section 2.1. With such a reminder of the spin-2 case in mind, in Section 2.2 we then move on to studying the asymptotic symmetry superalgebra of our favorite higher-spin model: that which is based on shs(1, 1), thus unveiling the supersymmetric $\mathcal{W}_\infty$-algebra which is found to govern the asymptotic dynamics of our theory. The latter case is treated quite explicitly and also encompasses in some sense the undeformed bosonic case. The extended corresponding asymptotic results are commented on at the end of the chapter, and a discussion of our results as well as comments on their relation to other topics are found in Chapter 3.

## 2.1 Asymptotics for the Spin-2 Connection

Our manifold $\mathcal{M}_3$ on which we integrate is assumed to have topology $\mathbb{R} \times \mathcal{D}_2$, where $\mathbb{R}$ is parametrized by the time coordinate $t \equiv x^0 \equiv x^t$ and $\mathcal{D}_2$ is our two-dimensional spatial manifold parametrized by $\theta \equiv x^1 \equiv x^\theta$ and $r \equiv x^2 \equiv x^r$, which we assume to have at least one boundary we focus the following asymptotic analysis on and which we call 'asymptotic infinity' or more loosely 'infinity'. This boundary will be assumed to correspond to $r \to \infty$.

We now describe the asymptotic form (boundary conditions) of our sl(2|$\mathbb{R}$)-connection $A_\mu$. We first start by recalling the Brown–Henneaux boundary conditions [67] in the metric formalism, and after commenting on how they were originally obtained we derive their translation in terms of the Chern–Simons gauge potential. We then proceed with the computation of the asymptotic symmetries in that picture and obtain the centrally-extended Virasoro algebra, on which we comment at the end of the section.

### 2.1.1 Boundary Conditions

The first step in the computation of asymptotic symmetries for any theory is the determination of the asymptotic behavior which we decide to impose on our fields. This is, in fact, far from being evident in general and we thus find it useful to recall below how the asymptotic conditions were first obtained in the metric formalism for Gravity, and shall then translate them to our Chern–Simons description, with which we work for the rest of this part.

**Metric Formalism**

As aforementioned, natural and consistent boundary conditions for three-dimensional pure gravity with negative cosmological constant were first given by J. D. Brown and M. Henneaux in [67] in the metric formalism, for which they read

$$\begin{aligned}
g_{tt} &= -\tfrac{r^2}{\ell^2} + \mathcal{O}(1), \\
g_{rr} &= \tfrac{\ell^2}{r^2} + \mathcal{O}(1/r^4), \\
g_{\phi\phi} &= r^2 + \mathcal{O}(1), \\
g_{tr} &= \mathcal{O}(1/r^3), \\
g_{t\phi} &= \mathcal{O}(1), \\
g_{r\phi} &= \mathcal{O}(1/r^3),
\end{aligned} \qquad (2.1)$$

where the $\mathcal{O}$ functions are allowed to depend on $t$ and $\theta$, and for the pure-AdS$_3$ case they take a specific form, which one can infer from Appendix A.1 where the AdS metric is given in the same coordinates. Note that the $\mathcal{O}(1/r^3)$ terms in the above asymptotics can be 'gauged away' so to yield simpler boundary conditions,[1] which are the ones we will use in the sequel. Now, specifying boundary conditions is always a delicate procedure, since these are not uniquely dictated by the theory one considers.[2] Therefore, we find it instructive to briefly recall here how the above boundary conditions were obtained.

The first thing to realize is that there is no obvious definition of a spacetime being 'asymptotically anti-de Sitter', nor is it obvious what we should expect or require from such a geometry [132]. A legitimate thing to wish for is that the asymptotic symmetry algebra contains the symmetry algebra of AdS, that is, so$(2,2) \simeq$ sl$(2) \oplus$ sl$(2)$, which amounts to say that acting with so$(2,2)$ on an 'asymptotically anti-de Sitter' spacetime should yield an 'asymptotically anti-de Sitter' spacetime (whatever the definition). Another legitimate requirement would be that global AdS be 'asymptotically anti-de Sitter', and a natural procedure for generating boundary conditions that could serve as a definition for being 'asymptotically anti-de Sitter' would then be to act on the asymptotic form of the global AdS metric with a general element of so$(2,2)$, which is precisely how the above boundary conditions were obtained.

Naively, one might expect them to yield so$(2,2)$ as the asymptotic symmetry algebra or, at least, that some refinement of this procedure would indeed produce it. However, as we will see, the algebra which arises then is not so$(2,2)$ but the conformal algebra in two dimensions (more precisely a central extension thereof), which admits so$(2,2)$ as a subalgebra, and it is also seen to be both difficult and unnatural to restrict the above boundary conditions so to shrink the asymptotic conformal algebra down to so$(2,2)$. The general conclusion is then that we should not try to define 'asymptotically something' spaces by the asymptotic symmetry algebra we expect it to have. Rather, we should try to generate boundary conditions in a way which implements some basic natural requirements and then compute the corresponding asymptotic algebra, allowing for surprising enlargements of the bulk symmetry algebra at asymptotic infinity. Let us also point out that a third basic requirement for the definition of 'asymptotically anti-de Sitter' geometries is that the fall-off conditions make the associated surface charges finite which, in fact, is the case for the above asymptotics. Finally, we also note that in dimension four such an enlargement of the bulk isometries at infinity does not occur [132].

---

[1] The 'gauge transformations' we are referring to here are the ones generated by the part of the Killing vectors with which there are no associated charges and whose generators in the canonical formalism vanish weakly, that is, the so-called 'proper' gauge transformations among the Killing symmetries [67].

[2] As M. Henneaux once put it, *boundary conditions is actually something of an art.*

Natural and consistent boundary conditions having been determined, we should now turn to computing the corresponding asymptotic symmetry algebra, i.e. the Poisson-bracket Lie algebra formed by the canonical generators corresponding to (infinitesimal) vector fields which preserve the form of the above boundary conditions under Lie transport. This is precisely what was carried out in [67]. The whole process, however, in addition to being rather long and complicated, is full with subtleties. Although the reading of the original paper is highly recommended to all interested readers, especially because of those subtleties it conveniently highlights, as announced we shall instead switch to the Chern–Simons formalism in which the derivation goes much more easily.

**Chern–Simons Translation**

In the gauge picture we work with the sl(2|$\mathbb{R}$)-connection $A_\mu$ of Section 1.1.2, so that the first thing we ought to do is translate the asymptotic form (2.1) into boundary conditions for $A$ (and $\tilde{A}$). Upon translating the above boundary conditions to the Chern–Simons formalism one finds, to leading order,[3]

$$A_+ = \mathcal{O}\big(\tfrac{1}{r}\big)\sigma^+ + r\sigma^-, \quad A_- = 0, \quad A_r = \big(\mathcal{O}\big(\tfrac{1}{r^3}\big) + \tfrac{1}{r}\big)\sigma^3, \qquad (2.2\text{a})$$

$$\tilde{A}_- = \mathcal{O}\big(\tfrac{1}{r}\big)\sigma^- + r\sigma^+, \quad \tilde{A}_+ = 0, \quad \tilde{A}_r = \big(\mathcal{O}\big(\tfrac{1}{r^3}\big) - \tfrac{1}{r}\big)\sigma^3, \qquad (2.2\text{b})$$

where we have changed coordinates for the spacetime, setting $x^\pm \equiv t \pm l\theta$, and we refer to Appendix B for details about the basis of sl(2|$\mathbb{R}$). Note that, because of Local Lorentz Invariance (see Subsection 1.1.1), the translation from the metric-like form to the above Chern–Simons expression is not unique and requires some guess work. One can easily check, however, that the above formulas reproduce the original Brown–Henneaux asymptotics (up to some terms which we can gauge away), which one does unambiguously.

---

[3] Note that the above fall-off conditions correspond the asymptotics (2.1) with the $\mathcal{O}(1/r^3)$ terms gauged away. We also note that the translation is not unique, because of the freedom to perform local Lorentz transformations (see Subsection 1.1.1).

Now, performing a gauge transformation conveniently allows us to eliminate the $r$ dependence from the above asymptotics (to leading order). From now on we mostly work with the first chiral copy only. The gauge generator is

$$\Omega \equiv \exp\left(\sigma^3 \ln(r)\right), \tag{2.3}$$

and acts via the global version of (1.42), that is,

$$A_i \to A'_i = \Omega \partial_i \Omega^{-1} + \Omega A_i \Omega^{-1}. \tag{2.4}$$

In order to perform that computation it is convenient to use the matrix representation of our generators for sl(2|$\mathbb{R}$), given in Appendix B, in which we have

$$\Omega = \begin{pmatrix} r^{\frac{1}{2}} & 0 \\ 0 & r^{-\frac{1}{2}} \end{pmatrix}, \tag{2.5}$$

and thus

$$A'_+ = \mathcal{O}(1)\sigma^+ + \sigma^-, \quad A'_- = 0, \quad A'_r = \mathcal{O}(\tfrac{1}{r^3})\sigma^3. \tag{2.6}$$

Therefore, the only asymptotically non-vanishing part of $A$ is given, in this gauge, by

$$A'_+|_{\text{LEADING ORDER}} \equiv (L\sigma^+ + \sigma^-) \equiv B, \tag{2.7}$$

where $L$ is the function forming the leading part of the $\mathcal{O}(1)$ term in $A'_+$, and which depends on $x^\pm$ only. Note that similar steps for the other chiral copy lead to the following non-vanishing piece

$$\tilde{A}'_-|_{\text{LEADING ORDER}} \equiv (\tilde{L}\sigma^- + \sigma^+) \equiv \tilde{B}. \tag{2.8}$$

### 2.1.2 Asymptotic Symmetry Algebra

Now that we have determined fall-off conditions for our Chern–Simons connection we proceed with the computation of asymptotic symmetries: first we determine the general form of the residual gauge parameter which preserves the asymptotics and then we extract the Poisson-bracket algebra therefrom. Again, we shall be mostly concerned with the first chiral copy, and the computation for the other one is really analogous. Let us also point out Reference [133], where the asymptotic symmetries of pure Gravity in the Chern–Simons formalism were first obtained.

**Residual Gauge Transformations at Asymptotic Infinity**

We now want to act with a general element[4] $\Lambda \in \text{sl}(2|\mathbb{R})$ on the above asymptotics for the connection $A_\mu$, require the transformed asymptotics to have the same form as the original ones and from that derive the conditions such a requirement yields on $\Lambda$ (if any). By 'the same form' as the original asymptotics we mean the transformed $B$ reads exactly as (2.7) except for the form of the functions multiplying the $\sigma^+$ generators, which is allowed to change (but we do not allow introducing $r$ dependence). Obviously, we also mean $A'_-$ and $A'_r$ remain zero (to leading order). Because the aforegiven asymptotics for $A_\mu$ have a very particular form, we expect this requirement of asymptotic gauge invariance of the boundary conditions to yield severe restrictions on the form of $\Lambda$.

The infinitesimal gauge transformation with parameter $\Lambda$ acts on $B$ by the adjoint action, i.e.

$$B \to B + \delta B \qquad (2.9)$$

with

$$\delta B = \partial_+ \Lambda + [B, \Lambda], \qquad (2.10)$$

where $\partial_+ \equiv \partial/\partial x^+$, and the same relation holds for $A'_-$ and $A'_r$. From the requirement that $\delta A'_-$ and $\delta A'_r$ be zero, taking into account that these are initially zero, the above relation states that $\Lambda$ cannot depend on either $x^-$ or $r$. Note that this already implies that no $r$ dependence will be introduced in $B$ by such a $\Lambda$ parameter, as required. With such a $\Lambda$ free of all $r$ (and $x^-$) dependence, our requirement of boundary conditions invariance amounts to ask for the coefficients of the $\sigma^-$ and $\sigma^3$ generators in $\delta B$ above to be zero. Note that we don't allow the appearance of the $\sigma^3$ generator, even though it is present in (2.7), because its coefficient in the later expression is just a fixed number and not a function.

Our goal is now to compute the expression (2.10) and derive the form that $\Lambda$ must have so that no $\sigma^-$ and $\sigma^3$ generators appear in it. As always, such requirements of asymptotic invariance of the boundary conditions will determine some of the components of $\Lambda$ in terms of the other, left-arbitrary ones. Let us denote the components in the following way:

$$\Lambda \equiv \Lambda^+ \sigma^+ + \Lambda^- \sigma^- + \Lambda^3 \sigma^3. \qquad (2.11)$$

First we note that

$$\delta B = (\partial_+ \Lambda^+ - L\Lambda^3)\sigma^+ + (\partial_+ \Lambda^- + \Lambda^3)\sigma^- + (\partial_+ \Lambda^3 + 2L\Lambda^- - 2\Lambda^+)\sigma^3, \qquad (2.12)$$

---

[4] Note that this $\Lambda$ has nothing to do whatsoever with the cosmological constant.

from which we deduce

$$\Lambda^+ = -\tfrac{1}{2}\partial_+^2 \Lambda^- + L\Lambda^-, \tag{2.13}$$

$$\Lambda^3 = -\partial_+ \Lambda^-. \tag{2.14}$$

Therefore, the conditions on $\Lambda \in \mathrm{sl}(2|\mathbb{R})$ translating the requirement that the form of the given boundary conditions (2.7) be preserved by infinitesimal gauge transformations with parameter $\Lambda$ (asymptotic invariance under residual gauge transformations) can be understood as leaving $\Lambda^-$ arbitrary, the other components being functions of the later (and of $B$). The asymptotic symmetries, i.e. the searched-for residual gauge transformations preserving the aforegiven boundary conditions are therefore spanned by $\Lambda$ elements of the form

$$\Lambda \equiv \left(-\tfrac{1}{2}\partial_+^2 \Lambda^- + L\Lambda^-\right)\sigma^+ + \Lambda^- \sigma^- - \partial_+ \Lambda^- \sigma^3. \tag{2.15}$$

**Generators of Asymptotic Symmetries**

We now want to compute the commutation relations for the algebra of asymptotic symmetries, which are spanned by $\Lambda$'s of the above form. The commutation relations we are talking about are the Poisson brackets among the generators of these asymptotic symmetries, and we thus need to identify these generators first. Actually, because there is only one left-arbitrary $\Lambda$ component, there will be only one generator of the asymptotic symmetry algebra.

In the above paragraph, our use of the word 'generator' is not the same as previously in this text. Let us be more specific: as usual, when we speak of the asymptotic symmetry algebra, what we are interested in is the algebra of transformations ('symmetries') on phase-space functions corresponding to the subset of the 'gauged' $\mathrm{sl}(2|\mathbb{R})$ which[5] is associated with the residual gauge symmetries at asymptotic infinity ($r \to \infty$). We thus want to study gauge transformations of phase-space functions at asymptotic infinity, which form an algebra we call asymptotic symmetry algebra and the basis elements of which we now also call 'generators', but those are of course functions and not elements of $\mathrm{sl}(2|\mathbb{R})$, so that we dare say confusion is unlikely.

---

[5] We say 'gauged' because an element of $\mathrm{sl}(2|\mathbb{R})$ is a linear combination of its generators with constant coefficients. Allowing the coefficients to be spacetime functions turns it into a gauge symmetry, but we shall still call the latter $\mathrm{sl}(2|\mathbb{R})$, writing e.g. $\Lambda \in \mathrm{sl}(2|\mathbb{R})$, but from the context it should be evident that the $\Lambda^i$ are functions.

The usual way to identify the generators of any symmetry algebra of transformations of phase-space functions corresponding to a given gauge symmetry of the action, like sl(2|$\mathbb{R}$), is the following: a gauge transformation with parameter $\Lambda \in$ sl(2|$\mathbb{R}$) acts on any phase-space function $\mathcal{O}$ through

$$\mathcal{O} \to \mathcal{O} + \delta\mathcal{O} \qquad (2.16)$$

with

$$\delta\mathcal{O} = \{\mathcal{O}, G[\Lambda]\}_{\text{PB}}, \qquad (2.17)$$

where $\{\cdot\,,\cdot\}_{\text{PB}}$ is the usual equal-time Poisson bracket providing phase-space functions with a Lie-type algebraic structure. According to the general principle of gauge theory, the functional $G[\Lambda]$ is (in our case) given by

$$G[\Lambda] \equiv \int d^2x \left(\Lambda^+ \mathcal{G}_+ + \Lambda^- \mathcal{G}_- + \Lambda^3 \mathcal{G}_3\right) + S_\infty, \qquad (2.18)$$

where $\mathcal{G}_{\pm,3}$ are the Chern–Simons–Gauss constraints of (1.51) (given in a different basis here) and $S_\infty$ is a boundary term at asymptotic infinity defined by the requirement that $G[\Lambda]$ must have well-defined functional derivatives with respect to the fields [134], i.e. $\delta_A G[\Lambda]$ contains only undifferentiated field variations under the given boundary conditions for $A_\mu$. To identify the generator, one must now compute the above expression, use if necessary that some $\Lambda$ components depend on the arbitrary one and then identify in $G[\Lambda]$ the function coefficient multiplying the arbitrary $\Lambda$ component, which is the searched-for generator.[6]

We thus want to find an explicit form for $G[\Lambda]$. By definition, the bulk term in (2.18) involving the constraints vanishes on the constraint surface of our theory ('on-shell'), on which $G[\Lambda]$ therefore reduces to the boundary term $S_\infty$, which we can compute by explicitly implementing the requirement that no differentiated field variations appear in $\delta G[\Lambda]$ and integrating by parts. From now on we will place ourselves on this constraint surface, so that we find

$$G[\Lambda] = S_\infty = \frac{k}{2\pi} \int d\theta\, \chi L, \qquad (2.19)$$

where we have renamed $\Lambda^- \equiv \chi$, as that component plays a special role. We thus see that the generator of our asymptotic symmetry algebra (the coefficient of $\chi$ in $G[\Lambda]$) is actually the non-trivial component of our asymptotic connection (boundary conditions) up to a factor of $k/2\pi$, and we recall that $k = l/4G$.

---

[6] The generators are really the Fourier modes contained in those function coefficients, but we will use the word generator also for the function coefficients themselves.

To be precise, the Poisson brackets are only defined for the 'full' generators (2.18), containing both the bulk and the boundary term. As explained hereabove, in the sequel we shall drop the bulk term. However, we shall still call our brackets Poisson brackets, which is formally wrong. The complete explanation of what we do is the following: the Poisson bracket of two well-defined generators is another, well-defined generator [103], which again has a bulk term and a boundary one. However, as we know the boundary part is really what is responsible for the asymptotic symmetries: a (well-defined) generator lacking such a boundary term would only transform phase-space functions in the interior of our spacetime. The important point is then that, as one can check, the surface term corresponding to the Poisson bracket of two generators only depends on the respective surface terms of the two generators that we have taken the bracket of.

**Remark** : one might also wonder about gauge fixings. If one fixes the gauge, the system of constraints (which now includes the gauge conditions) is no longer first class, and in particular the Chern–Simons constraints are no longer first class. One may thus set them to zero strongly and rightfully use only the boundary terms, but the Poisson bracket has now become a Dirac bracket, because of the gauge fixing [103]. Nonetheless, one can demonstrate that, because the system of original constraints is first class the Dirac bracket coincides with the Poisson one at least when considered on gauge-invariant observables (such as our generators).

This $L$ (up to a factor) is thus the generator of the asymptotic symmetries of phase-space functions under the given boundary conditions. Note that $G[\Lambda]$ turned out to be already expressed in terms of the arbitrary components of $\Lambda$ only and we needed not use any relation of the type (2.13). Those relations will have to be used, however, in the following Poisson-brackets computations.

**Poisson-Bracket Algebra**

We now aim at computing the Poisson bracket of two $L$ generators in order to recognize the asymptotic symmetry algebra we are dealing with. The 'trick' we will employ to compute these Poisson brackets is the following: the formula (2.17) holds of course for any phase-space function, and in particular for our generator $L$. However, because this generator is precisely the component (up to a factor) of the asymptotic form of our connection, we already know from (2.10) how it transforms under an infinitesimal gauge symmetry with parameter $\Lambda$, and we are therefore naturally led to exploit the equality
$$\{L, G[\Lambda]\}_{\text{PB}} = \partial_+ \Lambda^+ + [B, \Lambda]|_{\sigma^+}, \tag{2.20}$$
that is, using the relations (2.13),
$$\frac{k}{2\pi} \int d\theta' \chi(\theta') \{L(\theta), L(\theta')\}_{\text{PB}} = -\frac{1}{2}\partial^3 \chi + \partial(L\chi) + L\partial\chi \tag{2.21}$$
$$= \int d\theta' \delta(\theta' - \theta)\left(-\tfrac{1}{2}\partial^3\chi + \partial(L\chi) + L\partial\chi\right),$$

where we have made it explicit for the angular dependence at a fixed time (recall the $B$ components are functions of $x^\pm$ whereas the $\Lambda$ ones are functions of $x^+$ only), the right hand side of the first line depending only on $\theta$, whereas in the parenthesis of the second line everything depends on $\theta'$. Note that we have also renamed $\partial_+ \equiv \partial$. Let us point out that this equality is completely natural. Indeed, all the above equation means is that the boundary conditions are preserved under the action of the Poisson bracket (2.17) when treated as phase-space functions (as opposed to components of the connection), which is not only natural but of course also needed for consistency.

Now, because $\chi$ is arbitrary, integrating by parts the terms in the last member of the above equation allows us to read-off the Poisson brackets among the $L$'s. This yields, upon rescaling $L \to \frac{k}{2\pi} L$ (but we rename it $L$),
$$\{L(\theta), L(\theta')\}_{\text{PB}} = \frac{k}{4\pi}\delta'''(\theta - \theta') - \big(L(\theta) + L(\theta')\big)\delta'(\theta - \theta'), \tag{2.22}$$

where our derivatives '$\cdot$' are with respect to the arguments (this has nothing to do with $\theta'$, which is just a second angular variable). As expected, the above relations match the ones defining the conformal algebra or, rather, a central extension thereof: the Virasoro algebra!

Let us recast the asymptotic result obtained so far into Fourier modes. We will use so-called quantum-mechanical notations, setting
$$A(\theta) \equiv \frac{1}{2\pi}\sum_{n \in \mathbb{Z}} A_n e^{in\theta} \tag{2.23}$$

for any function of $\theta$ and using the correspondence $\{\cdot,\cdot\}_{\text{PB}} = -i[\cdot,\cdot]$, where the later is defined as the usual commutator on Fourier modes $A_n$, i.e. it is defined as the antisymmetrization of the product of modes. Let us recall that we shall be taking advantage of the mode expansion of the Dirac delta function

$$\delta(\theta - \theta') = \frac{1}{2\pi} \sum_{n \in \mathbb{Z}} e^{in(\theta - \theta')}, \tag{2.24}$$

so that its $k$-th derivative $\delta^{k]}(\theta - \theta')$, appearing in the above Poisson bracket is easily calculated in terms of modes.

A straightforward computation then shows that the Fourier mode analogue of the Poisson bracket (2.22) is

$$[\mathrm{L}_n, \mathrm{L}_m] = \tfrac{1}{2} k n^3 \delta_{n+m,0} + (n-m) \mathrm{L}_{n+m}, \tag{2.25}$$

where $\delta_{a,b}$ is the usual Kronecker symbol (equal to zero except for $a=b$, in which case it has value 1). The most used convention in the literature is that the coefficient of $n^3 \delta_{n+m,0}$ is equal to $\frac{c}{12}$, where $c$ is the central charge. This, combined with $k = \frac{\ell}{4\mathrm{G}}$ (which is a consequence of our normalization for the generators of the gauge algebra), leads to the well-known result

$$c = \frac{3\ell}{2\mathrm{G}}, \tag{2.26}$$

the celebrated Brown–Henneaux central charge at asymptotic infinity of pure Gravity with negative cosmological constant $\Lambda = -\frac{1}{\ell^2}$ [67]. Let us further note that the sl($2|\mathbb{R}$) algebra is a subalgebra of the above Virasoro algebra, corresponding to the generators $\{L_0, L_{\pm 1}\}$, up to the presence of the central charge. However, one can perform the following field redefinition:

$$L_0 \to L_0 - \tfrac{1}{4} k, \tag{2.27}$$

which turns the commutation relations into

$$[\mathrm{L}_n, \mathrm{L}_m] = \tfrac{1}{2} k n (n^2 - 1) \delta_{n+m,0} + (n-m) \mathrm{L}_{n+m}, \tag{2.28}$$

where we now explicitly see that the central charge vanishes for the $\{L_0, L_{\pm 1}\}$ sector. In fact, the above redefinition of $L_0$ has a deeper interpretation: indeed, the generator $L_0$ is a global charge, and the redefinition (2.27) is equivalent to requiring the redefined generator to vanish on the AdS$_3$ solution, as can be seen by looking at (C.22) — let us recall that the generators are always defined up to a constant. Thus, (half of) the isometry algebra of the bulk vacuum solution, sl($2|\mathbb{R}$), is indeed a subalgebra of the asymptotic symmetry algebra (provided we set the vacuum charges to zero), as it should be. Finally, we point out that the computation goes along the same lines for the second chiral copy, and one again finds a corresponding Virasoro algebra with same central charge.

### 2.1.3 Superconformal Virasoro Algebra

In Section 2.2 we shall be interested in computing the asymptotic symmetries of Higher-Spin Gravity models on AdS$_3$ spacetimes. More specifically we shall focus on supersymmetric theories which extend three-dimensional Supergravity with gauge algebra osp($N, 2|\mathbb{R}$) (for one chiral sector). Therefore, hereafter we quote the equivalent of the above pure-Gravity result for the case of extended osp($N, 2|\mathbb{R}$) Supergravity [60] (see also [135]).

As explained in Subsection 1.2.3, we can formulate the Supergravity we are interested in as a Chern–Simons action based on two copies of the osp($N, 2|\mathbb{R}$) superalgebra, in full analogy with the Einstein–Hilbert case. This means that, e.g. for $N = 1$ our connection one-form $\Gamma_\mu$ also has components along the $R^\pm$ generators of (1.52), and in the extended case it also has internal components along the $J_{ij}$ generators of Appendix B (then the $R^\pm$ generators further carry an internal index, so that they become the $R_i^\pm$ given therein). Accordingly, we should define the asymptotic behavior of these extra components — the fall-off conditions on the sl(2|$\mathbb{R}$) subsector are of course kept as in (2.2a). In [60] the methods previously applied to pure Gravity were used again to generate boundary conditions for the full osp($N, 2|\mathbb{R}$) connection, and for the first chiral copy of the generic-$N$ case they read

$$\Gamma'_+|_{\text{LEADING ORDER}} = \left(L(x^\pm)\sigma^+ + Q^i(x^\pm)R_i^+ + B^{ij}(x^\pm)J_{ij} + \sigma^-\right), \quad (2.29)$$

where we have already performed the gauge redefinition leading to (2.7). In the fermionic sector we thus impose a condition analogous to that for $A_\mu$, while in the internal sector we allow for any generator to appear asymptotically. In fact, the generators $R_i^\pm$ and $J_{ij}$ beyond sl(2|$\mathbb{R}$) come into irreducible representations of the sl(2|$\mathbb{R}$) subalgebra: the $R_i^\pm$'s form $N$ irreducible representations[7] of spin $\frac{3}{2}$ and the $J_{ij}$'s have spin 1 (see Appendix B). With this point of view in mind one sees that the above asymptotics correspond to allowing for one arbitrary function of $x^\pm$ for every highest-weight generator (the $R_i^+$'s and $\sigma^+$), except in the internal sector where we allow for one arbitrary function for all generators, because they have no spin. As we will see in Section 2.2, this is actually the logic we shall follow in order to determine boundary conditions for the higher-spin generators, which shall complement the above osp($N, 2|\mathbb{R}$) algebra so to make it a higher-spin algebra extending standard Supergravity.

---

[7] Recall that in our terminology the conformal spin of the generator is the spin of the sl(2|$\mathbb{R}$) representation it is in, and the spin of a generator is lifted by one unit with respect to the conformal spin.

The computation now proceeds as in the Gravity case, each arbitrary function in (2.29) giving rise to an asymptotic current, denoted with the same letter (up to a factor). Let us only quote the asymptotic symmetry result [60], which in the non-extended case reads

$$\{L(\theta), L(\theta')\}_{\text{PB}} = \frac{k}{4\pi}\delta'''(\theta - \theta') - \big(L(\theta) + L(\theta')\big)\delta'(\theta - \theta'), \quad (2.30\text{a})$$

$$\{L(\theta), Q(\theta')\}_{\text{PB}} = -\big(Q(\theta) + \tfrac{1}{2}Q(\theta')\big)\delta'(\theta - \theta'), \quad (2.30\text{b})$$

$$i\{Q(\theta), Q(\theta')\}_{\text{PB}} = -\frac{k}{\pi}\delta''(\theta - \theta') + 2L(\theta)\delta(\theta - \theta'). \quad (2.30\text{c})$$

Note that, in principle, there is a term proportional to $\delta(\theta - \theta')Q(\theta)Q(\theta')$ appearing in the right-hand side of the first line above, but it is zero because of the Grassmann parity of $Q$ (see Appendix B.3). In the extended case we also have commutators involving the currents $B^{ij}$ corresponding to the function-coefficients of $J_{ij}$ in (B.22), and for the sake of conciseness we refer to [60] for further details. Let us point out, however, an important feature of the extended case, which is that of quadratic terms in the $B$'s appearing in the right hand side of the above anticommutator of two (conformal) spin-$\tfrac{1}{2}$ currents. Indeed, in the extended case the last relation hereabove becomes, schematically:

$$i\{Q(\theta), Q(\theta')\}_{\text{PB}} = -\frac{k}{\pi}\delta''(\theta - \theta') + 2L(\theta)\delta(\theta - \theta') + \text{`}B \times B\text{'}. \quad (2.31)$$

These nonlinearities, which only appear in the extended case, spoil the standard Lie structure and the resulting relations form what is called a *nonlinear deformation* of a Lie algebra (in this case infinite dimensional). As we shall see in Section 2.2, the appearance of nonlinearities in the asymptotic commutation relations is also a feature common to higher-spin models, although not only for the extended versions thereof.

The above relations are recognized as the defining relations of the centrally-extended superconformal algebras in two dimensions, which are also called the super-Virasoro algebras. In Fourier modes we then obtain

$$[\text{L}_n, \text{L}_m] = \tfrac{1}{2}kn(n^2 - 1)\delta_{n+m,0} + (n - m)\text{L}_{n+m}, \quad (2.32\text{a})$$

$$[\text{L}_n, \text{Q}_m] = \big(\tfrac{1}{2}n - m\big)\text{Q}_{n+m}, \quad (2.32\text{b})$$

$$\{\text{Q}_n, \text{Q}_m\} = 2k\left(n^2 - \tfrac{1}{4}\right)\delta_{n+m,0} + 2\text{L}_{n+m}, \quad (2.32\text{c})$$

and similarly for the extended case. One could also wonder about the asymptotic symmetry algebra corresponding to the other extended supergravities, listed in Appendix B. In fact, it was shown in [60] that the structure of the symmetries at asymptotic infinity is the same for all types of supergravities, and only the detailed coefficients in the right hand sides notice the change. In Chapter 3 we shall also comment on higher-spin theories extending these other supergravities, but in the next section we focus on the orthosymplectic version.

As is well known, the Virasoro algebra is a subalgebra of the above superalgebra, which stems from the fact that sl$(2|\mathbb{R})$ is a subalgebra of osp$(N, 2|\mathbb{R})$. Moreover, one again sees that osp$(N, 2|\mathbb{R})$ is a subalgebra of the above superconformal algebra: in the above example of the non-extended case one easily sees that the relations close for the generators $\{Q_{\pm 1/2}, L_{0,\pm 1}\}$, for which the central charges also vanish (in (2.32) the redefinition (2.27) has already been performed). We also point out that the central charge in the Virasoro sector still has the standard Brown–Henneaux value (2.26), while in the fermionic sector we notice the appearance of a new central charge, which is nonetheless proportionally related to the Brown–Henneaux one (it is not parametrized independently). Finally, we note that according to the standard terminology [4], the $L$ generators indeed have spin 2 and the $Q$ ones have spin $\frac{3}{2}$, as the presence of the term proportional to $n$ (resp. $\frac{n}{2}$) indicates in the above adjoint action of $L$ on itself (resp. on $Q$), which one can also see at the level of the relations (2.30).

## 2.2 Asymptotic Symmetries Beyond Spin 2 and $\mathcal{W}_\infty$

In the previous section we have recalled the way in which one obtains the asymptotic symmetry algebra of pure Gravity and Supergravity in AdS$_3$, and this will now serve us as a guideline for computing the asymptotic symmetry algebra of three-dimensional higher-spin models. As indicated earlier in Section 1.2, we shall be interested in those theories which are based on the shs$(N, 2|\mathbb{R})$ algebras, which contain osp$(N, 2|\mathbb{R})$ as a subalgebra. More precisely, in the sequel we explicitly go through the computation of asymptotic symmetries in the shs$(1, 2|\mathbb{R}) \equiv$ shs$(1, 1)$ case and only comment on the extended results at the end of the section, relegating the corresponding computations to Appendix C, where we sketch the extended version of the procedure.

In Subsection 2.2.1 we first generalize the boundary conditions of the Supergravity case (2.29) to include higher spins, and in Subsections 2.2.2 and 2.2.3 we then proceed along the lines of Subsection 2.1.2 in order to obtain the corresponding asymptotic symmetry superalgebra, which is found to be a nonlinear deformation of some supersymmetric so-called $\mathcal{W}_\infty$-algebra, which we comment on in Chapter 3.

### 2.2.1 Higher-Spin Superconnection Fall-Off Conditions

As explained in Subsection 1.2.3, our Higher-Spin Theory is described by (two copies of) a Chern–Simons term whose gauge connection one-form $\Gamma_\mu$ takes values in the shs(1, 1) superalgebra, and the latter algebra is realized as the space of polynomials of all degrees in the $q_\alpha$'s (with suitable reality conditions) equipped with the Lie bracket (1.91), where the $\star$-product is given in (1.85). Moreover, a useful basis for such a space is that which is displayed in (1.90), and which we shall work with hereafter. With our normalization conventions for the Lie superbracket (1.91), the superalgebra shs(1, 1) is realized as the space of linear combinations of the basis elements $X_{(p,q)}$ of (1.90) with real coefficients. Let us, then, define the components of the shs$(1, 2|\mathbb{R})$-valued connection one-form $\Gamma_\mu$ as

$$\Gamma \equiv \sum_{p+q \,\in\, \mathbb{N}_0} \Gamma^{(p,q)} X_{(p,q)} \equiv \sum_{p+q \,\in\, \mathbb{N}_0} \Gamma_\mu^{(p,q)} X_{(p,q)} \mathrm{d}x^\mu, \qquad (2.33)$$

where $\mu = +, -, r$ is a three-dimensional spacetime index referring to the coordinates introduced in Section 2.1: $x^\pm \equiv t \pm \ell\theta$, $x^r \equiv r$, and we will work with AdS radius $\ell = 1$, so that $x^\pm \equiv t \pm \theta$. Let us also recall we don't consider the $p + q = 0$ sector, according to the sum in the above equation (see Subsection 1.2.3).

We now want to extend the fall-off conditions given in (2.29) to all the components of the above shs$(1, 2|\mathbb{R})$ connection, and recall that the osp$(1, 2|\mathbb{R})$ sector is encoded in the generators with $p + q = 1, 2$, which are proportional to the $R^\pm$ and $H, E, F$ generators of (1.52) (see Appendix B). However, a priori it is not clear how to extend the low-spin asymptotic behavior to the higher-spin sector. Inspired by the strategy used in [60] to generalize the boundary conditions of [67] on the sl$(2|\mathbb{R})$ sector to the osp$(1, 2|\mathbb{R})$ superalgebra,[8] we might want to carry out an analogous procedure here, namely acting on the osp$(N, 2|\mathbb{R})$ asymptotics with a general transformation of shs$(1, 2|\mathbb{R})$ and postulating the resulting form to be our boundary conditions for the full connection. However, we shall adopt a more 'heuristic' attitude, described hereafter, which was also used in the previous study of the bosonic case [1].

**Remark** : in fact, the 'strategy used in [60]' seemingly fails to generate the boundary conditions which we work with in the sequel (see below). Noticeably, such a procedure already fails in the case of pure Gravity in its Chern–Simons form, or at least it does not work in the most naive way.

---

[8] Note that this strategy was originally proposed in [132].

As explained in Subsection 2.1.1, determining boundary conditions may require some guess work and in the present case, much like in [1] we postulate the following form: as only $X_{(1,0)}$ and $X_{(2,0)}$ appear in the low-spin asymptotics of (2.29) (where we call them $R^+$ and $\sigma^+$), we postulate rather intuitively that the $X_{(n,0)}$ generators are the only ones appearing in the asymptotics for the full-fledged connection, that is,

$$\Gamma_+ \stackrel{r\to\infty}{\longrightarrow} \Delta_+ \equiv \Delta \quad \text{(at leading order)} \tag{2.34}$$

with

$$\begin{aligned}\Delta &= -X_{22} + \Delta^1(x^\pm)X_1 + \Delta^{11}(x^\pm)X_{11} + \Delta^{111}(x^\pm)X_{111} + \cdots \\ &= -X_{22} + \sum_{i\,\in\,\mathbb{N}_0} \Delta^{(i,0)}(x^\pm)X_{(i,0)},\end{aligned} \tag{2.35}$$

also setting $\Delta_-$ and $\Delta_r$ (the asymptotic forms of $\Gamma_-$ and $\Gamma_r$) to zero.[9] This condition has an algebraic justification, which is anticipated at the end of Subsection 2.1.3: only the highest-weight $X_{(n,0)}$ generators appear in the asymptotics! Indeed, let us again recall that under the sl$(2|\mathbb{R})$ sub-algebra of shs$(1,1)$, the higher-spin generators of degree $n$ form irreducible representations of spin $s = \frac{n}{2}$, and those are said to have spin $s+1$ or conformal spin $s$ (see Appendix B). The equivalent of (2.35) for the second chiral copy would evidently be

$$\begin{aligned}\tilde{\Delta} &= X_{11} + \tilde{\Delta}^2(x^\pm)X_2 + \tilde{\Delta}^{22}(x^\pm)X_{22} + \tilde{\Delta}^{222}(x^\pm)X_{222} + \cdots \\ &= X_{11} + \sum_{i\,\in\,\mathbb{N}_0} \tilde{\Delta}^{(0,i)}(x^\pm)X_{(0,i)},\end{aligned} \tag{2.36}$$

with $\tilde{\Delta} \equiv \tilde{\Delta}_-$ and $\tilde{\Delta}_+ = \tilde{\Delta}_r = 0$. As in the Gravity case, the latter asymptotic behavior is not the same as the one of the first chiral copy but it will, however, lead to the same asymptotic symmetry algebra, so that we shall again restrict our attention to the first chiral sector in the rest of this chapter.

### 2.2.2  Asymptotic Symmetries for Higher-Spin Currents

We now follow the steps presented in Subsection 2.1.2 for the case of Gravity, namely: we want to act with a general element $\Lambda \in \text{shs}(1,1)$ on the above asymptotics for the connection $\Gamma$, require the transformed asymptotics have the same form as the original ones and therefrom derive the conditions such requirement yields on $\Lambda$. The Poisson-bracket symmetry algebra is then extracted along the lines of Subsection 2.1.2. As we will see, the difference will lie in the determination of the conditions yielded on $\Lambda$. Again, by 'the same form' as the original asymptotics we mean the transformed $\Delta$ reads exactly as (2.35) except for the form of the functions multiplying the $X_{(n,0)}$ generators, which is allowed to change (but we do not allow introducing $r$ dependence). Obviously, we also mean $\Delta_-$ and $\Delta_r$ remain zero.

---

[9] Note that, as in [1], the whole of $\Gamma_-(r \to \infty)$ is zero while only the leading part $\Delta_r$ of $\Gamma_r(r \to \infty)$ is zero.

**Residual Gauge Symmetries**

The infinitesimal gauge transformation with parameter $\Lambda$ acts on $\Delta$ by the adjoint action, i.e.

$$\Delta \to \Delta + \delta\Delta \quad \text{with} \quad \delta\Delta = \partial_+ \Lambda + [\Delta, \Lambda]_\star, \tag{2.37}$$

where $\partial_+ \equiv \partial/\partial x^+$ and the same relation holds for $\Delta_-$ and $\Delta_r$. From the requirement that $\delta\Delta_-$ and $\delta\Delta_r$ be zero, taking into account that these are initially zero, the above relation states that $\Lambda$ cannot depend on either $x^-$ or $r$. Note that this already implies that no $r$ dependence will be introduced in $\Delta$ by such a $\Lambda$ parameter, as required. With such a $\Lambda$ free of all $r$ (and $x^-$) dependence, our requirement of boundary conditions invariance amounts to ask for all the coefficients of the generators with at least one index 2 in $\delta\Delta$ above to be zero. Note that we don't allow the appearance of the $X_{(0,2)}$ generator, even though it is present in (2.35), because its coefficient in it is just a fixed number and not a function. Also note that, although we should write our Lie superbracket as $[\,\cdot\,,\,\cdot\,\}_\star$ we shall often abuse the notation and denote it as a standard Lie bracket, as in the above equation. Furthermore, the $\star$ symbol in subscript shall often be dropped.

In order to further proceed let us first write $\Lambda$ in an 'ordered' way we will take advantage of in the following:

$$\Lambda \equiv \sum_{p+q \in \mathbb{N}_0} \Lambda^{(p,q)} X_{(p,q)} \tag{2.38}$$

$$= \sum_{i \in \mathbb{N}_0} \Lambda^{(0,i)} X_{(0,i)} + \sum_{i \geq 1} \Lambda^{(1,i-1)} X_{(1,i-1)} + \sum_{i \geq 2} \Lambda^{(2,i-2)} X_{(2,i-2)}$$

$$+ \sum_{i \geq 3} \Lambda^{(3,i-3)} X_{(3,i-3)} + \sum_{i \geq 4} \Lambda^{(4,i-4)} X_{(4,i-4)} + \cdots,$$

where we stress again that the components $\Lambda^{(p,q)}$ are functions of $x^+$ only and that we do not include the spin-1 (zero indices) generator. This rewriting will allow us to use the fact that $\Delta$ has a specific form, namely only the $X_{(n,0)}$ generators appear therein.

As aforesaid, our goal is now to compute the expression (2.37) and derive the form that $\Lambda$ must have so that no generators with indices equal to 2 appear therein. As in the low-spin case, such requirements of asymptotic invariance of the boundary conditions will determine some of the components of $\Lambda$ in terms of the other left-arbitrary ones. Now, if one is to determine all the commutation relations of the asymptotic algebra, one needs the complete form that these 'dependent' components take in terms of the arbitrary ones. However, in the following we shall only compute the Poisson-bracket relations among *some* of the (yet-to-be identified) generators of the asymptotic symmetries, hence we will only need the form of *some* of the dependent components of $\Lambda$. Therefore, we shall not carry out here the derivation of the form of all the dependent components of $\Lambda$. Rather, we shall confine ourselves to showing which ones are left arbitrary, for this we always need to know in order to identify the phase-space generators (see Subsection 2.1.2), and will indicate the procedure allowing to determine the other ones in terms of the latter.

Let us show what $\Lambda$ components can be thought of as being left arbitrary and at the same time sketch a procedure one can follow in order to compute the form of all the dependent $\Lambda$ components. First, let us give a name to the components of (2.37):

$$\delta\Delta \equiv \sum_{p+q\,\in\,\mathbb{N}_0} c^{(p,q)} X_{(p,q)} \equiv \sum_{p+q\,\in\,\mathbb{N}_0} \partial\Lambda^{(p,q)} X_{(p,q)} \,+\, \sum_{p+q\,\in\,\mathbb{N}_0} [\Delta,\Lambda]^{(p,q)} X_{(p,q)}, \tag{2.39}$$

where $\partial \equiv \partial_+$ and we sum on all possible generators for a priori all of them could appear. The requirement that the form of the asymptotics (2.35) be preserved by residual gauge transformations now evidently reads

$$c^{(p,q)} = 0 \quad \forall\, p \in \mathbb{N},\, q \in \mathbb{N}_0. \tag{2.40}$$

The strategy that will now allow us to determine which coefficients $\Lambda^{(p,q)}$ can be left arbitrary is to impose the above equations step by step in a certain order, namely

$$\begin{aligned} c^{(0,n)} &= 0 \quad \forall\, n \geq 1, \\ c^{(1,n-1)} &= 0 \quad \forall\, n \geq 2, \\ c^{(2,n-2)} &= 0 \quad \forall\, n \geq 3, \\ c^{(3,n-3)} &= 0 \quad \forall\, n \geq 4, \end{aligned} \tag{2.41}$$

etc.[10] Note that, as it should be, by following the above strategy we never set any $c^{(n,0)}$ coefficient equal to zero because of the lower bounds on $n$, which at each step are adjusted so to 'spare' those highest-weight generators, which obviously should not be set to zero because these are allowed (note the shift by one unit with respect to the lower bounds on the summing indices of (2.38), in which all components can a priori appear). Let us now apply the above strategy step by step and see how the structure of $\Lambda$ emerges. We further point out that all we are doing is consistently apply a generalized version of the procedure employed in [1] for the bosonic case.

**First Step:** one can easily convince oneself that the supercommutator in (2.37) only yields generators with all indices equal to 2 via $[\Delta, X_{(0,i)}\}$, so that using (B.37a) for $r = n$ odd ($m$ even) one finds

$$c^{(0,n)} = \partial \Lambda^{(0,n)} + n\Lambda^{(1,n-1)} + \sum_{i \in \mathbb{N}} \frac{(-)^i}{(2i+1)!} \Delta^{(2i+1,0)} \Lambda^{(0,n+2i+1)}$$
$$\equiv \partial \Lambda^{(0,n)} + n\Lambda^{(1,n-1)} + f^0\big(\Delta^{(i,0)}, \Lambda^{(0,j)}\big),$$
(2.42)

where the second term in the right hand side above comes from $[-X_{22}, \Lambda]$. From the expression above, we see that the condition $c^{(0,n)} = 0$ can be thought of as determining $\Lambda^{(1,n-1)}$ in terms of the coefficients $\Delta^{(i+1,0)}$, $\Lambda^{(0,n+i)}$ with $i \in \mathbb{N}$. Note that with our writing for $f^0$ it looks like a priori it depends on *all* of the $\Delta^{(i,0)}$ and $\Lambda^{(0,j)}$ functions, while it clearly does not. However, this will not matter much in the sequel and we thus keep this simple notation for $f^0$, as we will do for the other $f^i$'s below. This is also why we have not labeled $f^0$ by an index depending on $n$. More loosely, we will thus say that the $\Lambda^{(1,m)}$ coefficients are determined in terms of the $\Delta^{(i,0)}$, $\Lambda^{(0,j)}$ ones, $m \in \mathbb{N}$ (and similarly below).

Note that, even though the $f^i$'s contain an infinite number of terms (see equations above), one can check that each $f^j$ contains only a finite number of terms involving $\Lambda^{(0,j)}$ for a given $j$, so that in principle all Poisson brackets are unambiguously computable.

**Second Step:** one can also convince oneself that the supercommutator in (2.37) only yields generators with all indices equal to 2 except one

---

[10] So that, in a sense, we start with the 'worse': we begin by setting to zero the coefficients with all indices equal to 2, then the coefficients with all indices equal to 2 except one of them, then the coefficients with only indices '2' except two of them, ...

via $[\Delta, X_{(0,i)}\}$ and $[\Delta, X_{(1,j)}\}$, so that one finds

$$c^{(1,n-1)} = \partial \Lambda^{(1,n-1)} + (n-1)\Lambda^{(2,n-2)} + f^1\big(\Delta^{(i,0)}, \Lambda^{(0,j)}, \Lambda^{(1,k)}\big), \tag{2.43}$$

with a similar structure to the one found in the previous step. Note that we do not give the expression for $f^1$ as it is somewhat more involved than $f^0$ and we do not need it now. We see that the conditions of this step can be thought of as determining the $\Lambda^{(2,m)}$ coefficients in terms of the $\Delta^{(i,0)}, \Lambda^{(0,j)}, \Lambda^{(1,k)}$ ones, $m \in \mathbb{N}$. However, the $\Lambda^{(1,k)}$ coefficients where determined at the previous step in terms of the $\Delta^{(i,0)}, \Lambda^{(0,j)}$ coefficients so that this step, performed after the first one, really determines the $\Lambda^{(2,m)}$ coefficients in terms of the $\Delta^{(i,0)}, \Lambda^{(0,j)}$ ones only.

**Third Step:** now the supercommutator in (2.37) only yields generators with all indices equal to 2 except two via $[\Delta, X_{(0,i)}\}$, $[\Delta, X_{(1,j)}\}$ and $[\Delta, X_{(2,k)}\}$, so that one finds

$$c^{(2,n-2)} = \partial \Lambda^{(2,n-2)} + (n-2)\Lambda^{(3,n-3)} + f^2\big(\Delta^{(i,0)}, \Lambda^{(0,j)}, \Lambda^{(1,k)}, \Lambda^{(1,l)}\big), \tag{2.44}$$

which we can again see (by a reasoning similar to the one of the previous step) as determining the $\Lambda^{(3,m)}$ coefficients in terms of the $\Delta^{(i,0)}, \Lambda^{(0,j)}$ ones ($m \in \mathbb{N}$). The procedure continues on and on but we stop here, the important thing being the triangular pattern of the procedure, which has been made clear.

Note that, as pointed out before, the above procedure yields no conditions of the above kind on the $c^{(n,0)}$ coefficients, which is normal since these are allowed to appear in (2.37), and those we therefore need not constrain. Rather, they can be determined in terms of $\Delta^{(i,0)}$, $\Lambda^{(0,j)}$ and they themselves determine the allowed-for variation of $\Delta$ under the asymptotic gauge transformations through

$$c^{(n,0)} = \delta \Delta^{(n,0)} \equiv \partial \Lambda^{(n,0)} + [\Delta, \Lambda]^{(n,0)} \quad \forall\, n \in \mathbb{N}_0, \tag{2.45}$$

as it is obvious by considering

$$\delta \Delta = \sum_{i\,\in\,\mathbb{N}_0} \delta \Delta^{(i,0)} X_{(i,0)}, \tag{2.46}$$

which itself is trivially derived from (2.35).

It is now clear from the pattern emerging from the above procedure that the conditions on $\Lambda \in \text{shs}(1,2|\mathbb{R})$ translating the requirement that the form of the given boundary conditions (2.35) be preserved by infinitesimal gauge transformations with parameter $\Lambda$ (asymptotic invariance under residual gauge transformations) can be understood as leaving all the $\Lambda^{(0,n)}$ components (functions) arbitrary, the other ones being functions of the later and of the components of $\Delta$. Also, it is in principle possible to determine the form of these dependent components by following the above procedure, which we postpone. The asymptotic symmetries, i.e. the searched-for residual gauge transformations preserving the aforegiven boundary conditions are therefore spanned by $\Lambda$ elements of the form

$$\begin{aligned}\Lambda &= \sum_{i \in \mathbb{N}_0} \Lambda^{(0,i)} X_{(0,i)} + \sum_{p \in \mathbb{N}_0, q \in \mathbb{N}} \Lambda^{(p,q)}(\Delta^{(k,0)}, \Lambda^{(0,j)}) X_{(p,q)} \\ &\equiv \sum_{i \in \mathbb{N}_0} \Lambda^i X_{(0,i)} + \sum_{p \in \mathbb{N}_0, q \in \mathbb{N}} F^{(p,q)} X_{(p,q)},\end{aligned} \quad (2.47)$$

where the $F^{(p,q)} \equiv \Lambda^{(p,q)}$'s depend on the $\Lambda^j \equiv \Lambda^{(0,j)}$ (and $\Delta^i$) functions, which we have stressed by denoting them with the letter $F$. Note that we have also simplified the notation for the $\Lambda^{(0,j)}$ components, as those now play a special role (they are the arbitrary ones). Similarly, we will also simplify the notation for the components of $\Delta$: $\Delta^i \equiv \Delta^{(i,0)}$, as they will also be seen to play a special role hereafter.

We now want to compute the commutation relations for the superalgebra of asymptotic symmetries, which are spanned by $\Lambda$'s of the above form. The commutation relations we are talking about are the (super-)Poisson brackets of the generators of these asymptotic symmetries on phase-space functions, and we thus need to identify these generators first.

**Generators and Low-Spin Brackets**

Following the strategy outlined in Section 2.1, we recall that a gauge transformation with parameter $\Lambda \in \text{shs}(1,2|\mathbb{R})$ acts on any phase-space function $\mathcal{O}$ through

$$\mathcal{O} \to \mathcal{O} + \delta\mathcal{O} \quad \text{with} \quad \delta\mathcal{O} = \{\mathcal{O}, G[\Lambda]\}_{\text{PB}}, \quad (2.48)$$

where $\{\cdot,\cdot\}_{\text{PB}}$ is the Poisson bracket and the functional $G[\Lambda]$ is given by

$$G[\Lambda] \equiv \int d^2x \sum_{p+q \in \mathbb{N}_0} \Lambda^{(p,q)} \mathcal{G}_{(p,q)} \ + \ S_\infty \equiv \int d\theta dr \sum_{p+q \in \mathbb{N}_0} \Lambda^{(p,q)} \mathcal{G}_{(p,q)} \ + \ S_\infty, \quad (2.49)$$

where $\mathcal{G}_{(p,q)}$ are the Chern–Simons-Gauss constraints of the theory (they are defined by (1.51), where the Latin indices therein are now the $(p,q)$ indices) and $S_\infty$ is a boundary term at asymptotic infinity, again defined by the requirement that $G[\Lambda]$ must have well-defined functional derivatives with respect to $\Lambda$ [134]. To identify the generators, one computes the above expression, uses the dependency relations among the $\Lambda$ components and then identifies in $G[\Lambda]$ the function coefficients multiplying the arbitrary ones, which are the searched-for generators.

Going again to the on-shell surface (constraint surface) we find the explicit form

$$G[\Lambda] = S_\infty = -\frac{k}{2\pi}\int d\theta\Big(\sum_{n\,\in\,\mathbb{N}_0}\frac{i^{3n}}{n!}\Lambda^n\Delta^n\Big), \qquad (2.50)$$

where we have used (B.35), the $X_{(n,0)}$ generators being the only ones appearing in $\Delta$. In full analogy with the pure-Gravity case we thus see that the generators of our asymptotic symmetry algebra are actually the components of our asymptotic connection, up to a factor. In order not to make the following discussion too cumbersome and to get closer to the usual notation in the literature we set

$$N^n \equiv -\frac{k}{2\pi}\frac{i^{3n}}{n!}\Delta^n \equiv \alpha^n \Delta^n, \qquad (2.51)$$

so that

$$G[\Lambda] = \int d\theta \sum_{n\,\in\,\mathbb{N}_0} \Lambda^n N^n. \qquad (2.52)$$

These $N^n$ are the generators of the asymptotic symmetries of phase-space functions under the given boundary conditions. Note that $G[\Lambda]$ turned out to be already expressed in terms of the arbitrary components of $\Lambda$ only, and we needed not use any relation eventually derived from the strategy discussed in the previous subsection.

We now compute the Poisson brackets between the above $N^n$ generators in order to recognize the asymptotic symmetry superalgebra we are dealing with. The technique we employ is the same as in Section 2.1, that is, we use

$$\delta N^n = \alpha^n\big(\partial\Lambda^{(n,0)} + [\Delta,\Lambda]^{(n,0)}\big) \qquad (2.53)$$

in order to derive

$$\{N^n(\theta), \int d\theta'\big(\sum_{i\,\in\,\mathbb{N}_0}\Lambda^i(\theta')N^i(\theta')\big)\}_{\text{PB}} = \alpha^n\big(\partial\Lambda^{(n,0)} + [\Delta,\Lambda]^{(n,0)}\big), \qquad (2.54)$$

and we recall that the $\Delta$ components are functions of $x^\pm$ whereas the $\Lambda$ ones are functions of $x^+$ only, and the right hand side depends only on $\theta$. Now, the procedure carried out above determines the coefficient $c^{(n,0)}$ (which is the parenthesis in the right hand side of the above equality) in terms of $\Lambda^i$ and $\Delta^j$ (equivalently $N^j$), so that identifying the coefficients of the $\Lambda^i$ parameters on both sides of the equation (2.54) above makes it possible to read off the Poisson brackets

$$\{N^n(\theta), N^m(\theta')\}_{\text{PB}}, \quad n, m \in \mathbb{N}_0. \tag{2.55}$$

Note that this identification is possible only because the $\Lambda^i$ functions are arbitrary, and we need to use the strategy of the previous section in order to determine the form of the non-arbitrary $\Lambda$ components appearing in the right-hand side of (2.54) for whatever Poisson bracket we want to compute. We stress that, as it is clear from (2.54) and from the definition (2.39) of $c^{(n,0)}$, the expressions for $\{N^n(\theta), N^m(\theta')\}_{\text{PB}}$ are closed, i.e. they depend on the $N^i$ functions only. We also point out that some terms generated in the above Poisson brackets are nonlinear polynomials in the $N^i$ functions, which appear because of the nonlinearities in the $f^n$ functions, themselves introduced when we use the dependency relations for the non-arbitrary $\Lambda$ components. This means that the asymptotic symmetry algebra is actually not an algebra but a nonlinear deformation thereof. However, the Jacobi identity for the above Poisson bracket still holds, because they are Poisson brackets! We will speak of a nonlinear Lie algebra, and we comment on this point and others in Chapter 3.

We now turn to computing the expressions (2.55). First, let us confine ourselves to the $\mathfrak{osp}(1, 2|\mathbb{R})$ sector ($\Lambda$ and $\Delta$ are now truncated to belong to this subsuperalgebra). Applying the first three steps of the procedure of the previous subsection to this case one finds ($f^1$ is now easily computed)

$$c^{(0,1)} = 0 = \partial \Lambda^1 + \Lambda^{(1,0)} + \Delta^1 \Lambda^2, \tag{2.56a}$$

$$c^{(0,2)} = 0 = \partial \Lambda^2 + 2\Lambda^{(1,1)}, \tag{2.56b}$$

$$c^{(1,1)} = 0 = \partial \Lambda^{(1,1)} + \Lambda^{(2,0)} - i\Delta^1 \Lambda^1 + \Delta^2 \Lambda^2, \tag{2.56c}$$

which allows us to determine $\Lambda^{(1,0)}$, $\Lambda^{(1,1)}$ and $\Lambda^{(2,0)}$ in terms of $\Lambda^1$, $\Lambda^2$, $\Delta^1$ and $\Delta^2$:

$$\Lambda^{(1,0)} = -\partial \Lambda^1 - \Delta^1 \Lambda^2, \tag{2.57a}$$

$$\Lambda^{(1,1)} = -\tfrac{1}{2}\partial \Lambda^2, \tag{2.57b}$$

$$\Lambda^{(2,0)} = \tfrac{1}{2}\partial^2 \Lambda^2 + i\Delta^1 \Lambda^1 - \Delta^2 \Lambda^2. \tag{2.57c}$$

Using (2.54) together with the definition of $c^{(p,q)}$ as well as the super-commutation relations (B.34) for the $\text{osp}(1,2|\mathbb{R})$ sector one further finds

$$\{N^1, \int d\theta'(\Lambda^1 N^1 + \Lambda^2 N^2)\}_{\text{PB}} = \frac{ik}{2\pi}\big(\partial\Lambda^{(1,0)} + \Delta^1\Lambda^{(1,1)} + \Delta^2\Lambda^1\big), \tag{2.58a}$$

$$\{N^2, \int d\theta'(\Lambda^1 N^1 + \Lambda^2 N^2)\}_{\text{PB}} = \frac{k}{4\pi}\big(\partial\Lambda^{(2,0)} - 2i\Delta^1\Lambda^{(1,0)} + 2\Delta^2\Lambda^{(1,1)}\big), \tag{2.58b}$$

where everything depends on $\theta$ except for the functions in the integral, which evidently depend on $\theta'$. Now, using the expressions (2.57) for $\Lambda^{(1,0)}$, $\Lambda^{(1,1)}$ and $\Lambda^{(2,0)}$, the above brackets allow us to read off the Poisson brackets within the $\text{osp}(1,2|\mathbb{R})$ sector upon identifying the coefficients of the $\Lambda^1$, $\Lambda^2$ components in them. As was expected, the result matches with the centrally-extended superconformal algebra given in (2.30) upon renaming the low-spin generators as $N^1 \equiv Q$, $N^2 \equiv L$.

Let us point out that the above commutation relations, that we have shown to be valid *within* the $\text{osp}(1,2|\mathbb{R})$ sector, are also valid if we consider the whole $\text{shs}(1,2|\mathbb{R})$ algebra, that is, allowing higher spin components does not add any terms to the above relations, which one can easily convince oneself of.

### 2.2.3 Supersymmetric $\mathcal{W}_\infty$ Algebra

As was anticipated, the low-spin part of our asymptotic symmetry algebra reproduces the super-Virasoro commutation relations. Let us now explore 'higher' commutators, in order to gain insight into the structure of the asymptotic symmetries. We begin by computing the generic 'low-higher' Poisson brackets, that is, those involving an $L$ or $Q$ generator and a higher-spin one. We then compute the first 'higher-higher' brackets, namely those involving no low-spin generators. As the procedure leading to the Poisson-bracket algebra has been exposed several times in the above considerations we shall allow for some sketchiness in the sequel.

**Low-Higher Poisson Brackets**

Let us first compute $\{L(\theta), N^n(\theta')\}_{\text{PB}}$ and $\{Q(\theta), N^n(\theta')\}_{\text{PB}}$. We begin by the Poisson bracket for $L$, first deriving

$$\int d\theta' \sum_{n \in \mathbb{N}_0} \Lambda^n(\theta')\{L(\theta), N^n(\theta')\}_{\text{PB}} = \int d\theta' \delta(\theta' - \theta)\Big(\frac{k}{4\pi}\partial\Lambda^{(2,0)} + \sum_{l \in \mathbb{N}_0} (-)^l l N^l \Lambda^{(1,l-1)}\Big), \quad (2.59)$$

where everything in the parenthesis depends on $\theta'$ (the derivative also being with respect to $\theta'$). We thus need to find the $\Lambda^{(2,0)}$ and $\Lambda^{(1,i)}$ components in terms of the $\Lambda^j$ functions. These we find to be given by

$$\Lambda^{(1,n-1)} = -\frac{1}{n}\partial\Lambda^n + \delta_{1,|n|_2}\frac{2i\pi}{nk}\sum_{l \in \mathbb{N}} N^{2l+1}\Lambda^{n+2l+1}, \quad (2.60)$$

$$\Lambda^{(2,0)} = \frac{1}{2}\partial^2\Lambda^2 - \frac{2\pi}{k}\sum_{l \in \mathbb{N}_0} (-)^l l N^l \Lambda^l, \quad (2.61)$$

where we have used the previously derived relation $\Lambda^{(1,1)} = -\frac{1}{2}\partial\Lambda^2$ to later reinsert it in the expression for $\Lambda^{(2,0)}$, in order to find for the later an expression in terms of the arbitrary components only. Also note that we employ the notation $|j|_2$, meaning the value of $j$ modulo 2. Using the above expressions in (2.59) and identifying the coefficients of $\Lambda^n$ in the right-hand side of it now yields, for $n \geq 3$,

$$\{L(\theta), N^n(\theta')\}_{\text{PB}} = -\delta'(\theta - \theta')\left(N^n(\theta) + \tfrac{1}{2}nN^n(\theta')\right), \quad (2.62)$$

where we see that the current $N^n$ has conformal dimension $s = 1 + \frac{n}{2}$. We point out that, as in the case of $\{L(\theta), L(\theta')\}_{\text{PB}}$ where we had to use $Q(\theta)Q(\theta) = 0$, to derive the above formula we have used

$$\sum_{l=0}^{\lfloor \frac{n}{2} \rfloor - 1} N^{n-2l-1}(\theta)N^{2l+1}(\theta) = 0 \quad \forall n, \quad (2.63)$$

which in principle appears in the above Poisson bracket (for even $n$), multiplied by $\delta(\theta - \theta')$. Note that, as expected, the subspace of bosonic generators ($n$ even) forms an asymptotic symmetry subalgebra with respect to the above relation, and our results thus contain the results of [1] (observe the shift in $n$ operated in [1] in order to rewrite the commutator, which we have not performed here).

Let us now turn to computing our second 'low-higher' Poisson bracket, $\{Q(\theta), N^n(\theta')\}_{\text{PB}}$. The procedure is now assumed to be clear also for this kind of commutator, and we only quote the result, again forgoing the super-Virasoro sector:

$$i\{Q(\theta), N^n(\theta')\}_{\text{PB}} = +(-)^{n+1}(n+1)\delta(\theta - \theta')N^{n+1}(\theta)\delta_{|n|_2,1} \quad (2.64)$$
$$+ i(-)^{n+1}\delta'(\theta - \theta')\left(\tfrac{1}{n}N^{n-1}(\theta) + N^{n-1}(\theta')\right)\delta_{|n|_2,0}.$$

This is a completely new result in the sense that it does not appear at all in the bosonic analysis of [1], and in particular we notice that $Q$ indeed acts like the supercharge. Alternatively, upon particularizing to $n$ even or odd one may rewrite the above brackets more simply. We obtain:

$$\{Q(\theta), N^n(\theta')\}_{\text{PB}} = -\delta'(\theta - \theta')\left(\tfrac{1}{n}N^{n-1}(\theta) + N^{n-1}(\theta')\right) \quad (n\text{ even}), \tag{2.65a}$$

$$i\{Q(\theta), N^n(\theta')\}_{\text{PB}} = +(n+1)\delta(\theta - \theta')N^{n+1}(\theta) \quad (n\text{ odd}). \tag{2.65b}$$

The Fourier mode form of the above Poisson brackets is easily computed and turns out to be, for $s \geq 3$:

$$[\mathrm{L}_n, \mathrm{N}^s_m] = \left(\tfrac{1}{2}sn - m\right)\mathrm{N}^s_{n+m}, \tag{2.66a}$$

$$[\mathrm{Q}_n, \mathrm{N}^s_m\} = (s+1)\delta_{|s|_2,1}\mathrm{N}^{s+1}_{n+m} + \delta_{|s|_2,0}\left(n - \tfrac{1}{s}m\right)\mathrm{N}^{s-1}_{n+m}. \tag{2.66b}$$

**Higher-Higher Poisson Brackets and Extended Supersymmetry**

In order to gain further insight into the structure of our asymptotic symmetry superalgebra, we find it worth it to compute some of the 'higher-higher' Poisson brackets, i.e. brackets between two higher-spin generators (more than spin 2). Let us compute for example $\{N^3(\theta), N^3(\theta')\}_{\text{PB}}$ and $\{N^4(\theta), N^4(\theta')\}_{\text{PB}}$, the brackets among the spin-$\tfrac{5}{2}$ and spin-3 currents, $N^3 \equiv R$ and $N^4 \equiv M$. For the first one we find:

$$\{R(\theta), R(\theta')\}_{\text{PB}} = \frac{\alpha^3}{6}\delta'''' + \frac{\alpha^3}{12\alpha^6}(N^6(\theta) + N^6(\theta'))\delta - \frac{5\alpha^3}{6\alpha^2}(L(\theta) + L(\theta'))\delta''$$
$$+ \frac{3\alpha^3}{2(\alpha^2)^2}L(\theta)L(\theta')\delta + \frac{i\alpha^3}{6(\alpha^1)^2}Q(\theta)Q(\theta')\delta'$$
$$- \frac{\alpha^3}{3\alpha^2}(L'(\theta) - L'(\theta'))\delta', \tag{2.67}$$

and the second one turns out to be

$$\{M(\theta), M(\theta')\}_{\text{PB}} = \frac{\alpha^3}{24}\delta''''' - \frac{i\alpha^3}{2(\alpha^1)^2}(Q'(\theta)Q(\theta) + Q'(\theta')Q(\theta'))\delta'$$
$$+ \frac{2i\alpha^3}{3(\alpha^1)^2}Q(\theta)Q(\theta')\delta'' + \frac{\alpha^3}{6\alpha^6}(N^6(\theta) + N^6(\theta'))\delta'$$
$$- \frac{\alpha^3}{4\alpha^2}(L'(\theta) - L'(\theta'))\delta'' - \frac{5\alpha^3}{12\alpha^2}(L(\theta) + L(\theta'))\delta'''$$
$$+ \frac{\alpha^3}{(\alpha^2)^2}(L^2(\theta) + \tfrac{2}{3}L(\theta)L(\theta') + L^2(\theta'))\delta', \qquad (2.68)$$

and we recall that the numerical factors $\alpha^n$ are defined in (2.51). Note that, for the sake of readability, we have used the compact notation $\delta^{k\rfloor} \equiv \delta^{k\rfloor}(\theta - \theta')$.

As can be seen, the nonlinearities, which were absent in the 'low-higher' Poisson brackets, start appearing as soon as one considers 'higher-higher' commutators for conformal spins higher than 1 (spin higher than 2). Given the defining relations of a $\mathcal{W}$-algebra [4], it is clear from the above relations that our asymptotic symmetry algebra is a $\mathcal{W}_\infty$ algebra, but it is not clear a priori *which one* it is — as noted in [4], although the structure is quite rigid when only a finite number of currents are included, for an infinite number thereof one can have different structures. We shall dwell on these matters in Chapter 3, where we also formulate our other comments on the structure unveiled above and on its relation to the existing literature.

Before moving on to Chapter 3, where we discuss the results obtained so far, let us comment on the extended expressions corresponding to the $\mathcal{N} = 1$ relations above. In the $N \geq 2$ case the algebra $\text{shs}(1,1) \equiv \text{shs}(N,2|\mathbb{R})$ becomes $\text{shs}(N,2|\mathbb{R})$, and we thus need to generalize the boundary conditions of (2.35) to the case where our connection one-form $\Gamma_\mu$ lives in the extended higher-spin superalgebra. The details are relegated to Appendix C, where we exhibit again the procedure leading to the asymptotic symmetries, this time including extended indices. Let us mention, though, that the only added complication is the presence of the internal indices, labeling the Grassmann-odd oscillators $\psi_i$ of Subsection 1.2.3 and which account for the extended supersymmetry (see also Appendix B). Inspired by Reference [60] where, as explained in Subsection 2.1.3, the boundary conditions do not constrain the internal indices, and the extended version of the computation is really akin to the non-extended one.

Note that, for the sake of conciseness, in Appendix C we do not explicitly give the extended version of the above commutation relations, which could be obtained straightforwardly. However, we point out that the salient new features that arise are:

1. There are now fields $\Delta^{0;i_1,\cdots,i_N}$ of conformal dimension 1 (see Appendix C). These are the currents of the internal symmetry, and they form an affine subalgebra. Their brackets with the other generators reflect how these other generators transform under the internal symmetry.

2. In full analogy with the case of Supergravity, in the extended case we also see the appearance of additional nonlinearities involving the internal indices in the right-hand sides of the commutation relations, and a so-called Sugawara redefinition of the Virasoro generator $L$ must be performed, as already found in [60] for extended AdS$_3$ Supergravity.

3. There is more than a single current at each conformal dimension, and the degeneracy of each non-zero conformal weight is equal to $2^{N-1}$, while the degeneracy of conformal dimension 0 is $2^{N-1} - 1$ (internal currents). In particular, the Virasoro currents are no longer the only ones with conformal dimension 2.

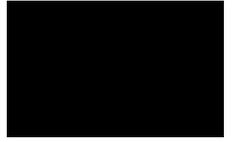

CHAPTER 3

# Discussion

Let us discuss our results and put them into perspective. Unless otherwise specified, the comments below apply equally well to the extended and non-extended case, and we often formulate them in the non-extended language.

**Summary of Results**

Exploiting the powerful realization of asymptotic symmetries as a Poisson-bracket algebra we have computed the asymptotic symmetry algebra of the three-dimensional Higher-Spin Theory based on the undeformed gauge superalgebra $\text{shs}(1,1)$, and we have highlighted the main differences which would endow the extended case for which the bulk gauge algebra is $\text{shs}(N,2|\mathbb{R})$. We find the symmetries at asymptotic infinity to be of the nonlinear, supersymmetric $\mathcal{W}_\infty$ type, also denoted nonlinear $s\mathcal{W}_\infty$, or even $s\hat{\mathcal{W}}_\infty$ [136]. We have explicitly computed the brackets among the spin-2 and spin-$\frac{3}{2}$ generators, which reproduce the superconformal Virasoro algebra with non-zero central charge(s). The bracket of the higher-spin current of spin $s$ with the spin-2 one as well as with the spin-$\frac{3}{2}$ current have also been explicitly worked out, and in Fourier modes they read, for $s > 2$,

$$[\text{L}_n, \text{N}_m^s] = \left(\tfrac{1}{2}sn - m\right) \text{N}_{n+m}^s, \tag{3.1}$$

$$\{\text{Q}_n, \text{N}_m^s\} = (s+1)\text{N}_{n+m}^{s+1} \qquad (n\,\text{odd}), \tag{3.2a}$$

$$[\text{Q}_n, \text{N}_m^s] = \left(n - \tfrac{1}{s}m\right)\text{N}_{n+m}^{s-1} \quad (n\,\text{even}). \tag{3.2b}$$

These are the defining features of a supersymmetric $\mathcal{W}_\infty$ algebra [4].

The nonlinearities start making their appearance at the level of the brackets involving two higher-spin generators, such as $\{R(\theta), R(\theta')\}_{\text{PB}}$ and $\{M(\theta), M(\theta')\}_{\text{PB}}$, which we have computed in (2.67) and (2.68). A priori, the s$\mathcal{W}_\infty$ superalgebra we have unveiled is thus of the nonlinear type, and we note that in the extended case the nonlinearities are present in the low-spin sector already [60]. There is an infinite number of Lie brackets we have not computed explicitly, so that we have not given all the structure constants. However, the procedure has been presented in great detail, thereby making the Lie bracket between any two asymptotic generators computable. We also note that, a priori, we have a central charge for each sector, and not only in the Virasoro subalgebra, which is a feature shared by the super-Virasoro algebra. To be precise, however, one really has only one central charge, in the sense that all central elements are parametrized by the value of the Brown–Henneaux one.

**Subalgebras and Truncations**

In order to gain insight into the structure of our s$\mathcal{W}_\infty$, let use explore its subalgebras and truncations. First of all, the super-Virasoro algebra is a subalgebra,[1] because we find the spin-2 and spin-$\frac{3}{2}$ generators to satisfy precisely the commutation relations of the superconformal algebra with Brown–Henneaux central charge in the Gravity sector. This might have been expected, since the $osp(1, 2|\mathbb{R})$ superalgebra is a subalgebra of $shs(1, 1)$, and we know the former yields the super-Virasoro algebra at asymptotic infinity [60] (and similarly for the extended case). However, the possibility of having nonlinearities in the right-hand sides of the commutation relations — in Appendix C.2.3 we prove that such is the generic situation — makes such expectations wrong in general. For example, as we shall comment on more below, although $hs(1, 1)$ is a subalgebra of $shs(1, 1)$, the $\mathcal{W}_\infty$ algebra found in [1] is not a subalgebra of our s$\mathcal{W}_\infty$. In light of these considerations, one might thus ask why it is that the super-Virasoro sector does form a subalgebra. The answer is that, by simple dimensional analysis, one can check that no nonlinear terms can appear in the right-hand side of the super-Virasoro commutators (in the non-extended case), except for $\delta(\theta - \theta')Q(\theta)Q(\theta')$, which can formally appear but is identically zero, as was noted in Subsection 2.1.3.

Let us then comment on the bosonic $\mathcal{W}_\infty$ algebra, found for the first time in [1] from the asymptotic symmetry standpoint. Having in mind the above reasoning, we now expect that the latter is not a subalgebra of our supersymmetric $\mathcal{W}_\infty$. This is indeed the case: because of the nonlinearities in the right-hand side of the commutators, the latter is not truly a subalgebra. However, the bosonic $\mathcal{W}_\infty$ algebra is nonlinear to start with, and one thus wonders what exactly spoils the 'subalgebra property' at infinity. One notices that the presence of nonlinearities allows, in particular, for fermionic generators (such as the supercharge $Q$) to appear on the right-hand side of a bracket between two bosonic ones, which would be impossible otherwise. In general, 'throwing away' generators spoils the Jacobi identity, but, because hs(1, 1) is a subalgebra of shs(1, 1), it might be felt that a 'miracle' should happen, namely that the Jacobi identity should hold good even with the nonlinear terms involving the fermionic generators brutally set to zero. This is precisely what happens. Note, however, that when comparing our spin-3 currents self bracket with the expression given in [1], we find a mismatch not only because of our nonlinearities involving the fermionic currents. This is because, in [1], the expression is given for the case when the hs(1, 1) algebra is truncated down to sl(3|$\mathbb{R}$).[2]

---

[1] In the extended case the analogous conclusion seems to be difficult to reach for a large number of supersymmetries.

Another natural question is that of $\mathcal{W}_N$ algebras, which have been obtained independently e.g. in [137] from sl($n|\mathbb{R}$) Chern–Simons models. Are they contained, in any sense of the word, in our structure? As explained in [1], the $\mathcal{W}_3$ nonlinear algebra of [138] is 'almost' a subalgebra of the bosonic $\mathcal{W}_\infty$, found therein. In fact, it is again not really a subalgebra, and one needs to brutally set to zero some nonlinearities in the right hand side of the bracket of two spin-3 ones for $\mathcal{W}_\infty$ in order to reproduce the $\mathcal{W}_3$ commutation relations. Therefore, as we just argued that $\mathcal{W}_\infty$ is not a subalgebra of s$\mathcal{W}_\infty$, $\mathcal{W}_3$ cannot possibly be a subalgebra of our supersymmetric s$\mathcal{W}_\infty$. It is clear, however, that by artificially setting to zero some terms in the right-hand side of the bracket (2.68) one reproduces the correct commutator for the $\mathcal{W}_3$ structure. What about $\mathcal{W}_N$ algebras for $N \geq 4$? As explained in [1], the procedure cannot be repeated, and the reason is that the 'miracle' which happens for $N = 3$, that sl(3|$\mathbb{R}$) can be obtained by truncating hs(1, 1), does not hold in general for sl($N|\mathbb{R}$). Therefore the corresponding asymptotic statement does not hold either, and one cannot truncate $\mathcal{W}_\infty$ to obtain $\mathcal{W}_N$ at $N = 4$ or greater. In the supersymmetric case we should expect a similar conclusion with respect to the supersymmetric $\mathcal{W}_3$ algebras, studied e.g. in [139] as asymptotic symmetries of models with sl($N|N-1$) gauge invariance. Namely, although the latter algebras (see also [140, 141]) are not true subalgebras of our s$\mathcal{W}_\infty$, it should be that by truncating artificially

---

[2] sl(3|$\mathbb{R}$) is *not* a subalgebra of hs(1, 1), but one can nevertheless truncate the latter to obtain the former [1].

the commutation relations obtained above one recovers the relations of s$\mathcal{W}_3$.

Let us now comment on the so-called *wedge* subalgebra of s$\mathcal{W}_\infty$. The bulk gauge algebra is the 'local version' of shs$(1,1)$ (the coefficients depend on spacetime coordinates). The vacuum solution is AdS$_3$, which can be seen to be left invariant by the rigid part of the gauge algebra, that is, by the constant gauge parameters $\Lambda \in$ shs$(1,1)$. By definition, such parameters close to shs$(1,1)$, which is thus the isometry of the vacuum. Now, one always expects the isometry part to be included in the asymptotic symmetries.[3] This is the case in our analysis, but it is not obvious, since the $\Lambda = \Lambda_0$ constant gauge parameter is *not* a particular case of the general relation among the components of the residual gauge parameters, found in (2.15). The reason is that the expression (2.15) corresponds to the background in a non-zero form, and in Appendix C.2 we detail this feature and explain that the exact background symmetries, when considered in the correct gauge, correspond to the part of the asymptotic generators forming the so-called *wedge* sector, that is, the generators $N_n^s$ with $|n| \leq s-1$. For example, in the low-spin sector it would correspond to $Q_{-1/2}, Q_{1/2}$ and $L_{-1}, L_0, L_1$, which can indeed be seen to close to the osp$(1,2|\mathbb{R})$ algebra, forming a subalgebra of the superconformal algebra.

The superconformal algebra is linear (it is a true algebra), so that the wedge sector forms a subalgebra thereof. However, in the higher-spin case of shs$(1,1)$, the wedge part does not form a subalgebra, again because of the nonlinearities. It is easy to see that the linear piece of the commutation relations indeed reproduces the shs$(1,1)$ superalgebra. Furthermore, the central charges all vanish when restricted to this sector. We thus say that the wedge sector closes to shs$(1,1)$ *up to nonlinearities*. This is the precise sense in which the exact background symmetries are embedded into the asymptotic symmetries. In fact, in order for the aforementioned identification to work properly one needs to further set the generators to zero when evaluated on the AdS connection. This is what is done in (2.27), where the rescaling of $L_0$ contributes to the central-charge piece so to make it vanish when restricted to the wedge. Note that, as explained in [143], it is really the fact that we are looking at an exact symmetry of the background which ensures this property. Let us further note that the existence of the wedge subalgebra as well as its relation with the exact symmetries is something that was noticed previously, e.g. in [121, 144], and the fact that it reproduces the exact symmetries is a general result known within the so-called Drinfeld–Sokolov reduction [145, 146] (see also [147]). Here we have provided a 'geometrical' proof within the asymptotic symmetry analysis.

---

[3] In recent investigations, asymptotic symmetries have been obtained among which one does not find all the bulk isometry generators [142].

Our last digression on subalgebras will be about *linear* $\mathcal{W}_\infty$ (super-) algebras. As we explain below, in the early days of $\mathcal{W}$-Symmetry, $\mathcal{W}_\infty$ algebras and superalgebras were constructed that were linear [121, 145, 148, 149] (see also [4] and references therein). Those are not subalgebras of our nonlinear structures, of course, because the latter are nonlinear. More importantly, the linear part of our commutation relations does not reproduce the linear brackets obtained previously (except inside the wedge). Interestingly, the cause of the latter discrepancy can be argued to be the presence of central charges in our realization of asymptotic symmetries. Indeed, without the latter, linear terms in the Jacobi identity for our Poisson-brackets would only receive contributions from the linear part of the commutation relations. This would mean that, upon removing all the nonlinear pieces from the right-hand sides of our commutators the Jacobi identity would boil down to its linear part. Now, previously-found linear $\mathcal{W}_\infty$'s were precisely obtained by solving the Jacobi identity for a linear Ansatz,[4] and the solutions were argued to be unique. Therefore, we would expect that throwing away all nonlinearities would reproduce precisely those structures. The presence of central charges crucially spoils this reasoning: the linear part of the Jacobi identity equation receives contributions from the nonlinear part of the commutation relations! The nonlinear pieces are essential for the closing of the Jacobi identity, and as we have explained there is a non-trivial interplay between nonlinear terms and central charges.

At this point, and given the added complications brought in by the nonlinearities, one might wonder whether quadratic nonlinear deformations are the only type of deformations that can occur. In Appendix C.2.3 we show that this is not the case: quadratic nonlinearities are not restricted by general arguments, and nor are higher-order ones. We further comment on the nonlinear terms in the paragraphs below.

**Deformations and Other Higher-Spin Theories**

In the main text we have been concerned with undeformed Higher-Spin Theory or, differently put, undeformed higher-spin algebras ($\lambda = \frac{1}{2}$). So-called *deformed* algebras also exist, at least in the bosonic case: they are noted hs[$\lambda$]. As explained in the main text, the supersymmetrization of the undeformed cases goes without much trouble — it is the case we have treated explicitly —, with or without extended supersymmetry, but the deformed bosonic algebras hs[$\lambda$] are less easily deformed (at least for $\mathcal{N} > 1$). Regardless of these issues, we argue that the deformed equivalent of $\mathcal{W}_\infty$ is also a nonlinear $\mathcal{W}_\infty$-algebra. This was checked explicitly in the bosonic case in [137, 147], and for the supersymmetric $\mathcal{N} = 1$ deformed case in [150]. This opens the door to a supersymmetric version of Minimal Model Holography, and it has been conjectured recently that the holographic equivalents thereof should be the so-called Kazama–Suzuki models in the large-$N$ limit [81, 83–86, 89, 90] (see also [6]).

---

[4] Another way of obtaining them was via the $N \to \infty$ limit of $\mathcal{W}_N$-algebras, at the same time performing some rescaling.

In this work we have investigated three-dimensional higher-spin supergravities extending Supergravity theories of the most standard type, namely those which are based on the osp($N, 2|\mathbb{R}$) $\oplus$ osp($N, 2|\mathbb{R}$) superalgebra. However, as stated in the text, one may think about higher-spin theories extending other types of (extended) supergravities, the list of which can be found in Appendix B.2.2. Indeed, it would in principle suffice to proceed again along the lines of the universally enveloping technique, which yields shs(1, 1) starting from osp(1, 2|$\mathbb{R}$) (for the undeformed case), but considering as the starting point one of the Lie superalgebras of Table B.2.2. This can always be done, but one might further wonder whether an oscillator realization is available for all cases. The answer is positive: as explained in [151, 152], all the superalgebras of AdS$_3$ extended Supergravity theories can be realized in terms of quadratic combinations of oscillators, along the lines used to build osp($N, 2|\mathbb{R}$) in the bulk of this work. One then obtains the associated Higher-Spin superalgebra by relaxing the condition of being quadratic in the oscillators, thereby allowing for higher-degree polynomials. The asymptotic superalgebra would again be of the s$\mathcal{W}_\infty$ type. Indeed, the general structure of the commutators would remain the same, so that at infinity one would recover the usual action of the Virasoro generators on the higher-spin ones. Furthermore, as we have explained above, the (super-)Virasoro sector would remain intact.

Finally, we also note that in three dimensions we can perform the same analysis for flat-space higher spins, thereby obtaining a different asymptotic symmetry algebra [153, 154], which should be related to some $\mathcal{W}$ structure by contraction.

**Comparison with Past Approaches and Nonlinearities**

The existence of algebras of the $\mathcal{W}_\infty$ type is not a novelty. In fact, in the late eighties and early nineties they were under intense investigation, which started with the seminal paper of Zamolodchikov [138], where the Virasoro algebra was first extended to include generators of higher conformal spin (for a review we refer to [4]). It is not their appearance, but rather their *re*appearance in the higher-spin context which is appealing, and which was first noticed in two triggering papers in 2010 [1, 2]. Accordingly, there is quite an amount of 'old' literature to be found about $\mathcal{W}$-algebras in general, dating from the turn of the last decade of the past century, and one should feel compelled to compare the structures found to govern the asymptotic dynamics of three-dimensional higher-spin models with those unveiled in the years following the publication of [138]. Indeed, although it is clear that our asymptotic symmetries form a s$\mathcal{W}_\infty$ algebra it is not clear a priori *which* s$\mathcal{W}_\infty$ structure that is (as explained in [4] when infinitely many currents are considered there is some freedom in determining the structure). We now attempt at carrying out such a comparison. Needless to say, the following comments should not be expected to be exhaustive in any way.

When investigating the older literature on $\mathcal{W}$-algebras, the first thing one notes is that the presence of nonlinearities in the right-hand side of the commutation relations is mostly confined to $\mathcal{W}_N$-like algebras, possessing only a finite number of higher-spin currents [4]. Indeed, in the finite-$N$ case such nonlinearities are unavoidable when going beyond the Virasoro sector, whereas they can be avoided by considering an infinite number of higher-spin currents. This can be intuitively understood in a simple manner: consider some (nonlinear) $\mathcal{W}_N$ algebra and declare any quadratic term in the right-hand side of the commutation relations to be some higher-spin current of higher conformal spin. In such a way one evidently linearizes the algebra, but a number of currents are added to the spectrum, and further commutation relations involving the latter are thus to be considered. The procedure, if it succeeds in the closing of all the commutation relations, never terminates and one ends up with a linear Lie algebra, but it now contains an infinite tower of higher-spin currents. This linearization by addition of an infinite number of currents was implemented by taking the limit $N \to \infty$ of $\mathcal{W}_N$-algebras, and it was noticed that upon performing non-trivial rescalings, the nonlinearities are lost in the limit.

Such observations do not underline an obstruction to building nonlinear $\mathcal{W}_\infty$-algebras — we just did it ! —, and if one takes the limit of $\mathcal{W}_N$-algebras in a naive way the nonlinearities are typically present. Nevertheless, the past literature on *nonlinear* $\mathcal{W}_\infty$-algebras is scarce, and only a few considerations can be found in [136, 155–166]. The simple reason for such a lack of interest is that of intricacy; nonlinear deformations of Lie algebras are much more hardly constructed and dealt with, and do not benefit from any mathematically well-developed theory allowing to tackle their structure. In particular, the construction of representations is made highly non-standard by the nonlinear terms, but perhaps another reason for the lack of work on nonlinear $\mathcal{W}_\infty$-like structures is the apparent absence of the following key property which they were expected to have: one hoped that some nonlinear $\mathcal{W}_\infty$ could be built such that it would contain all $\mathcal{W}_N$-algebras upon truncation or contraction (linear $\mathcal{W}_\infty$-algebras were not suited for that purpose because they could hardly reproduce the nonlinear part of the $\mathcal{W}_N$-brackets). Such so-called 'universal' algebras could be useful, but the conclusions of the program consisting in their construction were seemingly mitigated, regardless of Supersymmetry, although nontrivial relations were found to exist among different (linear and nonlinear) $\mathcal{W}_\infty$ structures [136]. One should note, though, that the quantum $\mathcal{W}_\infty$-algebra unveiled in [80] is universal precisely in this sense: it is nonlinear and it can be truncated to finitely-generated $\mathcal{W}_N$ algebras for any integer $N$.

Furthermore, if references dealing with nonlinear versions of $\mathcal{W}_\infty$-algebras are rare, the supersymmetric equivalents are even harder to find, and to our knowledge the exceptions include [166]. It is the author's opinion that such lack of investigation, thought of as originating in the difficulty in building them, adds importance to our result. Indeed, in our approach the algebra is guaranteed to satisfy the Jacobi identity (because it is a Poisson-bracket algebra), and any commutator can be straightforwardly computed. Moreover, as we have said above the presence of nonlinearities is rather independent of the particular bulk gauge algebra one chooses to work with. Now, given the variety of higher-spin algebras one can conceive of (one can at least universally envelop all the extended Supergravity algebras), this implies the existence of a vast landscape of nonlinear (supersymmetric) $\mathcal{W}_\infty$-like algebras. In a pinch, the asymptotic symmetry machinery, when viewed as a procedure for constructing consistent $\mathcal{W}$-algebras, seems rather powerful, if not convenient. The comparison of the structure obtained in this work with previously-found nonlinear, supersymmetric versions such as those of [166] should be interesting. Let us further note that, in mathematical terms, the asymptotic procedure for determining symmetry algebras is just a rephrasing of the well-known Drinfeld–Sokolov reduction procedure [4, 167].

**Other Related Topics**

Many more comments can be made on our findings, and relations with many other areas of physics exist. However, the scope of the present work must be finite and we thus refer the reader to [61], among others, for additional comments on black holes with $\mathcal{W}_\infty$ hair, String Theory embeddings, Self-Dual Gravity in dimension four, $\mathcal{W}$-strings, the Drinfeld–Sokolov reduction procedure and other topics (see also [6]).

# Part II

# Dimension D

*It is an error to believe that rigor is the enemy of simplicity.*
*On the contrary we find it confirmed by numerous examples that*
*the rigorous method is at the same time the simpler*
*and the more easily comprehended.*
*The very effort for rigor forces us to find out simpler methods of proof.*

David HILBERT

# Invitation

In Part I of this thesis we have lived in spacetime dimension three which, as often, has proved to be a fruitful yet workable laboratory for the study of Gravity and related issues [66]. However, as we have seen, the three-dimensional models lack some properties which crucially characterize higher-spin theories in dimension four or greater. The central difference is that of interactions: while in dimension $D \geq 4$ interacting higher spins are very much constrained [52], in three dimensions one has a whole variety of consistent, non-linear higher-spin theories to choose from. In order to fully understand the dynamics of particles with spin larger than two it is therefore mandatory to address the higher-spin problem in dimension four at least, which is what this second part of the text is devoted to.

Nowadays, most of the research investigating higher-spin couplings is carried out on AdS backgrounds where, among other things, one wishes to gain insight into the structure of the Vasiliev system and its three-dimensional conformal dual. Moreover, as the many no-go theorems for interacting higher spins concern flat spacetime [42–45, 168], it is natural to investigate constant-curvature backgrounds, where fully consistent gauge theories are known. Nonetheless, in the present work we shall be concerned with flat spacetime propagation and interaction of higher-spin gauge fields. On the one hand, despite the no-go results precluding the existence of a fully consistent theory of higher spins on flat spacetime (at least when locality and finiteness of the spectrum are insisted on), one feels inclined to exploring precisely to what extent that is true and what exactly the obstruction is. On the other hand, the perspective of an interplay between Higher-Spin Theory and String Theory — see e.g. [25, 169, 170] —, which contains (massive) higher spins and is under best control on Minkowski spacetimes, calls for an understanding of flat-space higher-spin interactions as refined as possible.

With the latter motivations in mind it seems one cannot dispose of fermions. Indeed, they are required by Supersymmetry, which in turn is instrumental in proving String Theory to be consistent. Also, besides Supersymmetry, fermions are present in nature (although they are massive), so that studying higher-spin theory with fermions should be interesting at any rate. An understanding of fermionic higher-spin gauge interactions in Minkowski spacetimes is thus necessary, and should complement the bosonic study, typically carried out first. In [63, 64] the electromagnetic and gravitational cubic couplings of higher-spin gauge fermions were analyzed in detail, thus partially filling a gap in the literature, which so far had been dealing primarily with bosons — with [25, 62, 171] among the exceptions. Those are the results which this second part is concerned with, namely, $1-\frac{n}{2}-\frac{n}{2}$ and $2-\frac{n}{2}-\frac{n}{2}$ gauge-invariant couplings for arbitrary $n$ in flat spacetime of generic dimension $D \geq 4$. All such couplings are derived in a systematic way in Chapter 5 and 6, and we obtain explicit, neat and off-shell forms for them in the metric-like formalism. The methods we will employ to achieve the classification and exhibition of our vertices are the so-called BRST-Antifield techniques, and they are recalled in Chapter 4.

Our results will be seen to be in agreement with the String Theory-inspired expressions obtained in [25, 171] as well as with the Light-Cone analysis of [62]. However, we derive our off-shell couplings in a totally independent manner; we take no input from any other work and make only minimal hypothesis. Also, in line with the various no-go theorems [42–45, 168], we prove the obstruction to cubic minimal coupling and also to making a theory involving non-minimal couplings consistent to second order in perturbation theory. More precisely, full consistency will be seen to be obstructed if one insists on locality and on the original spectrum. Finally, we comment on the link between the obtained interactions and various topics, such as $\mathcal{N}=2$ Supergravity, String Theory, massive higher spins, non-locality, AdS backgrounds, bosonic fields, etc. They are found, together with a summary of our results, in Chapter 7.

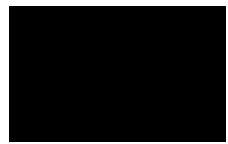

CHAPTER 4

# BRST Techniques

In Chapters 5 and 6, the techniques used to find out and classify interaction vertices are the so-called BRST–BV[1] ones. This formalism, which we originally owe to Bechi, Rouet, Stora, Tyutin as well as to Batalin and Vilkovisky was first discovered in a quantum setup [172–175], and it was realized only later on that one could use it also at the classical level to consistently search for deformations of given gauge theories [176–179], which is the application we are interested in and present below. Also, as shall be made clear, it is not the most general case that shall be recalled, but that which considers as the starting point a *free* theory. The literature on the BRST-Antifield reformulation of the deformation problem is nowadays somewhat extent, and includes in particular the very good review [180]. We also point out the algebraically and geometrically oriented lectures [181], the report [182] as well as the comprehensive book [103], which all go beyond the scope of the present introduction.

The following guide to the BRST-Antifield formulation is meant to be pedagogical. However, as it is only given here so that the reader can follow the next chapters, many demonstrations and historical considerations have been left out, so that the reader might sometimes feel it to be a little *ad hoc*. Nonetheless, although meant to be but a mere *vade mecum* for understanding the subsequent investigations, we have tried to make the naturalness of the formalism all the more salient by means of a step-by-step structure, further complemented by qualitative arguments. We hope this shall help the reader to grasp the essence of the BRST reformulation of the deformation problem as well as its beauty.

---

[1] Also called BRST-Antifield, or simply BRST — which we shall use interchangeably.

## 4.1 A Rationale for the Noether Procedure

Let us begin by some more introductory words. The skeptical or unfamiliar reader might be wondering what the need is for the formalism presented in Section 4.2. Indeed, if our goal is to find out and classify consistent vertices which may deform a given free theory, why do we not simply use the (more) familiar Noether method ? The answer is: because we can do better. Indeed, as we shall see the BRST–BV framework is basically a reformulation of the Noether procedure, but it does so in a way that presents many an advantage.[2] In a nutshell, the advantage is that the off-shell search for interactions is made more systematic by relating the consistent deformations, as well as the obstructions (including second-order ones), to cohomology classes of nilpotent operators acting on an enlarged space of fields. By so formalizing the problem one can use the powerful tools of homological theory — a well developed area of mathematics — which allow one to severely constrain the search. Also, as we shall see, many of the mathematical objects and cohomology classes introduced below actually have either a geometrical or a physical meaning.

Moreover, practically speaking an important upside is the possibility of addressing the problem of consistently deforming a free theory *backwards*. What we mean is the following: suppose one finds a tentative vertex, to be added to some free theory. If the said vertex is non-abelian, then it deforms the original, abelian gauge transformations (of the free theory), and the deformed ones thus contain terms of first order in the coupling constant when compared with the original ones. Then, one needs verify whether or not the deformed gauge transformations close to some deformed gauge algebra. If they do, then one needs to determine the algebraic structure thereof, and the procedure needs be repeated for every putative vertex. In the BRST setup, one does the exact opposite, namely, the formalism makes it very natural for one to start with the classification of the possible deformations of the gauge algebra. The advantage then is that the possible deformations of the gauge algebra are much constrained, and in our formalism those conditions are easily stated and translated to precise cohomological statements, so that we are assured to be exhaustive. Then, one needs only follow a systematic procedure — the so-called *consistency cascade* — based on solving cohomology equations in order to find out whether the tentative algebra deformation is consistent or not, and the obstructions to consistency are again related to precise cohomology classes. If the algebra deformation is consistent, the procedure also yields the corresponding gauge-symmetry and Lagrangian deformations.

---

[2] Although we try and argue our way to this point, it should nevertheless be said that this is merely the author's opinion, and others might disagree.

Another way of phrasing what the cohomological approach does for us is the following. The difficulty in the problem of constructing non-abelian vertices can be traced back to two simple facts. The first fact is that one poses two questions at the same time and seeks a common answer, that is, we look for a vertex which will possibly deform the Lagrangian and the gauge transformations at the same time. The second fact is that one is interested in gauge-invariant deformations but should also take into account the redundancy brought in by the possibility of performing field redefinitions, which is not necessarily a simple task. As we shall explain, the BRST reformulation cleverly deals with both these difficulties. On the one hand, instead of using the standard (free) action it constructs a so-called (free) *master action*, which on top of the original action contains terms with explicit information about the gauge transformations. That 'unified' object is what we then deform, thus naturally dealing at the same time with the problem of deforming the Lagrangian as well as the gauge transformations in a consistent way. On the other hand, and perhaps more importantly, a so-called BRST operator is constructed which implements at the same time the gauge transformations and the field redefinitions, and relates them to precise cohomology classes. Thus, one is ensured to have used all the freedom granted by field redefinitions when classifying different gauge-invariant vertices. One says that the field redefinitions are fully accounted for in this way when passing to the cohomology of gauge-invariant quantities.

Evidently, all these advantages come at a price, namely, that of having to introduce 'auxiliary' fields (so-called *antifields* and *ghosts*), thus much enlarging the original phase space. Such is, however, the philosophy: in this enlarged phase space, to be defined below, it will be possible to precisely reformulate many statements and properties such as gauge invariance and on-shell triviality into cohomological considerations. In particular, the usual properties of consistency and non-triviality will be related to the two familiar aspects of cohomological calculus: computing the kernel of a nilpotent operator as well as the trivial part therein. In a first approach, the formalism presented here may therefore seem excessive, but we nevertheless think it not only extremely useful but also, as aforementioned, quite natural in fact. As a final word let us further stress that the formalism is intrinsically off-shell in spirit, and no gauge-fixing is required nor implied.

## 4.2 BRST and Antifields

Let us now enter the details of the aforementioned scheme. As aforementioned, we restrict ourselves to presenting the reformulation in the case where the theory we start with is free, which is what we need in the next chapters, but in principle the same framework can be used to address the problem of deforming a theory which is not free — see e.g. [181]. We point out that the main added difficulty in the latter case[3] is that the BRST differential will no longer simply be the sum of the Kozul–Tate differential plus the longitudinal differential along the gauge orbits, to be defined below.

Another assumption we make is that of irreducibility of the free gauge theory, which again is the case we encounter when addressing free higher spins — see next chapters. This means that we start from a free, irreducible gauge theory of a collection of fields $\{\phi^i\}$, with $m$ gauge invariances

$$\delta_\varepsilon \phi^i \equiv R^i_\alpha \varepsilon^\alpha, \quad \alpha = 1, 2, \ldots, m, \tag{4.1}$$

which leave the free action $S^{(0)}[\phi^i]$ invariant. Evidently, the $R^i_\alpha$ may be (and usually are) differential operators. We shall present the formalism without any assumption of the dimensionality of spacetime, and such will also be the spirit of the following chapters, where interactions are classified in generic dimension $D$.

We shall now proceed to introducing seven steps of formalism, which shall set up in a simple way the essential tools we will need to address the consistency of the couplings. The seven steps are the following:

**Step 1:** Replacing Gauge Parameters by Ghosts ,

**Step 2:** Introducing Antifields to Source Gauge Variations,

**Step 3:** Implementing Gauge Variations via $\Gamma$ Operator,

**Step 4:** Implementing Field Redefinitions via $\Delta$ Operator,

**Step 5:** Combining $\Gamma$ and $\Delta$ into the BRST Differential $s$,

**Step 6:** Finding Deformations via Consistency Cascade ,

**Step 7:** Second-Order Consistency and the Antibracket.

---

[3] More precisely, the BRST differential will contain additional pieces when at least one of the following characterizes the starting theory: the gauge symmetries are not abelian or they form an *open* algebra [181].

In Step 1 and 2 we enlarge the phase space of our original fields $\phi^i$. Step 1 replaces gauge parameters by ghosts, which are added to the configuration space, whereas Step 2 introduces the so-called *antifields*, which source the gauge invariances in some generalized action called the *master action*. Then, in Step 3 and 4 we reformulate two important concepts in cohomological terms. The first one, dealt with in Step 3, is gauge invariance, and it will connect with Step 1. The second, addressed in Step 4, is field redefinitions, and it will relate to Step 2. Step 5 then combines both reformulations of these concepts into a single, unified operator: the BRST differential, conveniently implementing both the EoMs and the gauge symmetries. Once the correct operator has been identified, in Step 6 we explain how to search for consistent interactions in this formalism, exploiting the so-called *consistency cascade* and the possibility of classifying potential deformations at the level of the algebra. Finally, Step 7 is concerned with second-order consistency and quartic vertices. In the latter, we shall discover the so-called *antibracket*; a symplectic structure on our enlarged phase space which not only allows for an easier analysis of second-order consistency but also for a rather geometrical reformulation of the deformation problem in general, which we shall briefly touch upon.

As announced earlier, although we shall be concise and practical we shall nevertheless try to give the reader a feel of why we think this formalism is quite the natural one.

### Step 1: Replacing Gauge Parameters by Ghosts

The first step is merely a rewriting, almost a relabeling: each gauge parameter $\varepsilon^\alpha$ is replaced by a corresponding ghost field $\mathcal{C}^\alpha$, so that the gauge transformations (4.1) now read

$$\delta\phi^i \equiv R^i_\alpha \mathcal{C}^\alpha, \quad \alpha = 1, 2, \ldots, m. \tag{4.2}$$

However, $\mathcal{C}^\alpha$ is now declared to have the same algebraic symmetries but opposite Grassmann parity as $\varepsilon^\alpha$. This means that, e.g. if some gauge parameter is bosonic (as for example that of a spin-1 gauge field), the corresponding ghost is Grassmann odd, and vice versa. Calling them ghosts is therefore appropriate!

**Remark** : in this reminder we do not spell out spacetime indices, and the index $\alpha$ in $\mathcal{C}^\alpha$ is accounting for the different ghosts (replacing the different gauge parameters). As no assumption is made on the spin of the fields and their associated gauge parameters, it should be kept in mind that each of the latter can have spacetime indices, but we treat them all generically.

When we say that the ghosts have the same algebraic symmetries as the original fields they correspond to, we are referring to those spacetime indices.

The ghosts are now included in the phase space, thus enlarging that of the original fields, and all of them are sometimes collectively also called fields, which we denote by $\{\Phi^A\} \equiv \{\phi^i, \mathcal{C}^\alpha\}$. To be able to keep track of the nature of each of the fields we further introduce a grading, called the *pure ghost number*, defined to be 0 for the original fields and 1 for the ghosts:

$$\mathrm{pgh}(\phi^A) \equiv 0, \tag{4.3a}$$

$$\mathrm{pgh}(\mathcal{C}^\alpha) \equiv 1. \tag{4.3b}$$

The reason why we define the ghosts to be of opposite Grassmann parity to the corresponding original fields will be made clear in Step 3, where an operator $\Gamma$, implementing the gauge variations is built. Anticipating a little, and from a pragmatic standpoint, we might say that the parity properties of the ghosts are chosen so to make $\Gamma$ nilpotent of degree two, that is, $\Gamma^2 = 0$, which is the key property allowing us to define an associated cohomology (see below)

### Step 2: Introducing Antifields to Source Gauge Variations

We now want to define the so-called (free) *master action*, corresponding to the original, free action $S^{(0)}[\phi^i]$. As we anticipated in the previous section, the point is to build some generalized action which, on top of containing information about the (original) Lagrangian will also contain explicit information[4] about the gauge transformations. It is such a free master action, denoted $S_0$ (note the subtle change in notations), which we will try to deform later on. The idea is the following: in addressing the deformation problem for non-abelian vertices, one is not looking for deformations of the Lagrangian invariant under the original gauge transformations. Rather, one needs to allow for the gauge transformations to get deformed too. One virtue of the master action is that it contains explicit information about both these aspects, and the deformation problem, when formulated in terms of it, will automatically take into account both these features in a

---

[4] We use the word *explicit* for the following reason: given some standard (free) action one can always work out the corresponding gauge symmetries, so that this information is already contained in the action functional, although in an implicit manner.

way which ensures consistency and exhaustivity. The free master action is defined as follows:

$$S_0 \equiv S^{(0)}[\phi^i] + \int \mathrm{d}^D x \, \phi_i^* R_\alpha^i \mathcal{C}^\alpha, \qquad (4.4)$$

where the $\phi_i^*$ are the so-called *antifields*, which are seen to play the role of sources for the gauge variations in the master action.

The configuration space is thus further enlarged by introducing, for each field (original fields and ghosts), an antifield $\Phi_A^*$, which is again defined to have the same algebraic properties as $\Phi^A$ but opposite Grassmann parity (which correctly makes the above master action Grassmann even). Thus, our phase space now is given by $\{\Phi^A, \Phi_A^*\}$, where $\{\Phi_A^*\} \equiv \{\phi_i^*, \mathcal{C}_\alpha^*\}$. Note that, in Step 1, the gauge parameters are *replaced* by ghosts and those are then added to the phase space. Here, rather, we *supplement* the phase space with antifields corresponding to the fields.

Note that the antighosts do not enter the above master action, and at this stage one can think of them as being added too for the sake of democracy — their role will be clarified in the sequel. As for the antifields, besides sourcing the gauge variations in the master action, in Step 4 we shall see that they have another role to play. However, for the moment let us be content with the explicit presence of information about the gauge symmetries in the master action. Our simple rule will be that, whatever multiplies the antifields in the master action is the gauge transformation of the corresponding field. This will be of much use when deforming the free theory.

**Remark** : a word of caution about the interpretation of the above action should be added. For the unfamiliar reader, it might be tempting to consider the antifields and the ghosts as auxiliary fields in the usual sense of Field Theory. The corresponding paradigm is that the EoMs which follow from variating the action with respect to those auxiliary fields allow one to solve for them, hence integrating them out when plugging their algebraic expressions (in terms of the dynamical fields) back into the action. However, such is not the way in which one should understand the master action. Rather, the latter is really a tool, allowing to keep gauge invariance (and other things, as will be explained) under control, and should by no means be thought of as a standard action.

Finally, as we have introduced new (anti-)fields we need a new grading to keep track of who is who in the new, enlarged configuration space of all the fields and antifields. We thus define the *antighost number* as

$$\mathrm{agh}(\Phi^A) \equiv 0, \tag{4.5a}$$

$$\mathrm{agh}(\Phi_A^*) \equiv \mathrm{pgh}(\Phi^A) + 1. \tag{4.5b}$$

Also, we need to extend the definition of the pure ghost number to the antifields, and the correct definition is $\mathrm{pgh}(\Phi_A^*) \equiv 0$.

### Step 3: Implementing Gauge Variations via $\Gamma$ Operator

As explained at the beginning of the present section, this third step has to do with Step 1. What is done is to define an operator, $\Gamma$, implementing the (free) gauge variations on our enlarged phase space. The definition is the following:

$$\Gamma \phi^i \equiv R_\alpha^i \mathcal{C}^\alpha, \tag{4.6}$$

that is, $\Gamma$ is the 'longitudinal derivative along the gauge orbits' [103]. Now, from the above definition one sees that $\Gamma$ must be Grassmann odd, and from there it is found that $\Gamma^2 = 0$, and hence that $\Gamma \mathcal{C}^\alpha = 0$. To finish implementing the gauge variation, one needs to further define the action of $\Gamma$ on the antifields and the antighosts, and the correct definition is $\Gamma \Phi_A^* \equiv 0$. We point out that one can also check the nilpotency of $\Gamma$ by acting twice on any of the fields and directly obtaining zero.

As for every nilpotent operator (of degree two), it is natural to consider the *cohomology* of $\Gamma$, $\mathrm{H}(\Gamma)$, that is,

$$\mathrm{H}(\Gamma) \equiv \{X \in \text{phase space} \mid \Gamma X = 0, \, X \neq \Gamma Y\}. \tag{4.7}$$

The physical interpretation is clear: the cohomology of $\Gamma$ is the set of gauge-invariant combinations that are not themselves gauge variations of something else ('pure gauge', one might say). With this definition we are really beginning our formalization of the properties that are crucial for us. Indeed, all the above definition does is formalize the definition of what 'being an observable' means — to be refined in Step 5. We cannot possibly stress enough that this approach is at the core of the present reformulation, and in fact it is precisely the idea of associating physical quantities with cohomological classes of nilpotent operators which the BRST approach put forward for the first time [172–174]. Note that $\Gamma$ has pure ghost number equal to 1 but leaves the antighost number unchanged.

Before moving on to Step 4 we should add a few words about terminology and notation. In the language of cohomology, a combination which is annihilated by some operator is said to be *closed*, and one that can be expressed as the application of the operator to some other quantity is said to be *exact*. An equivalent way of phrasing things is to declare any closed object a *cocycle* and any exact one a *coboundary*. We shall switch back and forth between both terminologies, although we prefer the former. The cohomology of $\Gamma$, for example, will be said to be the space of all $\Gamma$-closed combinations which are not $\Gamma$-exact. Also, a $\Gamma$-exact element shall sometimes be said to be *trivial* in the cohomology, and therefore we shall sometimes drift towards the standard abuse of terminology according to which 'being in the cohomology' is understood as being $\Gamma$-closed and $\Gamma$-exactness is expressed as being trivial in $H(\Gamma)$. We believe, however, that confusion is unlikely for the attentive reader.

Let us also comment on the content of $H(\Gamma)$. Indeed, the reader may now be wondering about the fact that, manifestly, on top of familiar gauge invariant quantities built out of the original fields (such as the curvatures, when our original fields are photons or gravitons), quantities such as combinations of the antifields and antighosts also belong to the cohomology of $\Gamma$. As such, and given our interpretation of it, those quantities — e.g. $\phi_i^*$ itself — should be declared to be gauge invariant objects or 'observables', but they have no clear interpretation in terms of our familiar fields. Therefore, when we shall solve for the first-order deformation of the free master action by requiring it to be gauge invariant (up to field redefinitions), those quantities will be part of the solution. However, as we have seen, the master action gives a role to the antifields as well, and hence those combinations containing them shall simply be the part of the deformation which concerns the gauge symmetries, whereas the quantities containing the original fields only shall be the deformations of the original, free Lagrangian. Furthermore, as shall be made explicit below, such a first order deformation of the free master action will have to obey certain restrictions at the level of the gradings we have introduced, so that *any* quantity shall not even be acceptable a priori.

With this being said, for the sake of completeness we should probably comment on the role of the antighosts, but we shall simply anticipate here that, much like the terms proportional to the antifields in the master action are understood as being the gauge symmetries, the terms multiplying the antighosts shall be seen to correspond to the deformations of the gauge algebra! Note that this fits quite nicely with the fact that the *free* master action $S_0$ does not contain any such terms, for the algebra of the free theory is abelian.

**Step 4: Implementing Field Redefinitions via $\Delta$ Operator**

Step 3 had to do with Step 1, and the present step has to do with Step 2, that is, with the antifields. Indeed, so far we have dealt with the gauge invariance, and we further need to address the freedom granted by field redefinitions. Again, this will be done by the introduction of an odd operator, this time named $\Delta$. The possibility of field-redefining our deformations may sound like one which can be handled easily even in the standard approach, but it is not, and although one may like to think of $\Gamma$ as being the main ingredient of the ultimate BRST differential $s = \Gamma + \Delta$ (see Step 5), the inclusion of $\Delta$ is in fact crucial. It will ensure that no such redundancy is left unconsidered when passing to the cohomology of $s$, describing our gauge invariant observables.

In order to properly deal with fields redefinitions we again follow the fruitful paradigm according to which 'physical' combinations should be associated with elements of the cohomology of our nilpotent operator $\Delta$. More precisely, this means that the quantities we wish to consider as redundant (defining some equivalence class) should be associated with trivial elements of the cohomology. The analogy with the previous step, concerned with gauge invariance, is thus clear: in the same way as $\Gamma$-exact objects are gauge variations of something else, which are the redundancies corresponding to the possibility of performing gauge transformations; $\Delta$-exact objects will be associated with combinations which are on-shell zero, that is, with field redefinitions. We thus would like to define the action of $\Delta$ on our phase space in a way such that the EoMs for our original fields are equal to $\Delta$-variations of something else. Actually, the correct definition is:

$$\Delta \phi_i^* \equiv \mathrm{E}_i(\phi^j), \qquad (4.8)$$

where $\mathrm{E}_i$ is defined to be precisely the EoM for $\phi^i$, that is, $\delta S^{(0)} \equiv \mathrm{E}_i \delta \phi^i$. Moreover, starting from the above definition, the Bianchi identities (zeroth-order Noether identities) endowing the EoMs are easily seen to enforce the following relations:

$$\Delta(R^i_\alpha \phi_i^*) = 0. \qquad (4.9)$$

Then, it can be shown that having objects such as $R^i_\alpha \phi_i^*$ in the cohomology of $\Delta$ leads to inconsistencies in the formalism [181]. However, the cure to this problem is obvious: if an object is $\Delta$-closed and we wish to exclude it from the cohomology, the way out is to make it $\Delta$-exact. Accordingly, we define

$$R^i_\alpha \phi_i^* \equiv \Delta C^*_\alpha. \qquad (4.10)$$

Our antighosts finally get to play a role! Note that, from the relation (4.8) one concludes that $\Delta$ is also Grassmann odd, just like $\Gamma$, and its nilpotency is thus again guaranteed (and can be checked explicitly on any field).

The last definitions may have come as being a little *ad hoc* for the reader, and we now comment on them. We begin with the last equation hereabove. As we have pointed out, the necessity of having $R^i_\alpha \phi^*_i$ be $\Delta$-exact comes from rather involved considerations about the consistency of the formalism [103]. However, one might ask whether other options than setting it to $\Delta C^*_\alpha$ exist. The obvious answer to this is that, in fact, it is only for the sake of democracy that the antighosts have been introduced before, and as a matter of fact we have never needed them so far. Thus, one might think of the above relation in the following way: we are just adding the antighosts to the game (assuming they had not been added before) in order to express $R^i_\alpha \phi^*_i$ as a $\Delta$-exact quantity, mush in the same way as we simply added the antifields because we needed an object that could source the gauge variations in the master action (see Step 2). Also, let us point out that, given the index structure of the fields at hand, there was no other option one could have thought of (this time considering the antighosts as introduced beforehand).

Similarly, although the need for the EoMs to be $\Delta$-exact is clear enough, we have not justified why the antifields are precisely the ones appearing in (4.8). The 'sloppy' answer to the previous question still applies, namely, arguing that there is no other object we could have used. There is, of course, a deeper answer behind the scenes but it is beyond the scope of this presentation, and we thus refer again to [181] for further information.

The action of the $\Delta$ operator on the rest of the original fields is derived by acting with it on (4.8), which yields $\Delta \phi^i = 0$, and the action on the ghosts then follows from applying $\Delta$ to (4.6) and noticing that $\Delta$ anticommutes with $\Gamma$, for they are both Grassmann odd. We are ready to formulate our second cohomology, that of $\Delta$:

$$\mathrm{H}(\Delta) = \{X \in \text{phase space} \mid \Delta X = 0,\ X \neq \Delta Y\}. \qquad (4.11)$$

The physical meaning is, again, quite clear: the cohomology of $\Delta$ is the space of all quantities built out of the $\Phi^A$ which are not themselves the $\Delta$-variation of something else, namely, which are not proportional to the EoMs (equivalently, which are not field redefinitions, or on-shell zero). We are making progress towards a complete and refined formalization of what we mean by an observable: a non-trivial, gauge invariant quantity identified with others up to field redefinitions and gauge variations. Evidently, the correct operator which will compute for us the 'physical' cohomology will need to combine both $\Delta$ and $\Gamma$. That operator is called the BRST differential, and is formally introduced in Step 5 below.

Finally, before going to the next step, let us note that as far as the gradings are concerned, $\Delta$ has $\mathrm{agh}(\Delta) = -1$ and $\mathrm{pgh}(\Delta) = 0$, as is easily deduced from its action on the various fields and antifields.

### Step 5: Combining $\Gamma$ and $\Delta$ Into the BRST Differential $s$

With the above considerations in mind we can finally construct our ultimate nilpotent operator: the BRST differential $s$. The aim is that its cohomology should correctly describe the notion of an observable. Differently put, $\mathrm{H}(s)$ should be associated with the space of inequivalent, gauge-invariant and non-trivial deformations $S_1$ of the free master action $S_0$. The definition which computes the correct cohomology is the obvious one, namely:

$$s \equiv \Gamma + \Delta, \qquad (4.12)$$

and one can again check its nilpotency, either by noticing that $\Gamma$ and $\Delta$ anticommute or directly. Its cohomology,

$$\mathrm{H}(s) = \{X \in \text{phase space} \mid sX = 0,\ X \neq sY\}, \qquad (4.13)$$

is exactly the one computing all the consistent deformations up to field redefinitions. Indeed, it is the space of combinations which are on-shell gauge invariant but which are not themselves the gauge variation of something else or a field redefinition of something else, as follows from simply investigating the cohomology conditions in light of the decomposition $s = \Gamma + \Delta$.

Let us now consider a consistent deformation of the free master action into some deformed master action $S$, that is,

$$S = S_0 + gS_1 + g^2 S_2 + \cdots, \qquad (4.14)$$

where the ellipses stand for higher order deformations. In the generic case where the deformation is possibly non-abelian, the gauge transformations also get deformed in such a perturbative way, and $S_0$ is assumed to be invariant under the zeroth-order gauge transformations. As aforementioned, we shall primarily address the problem of first-order deformations. Now, as it is well known and also easy to check, for the first order piece $S_1$ the requirement of perturbative gauge invariance is really that of being invariant under the *free* gauge symmetries — this is equally true whether one considers the master action or the original action. Thanks to this fact, that we have implicitly used until now, we can analyze the problem of finding out and classifying the consistent deformations by means of the BRST differential $s$, which implements all at once the necessary requirements of (free) gauge invariance and (free) on-shell triviality. The condition our first-order deformation of the free master action must satisfy is then simply expressed in purely cohomological terms:

$$S_1 \in \mathrm{H}(s). \tag{4.15}$$

Finding all the consistent first-order deformations up to equivalency is thus tantamount to computing the cohomology of $s$, which is a well-defined mathematical problem. However, we shall now proceed to introducing one last refinement of the cohomology we want to compute, and that is the one of partial integration. Indeed, we shall always assume that the deformations we seek are local,[5] that is, they are spacetime integrals of functionals of our phase space variables (the fields, antifields, ... ) and of derivatives thereof, provided the derivatives appear up to finite order only. Our notation goes

$$S_1 \equiv \int a. \tag{4.16}$$

Therefore, our problem can be (and will be) reformulated in terms of $a$, namely, at the level of the (master) Lagrangian instead. Consequently, provided we are interested in the local dynamics only, to which boundary terms in the action never contribute, we have the freedom of performing integrations by parts. Therefore, the relevant cohomology is not $\mathrm{H}(s)$ but, rather, the cohomology of *s modulo d*, noted $\mathrm{H}(s|\mathrm{d})$, which is defined as $\mathrm{H}(s)$ but with the extra freedom of performing partial integrations when computing the BRST variations. The ultimate condition that our deformation must satisfy then reads

$$a \in \mathrm{H}(s|\mathrm{d}), \tag{4.17}$$

which is both necessary and sufficient. On top of this condition, our deformation $a$ might of course be required 'by hand' to preserve certain global symmetries, such as Lorentz invariance or parity.

---

[5] The fact that locality is compatible with the formalism was an issue at the time, which the work [183] cleared out, giving a positive answer to the question of compatibility.

Before we can start analyzing the above condition in detail there is one more grading we need to introduce, namely, the *total ghost number* (or simply the *ghost number*), equally defined on all fields as the pure ghost number minus the antighost number. That we do so now is not an accident. Indeed, the BRST differential $s$ does not have neither definite pure ghost number nor definite antighost number, as is inferred from the properties of $\Gamma$ and $\Delta$. The correct quantum number which keeps track of the action of $s$ is the (total) ghost number, and in fact we find $\text{gh}(s) = 1$. The ghost number of the various fields and antifields are straightforwardly computed, and also given in Table 4.2 at the end of the section, together with the action of the various operators on the various fields and antifields.

With our last quantum number, the total ghost number at hand, it is time we mention a condition on the deformation we have been neglecting so far, and it is that of the quantum numbers which it must have. Firstly, let us note a simple yet important fact: just like $s$ the free master action does not have definite pure ghost or antighost number, but it has total ghost number 0, which the reader shall easily verify. Note that this further indicates that the BRST differential is the right operator to consider when deforming the free master action but, more importantly, it means that $S_1$, and hence $a$ must have total ghost number 0. While this condition has been derived almost trivialy, its content is nevertheless not empty, and it will much restrict the possible ingredients one may use to build a tentative deformation term. Also, as we shall see the elements of $\text{H}(s|\text{d})$ with higher ghost number shall also enter the game at some point. The subset of $\text{H}(s|\text{d})$ having ghost number equal to $k$ is called the cohomology at ghost number $k$ and it is denoted by $\text{H}^k(s|\text{d})$. The final word about the deformation is thus the following:

$$\boxed{a \in \text{H}^0(s|\text{d})}. \tag{4.18}$$

This condition repackages all the requirements of consistency and non-triviality of the deformation

### Step 6: Finding Deformations via Consistency Cascade

Having formulated in a precise way the cohomology we wish to compute, we now explore how to do so in a clever way. An obvious thing to do is to use the gradings we have introduced to further inspect the problem. It is found that the antighost number is most useful in doing so, and the main reason for that is the following theorem: let

$$a = a_0 + a_1 + a_2 + \cdots, \tag{4.19}$$

where $\operatorname{agh}(a_i) \equiv i$ (note negative antighost numbers cannot occur). The theorem, proved in [178] under very generic assumptions, states that $a_i = 0\ \forall\, i > 2$. This result is, in general, very strong and as we shall see below it will be crucial in being able to analyze the deformation in a systematic way. We point out, though, that for cubic deformations the theorem is trivially proved, in the sense that there is no $\operatorname{gh}\# = 0$ combination of three of our fields and antifields of antighost number higher than 2, as one can directly observe by considering the various quantum numbers we have assigned each of our fields. In the sequel we shall confine ourselves to cubic deformations.

**Remark** : as we are addressing first-order deformations anyway, the reader might wonder what it means to further confine ourselves to cubic deformations. Could one think of e.g. quartic first-order deformations ? In fact, although this situation never arises in physics it is nevertheless a logical possibility. In [184] it has been proved that such deformations never occur, up to spin 5 and argued to be true for all spins. In the present approach we shall not prove the analogous result and simply make the assumption that our first-order deformations are cubic, further arguing this hypothesis to be very reasonable.

The above result is more useful than it might seem at first glance. First of all, the interpretation of the three pieces appearing in the above decomposition of $a$ is much clear: $a_0$ is the deformation of the Lagrangian, $a_1$ is the deformation of the gauge symmetries and $a_2$ is the deformation of the gauge algebra ! This can be verified by noticing that $a_0$ contains only the original fields ($\operatorname{agh}\# = 0$), $a_1$ contains one original field, one ghost and one antifield ($\operatorname{agh}\# = 1$) and $a_2$ contains two ghosts and one antighost ($\operatorname{agh}\# = 2$) — and once again recalling that $a$ has total ghost number zero.

Secondly, with the above decomposition in mind the cohomology condition
$$sa + \mathrm{d}(...) = (\Gamma + \Delta)(a_0 + a_1 + a_2) + \mathrm{d}(...) = 0, \qquad (4.20)$$
when analyzed antighost number by antighost number, gives rise to three independent conditions:

$$\Gamma a_2 \doteq 0, \qquad (4.21\mathrm{a})$$
$$\Delta a_2 + \Gamma a_1 \doteq 0, \qquad (4.21\mathrm{b})$$
$$\Delta a_1 + \Gamma a_0 \doteq 0. \qquad (4.21\mathrm{c})$$

These conditions form what is known as the *consistency cascade*, and we have used a new notation: '$\doteq$' is understood as the standard equality up to total derivatives. Moreover, another general theorem [178] teaches us that one can always assume $\Gamma a_2 = 0$, which is stronger than the same condition up to total derivatives.

The above consistency cascade will be our main tool in finding out consistent deformations, and it is worth commenting on. The equation (4.21c) for $a_0$ is familiar: it expresses the fact that the deformation of the Lagrangian, $a_0$, is invariant up to field redefinitions $\Delta a_1$ and total derivatives. The two remaining equations are less easily interpreted, but their role is to ensure first-order consistency of the deformation of the Lagrangian. Intuitively, the situation is clear: (4.21b) involves $a_1$ and $a_2$, and is thus ensuring that the deformation of the gauge symmetries induced by $a_0$ closes to a gauge-algebra deformation $a_2$. Then, Equation (4.21a) ensures consistency of the gauge-algebra deformation $a_2$. In fact, one can check that (4.21b) is the first-order projection of the condition that the gauge symmetries close to some algebra and (4.21b) is a consistency condition for the gauge-algebra deformation, again projected to first order in the deformation.

Having established the above consistency cascade, we are in principle ready to attack the problem of computing $H^0(s|d)$. However, as mentioned at the beginning of this section, the BRST formalism will allow us to tackle that problem *backwards*. This means that, instead of classifying the $a_0$'s satisfying Equation (4.21c) and then working our way up the consistency cascade, we shall rather classify the $a_2$ satisfying Equation (4.21a) and from there make progress all the way down to the corresponding, consistent $a_0$. Such is the power of the BRST framework: we classify the consistent gauge-algebra deformations first and from there extract, by solving the consistency cascade (first for $a_1$ and then for $a_0$), the corresponding Lagrangian deformations. In this fashion the search for consistent deformations is rendered systematic and involves only the solving of precise cohomology equations. Also note that, by construction, a byproduct of this method is that the gauge-symmetry and gauge-algebra deformations corresponding to some found $a_0$ are readily available.

Let us discuss non-abelian vertices first. The strategy is the following: classify all the $a_2$ satisfying Equation (4.21a). Take a linear combination of all of them with arbitrary coefficients and plug it into Equation (4.21b). Then solve Equation (4.21b) for $a_1$. Finally, plug the found $a_1$ into Equation (4.21c) and solve for $a_0$: it is the non-abelian deformation of the Lagrangian we were looking for.

Actually, the classification of $a_2$ is further constrained by an equivalence relation. Indeed, two different $a_2$, both satisfying Equation (4.21a), might yield the same $a_0$. This simply stems from the fact that, so far, in addressing the computation of $\mathrm{H}^0(s|\mathrm{d})$ we have only analyzed the condition that $a$ should be $s$-closed modulo d, and in (4.20) we have expanded it in antighost number. The condition of non-triviality in the cohomology should also be taken into account, and this means that the $a$ candidates are defined up to the equivalence relation given by the addition of $s$-exact terms modulo d, that is, terms of the form $sm + \mathrm{d}n$. Now, as the reader shall easily verify upon recalling that $m$ and $n$ should also have total ghost number zero (and also stop at antighost number 2), such an equivalence relation yields the following three equivalence relations ($\sim$) for the different components of $a$:

$$a'_2 \sim a_2 + \Gamma b_2 + \mathrm{d}c_2, \tag{4.22a}$$
$$a'_1 \sim a_1 + \Delta b_1 + \mathrm{d}c_1, \tag{4.22b}$$
$$a'_0 \sim a_0 + \Delta b_0 + \mathrm{d}c_0. \tag{4.22c}$$

This means that, when listing all the $a_2$ satisfying Equation (4.21a) we do so up to the above equivalency, and indeed one straightforwardly checks that two $a_2$'s differing by $\Gamma$-exact terms modulo d yield the same $a_0$, if any. Those would thus be two equivalent ways of writing down the gauge-algebra deformation induced by some given $a_0$.

We now address abelian vertices, that is, deformations $a$ for which the $a_2$ part is trivial, i.e. $\Gamma$-exact up to total derivatives. Now comes of use another theorem: when $a_2$ is trivial one can always choose it to be zero, and hence the consistency cascade starts one step lower, with Equation (4.21b) at $a_2 = 0$, that is, $\Gamma a_1 \doteq 0$ [179]. Even better, the theorem further guarantees that one can chose $a_1$ to be exactly $\Gamma$-closed, and not only modulo d. The abelian vertices which nonetheless deform the gauge transformations are thus found by classifying all the inequivalent $a_1$'s which are $\Gamma$-cocycles. Again, two equivalent $a_1$'s, differing by $\Delta$-exact terms up to total derivatives, are seen to yield the same $a_0$.

Last of all we address the 'completely' abelian vertices, namely, those that not only preserve the gauge algebra but also leave the gauge transformations undeformed. This kind of deformations will have zero $a_2$ and trivial $a_1$, that is, the $a_1$ piece will be $\Delta$-exact modulo d. In that case one can evidently remove $a_1$ so to be left with only Equation (4.21c) at $a_1 = 0$, to be solved for $a_0$, i.e. $\Gamma a_0 \doteq 0$, which is to be solved in light of the equivalence relation for $a_0$, simply given by field redefinitions and total derivatives.

We are almost at the end of our step-by-step guide to the BRST-Antifield reformulation of the deformation problem (in the free, irreducible case), and shall now make some comments about it. One point worth highlighting is the crucial role played by the cohomology of $\Gamma$ at antighost number 2, which is indeed the one computing the inequivalent $a_2$ candidates in the non-abelian case.[6]

Also, there is a subtlety in the non-abelian case which we should comment on right away, and which we have overlooked so far in order not to crowd our first approach of the consistency cascade, but which is nonetheless an important point. Let us consider some $\Gamma$-closed $a_2$ and plug it into Equation (4.21b) in order to solve for $a_1$. The subtlety is the following: the solution for $a_1$, if it exists, is in fact defined up to $\Gamma$-closed terms only. Indeed, if two $a_1$'s differ by $\Gamma$-closed terms they will correspond to the same $a_2$. The solution $a_1$, if it exists, is then usually denoted as

$$a_1 = \hat{a}_1 + \tilde{a}_1, \tag{4.23}$$

where the $\Gamma$-closed term $\tilde{a}_1$ is called the ambiguity and the non-ambiguous piece $\hat{a}_1$ is the solution found to solve Equation (4.21b) for our candidate $a_2$. Now comes the complication: when plugging the above $a_1$ into the last consistency equation, $\Delta a_1 + \Gamma a_0 \doteq 0$, it might be that a solution for $a_0$ only exists for some ambiguity $\tilde{a}_1$, and in general that is the case. In fact, this is the way the ambiguity is fixed. The computational intricacy is then that, in general, it might be difficult to either guess the correct ambiguity or express it into its most general form to then plug it into the last equation. For non-abelian vertices the situation is thus usually the following: determining whether a candidate $a_2$ has a corresponding $a_1$ is usually not extremely difficult but, if there is such an $a_1$, finding out the correct ambiguity (or establishing that no $a_0$ solves the last equation for the found $a_1$) can be tricky, and it is the most non-trivial part of the procedure. Although this part of the procedure is still 'systematic', in the sense that one can just try the most general $\Gamma$-closed $\tilde{a}_1$ (with correct quantum numbers) and then plug it into $\Delta a_1 + \Gamma a_0 \doteq 0$ in order to solve for $a_0$, it is nonetheless the least algorithmic part of the work.

Finally, before addressing second-order consistency in the next step, let us comment on a procedural point, having to do with abelian vertices. For the latter, the equation to solve is $\Gamma a_0 \doteq 0$, as we have just seen. However, it might not be the easiest one to solve and, as it usually turns out, it is easier to allow for $\Delta$-exact terms in $\Gamma a_0$. It is then easier to find the inequivalent $a_0$'s satisfying an equation of the form $\Gamma a_0 \doteq \Delta(...)$. However, the corresponding $a_1$ is then in general not equal to zero, but one can check that, because $a_1$ is trivial, it can always be canceled by the addition of $\Delta$-exact terms in $a_0$. By performing field redefinitions at the level of our vertex one can thus render manifest the invariance of it up to total derivatives only (or do the inverse thing). Differently put, depending on the chosen 'representation' for our vertex, the absence of deformation of the gauge transformations (triviality of $a_1$) may appear explicitly ($a_1 = 0$) or not ($a_1 \neq 0$).

---

[6] Actually, as has been emphasized, one can choose $a_2$ to be strictly $\Gamma$-closed, and not only $\Gamma$-closed modulo d.

**Step 7: Second-Order Consistency and the Antibracket**

Everything we have mentioned so far had to do with first-order consistency. In the completely abelian case, when only the Lagrangian is deformed, no quartic or higher-order terms are needed and the consistency is automatic to all orders in perturbation theory. However, as is well known, in the non-abelian case (and in the 'intermediate' case too) the situation is different; either the vertex is consistent to second order only up to the addition of a quartic term, $S_2$, or it is obstructed. In general, determining whether a non-abelian vertex is obstructed or not and, in the latter case, determining the quartic term that needs be added in order to render the theory fully consistent is rather tedious. However, it is remarkable that the BRST-Antifields also provides one with just the right tool to deal with this issue, and that tool is called the *antibracket*, which we now introduce.

One defines the following odd, symplectic structure on the space of functionals of our fields and antifields:

$$(X, Y) \equiv \frac{d^R X}{d\Phi^A} \frac{d^L Y}{d\Phi^*_A} - \frac{d^R X}{d\Phi^*_A} \frac{d^L Y}{d\Phi^A}. \tag{4.24}$$

This definition gives $(\Phi^A, \Phi^*_B) = \delta^A_B$, which is real. Because a field and its antifield have opposite Grassmann parity, it follows that if $\Phi^A$ is real, $\Phi^*_B$ must be purely imaginary, and vice versa. Note that the antibracket satisfies the graded Jacobi identity.

To understand the usefulness of the antibracket, we first note the following peculiar and a priori anodyne fact: the action of the free BRST differential $s$ can be rephrased as taking the antibracket with the free master action $S_0$, that is,

$$sF = (S_0, F), \tag{4.25}$$

for any phase-space functional $F$. For the reader unfamiliar with the BRST formulation in the context of quantizing gauge theories, this relation might come as a surprise, even after proving it to be true (which is of little difficulty). In fact, there is even more to it as we shall see and the antibracket will shed a whole new light on the framework we have introduced so far. Indeed, one can actually prove the equivalent of the above statement for the fully deformed theory too! Let us denote the completely deformed BRST differential by $\mathfrak{s}$, so that[7]

$$\mathfrak{s} \equiv s + s_1 + s_2 + \cdots, \qquad (4.26)$$

where for example $s_1$ is the sum of some $\Gamma_1$, implementing the deformed piece brought in by $a_1$, and some $\Delta_1$, which implements the contribution to the free EoMs induced by $a_0$. The full statement is then

$$\mathfrak{s}F = (S, F), \qquad (4.27)$$

which can be seen to hold by virtue of the Noether identities and the higher-order gauge-structure equations [181].

We are now ready to formulate an equation which has been the cornerstone of the BRST approach to Gauge Theory. Indeed, the full master action $S$ is invariant under the full BRST differential $\mathfrak{s}$ which, by virtue of the above relation reads

$$(S, S) = 0. \qquad (4.28)$$

It is the so-called (classical) *master equation*, which contains all the information about the Noether identities and the higher-order gauge structure equations. It remarkably repackages all the conditions defining a fully consistent deformation into a 'single', geometrical equation. This structure allows for a rephrasal of many a property. For example, in this way one can see the nilpotency of $\mathfrak{s}$ as a mere consequence of the graded Jacobi identity for the antibracket. Furthermore, let us also point out that the above odd structure is somehow related to the more familiar Poisson bracket, and to other structures as well [185], but in the present reminder we shall not dwell on these interesting questions.

To see how this new structure helps addressing the problem of second-order consistency let us split the master equation above in terms of the coupling constant $g$ by inserting in it the perturbative expression (4.14).

---

[7] Note that, to be homogeneous in our use of notation we should have called the full BRST operator $s = s_0 + s_1 + \cdots$, but as the zeroth-order part is the most often used piece we have chosen to be more economical.

The first orders give us

$$(S_0, S_0) = 0, \tag{4.29a}$$
$$(S_0, S_1) = 0, \tag{4.29b}$$
$$(S_1, S_1) = -2(S_0, S_2). \tag{4.29c}$$

The first equation hereabove is satisfied by assumption: it can be rewritten as $sS_0 = 0$, which is simply the statement of invariance of the free master action under the (free) BRST differential. The second equation translates to $sS_1 = 0$, which is the integrated version of the cohomological condition written down in (4.18). As for the third one, it expresses in a compact way the condition that $S_1$ must satisfy so to be consistent at second order, where it is completed by a quartic term $S_2$. It determines whether or not, in a local theory, a consistent first-order deformation gets obstructed at the second order. One thus sees that the second-order consistency is controlled by the local cohomology group $\mathrm{H}^1(s)$, for $(S_1, S_1)$ can be seen to be of ghost number 1. More precisely, one easily checks that $(S_1, S_1)$ is BRST-closed (by the graded Jacobi identity), and what the third equation hereabove does is to further require it to be trivial in $\mathrm{H}^1(s)$. Keeping in mind that $s$ annihilates $(S_1, S_1)$, one may thus rewrite the second order consistency condition as

$$(S_1, S_1) \notin \mathrm{H}^1(s). \tag{4.30}$$

Moreover, it can be shown that $\mathrm{H}^1(s)$ is also the cohomology group controlling higher-order deformations. However, more often than not in gauge theory a deformation is either fully consistent, that is, to all orders, or consistent only at the cubic level, and hence fails to satisfy the above requirement. Note that in the above condition it is truly the cohomology of $s$ which is used, not the cohomology modulo d, so that one should expect strong conditions to arise from it.

Now, the introduction of the antibracket structure for the sole purpose of addressing the second order consistency may seem excessive to the reader. However, as we shall see, the second order problem will be straightforwardly solved upon using the above condition (see next chapters). But there is (even) more to the antibracket, as we have mentioned already. Indeed, with such a symplectic structure one can actually reformulate the whole problem of deformation of the free master action. Indeed, as we have seen the free master action does satisfy the master equation (4.28), and the full deformation should also fulfill it. The problem of consistently deforming a free theory can thus be reformulated as the problem of deforming the solution $S_0$ to the master equation, and this allows for a different mathematical approach to the problem, which has proved much useful [186]. More generically speaking, the BRST reformulation presented here has allowed for a systematic study of many aspects of Gauge Theory, as for example that of [187], where the above techniques are used to discard as inconsistent theories involving a colored graviton. Let us also point out the work [188], where the simplest four-dimensional Supergravity is proved to be unique by making use of the formalism presented here.

We end this section with a reminder of the quantum numbers for our fields and antifields as well as the action of the different operators on them. In the free, irreducible case of interest to us they read as follows.

Table 4.1: Properties of the Various Fields, Antifields and Operators

| $Z$ | $\Gamma(Z)$ | $\Delta(Z)$ | $\mathrm{pgh}(Z)$ | $\mathrm{agh}(Z)$ | $\mathrm{gh}(Z)$ | $\epsilon(Z)$ |
|---|---|---|---|---|---|---|
| $\phi^i$ | $R^i_\alpha \mathcal{C}^\alpha$ | 0 | 0 | 0 | 0 | 0 |
| $\mathcal{C}^\alpha$ | 0 | 0 | 1 | 0 | 1 | 1 |
| $\phi^*_i$ | 0 | $\mathrm{E}_i[\phi^j]$ | 0 | 1 | $-1$ | 1 |
| $\mathcal{C}^*_\alpha$ | 0 | $R^i_\alpha \phi^*_i$ | 0 | 2 | $-2$ | 0 |

## 4.3 QED: a Pedagogical Example

The BRST-Antifield reformulation of the interaction problem has been introduced in the previous section for free, irreducible theories. Although we have tried to make such an introduction as pedagogical as possible, we believe the explicit treatment of a familiar example could be of use to the reader, before moving to the next chapters where we deal with higher-spin fields. Indeed, in Chapter 5 we begin by studying the electromagnetic couplings of a spin-$\frac{3}{2}$ Rarita–Schwinger field, where some features typical of higher-spin couplings will already appear, and in the present section we thus address the more familiar setup of QED. In fact, although the term 'higher spin' is usually understood as referring to spins greater than two, the spin $\frac{3}{2}$ is somewhat exceptional: it is somewhat 'standard' regarding its gravitational coupling (found in Supergravity) but it counts as 'higher' when its electromagnetic coupling is addressed. These considerations are expanded on in the corresponding chapters below. Also, to be completely honest we should point out that the following example of QED is perhaps a little treacherous, in the sense that there is no gauge invariance for the fermion and hence the problem of building consistent interactions becomes much, much simpler as we shall see. We believe it is nevertheless a good place to start applying our formalism, and it is certainly interesting, if only because it provides a point to compare our forthcoming higher-spin study with.

Let us, then, construct all the off-shell $1$–$\frac{1}{2}$–$\frac{1}{2}$ cubic vertices by employing our beloved BRST–BV cohomological methods. Our assumptions are Lorentz invariance, Parity invariance and locality. The starting point is the free theory, which contains a photon $A_\mu$ and a massless electron field $\psi$, described by the action

$$S^{(0)}[A_\mu, \psi] = \int d^D x \left(-\tfrac{1}{4} F_{\mu\nu}^2 - i\bar{\psi}\slashed{\partial}\psi\right), \tag{4.31}$$

which enjoys the abelian gauge invariance

$$\delta_\lambda A_\mu = \partial_\mu \lambda, \tag{4.32}$$

and no gauge invariance for $\psi$.

For the Grassmann-even bosonic gauge parameter $\lambda$, we introduce the Grassmann-odd bosonic ghost $C$, and no ghost corresponding to $\psi$ is introduced for the latter enjoys no gauge invariance.[8] Therefore, the set of fields becomes

$$\Phi^A = \{A_\mu, C, \psi\}. \tag{4.33}$$

For each of these fields, we introduce an antifield with the same algebraic symmetries in its indices but opposite Grassmann parity. The set of antifields thus reads

$$\Phi_A^* = \{A^{*\mu}, C^*, \bar{\psi}^*\}. \tag{4.34}$$

Now we construct the free master action $S_0$, which is an extension of the original gauge-invariant action (5.1) by terms involving ghosts and antifields. Explicitly,

$$S_0 = \int d^D x \left(-\tfrac{1}{4} F_{\mu\nu}^2 - i\bar{\psi}\slashed{\partial}\psi + A^{*\mu}\partial_\mu C\right). \tag{4.35}$$

Notice how the antifields appear as sources for the 'gauge' variations, with gauge parameters replaced by corresponding ghosts. It is easy to verify that (5.5) indeed solves the master equation $(S_0, S_0) = 0$. The different gradings and Grassmann parity of the various fields and antifields, along

Table 4.2: Properties of the Various Fields & Antifields ($n = 1$)

| $Z$ | $\Gamma(Z)$ | $\Delta(Z)$ | $\mathrm{pgh}(Z)$ | $\mathrm{agh}(Z)$ | $\mathrm{gh}(Z)$ | $\epsilon(Z)$ |
|---|---|---|---|---|---|---|
| $A_\mu$ | $\partial_\mu C$ | $0$ | $0$ | $0$ | $0$ | $0$ |
| $C$ | $0$ | $0$ | $1$ | $0$ | $1$ | $1$ |
| $A^{*\mu}$ | $0$ | $-\partial_\nu F^{\mu\nu}$ | $0$ | $1$ | $-1$ | $1$ |
| $C^*$ | $0$ | $-\partial_\mu A^{*\mu}$ | $0$ | $2$ | $-2$ | $0$ |
| $\psi$ | $0$ | $0$ | $0$ | $0$ | $0$ | $1$ |
| $\bar{\psi}^*$ | $0$ | $-i\bar{\psi}\overleftarrow{\partial\!\!\!/}$ | $0$ | $1$ | $-1$ | $0$ |

with the action of $\Gamma$ and $\Delta$ on them, are given in Table 4.3 below.

The cohomology of $\Gamma$ is isomorphic to the space of functions of

- The undifferentiated ghost $C$,
- The antifields $\{A^{*\mu}, C^*, \bar{\psi}^*\}$ and their derivatives,
- The curvature $F_{\mu\nu}$ and its derivatives,
- The field $\psi$.

Let us now classify the consistent couplings. We start by the non-abelian ones. In fact, it is easily seen that there can be no consistent non-abelian (cross) coupling in the present setup. Indeed, the construction of a candidate $a_2$ involving either $\psi$ or $\bar{\psi}^*$ fails at the level of the quantum numbers already, as can be derived by looking at the above table and further recalling that $a_2$ must be a cubic combination of total ghost number zero and antighost number two. From those considerations only, one sees that any $a_2$ should be of the schematic form ghost × ghost × antighost, and as we have no ghost corresponding to $\psi$ (because it enjoys no gauge invariance) one can only construct self-coupling $a_2$ candidates for $A_\mu$. We are not interested in those (which lead to the familiar Yang–Mills cubic term [189] when the photon is colored), and wish to look at cross couplings only. Therefore, the non-abelian case is covered, and we have found no consistent such couplings.

Let us now address the abelian couplings. As mentioned in the previous section, we should first investigate vertices which do not deform the gauge algebra but nevertheless deform the gauge transformations, and then move on to 'completely' abelian ones, namely those vertices which deform only the Lagrangian. In fact, in Section 5.4 we prove that the 'in-between' case cannot occur for the electromagnetic coupling of higher-spin gauge fermions. However, the present setup evades that theorem because of the absence of a ghost for the fermion field, and we shall find hereafter that there exists indeed a coupling which deforms the gauge transformations but is nevertheless abelian.

---

[8] Here is where our setup is a little misleading in illustrating the BRST methods, for only one ghost needs be introduced, and that will drastically simplify certain aspects.

In order to search for the latter couplings with trivial $a_2$ but non-trivial $a_1$, let us classify the possible gauge-transformations deformations. They should have antighost number 1 and total ghost number zero, and further be cubic in our fields and antifields. Evidently, it should also be Lorentz invariant and have all spinor indices contracted (and we further require it to preserve Parity). The only zero-derivatives such combination is easily concluded to be

$$a_1 = g\bar{\psi}^*\psi C, \tag{4.36}$$

If such a gauge-symmetry deformation indeed does not deform the gauge algebra, it should satisfy $\Gamma a_1 \doteq 0$ (see previous section), which indeed it does. If it corresponds to a vertex, it must also be such that $\Delta a_1 + \Gamma a_0 \doteq 0$. One easily realizes that, only when the coupling constant $g$ is imaginary does the above deformation get lifted to a Lagrangian vertex. Indeed, making use of partial integration the cohomology equation is easily solved for $a_0$, which is found to be

$$a_0 = ig\bar{\psi}\slashed{A}\psi. \tag{4.37}$$

One should now investigate the fate of $a_1$ candidates containing derivatives. However, those are immediately ruled out as trivial. To see it, let us first make clear that the antifield in $a_1$ can always be assumed to be undifferentiated, as $a_1$ is defined up to total derivatives only. Now, if a derivative acts on the ghost, it would produce the gauge variation of $A_\mu$, which is by definition a $\Gamma$-exact object, and because $\Gamma$ does not act on $\psi$ nor $\psi^*$, this situation would correspond to a $\Gamma$-exact $a_1$ (one can pull out $\Gamma$ to make it act on the whole $a_1$ above), and this would correspond to a $\Delta$-exact $a_0$, which is trivial (see Step 6 of the previous Section). If, on the other hand, a derivative acts on $\psi$, the following argument can be used: this derivative cannot come alone (because of Lorentz invariance), and there are no indices on the involved fields, so that it must be contracted either with another derivative or with a $\gamma$-matrix. Now, because no derivatives can act on the ghost (see above) and because $\square = \slashed{\partial}\slashed{\partial}$, both these situations give rise to the EoMs in $a_1$, that is, to $\Delta\psi^* = -i\psi^*\overleftarrow{\slashed{\partial}}$ or combinations thereof. This does not generically make the whole deformation $\Delta$-exact, because the $\Delta$-operator acts on $\bar{\psi}^*$, but one can check that the cohomology equation $\Delta a_1 + \Gamma a_0 \doteq 0$ is then either not satisfied ($g \in \mathbb{C} \setminus \mathbb{R}$), or it is satisfied ($g \in \mathbb{R}$) but the resulting $a_0$ is $\Delta$-exact. Considering even more derivatives only makes it worse, and we thus conclude that there is only one vertex (4.37) which deforms the gauge transformations, and it has zero derivatives.

Our search is now narrowed down to the couplings which preserve the gauge transformations. The only part of the deformation that we need care about is thus $a_0$, and the $a_1$ piece can always be chosen to be zero (see previous section). We are left with the equation $\Gamma a_0 \doteq 0$ to be solved, and as explained in the steps above we shall alternatively use the weaker equation $\Gamma a_0 \doteq \Delta(...)$. We start with vertices containing no spacetime derivatives. We directly see that the only Lorentz-invariant possibility is

$$a_0 = \bar{\psi} \slashed{A} \psi, \tag{4.38}$$

which obeys $\Gamma a_0 \doteq \Delta(...)$. This is the vertex we have already found above. Note that, if we had not found this vertex previously, we could be tempted to conclude that the latter is completely abelian. However, because we have used the 'weaker equation' $\Gamma a_0 \doteq \Delta(...)$ here, that is not guaranteed, and in this instance it is of course not true. We thus take this opportunity to recall that, when using the 'weaker equation', one should check whether the corresponding $a_1$ is trivial, that is, whether it can be canceled by field redefinitions at the level of the Lagrangian.

We then address the vertices containing one derivative. An obvious possibility is the term built in terms of the curvatures (in this case there is only one curvature, namely that of $A_\mu$, for the fermion has no spacetime indices):

$$a_0 = \bar{\psi} \slashed{F} \psi, \tag{4.39}$$

which is strictly gauge invariant (not even modulo d). The only other Lorentz-invariant combination with one derivative is $\partial \cdot A \, \bar{\psi}\psi$, but it is easily seen to violate the consistency equation $\Gamma a_0 \doteq 0$. Furthermore, it is easily proved that there are no higher-derivative candidates, for all such Lorentz-invariant combinations would be on-shell trivial up to partial integration, as the reader shall easily convince himself of.

Let us comment on the nature of the vertices. The last one is clearly the Born–Infeld-like one, namely, it is a product of curvatures and is strictly gauge invariant. The other one is seen to be different: whatever field redefinition we perform on it the best we can do is bring it to a form in which it is on-shell gauge invariant modulo d. That vertex is also the one which completes the free kinetic term for the fermion, turning it into the familiar expression involving the covariant derivative: $\bar{\psi}\slashed{D}\psi$, with $D_\mu \equiv \partial_\mu - igA_\mu$. The said cubic coupling is thus the one resulting from covariantizing the derivatives in the fermion kinetic term, namely, the so-called minimal coupling, which in the present setup has zero derivatives.

As we shall see in Chapter 5, for higher-spin fields the situation is very different, and in fact minimal couplings thereof will never exist on their own. Such a feature of higher-spin interactions is actually generic on flat spacetimes, and as we will see it also holds true for gravitational couplings, studied in Chapter 6. Furthermore, in the following study of higher-spin interactions with Electromagnetism and Gravity, there will be no couplings which deform the gauge transformations but not the gauge algebra (see Section 5.4 and Appendix D.4). Although simple, for future comparison it is useful to summarize the results of the present section, and we give them in the table hereafter.

Table 4.3: Summary of $1-\frac{1}{2}-\frac{1}{2}$ Vertices

| # of derivatives | Vertex | Nature | Exists in |
|---|---|---|---|
| 0 | $\bar{\psi}\slashed{A}\psi$ | $\frac{1}{2}$-Abelian | $D \geq 4$ |
| 1 | $\bar{\psi}\slashed{F}\psi$ | Abelian | $D \geq 4$ |

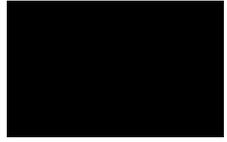

CHAPTER 5

# Electromagnetic Vertices

In this chapter, we consider the coupling of a massless fermion of arbitrary spin to a U(1) gauge field, in flat spacetime of dimension $D \geqslant 4$. We do not consider mixed-symmetry fields, and restrict our attention to totally symmetric Dirac fermions $\psi_{\mu_1\ldots\mu_n}$ of spin $s = n + \frac{1}{2}$. For these fields, we employ the powerful machinery of the BRST–BV cohomological methods (recalled in the previous chapter) to construct systematically consistent interaction vertices, with the underlying assumptions of locality, Poincaré invariance and conservation of Parity, and without relying on other methods. The would-be off-shell $1-s-s$ cubic vertices will complement their bosonic counterparts constructed in [184].

The organization of the chapter is as follows. We construct consistent off-shell $1-s-s$ vertices in the following three Sections. In particular, Section 5.1 considers the massless spin-$\frac{3}{2}$ field, while Section 5.2 pertains to $s = \frac{5}{2}$, and Section 5.3 generalizes the results, rather straightforwardly, to arbitrary spin $s = n + \frac{1}{2}$. Then, in Section 5.4, we prove an interesting property of the vertices under study: an abelian $1-s-s$ vertex, i.e. a $1-s-s$ vertex that does not deform the original abelian gauge algebra never deforms the gauge transformations. Finally, Section 5.5 investigates whether there are obstructions to the existence of second-order deformations corresponding to the non-abelian vertices, i.e. if they are consistent beyond the cubic order. Our comments are found, together with those concerning Chapter 6 about gravitational couplings, in Chapter 7. Various appendices present some useful technical details, much required for the following calculations.

## 5.1 Electromagnetic Coupling of Spin 3/2

In this section we construct parity-preserving off-shell $1-\frac{3}{2}-\frac{3}{2}$ vertices by employing the BRST–BV cohomological methods. The spin-$\frac{3}{2}$ system is simple enough so that one can implement the BRST deformation scheme with ease, while it captures many non-trivial features that could serve as guidelines as one moves on to higher spins. As the formalism has been recalled in Chapter 4 and, moreover, the simple example of QED has been treated explicitly in Section 4.3, we shall introduce the BRST elements in a rather straightforward way hereafter.

The starting point is the free theory [190], which contains a photon $A_\mu$ and a massless Rarita–Schwinger field $\psi_\mu$, described by the action

$$S^{(0)}[A_\mu, \psi_\mu] = \int \mathrm{d}^D x \left( -\tfrac{1}{4} F_{\mu\nu}^2 - i\bar{\psi}_\mu \gamma^{\mu\nu\rho} \partial_\nu \psi_\rho \right), \tag{5.1}$$

which enjoys the two abelian gauge invariances:

$$\delta_\lambda A_\mu = \partial_\mu \lambda, \qquad \delta_\varepsilon \psi_\mu = \partial_\mu \varepsilon. \tag{5.2}$$

For the Grassmann-even bosonic gauge parameter $\lambda$, we introduce the Grassmann-odd bosonic ghost $C$. Corresponding to the Grassmann-odd fermionic gauge parameter $\varepsilon$, we have the Grassmann-even fermionic ghost $\xi$. Therefore, the set of fields becomes

$$\Phi^A = \{A_\mu, C, \psi_\mu, \xi\}. \tag{5.3}$$

For each of these fields, we introduce an antifield with the same algebraic symmetries in its indices but opposite Grassmann parity. The set of antifields is

$$\Phi^*_A = \{A^{*\mu}, C^*, \bar{\psi}^{*\mu}, \bar{\xi}^*\}. \tag{5.4}$$

Now we construct the *free* master action $S_0$, which is an extension of the original gauge-invariant action (5.1) by terms involving ghosts and antifields. Explicitly,

$$S_0 = \int \mathrm{d}^D x \left( -\tfrac{1}{4} F_{\mu\nu}^2 - i\bar{\psi}_\mu \gamma^{\mu\nu\rho} \partial_\nu \psi_\rho + A^{*\mu} \partial_\mu C + (\bar{\psi}^{*\mu} \partial_\mu \xi - \partial_\mu \bar{\xi} \psi^{*\mu}) \right), \tag{5.5}$$

where the antifields appear as sources for the gauge variations, with gauge parameters replaced by the corresponding ghosts. It is easy to verify that (5.5) indeed solves the master equation $(S_0, S_0) = 0$. The different gradings and Grassmann parity of the various fields and antifields, along with the action of $\Gamma$ and $\Delta$ on them, are given in Table 5.1.

Table 5.1: Properties of the Various Fields & Antifields ($n = 1$)

| $Z$ | $\Gamma(Z)$ | $\Delta(Z)$ | $\mathrm{pgh}(Z)$ | $\mathrm{agh}(Z)$ | $\mathrm{gh}(Z)$ | $\epsilon(Z)$ |
|---|---|---|---|---|---|---|
| $A_\mu$ | $\partial_\mu C$ | $0$ | $0$ | $0$ | $0$ | $0$ |
| $C$ | $0$ | $0$ | $1$ | $0$ | $1$ | $1$ |
| $A^{*\mu}$ | $0$ | $-\partial_\nu F^{\mu\nu}$ | $0$ | $1$ | $-1$ | $1$ |
| $C^*$ | $0$ | $-\partial_\mu A^{*\mu}$ | $0$ | $2$ | $-2$ | $0$ |
| $\psi_\mu$ | $\partial_\mu \xi$ | $0$ | $0$ | $0$ | $0$ | $1$ |
| $\xi$ | $0$ | $0$ | $1$ | $0$ | $1$ | $0$ |
| $\bar{\psi}^{*\mu}$ | $0$ | $-\frac{i}{2}\bar{\Psi}_{\alpha\beta}\gamma^{\alpha\beta\mu}$ | $0$ | $1$ | $-1$ | $0$ |
| $\bar{\xi}^*$ | $0$ | $\partial_\mu \bar{\psi}^{*\mu}$ | $0$ | $2$ | $-2$ | $1$ |

For the spin-$\frac{3}{2}$ field the Fronsdal tensor is

$$\mathcal{S}_\mu = i\left[\slashed{\partial}\psi_\mu - \partial_\mu \slashed{\psi}\right] = -i\gamma^\nu \Psi_{\mu\nu}, \tag{5.6}$$

i.e. the $\gamma$-trace of the curvature. The cohomology of $\Gamma$ is isomorphic to the space of functions of

- The undifferentiated ghosts $\{C, \xi\}$,
- The antifields $\{A^{*\mu}, C^*, \bar{\psi}^{*\mu}, \bar{\xi}^*\}$ and their derivatives,
- The curvatures $\{F_{\mu\nu}, \Psi_{\mu\nu}\}$ and their derivatives,

and we recall that this cohomology is discussed in detail in Appendix E.

Let us now address the deformations of the above free master action. In Subsection 5.1.1 we address non-abelian vertices, while in Subsection 5.1.2 we turn to the abelian ones.

### 5.1.1 Non-Abelian Vertices

The non-abelian vertices are the ones that deform the gauge algebra, that is, the ones for which to corresponding $a_2$ is non-trivial. As explained in Chapter 4, we shall start by classifying all such gauge-algebra deformations, on the basis of quantum numbers and non-triviality, and shall then work our way down the consistency cascade (4.21) in order to find the corresponding $a_1$ and $a_0$, if any.

**Gauge-Algebra Deformation**

Let us consider, for the first-order deformation, the most general form of $a_2$ — the term with $\mathrm{agh}\# = 2$, that contains information about the deformation of the gauge algebra. $a_2$ must satisfy $\Gamma a_2 = 0$, and be Grassmann even with $\mathrm{gh}(a_2) = 0$. Besides, we require that $a_2$ be a parity-even Lorentz scalar. Then, the most general possibility is

$$a_2 = -g_0 C \left( \bar{\xi}^* \xi + \bar{\xi} \xi^* \right) - g_1 C^* \bar{\xi} \xi, \tag{5.7}$$

which is a linear combination of two independent terms: one contains *both* the bosonic ghost $C$ and the fermionic ghost $\xi$, while the other contains only $\xi$ but *not* $C$. The former one potentially gives rise to minimal coupling, while the latter could produce dipole interaction. This can be understood by first noting that the corresponding Lagrangian deformation, $a_0$, is obtained through the consistency cascade (4.21a)–(4.21c). From the action of $\Gamma$ and $\Delta$ on the fields and antifields, it is then easy to see that the respective $a_0$ would contain no derivatives and one derivative respectively.

**Deformation of Gauge Transformations**

Next, we would like to see if $a_2$ can be lifted to certain $a_1$, i.e. with the given $a_2$, if one could solve (4.21b) for some $a_1$. Indeed, one finds that [1]

$$\begin{aligned}
\Delta a_2 &= + g_0 C \left[ (\partial_\mu \bar{\psi}^{*\mu}) \xi - \bar{\xi}(\partial_\mu \psi^{*\mu}) \right] + g_1 (\partial_\mu A^{*\mu}) \bar{\xi} \xi \\
&= - g_0 \left[ \bar{\psi}^{*\mu} \partial_\mu (C\xi) - \partial_\mu (C\bar{\xi}) \psi^{*\mu} \right] - g_1 A^{*\mu} \partial_\mu (\bar{\xi}\xi) + \mathrm{d}(\ldots) \\
&= - \Gamma \left[ g_0 (\bar{\psi}^{*\mu} \psi_\mu + \bar{\psi}_\mu \psi^{*\mu}) C + g_0 (\bar{\psi}^{*\mu} A_\mu \xi - \bar{\xi} A_\mu \psi^{*\mu}) \right. \\
&\quad \left. + g_1 A^{*\mu} (\bar{\psi}_\mu \xi - \bar{\xi} \psi_\mu) \right] + \mathrm{d}(\ldots).
\end{aligned} \tag{5.8}$$

Therefore, in view of Eq. (4.21b), one must have

$$a_1 = g_0 \left[ \bar{\psi}^{*\mu} (\psi_\mu C + \xi A_\mu) + \mathrm{h.c.} \right] + g_1 A^{*\mu} (\bar{\psi}_\mu \xi - \bar{\xi} \psi_\mu) + \tilde{a}_1, \quad \Gamma \tilde{a}_1 = 0, \tag{5.9}$$

where the ambiguity, $\tilde{a}_1$, belongs to the cohomology of $\Gamma$. Its most general form will be

$$\tilde{a}_1 = \left[ \bar{\psi}^{*\mu} X_{\mu\nu\rho} \Psi^{\nu\rho} \right] C + \left[ \bar{\psi}^{*\mu} Y_{\mu\nu\rho} F^{\nu\rho} + \bar{\Psi}^{\mu\nu} Z_{\mu\nu\rho} A^{*\rho} \right] \xi + \mathrm{h.c.}, \tag{5.10}$$

where $X, Y$ and $Z$ may contain derivatives and spinor indices.

---

[1] Here one also needs $\Delta \xi^* = -\partial_\mu \psi^{*\mu}$, $\Gamma \bar{\psi}_\mu = -\partial_\mu \bar{\xi}$, which follow from Table 5.1.

**Lagrangian Deformation**

We note that $\Delta a_1$ must be $\Gamma$-closed modulo d, since

$$\Gamma(\Delta a_1) = \Delta(-\Gamma a_1) = \Delta\left[\Delta a_2 + \mathrm{d}(...)\right] = \mathrm{d}(...). \tag{5.11}$$

Condition (4.21c), however, requires that $\Delta a_1$ be $\Gamma$-exact modulo d. The $\Delta$-variation of neither of the unambiguous pieces in $a_1$ is $\Gamma$-exact modulo d, and the non-trivial part must be killed by $\Delta \tilde{a}_1$, if (4.21c) is to hold at all. But such a cancellation is impossible for the first piece, i.e. for the would-be minimal coupling, simply because $\tilde{a}_1$ contains too many derivatives. Therefore, minimal coupling is ruled out, and we must set $g_0 = 0$. Thus, we have

$$\Delta a_1 = -\Gamma(g_1 \bar{\psi}_\mu F^{\mu\nu} \psi_\nu) - \tfrac{1}{2} g_1 F^{\mu\nu}(\bar{\Psi}_{\mu\nu}\xi - \bar{\xi}\Psi_{\mu\nu}) + \Delta \tilde{a}_1 + \mathrm{d}(...). \tag{5.12}$$

The second term on the right-hand side is in the cohomology of $\Gamma$ modulo d, and must be canceled by $\Delta \tilde{a}_1$. To see if this is possible or not, we make use of the identity

$$\eta^{\mu\nu|\alpha\beta} \equiv \tfrac{1}{2}\left(\eta^{\mu\alpha}\eta^{\nu\beta} - \eta^{\mu\beta}\eta^{\nu\alpha}\right) = \tfrac{1}{2}\gamma^{\mu\nu}\gamma^{\alpha\beta} - 2\gamma^{[\mu}\eta^{\nu][\alpha}\gamma^{\beta]} - \tfrac{1}{2}\gamma^{\mu\nu\alpha\beta} \tag{5.13}$$

to rewrite the term as

$$\begin{aligned}
F^{\mu\nu}(\bar{\Psi}_{\mu\nu}\xi - \bar{\xi}\Psi_{\mu\nu}) =& +\tfrac{1}{2}\left(\bar{\Psi}\!\!\!/\,F\!\!\!\!/\, - 4\bar{\Psi}_{\mu\nu}\gamma^\mu F^{\nu\rho}\gamma_\rho\right)\xi \\
& -\tfrac{1}{2}\bar{\xi}\left(F\!\!\!\!/\,\Psi\!\!\!/\, - 4\gamma_\mu F^{\mu\alpha}\gamma^\beta \Psi_{\alpha\beta}\right) \\
& -\tfrac{1}{2}\left(\bar{\Psi}_{\mu\nu}\gamma^{\mu\nu\alpha\beta}F_{\alpha\beta}\xi - \bar{\xi}F_{\mu\nu}\gamma^{\mu\nu\alpha\beta}\Psi_{\alpha\beta}\right) \\
=& +\tfrac{1}{2}\left(\bar{\Psi}\!\!\!/\,F\!\!\!\!/\, - 4\bar{\Psi}_{\mu\nu}\gamma^\mu F^{\nu\rho}\gamma_\rho\right)\xi \\
& -\tfrac{1}{2}\bar{\xi}\left(F\!\!\!\!/\,\Psi\!\!\!/\, - 4\gamma_\mu F^{\mu\alpha}\gamma^\beta \Psi_{\alpha\beta}\right) \\
& +\Gamma\left(\bar{\psi}_\mu \gamma^{\mu\nu\alpha\beta} F_{\alpha\beta}\psi_\nu\right) + \mathrm{d}(...).
\end{aligned} \tag{5.14}$$

Notice that we have rendered the second line in the first step $\Gamma$-exact modulo d by virtue of the Bianchi identity $\partial_{[\mu} F_{\nu\rho]} = 0$. We now plug Eq. (5.14) into (5.12) and obtain

$$\begin{aligned}
\Delta a_1 =& -\Gamma(g_1 \bar{\psi}_\mu F^{+\mu\nu}\psi_\nu) + \Delta \tilde{a}_1 + d(...) \\
& -\tfrac{1}{4}g_1 \left[\left(\bar{\Psi}\!\!\!/\,F\!\!\!\!/\, - 4\bar{\Psi}_{\mu\nu}\gamma^\mu F^{\nu\rho}\gamma_\rho\right)\xi - \bar{\xi}\left(F\!\!\!\!/\,\Psi\!\!\!/\, - 4\gamma_\mu F^{\mu\alpha}\gamma^\beta \Psi_{\alpha\beta}\right)\right].
\end{aligned} \tag{5.15}$$

Now, the most important point is that, the terms in the second line of the above expression are $\Delta$-exact, so that it is consistent to set

$$\Delta \tilde{a}_1 = \tfrac{1}{4}g_1 \left[\left(\bar{\Psi}\!\!\!/\,F\!\!\!\!/\, - 4\bar{\Psi}_{\mu\nu}\gamma^\mu F^{\nu\rho}\gamma_\rho\right)\xi - \bar{\xi}\left(F\!\!\!\!/\,\Psi\!\!\!/\, - 4\gamma_\mu F^{\mu\alpha}\gamma^\beta \Psi_{\alpha\beta}\right)\right]. \tag{5.16}$$

This is tantamount to setting

$$\tilde{a}_1 = ig_1\left(\bar{\psi}^{*\mu}\gamma^\nu F_{\mu\nu} - \tfrac{1}{2(D-2)}\bar{\psi}^* \slashed{F}\right)\xi + \text{h.c.}, \qquad (5.17)$$

which, of course, is in the cohomology of $\Gamma$. Eq. (5.15) then reduces to

$$\Delta a_1 = -\Gamma(g_1 \bar{\psi}_\mu F^{+\mu\nu}\psi_\nu) + \mathrm{d}(...), \qquad (5.18)$$

so that we have a consistent Lagrangian deformation $a_0$. To summarize, we have

$$a_0 = g_1\bar{\psi}_\mu F^{+\mu\nu}\psi_\nu, \quad a_1 = g_1 A^{*\mu}(\bar{\psi}_\mu \xi - \bar{\xi}\psi_\mu) + \tilde{a}_1, \quad a_2 = -g_1 C^*\bar{\xi}\xi, \qquad (5.19)$$

and minimal coupling is ruled out. It may be englightening to point out that, when one writes down the minimal coupling (by covariantizing ordirary partial derivatives in the kinetic term), what fails regarding gauge invariance is the gauge variation of the putative vertex with respect to the fermion gauge parameter: it does not vanish on the free equations of motion, because it is proportional to $F^{\mu\nu}$ and not to its divergence.[2] This fits with the obstruction as seen from the consistency cascade perspective — see comments earlier in the text.

### 5.1.2 Abelian Vertices

Now that we have exhausted all the possibilities for $a_2$, any other vertex can only have a trivial $a_2$. In this case, as we will show in Section 5.4, one can always choose to write the vertex as the photon field $A_\mu$ contracted with a gauge-invariant current $j^\mu$:

$$a_0 = j^\mu A_\mu, \qquad \Gamma j^\mu = 0, \qquad (5.20)$$

where the divergence of the current is $\Delta$-exact:

$$\partial_\mu j^\mu = \Delta M, \qquad \Gamma M = 0, \qquad (5.21)$$

so that one has $a_1 = MC$. If, however, $M$ happens to be $\Delta$-exact modulo d in the space of invariants, one can add a $\Delta$-exact term in $a_0$, so that the new current is identically conserved [177, 178]. In the latter case, the vertex does not deform the gauge symmetry at all (see Section 5.4).

---

[2] The gauge variation with respect to the photon gauge parameter is of course zero on the free equations of motion because the vertex is obtained by covariantizing partial derivatives with help of the vector field, precisely.

Now the most general vertex of the form (5.20) contains the current

$$j^\lambda = \bar\Psi_{\mu\nu}\, X^{\mu\nu\alpha\beta\lambda}\, \Psi_{\alpha\beta}, \tag{5.22}$$

whose divergence is required to obey the condition (5.21). Here $X$ may contain Dirac matrices as well as derivatives. It is not difficult to see that, if $X$ contains more than one derivative, $a_0$ is $\Delta$-exact modulo d, i.e. trivial. First, if $X$ contains the Laplacian, $\Box$, the contribution is always $\Delta$-exact, by the EoM $\Box\Psi_{\mu\nu}=0$. We can also forgo the Dirac operator, $\slashed\partial$, because by using the relation $\slashed\partial\gamma^\mu = 2\partial^\mu - \gamma^\mu \slashed\partial$ one can always make $\slashed\partial$ act on the curvature to get $\Delta$-exact terms, thanks to the EoM $\slashed\partial\Psi_{\mu\nu}=0$. Therefore, any derivative contained in $X^{\mu\nu\alpha\beta\lambda}$ must carry one of the five indices. Given the EoM $\partial^\mu\Psi_{\mu\nu}=0$, the antisymmetry of the field strength $\Psi_{\mu\nu}$, and the commutativity of ordinary derivatives, the only potentially non-trivial way to have more than one derivative is

$$a_0 = \bigl(\bar\Psi_{\mu\alpha}\overleftarrow\partial_\nu\, \gamma^\lambda\, \partial^\mu \Psi^{\alpha\nu}\bigr) A_\lambda. \tag{5.23}$$

However, algebraic manipulations show that this vertex is actually $\Delta$-exact modulo d, i.e. trivial. To see this, we use $\Psi^{\alpha\nu} = \partial^\alpha\psi^\nu - \partial^\nu\psi^\alpha$ to rewrite (5.23) as

$$a_0 = \bigl(\bar\Psi_{\mu\alpha}\overleftarrow\partial_\nu\, \gamma^\lambda\, \partial^\mu\partial^\alpha\psi^\nu - \tfrac{1}{2}\bar\Psi_{\mu\alpha}\overleftarrow\partial_\nu\, \gamma^\lambda\, \partial^\nu \Psi^{\mu\alpha}\bigr) A_\lambda.$$

While the first term is identically zero, in the second term one can use the so-called 3-box rule, $2\partial_\mu X \partial^\mu Y = \Box(XY) - X(\Box Y) - (\Box X)Y$, so that

$$a_0 = -\tfrac{1}{4}\bigl[\Box\bigl(\bar\Psi_{\mu\alpha}\,\gamma^\lambda\,\Psi^{\mu\alpha}\bigr) - \bigl(\Box\bar\Psi_{\mu\alpha}\bigr)\,\gamma^\lambda\,\Psi^{\mu\alpha} - \bar\Psi_{\mu\alpha}\,\gamma^\lambda\,\bigl(\Box\Psi^{\mu\alpha}\bigr)\bigr] A_\lambda.$$

In the above, the last two terms are $\Delta$-exact, whereas in the first term a double integration by parts gives $\Box A_\lambda$, which is equal to $\partial_\lambda(\partial\cdot A)$ by the photon EoM. One is then left with

$$a_0 = -\tfrac{1}{4}\bigl(\bar\Psi_{\mu\alpha}\,\gamma^\lambda\,\Psi^{\mu\alpha}\bigr)\,\partial_\lambda(\partial\cdot A) + \Delta\text{-exact} + \mathrm{d}(...).$$

Now, upon integrating by parts w.r.t. $\partial_\lambda$, this indeed becomes $\Delta$-exact modulo d:

$$a_0 = \bigl(\bar\Psi_{\mu\alpha}\overleftarrow\partial_\nu\, \gamma^\lambda\, \partial^\mu \Psi^{\alpha\nu}\bigr) A_\lambda = \Delta\text{-exact} + \mathrm{d}(...). \tag{5.24}$$

The only possibilities are therefore that $X$ contains either no derivatives or one derivative. In the former case, we have the candidate $X^{\mu\nu\alpha\beta\lambda} = -2\eta^{\mu\nu|\alpha\beta}\gamma^\lambda$, which gives

$$M = -4i\bar{\Psi}_{\mu\nu}\partial^\mu\left(\psi^{*\nu} - \tfrac{1}{D-2}\gamma^\nu\psi^*\right) - \text{h.c.}, \quad (5.25)$$

which is obviously gauge invariant: $\Gamma M = 0$. However, explicit computation easily shows that $M$ is actually $\Delta$-exact modulo d. Therefore, one can render the current identically conserved by adding a $\Delta$-exact term to it. In fact, in view of identity (5.13), our candidate $j^\mu$ is

$$j^\mu = \tfrac{1}{2}\bar{\Psi}_{\mu\nu}\left(\gamma^{\mu\nu\alpha\beta}\gamma^\lambda + \gamma^\lambda\gamma^{\mu\nu\alpha\beta}\right)\Psi_{\alpha\beta} + \Delta\text{-exact}. \quad (5.26)$$

Then, it is clear from the identity

$$\tfrac{1}{2}\gamma^{\mu\nu\alpha\beta}\gamma^\lambda + \tfrac{1}{2}\gamma^\lambda\gamma^{\mu\nu\alpha\beta} = \gamma^{\mu\nu\alpha\beta\lambda} \quad (5.27)$$

that our 2-derivatives vertex is actually off-shell equivalent[3] ($\approx$) to

$$a_0 \approx \left(\bar{\Psi}_{\mu\nu}\,\gamma^{\mu\nu\alpha\beta\lambda}\,\Psi_{\alpha\beta}\right)A_\lambda. \quad (5.28)$$

This vertex does not deform the gauge symmetry, and is gauge invariant up to a total derivative only. Note that the vertex does not exist in $D = 4$, because of the presence of $\gamma^{\mu\nu\alpha\beta\lambda}$. This is in complete agreement with Metsaev's results [62].

Finally, we are left with the possibility of having just one derivative in $X$, which would correspond to a 3-derivatives vertex. The only candidate is $X^{\mu\nu\alpha\beta\lambda} = \tfrac{1}{2}\eta^{\mu\nu|\alpha\beta}\overrightarrow{\partial}^\lambda$, which is equivalent to $-\tfrac{1}{4}\gamma^{\mu\nu\alpha\beta}\overrightarrow{\partial}^\lambda$ up to $\Delta$-exact terms, thanks to the identity (5.13). We have

$$a_0 = \tfrac{1}{2}\left(\bar{\Psi}_{\mu\nu}\,\eta^{\mu\nu|\alpha\beta}\,\overrightarrow{\partial}^\lambda\,\Psi_{\alpha\beta}\right)A_\lambda = \tfrac{1}{2}\left(\bar{\Psi}_{\mu\nu}\partial^\lambda\Psi^{\mu\nu} - \bar{\Psi}_{\mu\nu}\overleftarrow{\partial}^\lambda\Psi^{\mu\nu}\right)A_\lambda. \quad (5.29)$$

In this case too, our candidate current reduces on-shell to an identically conserved one, so that the vertex actually does not deform the gauge symmetry. To see it we use the Bianchi identity $\partial^\lambda\Psi^{\mu\nu} = -\partial^\mu\Psi^{\nu\lambda} + \partial^\nu\Psi^{\mu\lambda}$ to write the vertex as

$$a_0 = \left(-\bar{\Psi}_{\mu\nu}\partial^\mu\Psi^{\nu\lambda} + \bar{\Psi}_\nu{}^\lambda\overleftarrow{\partial}_\mu\Psi^{\mu\nu}\right)A_\lambda.$$

Then, thanks to the EoM $\partial^\mu\Psi_{\mu\nu} = 0$, up to $\Delta$-exact terms, the current reduces to the total derivative of a fermion bilinear, which is identically conserved:

$$a_0 \approx 2\partial_\nu\left(\bar{\Psi}_\alpha{}^{[\mu}\Psi^{\nu]\alpha}\right)A_\mu. \quad (5.30)$$

---

[3] What we mean is that our proof of equivalence is off-shell, and did not require using the equations of motion nor any gauge fixing condition.

Lastly, upon integration by parts the above expression is seen to be just a 3-curvatures term (Born–Infeld type),

$$a_0 \approx \bar{\Psi}_{\mu\alpha} \Psi^{\alpha}_{\ \nu} F^{\mu\nu}. \tag{5.31}$$

This exhausts all possible $1$–$\tfrac{3}{2}$–$\tfrac{3}{2}$ vertices. In Chapter 7, Table 7 presents a summary of our vertices.

Let us parenthetically comment about the nature of the abelian vertices. As it turned out, the vertices that do not deform the gauge algebra do not deform the gauge transformations either. In other words, if $a_2$ is trivial, so is $a_1$. This is not accidental at all: in fact, in Section 5.4 we are going to show that, for a massless particle of arbitrary spin $s = n + \tfrac{1}{2}$ coupled to a U(1) vector field, the cubic couplings that do not deform the gauge algebra actually do not deform the gauge transformations either, and hence only deform the Lagrangian.

## 5.2 Massless Spin 5/2 Coupled to Electromagnetism

Now we move on to constructing parity-preserving off-shell cubic vertices for a spin-$\tfrac{5}{2}$ gauge field, which is a symmetric rank-2 tensor-spinor $\psi_{\mu\nu}$. The original free action is

$$S^{(0)}[A_\mu, \psi_{\mu\nu}] = \int \mathrm{d}^D x \left[ -\tfrac{1}{4} F^2_{\mu\nu} - \tfrac{1}{2} \left( \bar{\psi}_{\mu\nu} \mathcal{R}^{\mu\nu} - \bar{\mathcal{R}}^{\mu\nu} \psi_{\mu\nu} \right) \right], \tag{5.32}$$

where the tensor $\mathcal{R}^{\mu\nu}$ is related to the spin-$\tfrac{5}{2}$ Fronsdal tensor $\mathcal{S}^{\mu\nu}$ as follows:

$$\mathcal{R}^{\mu\nu} = \mathcal{S}^{\mu\nu} - \gamma^{(\mu} \not{\mathcal{S}}^{\nu)} - \tfrac{1}{2} \eta^{\mu\nu} \mathcal{S}', \qquad \mathcal{S}' \equiv \mathcal{S}^\mu_\mu. \tag{5.33}$$

Here the photon gauge invariance is as usual, while the fermionic part is gauge invariant under a *constrained* vector-spinor gauge parameter, $\varepsilon_\mu$,

$$\delta_\varepsilon \psi_{\mu\nu} = 2 \partial_{(\mu} \varepsilon_{\nu)}, \qquad \not{\varepsilon} = 0. \tag{5.34}$$

Then, the corresponding Grassmann-even fermionic ghost, $\xi_\mu$, must also be $\gamma$-traceless:

$$\not{\xi} = 0, \tag{5.35}$$

and so will be its antighost. The set of fields and antifields under study are given below.

$$\Phi^A = \{A_\mu, C, \psi_{\mu\nu}, \xi_\mu\}, \qquad \Phi^*_A = \{A^{*\mu}, C^*, \bar{\psi}^{*\mu\nu}, \bar{\xi}^{*\mu}\}. \tag{5.36}$$

The *free* master action, $S_0$, takes the form

$$S_0 = \int \mathrm{d}^D x \big[ -\tfrac{1}{4} F_{\mu\nu}^2 - \tfrac{1}{2} \big( \bar{\psi}_{\mu\nu} \mathcal{R}^{\mu\nu} - \bar{\mathcal{R}}^{\mu\nu} \psi_{\mu\nu} \big) + A^{*\mu} \partial_\mu C \\ + 2(\bar{\psi}^{*\mu\nu} \partial_\mu \xi_\nu - \partial_\mu \bar{\xi}_\nu \psi^{*\mu\nu}) \big]. \tag{5.37}$$

The properties of the various fields and antifields are given in Table 5.2. Note that the spin-$\tfrac{5}{2}$ curvature tensor is the 2-curl (see Appendix E for its properties),

$$\Psi_{\mu_1 \nu_1 | \mu_2 \nu_2} = [\partial_{\mu_1} \partial_{\mu_2} \psi_{\nu_1 \nu_2} - (\mu_1 \leftrightarrow \nu_1)] - (\mu_2 \leftrightarrow \nu_2). \tag{5.38}$$

The cohomology of $\Gamma$ is isomorphic to the space of functions of (again see Appendix E)

- The undifferentiated ghosts $\{C, \xi_\mu\}$ and the $\gamma$-traceless part of the 1-curl of the spinorial ghost, $\xi^{(1)}_{\mu\nu} = 2\partial_{[\mu} \xi_{\nu]}$,

- The antifields $\{A^{*\mu}, C^*, \bar{\psi}^{*\mu\nu}, \bar{\xi}^{*\mu}\}$, and their derivatives,

- The curvatures $\{F_{\mu\nu}, \Psi_{\mu_1\nu_1|\mu_2\nu_2}\}$, and their derivatives,

- The Fronsdal tensor $\mathcal{S}_{\mu\nu}$, and its symmetrized derivatives.

Table 5.2: Properties of the Various Fields & Antifields ($n = 2$)

| $Z$ | $\Gamma(Z)$ | $\Delta(Z)$ | pgh$(Z)$ | agh$(Z)$ | gh$(Z)$ | $\epsilon(Z)$ |
|---|---|---|---|---|---|---|
| $A_\mu$ | $\partial_\mu C$ | 0 | 0 | 0 | 0 | 0 |
| $C$ | 0 | 0 | 1 | 0 | 1 | 1 |
| $A^{*\mu}$ | 0 | $-\partial_\nu F^{\mu\nu}$ | 0 | 1 | $-1$ | 1 |
| $C^*$ | 0 | $-\partial_\mu A^{*\mu}$ | 0 | 2 | $-2$ | 0 |
| $\psi_{\mu\nu}$ | $2\partial_{(\mu} \xi_{\nu)}$ | 0 | 0 | 0 | 0 | 1 |
| $\xi_\mu$ | 0 | 0 | 1 | 0 | 1 | 0 |
| $\bar{\psi}^{*\mu\nu}$ | 0 | $\bar{\mathcal{R}}^{\mu\nu}$ | 0 | 1 | $-1$ | 0 |
| $\bar{\xi}^{*\mu}$ | 0 | $2\partial_\nu \bar{\chi}^{*\mu\nu}$ | 0 | 2 | $-2$ | 1 |

Note that, because the spin of the fermion field has increased by one unit, the divergence $\partial_\nu \mathcal{R}^{\mu\nu}$ is no longer zero, but is proportional to $\gamma^\mu$.[4] Because of this, when $\Delta$ acts on the fermionic antighost $\bar{\xi}^{*\mu}$, the result is more than a simple divergence of the antifield $\bar{\psi}^{*\mu\nu}$. This property, detailed in Appendix E, will not be crucial for the following electromagnetic couplings but will play an important role in their gravitational counterpart, studied in Chapter 6. Explicitly,

$$\Delta \bar{\xi}^{*\mu} = 2\partial_\nu \bar{\chi}^{*\mu\nu}, \qquad \bar{\chi}^{*\mu\nu} \equiv \bar{\psi}^{*\mu\nu} - \tfrac{1}{D}\bar{\slashed{\psi}}^{*\nu}\gamma^\mu. \tag{5.39}$$

---

[4] The action is still gauge invariant, however, thanks to the $\gamma$-tracelessness of the gauge parameter $\varepsilon_\mu$.

### 5.2.1 Non-Abelian Vertices

The set of all possible non-trivial $a_2$'s falls into two subsets: Subset 1 contains *both* the bosonic ghost $C$ and the fermionic ghost $\xi_\mu$, while Subset 2 contains $\xi_\mu$ but *not* $C$.

**Subset 1:** $\left\{ C\left(\bar{\xi}^*_\mu \xi^\mu + \bar{\xi}_\mu \xi^{*\mu}\right),\, C\left(\bar{\xi}^{*(1)}_{\mu\nu} \xi^{(1)\mu\nu} + \bar{\xi}^{(1)}_{\mu\nu} \xi^{*(1)\mu\nu}\right) \right\}$,

**Subset 2:** $\left\{ C^* \bar{\xi}_\mu \xi^\mu,\, C^* \bar{\xi}^{(1)}_{\mu\nu} \xi^{(1)\mu\nu} \right\}$.

One can easily verify that other possible rearrangements of derivatives or other possible contractions of the indices, e.g. by $\gamma$-matrices, all give trivial terms thanks to the $\gamma$-tracelessness of the fermionic ghost and its antighost. Here, the term $C\bar{\xi}^*_\mu \xi^\mu$ corresponds to potential minimal coupling and the other candidate $a_2$'s to higher-derivative interactions.

To see which of the $a_2$'s can be lifted to some $a_1$, let us solve Eq. (4.21b). Now, a computation similar to what leads one from (5.7) to (5.9) shows that both elements of Subset 1 above enjoy such a lift, thanks to the relations (E.49)–(E.50) among others. Explicitly,

$$a_2 = \begin{cases} C\,\bar{\xi}^*_\mu \xi^\mu \\ C\,\bar{\xi}^{*(1)}_{\mu\nu} \xi^{(1)\mu\nu} \end{cases} \rightarrow \quad a_1 = \begin{cases} -\bar{\psi}^{*\mu\nu}\left(\psi_{\mu\nu}C + 2\xi_\mu A_\nu\right) + \tilde{a}_1 \\ -\bar{\psi}^{*(1)\mu\nu\|\rho}(\psi^{(1)}_{\mu\nu\|\rho}C + 2\xi^{(1)}_{\mu\nu} A_\rho) + \tilde{a}_1, \end{cases} \tag{5.40}$$

and similarly for the hermitian conjugate terms. Here $\tilde{a}_1$ is the usual ambiguity, satisfying $\Gamma \tilde{a}_1 = 0$.

To see whether these could be further lifted to $a_0$'s, we write

$$\Delta a_1 = \begin{cases} -\bar{\mathcal{R}}^{\mu\nu}\left(\psi_{\mu\nu}C + 2\xi_\mu A_\nu\right) + \Delta\tilde{a}_1 \\ -\bar{\mathcal{R}}^{(1)\mu\nu\|\rho}(\psi^{(1)}_{\mu\nu\|\rho}C + 2\xi^{(1)}_{\mu\nu} A_\rho) + \Delta\tilde{a}_1. \end{cases} \tag{5.41}$$

It is important to notice that, up to total derivatives, the $\Delta a_1$'s have an expansion in the basis of *undifferentiated* ghosts, $\omega_I = \{C, \xi_\mu\}$. Because $\Gamma(\Delta \tilde{a}_1) = -\Delta(\Gamma \tilde{a}_1) = 0$, the coefficients $\alpha^I$ in the expansion of the ambiguity will be $\Gamma$-cocycles, i.e. they will be 'invariant polynomials'. Clearly, this is not the case for the unambiguous pieces, which are only cocycles of $\mathrm{H}(\Gamma|\mathrm{d})$.[5] However, their expansion coefficients $\beta^I$ are such that their $\Gamma$-variation is not the total derivative *of a* $\Gamma$-exact object. Schematically,

$$\Delta a_1 = \left(\alpha^I + \beta^I\right)\omega_I + \mathrm{d}(...), \qquad \Gamma \alpha^I = 0, \qquad \Gamma \beta^I \neq \mathrm{d}\Gamma(...). \quad (5.42)$$

Now, $\Gamma a_0$ is a pgh$\# = 1$ object that can be expanded, up to a total derivative, in the basis of $\{\partial_\mu C, \partial_{(\mu}\xi_{\nu)}\}$. Then, obviously, one can also expand it in the undifferentiated ghosts $\omega_I$:

$$\Gamma a_0 = -\left(\partial \cdot J\right)^I \omega_I + \mathrm{d}(...). \quad (5.43)$$

Let us then plug the respective expansions (5.42) and (5.43) for $\Delta a_1$ and $\Gamma a_0$ into the consistency condition (4.21c), and subsequently take a functional derivative w.r.t. $\omega_I = \{C, \xi_\mu\}$, finding

$$\alpha^I + \beta^I = \partial \cdot J^I = \mathrm{d}(...). \quad (5.44)$$

But, if this is true, then $\Gamma(\alpha^I + \beta^I) = \mathrm{d}\Gamma(...)$, which is in direct contradiction with the properties of $\alpha^I$ and $\beta^I$ given in (5.42) above. The conclusion is that none of the $a_2$'s in Subset 1 can be lifted all the way to $a_0$. It is important to notice that this obstruction originates from the very nature of the $a_2$'s themselves, namely each of them contains *both* the ghosts.

**Remark** : For the would-be minimal coupling, the impossibility can also be seen as a consequence of $\alpha^I$ containing too many derivatives compared to $\beta^I$, which is the argument we have used for spin $\frac{3}{2}$.

For Subset 2, the analysis simplifies because only the second term therein, with the maximum number of derivatives, can be lifted to an $a_1$. For the other term we have

$$\Delta\left(C^* \bar{\xi}_\nu \, \xi^\nu\right) = A^{*\mu}\left(\bar{\xi}^\nu \partial_\mu \xi_\nu + \partial_\mu \bar{\xi}_\nu \, \xi^\nu\right) + \mathrm{d}(...), \quad (5.45)$$

and because one can write $\partial_\mu \xi_\nu = \partial_{[\mu}\xi_{\nu]} + \partial_{(\mu}\xi_{\nu)}$, which is the sum of a non-trivial and a trivial element in the cohomology of $\Gamma$, the right-hand side of the above equation cannot be $\Gamma$-exact modulo d. Hence the candidate $C^*\bar{\xi}_\mu \, \xi^\mu$ is ruled out. However, for the second term of Subset 2 one finds that

$$\Delta\left(C^* \bar{\xi}^{(1)}_{\mu\nu} \, \xi^{(1)\mu\nu}\right) = A^{*\rho}\left(\bar{\xi}^{(1)\mu\nu} \partial_\rho \xi^{(1)}_{\mu\nu} + \partial_\rho \bar{\xi}^{(1)}_{\mu\nu} \, \xi^{(1)\mu\nu}\right) + \mathrm{d}(...) \quad (5.46)$$
$$= \Gamma\left[A^{*\rho}\left(\bar{\psi}^{(1)}_{\mu\nu\|\,\rho}\, \xi^{(1)\mu\nu} - \bar{\xi}^{(1)\mu\nu}\, \psi^{(1)}_{\mu\nu\|\,\rho}\right)\right] + \mathrm{d}(...),$$

---

[5] But still, because of Eq. (5.11) one must have $\beta^I \omega_I \in \mathrm{H}^1(\Gamma|\mathrm{d})$, and indeed this is the case.

thanks to the relation (E.50). Thus, indeed, $C^*\bar{\xi}^{(1)}_{\mu\nu}\xi^{(1)\mu\nu}$ gets lifted to an $a_1$:

$$a_2 = C^*\bar{\xi}^{(1)}_{\mu\nu}\xi^{(1)\mu\nu} \ \to \ a_1 = -A^{*\rho}\big(\bar{\psi}^{(1)}_{\mu\nu\|\rho}\xi^{(1)\mu\nu} - \bar{\xi}^{(1)\mu\nu}\psi^{(1)}_{\mu\nu\|\rho}\big) + \tilde{a}_1. \tag{5.47}$$

To see if this $a_1$ can be lifted to an $a_0$, we compute its $\Delta$ variation,

$$\Delta a_1 = \Gamma\big(\bar{\psi}^{(1)}_{\alpha\beta\|\mu}F^{\mu\nu}\psi^{(1)\alpha\beta\|}{}_\nu\big) + \tfrac{1}{2}F_{\mu\nu}\big(\bar{\Psi}^{\mu\nu|\alpha\beta}\xi^{(1)}_{\alpha\beta} - \bar{\xi}^{(1)}_{\alpha\beta}\Psi^{\mu\nu|\alpha\beta}\big)$$
$$+ \Delta\tilde{a}_1 + \mathrm{d}(...). \tag{5.48}$$

This equation bears striking resemblance with its spin-$\tfrac{3}{2}$ counterpart, Eq. (5.12). We recall that, in the latter, cancellation of non-$\Gamma$-exact terms was possible by the insertion of identity (5.13) in the contraction of curvatures, the Bianchi identity $\partial_{[\mu}F_{\nu\rho]}=0$ and the fermion EoMs in terms of the curvature, $\gamma^\mu\Psi_{\mu\nu}=0, \gamma^{\mu\nu}\Psi_{\mu\nu}=0$. In the present case as well, as shown in Appendix E, the fermion EoMs can be written as the $\gamma$-traces of the curvature, $\gamma^\mu\Psi_{\mu\nu|\alpha\beta}=0, \gamma^{\mu\nu}\Psi_{\mu\nu|\alpha\beta}=0$. Therefore, the non-$\Gamma$-exact terms from the unambiguous piece in the above expression can indeed be canceled by the $\Delta$ variation of a $\Gamma$-closed ambiguity,

$$\Delta\tilde{a}_1 = -\tfrac{1}{4}\big(\bar{\slashed{\Psi}}_{\alpha\beta}\slashed{F} - 4\bar{\Psi}_{\mu\nu|\alpha\beta}\gamma^\mu F^{\nu\rho}\gamma_\rho\big)\xi^{(1)\alpha\beta} + \text{h.c.} \tag{5.49}$$

We thus have a lift all the way to $a_0$, the latter being a 3-derivatives non-abelian vertex,

$$a_0 = -\bar{\psi}^{(1)}_{\alpha\beta\|\mu}F^{+\mu\nu}\psi^{(1)\alpha\beta\|}{}_\nu. \tag{5.50}$$

### 5.2.2 Abelian Vertices

All the statements (5.20)–(5.21) still hold good in this case, and the current in the vertex $a_0 = j^\mu A_\mu$ is an invariant polynomial, which takes the most general form

$$j^\lambda = \bar{\Psi}_{\mu_1\nu_1|\mu_2\nu_2}X^{\mu_1\nu_1\alpha_1\beta_1\lambda\mu_2\nu_2\alpha_2\beta_2}\Psi_{\alpha_1\beta_1|\alpha_2\beta_2}. \tag{5.51}$$

Notice that the Fronsdal tensor, although allowed in principle, cannot appear in the current simply because it would render the vertex $\Delta$-exact. In view of the spin-$\tfrac{5}{2}$ EoMs and the symmetry properties of the field strength, one can show, like in Subsection 5.1.2, that $X^{\mu_1\nu_1\alpha_1\beta_1\lambda\mu_2\nu_2\alpha_2\beta_2}$ can contain at most one derivative, which must carry one of the indices.

When $X$ does not contain any derivative, the corresponding vertex will contain 4. In this case, we have the candidate

$$X^{\mu_1\nu_1\alpha_1\beta_1\lambda\mu_2\nu_2\alpha_2\beta_2} = -2\eta^{\mu_1\nu_1|\alpha_1\beta_1}\eta^{\mu_2\nu_2|\alpha_2\beta_2}\gamma^\lambda, \tag{5.52}$$

but again, the identities (5.13) and (5.27) tell us that the resulting vertex deforms nothing. It reads:

$$a_0 \approx \left(\bar{\Psi}_{\mu_1\nu_1|\mu_2\nu_2}\gamma^{\mu_1\nu_1\alpha_1\beta_1\lambda}\Psi_{\alpha_1\beta_1|}{}^{\mu_2\nu_2}\right)A_\lambda. \tag{5.53}$$

Finally, the 1-derivative candidate is

$$X^{\mu_1\nu_1\alpha_1\beta_1\lambda\mu_2\nu_2\alpha_2\beta_2} = \tfrac{1}{2}\eta^{\mu_1\nu_1|\alpha_1\beta_1}\eta^{\mu_2\nu_2|\alpha_2\beta_2}\overrightarrow{\partial}^\lambda, \tag{5.54}$$

which is equivalent to a 5-derivatives and 3-curvatures term (Born–Infeld),

$$a_0 \approx \bar{\Psi}_{\mu_1\nu_1|\mu_2\rho}\Psi^{\mu_1\nu_1|\rho}{}_{\nu_2}F^{\mu_2\nu_2}. \tag{5.55}$$

In Table 7 of Chapter 7 we present a summary table for all possible $1-\tfrac{5}{2}-\tfrac{5}{2}$ vertices.

## 5.3 Arbitrary-Spin Couplings

The set of fields and antifields in this case is given by

$$\Phi^A = \{A_\mu, C, \psi_{\mu_1\ldots\mu_n}, \xi_{\mu_1\ldots\mu_{n-1}}\}, \tag{5.56a}$$
$$\Phi_A^* = \{A^{*\mu}, C^*, \bar{\psi}^{*\mu_1\ldots\mu_n}, \bar{\xi}^{*\mu_1\ldots\mu_{n-1}}\}. \tag{5.56b}$$

For $n > 2$, there is an additional constraint that the field-antifield are triply $\gamma$-traceless:

$$\slashed{\psi}'_{\mu_1\mu_3\ldots\mu_{n-3}} = 0, \qquad \slashed{\bar{\psi}}^{*\prime}_{\mu_1\mu_3\ldots\mu_{n-3}} = 0, \tag{5.57}$$

where prime denotes the trace w.r.t the Minkowski metric. Besides, the rank-$(n-1)$ fermionic ghost and its antighost are $\gamma$-traceless:

$$\slashed{\xi}_{\mu_1\ldots\mu_{n-2}} = 0, \qquad \slashed{\bar{\xi}}^*_{\mu_1\ldots\mu_{n-2}} = 0. \tag{5.58}$$

The properties of the various fields and antifields are given in Table 5.3 below, and we also recall that the antifield $\bar{\chi}^{*\mu_1\ldots\mu_n}$ is given by Eqs. (E.34)–(E.35).

The rank-$n$ tensor-spinor $\mathcal{R}_{\mu_1\ldots\mu_n}$ appearing in the spin-$s$ EoMs is an arbitrary-spin generalization of (5.33); it is related to the Fronsdal tensor by

$$\mathcal{R}_{\mu_1\ldots\mu_n} = \mathcal{S}_{\mu_1\ldots\mu_n} - \tfrac{1}{2}n\,\gamma_{(\mu_1}\slashed{\mathcal{S}}_{\mu_2\ldots\mu_n)} - \tfrac{1}{4}n(n-1)\,\eta_{(\mu_1\mu_2}\mathcal{S}'_{\mu_3\ldots\mu_n)}. \tag{5.59}$$

The cohomology of $\Gamma$ has already been given in the lower-spin cases treated in Section 5.2 and 5.1, with the details appearing in Appendix E.

Table 5.3: Properties of the Various Fields & Antifields ($\forall\, n$)

| $Z$ | $\Gamma(Z)$ | $\Delta(Z)$ | pgh$(Z)$ | agh$(Z)$ | gh$(Z)$ | $\epsilon(Z)$ |
|---|---|---|---|---|---|---|
| $A_\mu$ | $\partial_\mu C$ | $0$ | $0$ | $0$ | $0$ | $0$ |
| $C$ | $0$ | $0$ | $1$ | $0$ | $1$ | $1$ |
| $A^{*\mu}$ | $0$ | $-\partial_\nu F^{\mu\nu}$ | $0$ | $1$ | $-1$ | $1$ |
| $C^*$ | $0$ | $-\partial_\mu A^{*\mu}$ | $0$ | $2$ | $-2$ | $0$ |
| $\psi_{\mu_1\ldots\mu_n}$ | $n\partial_{(\mu_1}\xi_{\mu_2\ldots\mu_n)}$ | $0$ | $0$ | $0$ | $0$ | $1$ |
| $\xi_{\mu_1\ldots\mu_{n-1}}$ | $0$ | $0$ | $1$ | $0$ | $1$ | $0$ |
| $\bar\psi^{*\mu_1\ldots\mu_n}$ | $0$ | $\bar{\mathcal{R}}^{\mu_1\ldots\mu_n}$ | $0$ | $1$ | $-1$ | $0$ |
| $\bar\xi^{*\mu_1\ldots\mu_{n-1}}$ | $0$ | $2\partial_{\mu_n}\bar\chi^{*\mu_1\ldots\mu_n}$ | $0$ | $2$ | $-2$ | $1$ |

Then, one can immediately write down the set of all possible non-trivial $a_2$'s, and again, they fall into two subsets: Subset 1 contains *both* the bosonic ghost $C$ and the fermionic ghost $\xi_{\mu_1\ldots\mu_{n-1}}$, while Subset 2 contains only $\xi_{\mu_1\ldots\mu_{n-1}}$.

**Subset 1:** $\left\{C\,\bar\xi^{*(m)}_{\mu_1\nu_1|\ldots|\mu_m\nu_m\|\nu_{m+1}\ldots\nu_{n-1}}\xi^{(m)\mu_1\nu_1|\ldots|\mu_m\nu_m\|\nu_{m+1}\ldots\nu_{n-1}}\right\}$,

**Subset 2:** $\left\{C^*\bar\xi^{(m)}_{\mu_1\nu_1|\ldots|\mu_m\nu_m\|\nu_{m+1}\ldots\nu_{n-1}}\xi^{(m)\mu_1\nu_1|\ldots|\mu_m\nu_m\|\nu_{m+1}\ldots\nu_{n-1}}\right\}$.

Note that in Subset 1 the hermitian conjugate of the displayed expression should be added too, and we have given the subsets for $0 \leq m \leq n-1$. As a straightforward generalization of the spin-$\tfrac{5}{2}$ case, one finds that each element in Subset 1 gets lifted to an $a_1$:

$$\begin{aligned}a_1 =\ &-\bar\psi^{*(m)\mu_1\nu_1|\ldots|\mu_m\nu_m\|\nu_{m+1}\ldots\nu_n}\psi^{(m)}_{\mu_1\nu_1|\ldots|\mu_m\nu_m\|\nu_{m+1}\ldots\nu_n}C + \tilde a_1 \\ &- n\,\bar\psi^{*(m)\mu_1\nu_1|\ldots|\mu_m\nu_m\|\nu_{m+1}\ldots\nu_n}\xi^{(m)}_{\mu_1\nu_1|\ldots|\mu_m\nu_m\|(\nu_{m+1}\ldots\nu_{n-1}}A_{\nu_n)} + \text{h.c.}\end{aligned}\qquad(5.60)$$

Now, one can compute $\Delta a_1$ and expand it in the basis of pgh$\#=1$ objects in the cohomology of $\Gamma$, namely

$$\omega_I = \left\{C,\ \xi^{(m)}_{\mu_1\nu_1|\ldots|\mu_m\nu_m\|\nu_{m+1}\ldots\nu_{n-1}}\ \big|\ 0\leq m\leq n-1\right\}.\qquad(5.61)$$

Upon comparing the expansion coefficients for the unambiguous piece and the ambiguity $\tilde a_1$, again one can conclude that none of these $a_1$'s can be lifted to an $a_0$. On the other hand, for the elements of Subset 2, one notices that

$$\Delta\!\left(C^*\bar\xi^{(m)}\cdot\xi^{(m)}\right) = A^{*\nu}\bar\xi^{(m)}\cdot\partial_\nu\xi^{(m)} + \text{h.c.} + \text{d}(\ldots),\qquad(5.62)$$

where we have used the notation

$$\bar{\xi}^{(m)} \cdot \xi^{(m)} \equiv \bar{\xi}^{(m)}_{\mu_1\nu_1|...|\mu_m\nu_m\|\nu_{m+1}...\nu_{n-1}} \xi^{(m)\mu_1\nu_1|...|\mu_m\nu_m\|\nu_{m+1}...\nu_{n-1}}. \quad (5.63)$$

Then, in view of Eqs. (E.49) and (E.50), it is clear that the right-hand side of (5.62) is $\Gamma$-exact modulo d *only* for $m = n - 1$, which rules out, in particular, the would-be minimal coupling corresponding to $m = 0$. Therefore, one is left with the lift:

$$a_1 = -A^{*\nu_n}\,\bar{\psi}^{(n-1)}_{\mu_1\nu_1|...|\mu_{n-1}\nu_{n-1}\|\nu_n}\,\xi^{(n-1)\mu_1\nu_1|...|\mu_{n-1}\nu_{n-1}} + \text{h.c.} + \tilde{a}_1, \quad (5.64)$$

whose $\Delta$-variation is given by

$$\begin{aligned}\Delta a_1 = {}& \Gamma\big(\bar{\psi}^{(n-1)}_{\mu_1\nu_1|...|\mu_{n-1}\nu_{n-1}\|\,\mu_n} F^{\mu_n}{}_{\nu_n}\,\psi^{(n-1)\mu_1\nu_1|...|\mu_{n-1}\nu_{n-1}\|\,\nu_n}\big) \\ & + \tfrac{1}{2} F^{\mu_n\nu_n}\big(\bar{\Psi}_{\mu_1\nu_1|...|\mu_n\nu_n}\,\xi^{(n-1)\mu_1\nu_1|...|\mu_{n-1}\nu_{n-1}} + \text{h.c.}\big) \\ & + \Delta\tilde{a}_1 + \mathrm{d}(...).\end{aligned} \quad (5.65)$$

In view of Eq. (5.12) and (5.48), pertaining respectively to the spin-$\frac{3}{2}$ and spin-$\frac{5}{2}$ cases, and the subsequent steps, we realize that it is possible to cancel the non-$\Gamma$-exact terms in the above expression by inserting identity (5.13) in the contraction of curvatures, thanks to the Bianchi identity $\partial_{[\mu} F_{\nu\rho]} = 0$, and to the fermion EoMs in terms of curvatures (see Appendix E), $\gamma^{\mu_1}\Psi_{\mu_1\nu_1|...|\mu_n\nu_n} = 0$, $\gamma^{\mu_1\nu_1}\Psi_{\mu_1\nu_1|...|\mu_n\nu_n} = 0$. In other words, $\Delta a_1$ is rendered $\Gamma$-exact modulo d by an appropriate choice of the ambiguity $\tilde{a}_1$, just like in the previous examples, so that one finally has

$$a_0 = -\bar{\psi}^{(n-1)}_{\mu_1\nu_1|...|\mu_{n-1}\nu_{n-1}\|\mu_n} F^{+\mu_n}{}_{\nu_n}\psi^{(n-1)\mu_1\nu_1|...|\mu_{n-1}\nu_{n-1}\|\nu_n}, \quad (5.66)$$

which is the $(2n - 1)$-derivatives non-abelian vertex containing the $(n - 1)$-curl of the fermionic field.

For an abelian vertex $a_0 = j^\mu A_\mu$, the gauge-invariant current does not contain the Fronsdal tensor nor its derivatives, since their presence would render the vertex $\Delta$-exact. Again, non-triviality of the abelian deformation allows for only two possible values for the number of derivatives in the vertex: $2n$ and $2n+1$. The off-shell vertices can be obtained in exactly the same way as for spins $\frac{3}{2}$ and $\frac{5}{2}$, considered in Subsections 5.1.2 and 5.2.2 respectively. A summary of all possible $1-s-s$ vertices is given in Table 7 in Chapter 7.

## 5.4 Abelian Vertices and Gauge Symmetries

In the present section we prove an important result: abelian vertices also preserve the gauge symmetries. Differently put, this means that a given vertex either deforms both the gauge symmetries and the gauge algebra or none of them (hence deforming only the Lagrangian). This property has been verified explicitly in the previous sections dealing with the various electromagnetic couplings, and here we prove it in full generality. We note, however, that the result demonstrated below is valid in our context only (that is, for our set of fields, etc.), and examples exist in the literature of 'in-between' vertices, deforming the gauge symmetries but not the gauge algebra; see e.g. [191], or [192] for a treatment of the Freedman–Townsend model within the BRST formulation.

Abelian vertices are those that do not deform the gauge algebra, i.e. they can only have a trivial $a_2$. As mentioned in Chapter 4, for such a vertex it is always possible to choose $a_1$ to be $\Gamma$-closed [177, 178]:

$$\Gamma a_1 = 0, \tag{5.67}$$

and if it gets lifted to an $a_0$, one has the cocycle condition (4.21c),

$$\Delta a_1 + \Gamma a_0 + \mathrm{d} b_0 = 0. \tag{5.68}$$

Now, for the $1-s-s$ vertices under study, one can always write a vertex as the photon field $A_\mu$ contracted with a current $j^\mu$, which is a fermion bilinear:

$$a_0 = j^\mu A_\mu, \tag{5.69}$$

and one can always choose the current such that it satisfies

$$\Gamma j^\mu = 0, \qquad \partial_\mu j^\mu = \Delta M, \qquad \Gamma M = 0. \tag{5.70}$$

To see this, let us note that the $a_1$ corresponding to (5.69) has the general form

$$a_1 = MC + \left(\bar{P}_{\mu_1\ldots\mu_{n-1}}\xi^{\mu_1\ldots\mu_{n-1}} - \bar{\xi}_{\mu_1\ldots\mu_{n-1}}P^{\mu_1\ldots\mu_{n-1}}\right) + a'_1, \tag{5.71}$$

where $M$ and $P_{\mu_1\ldots\mu_{n-1}}$ belong to H($\Gamma$), with pgh$\# = 0$, agh$\# = 1$, and $a'_1$ stands for expansion terms in the ghost-curls. Given (5.69) and (5.71), the condition (5.68) reads

$$\Gamma\left(j^\mu A_\mu\right) + \Delta MC + \left(\Delta\bar{P}_{\mu_1\ldots\mu_{n-1}}\xi^{\mu_1\ldots\mu_{n-1}} - \bar{\xi}_{\mu_1\ldots\mu_{n-1}}\Delta P^{\mu_1\ldots\mu_{n-1}}\right)$$
$$+ \Delta a'_1 + \mathrm{d} b_0 = 0. \tag{5.72}$$

Now, $P_{\mu_1...\mu_{n-1}}$ consists of two kinds of terms: one contains the antifield $A^{*\mu}$ and its derivatives, and the other contains the antifield $\psi^{*\nu_1...\nu_n}$ and its derivatives. The former one also contains (derivatives of) the Fronsdal tensor $\mathcal{S}_{\nu_1...\nu_n}$, or (derivatives of) the curvature $\Psi_{\rho_1\nu_1|...|\rho_n\nu_n}$, while the latter one contains (derivatives of) the electromagnetic field strength $F_{\mu\nu}$. One can choose to get rid of derivatives on $A^{*\mu}$ and $F_{\mu\nu}$ by using the Leibniz rule,

$$P_{\mu_1...\mu_{n-1}} = A^{*\mu}\big(\vec{P}^{(\mathcal{S})\nu_1...\nu_n}_{\mu,\,\mu_1...\mu_{n-1}} \mathcal{S}_{\nu_1...\nu_n} + \vec{P}^{(\Psi)\rho_1\nu_1|...|\rho_n\nu_n}_{\mu,\,\mu_1...\mu_{n-1}} \Psi_{\rho_1\nu_1|...|\rho_n\nu_n}\big)$$
$$+ F^{\mu\nu}\big(\vec{P}^{(\psi^*)\nu_1...\nu_n}_{\mu\nu,\,\mu_1...\mu_{n-1}} \psi^*_{\nu_1...\nu_n}\big) + \partial^{\mu_n} p_{\mu_1...\mu_n}, \qquad (5.73)$$

where $\Gamma p_{\mu_1...\mu_n} = 0$ and the $\vec{P}$'s are operators acting to the right. Notice that the quantity in the parentheses in the first line is both $\Gamma$-closed and $\Delta$-exact [6]. Now, one can take the $\Delta$-variation of (5.73), and then add a total derivative in order to cast it in the form

$$\Delta P_{\mu_1...\mu_{n-1}} = \tfrac{1}{2} F^{\mu\nu} \Delta Q_{[\mu\nu],\,\mu_1...\mu_{n-1}} + \partial^{\mu_n} \Delta q_{\mu_1...\mu_n}, \qquad (5.74)$$

where $\Gamma Q_{[\mu\nu],\,\mu_1...\mu_{n-1}} = 0$, $\Gamma q_{\mu_1...\mu_n} = 0$. Therefore, we have

$$\bar{\xi}_{\mu_1...\mu_{n-1}} \Delta P^{\mu_1...\mu_{n-1}} = A_\mu \Delta\big[\partial_\nu\big(\bar{\xi}_{\mu_1...\mu_{n-1}} Q^{[\mu\nu],\,\mu_1...\mu_{n-1}}\big)\big] \qquad (5.75)$$
$$- \bar{\xi}_{\mu_1...\mu_{n-1}} \overleftarrow{\partial}_{\mu_n} \Delta q^{\mu_1...\mu_n} + \mathrm{d}(...).$$

The second term on the right side is $\Gamma$-closed, and can be broken as a $\Gamma$-exact piece plus terms involving the ghost-curls. The latter can always be canceled in the cocycle condition (5.72) by appropriately choosing the similar terms coming from $a'_1$. Thus,

$$\Delta MC + \Gamma\big[j^\mu A_\mu + \tfrac{1}{n}\Delta\big(\bar{\psi}_{\mu_1...\mu_n} q^{\mu_1...\mu_n} + \text{h.c.}\big)\big] \qquad (5.76)$$
$$- A_\mu \Delta\big[\partial_\nu\big(\bar{\xi}_{\mu_1...\mu_{n-1}} Q^{[\mu\nu],\,\mu_1...\mu_{n-1}}\big) + \text{h.c.}\big] + \mathrm{d}(...) = 0.$$

The $\Delta$-exact term added to the original vertex $j^\mu A_\mu$ is trivial, and therefore can be dropped. Now we are left with

$$A_\mu\big[\Gamma j^\mu - \Delta\big(\partial_\nu\big(\bar{\xi}_{\mu_1...\mu_{n-1}} Q^{[\mu\nu],\,\mu_1...\mu_{n-1}}\big) + \text{h.c.}\big)\big] + (\Delta M - \partial_\mu j^\mu) C \doteq 0, \qquad (5.77)$$

---

[6] $\Delta$-exactness of the first term is manifest, while in the second, the presence of the curvature admits only $\Delta$-exact terms, like its own $\gamma$-traces and divergences (see Appendix E).

and we point out that we have started to use the notation '$\doteq$' to denote equality up to d-exact terms. Then, taking a functional derivative w.r.t. $C$ produces part of the sought-after conditions (5.70):

$$\partial_\mu j^\mu = \Delta M, \qquad \Gamma M = 0, \tag{5.78}$$

while the functional derivative w.r.t. $A_\mu$ gives

$$\Gamma j^\mu = \partial_\nu \big(\bar{\xi}_{\mu_1...\mu_{n-1}} \Delta Q^{[\mu\nu],\,\mu_1...\mu_{n-1}}\big) + \text{h.c.}, \quad \Gamma Q^{[\mu\nu],\,\mu_1...\mu_{n-1}} = 0. \tag{5.79}$$

The expression for $\Gamma j^\mu$ has to be $\Gamma$-exact. This demands that $\partial_\nu Q^{[\mu\nu],\,\mu_1...\mu_{n-1}}$ be $\Delta$-closed and that $Q^{[\mu\nu],\,\mu_1...\mu_{n-1}}$ have the interchange symmetry $\nu \leftrightarrow \mu_i$, $i=1,...,n-1$. It follows that

$$j^\alpha = \tilde{j}^\alpha + \Delta\big(\tfrac{1}{n}\,\bar{\psi}_{\mu_1...\mu_n} Q^{[\alpha\mu_1],\,\mu_2...\mu_n} + \text{h.c.}\big), \qquad \Gamma \tilde{j}^\alpha = 0. \tag{5.80}$$

Therefore, by field redefinitions, the current can always be made gauge invariant:

$$\Gamma j^\mu = 0. \tag{5.81}$$

This completes the proof of (5.70), and from (5.68) one then obtains the lift:

$$a_1 = MC. \tag{5.82}$$

Now we will show that $M$ must be $\Delta$-exact modulo the derivative of a $\Gamma$-cocycle.. We recall that $M$ belongs to the cohomology of $\Gamma$, with pgh$\#=0$, agh$\#=1$. It will contain (derivatives of) the fermionic antifield and (derivatives of) the Fronsdal tensor $\mathcal{S}_{\nu_1...\nu_n}$ or the curvature $\Psi_{\rho_1\nu_1|...|\rho_n\nu_n}$. However, one can again choose to have no derivatives of the antifield by using the Leibniz rule, yielding the following general form for $M$:

$$M = \bar{\psi}^{*\mu_1...\mu_n}\big(\vec{M}^{(\mathcal{S})\nu_1...\nu_n}_{\mu_1...\mu_n} \mathcal{S}_{\nu_1...\nu_n} + \vec{M}^{(\Psi)\rho_1\nu_1|...|\rho_n\nu_n}_{\mu_1...\mu_n} \Psi_{\rho_1\nu_1|...|\rho_n\nu_n}\big) \\ + \partial^\mu m_\mu - \text{h.c.}, \tag{5.83}$$

where $\Gamma m_\mu = 0$ and the operators $\vec{M}$'s act to the right. Now, the first term in the parentheses is manifestly $\Delta$-exact, while the second one must contain either a $\gamma$-trace and or a divergence of the curvature, which are $\Delta$-exact too (see Appendix E). Therefore, $M$ must be $\Delta$-exact modulo $d\Gamma(...)$. This means that $a_1$, given in (5.82), can be rendered trivial by adding a $\Delta$-exact piece in $a_0$, and so the vertex will be gauge invariant up to a total derivative:

$$\Gamma a_0 + db_0 = 0. \tag{5.84}$$

In other words, one can always add a $\Delta$-exact term in $a_0$ so that the new current is identically conserved [177–179]:

$$j^\mu \to j'^\mu = j^\mu + \Delta k^\mu = \partial_\nu \mathcal{A}^{\mu\nu}, \qquad \mathcal{A}^{\mu\nu} = -\mathcal{A}^{\nu\mu}. \tag{5.85}$$

Thus we have proved that no abelian vertex deforms the gauge transformations.

## 5.5 Second-Order Consistency

The cubic vertices have all been classified in previous sections. Let us now turn to the analysis of the second-order consistency of the found first-order interactions. From Chapter 4 we recall that consistent second-order deformation requires $(S_1, S_1)$ to be $s$-exact:

$$(S_1, S_1) = -2sS_2 = -2\Delta S_2 - 2\Gamma S_2. \tag{5.86}$$

For abelian vertices, this antibracket is zero, so that the first-order deformations always go unobstructed. Non-abelian vertices, however, are more interesting in this respect.

We can see that there is an obstruction for the non-abelian vertices we have obtained, which do *not* obey the above condition. We prove our claim by contradiction. If Eq. (5.86) holds, then the most general form of the antibracket evaluated at zero antifields is

$$(S_1, S_1)|_{\Phi_A^* = 0} = \Delta M + \Gamma N, \tag{5.87}$$

where $M = -2\,[S_2]_{\mathcal{C}_\alpha^* = 0}$ and $N = -2\,[S_2]_{\Phi_A^* = 0}$. Note that $M$ is obtained by setting to zero *only* the antighosts in $S_2$. Furthermore, the equality (5.87) holds precisely because $S_2$ is linear in the antifields. The $\Gamma$-variation of the above condition is therefore $\Delta$-exact:

$$\Gamma(S_1, S_1)|_{\Phi_A^* = 0} = \Gamma \Delta M = -\Delta\left(\Gamma M\right). \tag{5.88}$$

It is relatively easier to compute the left-hand side of (5.88) for our non-abelian vertices. For spin $\tfrac{3}{2}$ we recall that

$$a_2 = -C^*\bar\xi\xi, \quad a_1 = A^{*\mu}(\bar\psi_\mu \xi - \bar\xi \psi_\mu) + \tilde a_1, \quad a_0 = \bar\psi_\mu F^{+\mu\nu}\psi_\nu,$$
$$\tilde a_1 = i\bigl(\bar\psi^{*\mu}\gamma^\nu F_{\mu\nu} - \tfrac{1}{2(D-2)}\bar\psi^* \slashed{F}\bigr)\xi + \text{h.c.} \tag{5.89}$$

To compute the antibracket of $S_1 = \int (a_2 + a_1 + a_0)$ with itself, we notice that a field-antifield pair shows up only in $\int a_1$, and between $\int a_0$ and $\int a_1$, so that it reduces to

$$(S_1, S_1) = 2\bigl(\int a_0, \int a_1\bigr) + \bigl(\int a_1, \int a_1\bigr). \tag{5.90}$$

Now, the second antibracket on the right-hand side necessarily contains antifields, while the first one will not contain any. Thus we have

$$(S_1, S_1)|_{\Phi_A^* = 0} = 2\Big(\int a_0, \int a_1\Big). \tag{5.91}$$

Notice that, while the unambiguous piece in $a_1$ contains the antifield $A^{*\mu}$, the ambiguity, $\tilde{a}_1$, contains instead the antifield $\bar{\psi}^{*\mu}$. Correspondingly, $\big(\int a_0, \int a_1\big)$ will contain two distinct kinds of pieces: 4-Fermions terms and Fermion bilinears. Explicitly,

$$\begin{aligned}(S_1, S_1)|_{\Phi_A^* = 0} &= \int \mathrm{d}^D x \big[4(\bar{\psi}_\mu \xi - \bar{\xi}\psi_\mu)\, \partial_\nu \big(\bar{\psi}^{[\mu}\psi^{\nu]} + \tfrac{1}{2}\bar{\psi}_\alpha \gamma^{\mu\nu\alpha\beta}\psi_\beta\big)\big] \\ &\quad + \int \mathrm{d}^D x \big[i\bar{\psi}_\mu F^{+\mu\nu}\big(2\gamma^\rho F_{\nu\rho} - \tfrac{1}{(D-2)}\gamma_\nu \slashed{F}\big)\xi + \text{h.c.}\big].\end{aligned} \tag{5.92}$$

Now, if the vertex is unobstructed and Eq. (5.88) holds, the $\Gamma$-variation of each of these terms should independently be $\Delta$-exact. Let us then consider the Fermion bilinears appearing in the second line of the above expression, originating from $\big(\int a_0, \int \tilde{a}_1\big)$. It is easy to see that their $\Gamma$-variation is *not* $\Delta$-exact. We conclude that the non-abelian $1-\tfrac{3}{2}-\tfrac{3}{2}$ vertex gets obstructed beyond the cubic order. The proof for arbitrary spin is very similar.

Our non-abelian electromagnetic couplings are therefore inconsistent in a complete theory under the assumptions of locality. Another underlying assumption is that of the spectrum: as we know, adding dynamical degrees of freedom might change the above conclusion, and these matters are discussed in Chapter 7, together with other issues.

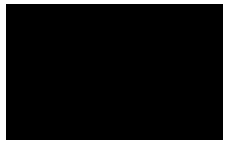

CHAPTER 6

# Gravitational Couplings

In Chapter 5 we have constructed all the consistent cubic vertices of the form $1-s-s$, and we now address the perhaps more interesting generalization of such a result to spin-2 couplings. That is, we study couplings of the form $2-s-s$, and the employed methods are again those of BRST–BV, reviewed in Chapter 4. As we shall see, the gravitational interactions are substantially more difficult to analyze than the electromagnetic ones. Interestingly, this can be thought of as stemming from the crucial difference between the photon EoMs and the graviton ones: the former are expressed as the divergence of a gauge-invariant tensor, $\partial_\mu F^{\mu\nu}$, while the latter are not (the Einstein tensor $G_{\mu\nu}$ is of course gauge invariant but it is not the divergence, or even the derivative of a gauge-invariant object). Nevertheless, we manage to obtain fully off-shell and neat expressions for all the couplings, and second-order consistency of our non-abelian vertices is again seen to require either a wider spectrum or non-locality.

The chapter is organized as follows: in Section 6.1 we consider in great detail $s = \frac{5}{2}$, which serves as a prototype for the arbitrary-spin case. Treating the gauge-algebra-deforming/preserving cases separately, we explicitly construct all the $2-\frac{5}{2}-\frac{5}{2}$ vertices, and cast them into various off-shell forms to make some desired properties manifest. Section 6.2 is then a straightforward arbitrary-spin generalization that mimics the spin-$\frac{5}{2}$ setup. The proof that our non-abelian vertices face obstructions in a local theory beyond the cubic order is, this time, relegated to Appendix D.5, and other appendices again supplement the main text to provide useful technical details. A discussion of the results, together with those of Chapter 5, is relegated to Chapter 7.

## 6.1 Gravitational Coupling of Spin 5/2

In this Section we construct parity-preserving off-shell $2-\frac{5}{2}-\frac{5}{2}$ vertices by employing the BRST-BV cohomological methods. The starting point is the free theory, which contains a graviton field $h_{\mu\nu}$ and a massless spin-$\frac{5}{2}$ tensor-spinor field $\psi_{\mu\nu}$, described by the action

$$S^{(0)}[h_{\mu\nu},\psi_{\mu\nu}] = \int d^D x \left[ G^{\mu\nu} h_{\mu\nu} + \tfrac{1}{2} \left( \bar{\mathcal{R}}^{\mu\nu} \psi_{\mu\nu} - \bar{\psi}_{\mu\nu} \mathcal{R}^{\mu\nu} \right) \right], \qquad (6.1)$$

which enjoys two abelian gauge invariances

$$\delta_\lambda h_{\mu\nu} = 2\partial_{(\mu}\lambda_{\nu)}, \qquad \delta_\varepsilon \psi_{\mu\nu} = 2\partial_{(\mu}\varepsilon_{\nu)}, \quad \text{with } \not{\varepsilon} = 0. \qquad (6.2)$$

For the Grassmann-even bosonic gauge parameter $\lambda_\mu$, we introduce the Grassmann-odd bosonic ghost $C_\mu$, and corresponding to the Grassmann-odd fermionic gauge parameter $\varepsilon_\mu$, we have the Grassmann-even fermionic ghost $\xi_\mu$, which is of course $\gamma$-traceless. Therefore, the set of fields becomes

$$\Phi^A = \{h_{\mu\nu}, C_\mu, \psi_{\mu\nu}, \xi_\mu\}. \qquad (6.3)$$

For each of these fields, we introduce an antifield with the same algebraic symmetries in its indices but opposite Grassmann parity, the set of which is

$$\Phi^*_A = \{h^{*\mu\nu}, C^{*\mu}, \bar{\psi}^{*\mu\nu}, \bar{\xi}^{*\mu}\}. \qquad (6.4)$$

Now we construct the free master action $S_0$, which is an extension of the original gauge-invariant action (6.1) by terms involving ghosts and antifields. Explicitly,

$$S_0 = \int d^D x \big[ G^{\mu\nu} h_{\mu\nu} + \tfrac{1}{2} \left( \bar{\mathcal{R}}^{\mu\nu} \psi_{\mu\nu} - \bar{\psi}_{\mu\nu} \mathcal{R}^{\mu\nu} \right) \qquad (6.5)$$
$$- 2h^{*\mu\nu}\partial_\mu C_\nu + (\bar{\psi}^{*\mu\nu}\partial_\mu \xi_\nu - \partial_\mu \bar{\xi}_\nu \psi^{*\mu\nu}) \big],$$

which is easily seen to satisfy the master equation $(S_0, S_0) = 0$. Again, the antifields appear as sources for the 'gauge' variations, with the gauge parameters replaced by the corresponding ghosts. We spell out in Table 6.1 below the different gradings and Grassmann parity of the various fields and antifields, along with the action of $\Gamma$ and $\Delta$ on them.

For the spin-$\frac{5}{2}$ field, the Fronsdal tensor is $\mathcal{S}_{\mu\nu} = i(\not{\partial}\psi_{\mu\nu} - 2\partial_{(\mu}\not{\psi}_{\nu)})$, and it is related to the original EoMs via

$$\mathcal{R}^{\mu\nu} = \mathcal{S}^{\mu\nu} - \gamma^{(\mu}\not{\mathcal{S}}^{\nu)} - \tfrac{1}{2}\eta^{\mu\nu}\mathcal{S}', \qquad \mathcal{S}' \equiv \mathcal{S}^\mu_\mu. \qquad (6.6)$$

Table 6.1: Properties of the Various Fields & Antifields ($n = 2$)

| $Z$ | $\Gamma(Z)$ | $\Delta(Z)$ | $\mathrm{pgh}(Z)$ | $\mathrm{agh}(Z)$ | $gh(Z)$ | $\epsilon(Z)$ |
|---|---|---|---|---|---|---|
| $h_{\mu\nu}$ | $2\partial_{(\mu}C_{\nu)}$ | 0 | 0 | 0 | 0 | 0 |
| $C_\mu$ | 0 | 0 | 1 | 0 | 1 | 1 |
| $h^{*\mu\nu}$ | 0 | $G^{\mu\nu}$ | 0 | 1 | $-1$ | 1 |
| $C^{*\mu}$ | 0 | $-2\partial_\nu h^{*\mu\nu}$ | 0 | 2 | $-2$ | 0 |
| $\psi_{\mu\nu}$ | $2\partial_{(\mu}\xi_{\nu)}$ | 0 | 0 | 0 | 0 | 1 |
| $\xi_\mu$ | 0 | 0 | 1 | 0 | 1 | 0 |
| $\bar\psi^{*\mu\nu}$ | 0 | $\bar{\mathcal{R}}^{\mu\nu}$ | 0 | 1 | $-1$ | 0 |
| $\bar\xi^{*\mu}$ | 0 | $2\partial_\nu \bar\chi^{*\mu\nu}$ | 0 | 2 | $-2$ | 1 |

As already stated in the previous chapter, an important property is that the divergence $\partial_\nu \mathcal{R}^{\mu\nu}$ is not zero, unlike that of the Einstein tensor, but is proportional to $\gamma^\mu$. The details are found in Appendix E, and they will affect the following computations non-trivialy so that we give, explicitly,

$$\Delta \bar\xi^{*\mu} = 2\partial_\nu \bar\chi^{*\mu\nu}, \qquad \bar\chi^{*\mu\nu} \equiv \bar\psi^{*\mu\nu} - \tfrac{1}{D}\bar{\slashed\psi}^{*\nu}\gamma^\mu. \qquad (6.7)$$

The cohomology of $\Gamma$ is isomorphic to the space of functions of

- The undifferentiated ghosts $\{C_\mu, \xi_\mu\}$, the 1-curl of the bosonic ghost $\mathfrak{C}_{\mu\nu}$ and the $\gamma$-traceless part of the 1-curl of the fermionic ghost $\xi_{\mu\nu}$,

- The antifields $\{h^{*\mu\nu}, C^{*\mu}, \bar\psi^{*\mu\nu}, \bar\xi^{*\mu}\}$ and their derivatives,

- The curvatures $\{R_{\mu\nu\rho\lambda}, \Psi_{\mu\nu|\rho\lambda}\}$ and their derivatives,

- The Fronsdal tensor $\mathcal{S}_{\mu\nu}$ and its symmetrized derivatives.

Note that, because of the spin of the graviton (one unit higher than that of the photon), the subsequent demonstrations lead to a proliferation of curls, and we have therefore adopted a simplified notation for this chapter:

$$\mathfrak{h}_{\mu\nu||\rho} \equiv h^{(1)}_{\mu\nu||\rho}, \quad \mathfrak{C}_{\mu\nu} \equiv C^{(1)}_{\mu\nu} \quad \psi_{...} \equiv \psi^{(m)}_{...} \quad \xi_{...} \equiv \xi^{(1)}_{...}. \qquad (6.8)$$

Let us also point out, finally, that we shall continue using the symbol $\doteq$ to denote equality up to total derivatives.

### 6.1.1 Non-Abelian Vertices

Non-abelian vertices are those that deform the gauge algebra. They correspond to deformations of the master action with nontrivial terms at agh# $= 2$. In other words, $a_2$ is a nontrivial element in H($\Gamma|$d) (see Chapter 4). Notice that $a_2$ is Grassmann even, hermitian and has gh$(a_2) = 0$. Besides, we require that $a_2$ be a parity-even Lorentz scalar.

It is then clear that any $a_2$ will consist of a single antighost and two ghost fields. Let us note that two $a_2$'s are equivalent iff they differ by $\Gamma$-exact terms modulo total derivatives. Without loss of generality, we can thus choose the antighost to be undifferentiated. Furthermore, any derivative acting on the ghost fields $\{C_\mu, \xi_\mu\}$ can be realized as a 1-curl $\{\mathfrak{C}_{\mu\nu}, \xi_{\mu\nu}\}$ up to irrelevant $\Gamma$-exact terms (see Appendix E). Because the derivative of a ghost-curl is $\Gamma$-exact, a nontrivial $a_2$ can never contain more than 2 derivatives, which already poses an upper bound of 3 on the number of derivatives in a non-abelian vertex.

To be more explicit, let us write down all the inequivalent $a_2$'s. In view of the actions of $\Gamma$ and $\Delta$ on various (anti)fields, given any $a_2$, the consistency cascade (4.21) unambiguously counts the number of derivatives $p$ contained in the corresponding vertex $a_0$. Thus we can classify $a_2$'s based on the value of their corresponding $p$. Also, the set of all possible nontrivial $a_2$'s again falls into two subsets: Subset 1 contains the bosonic antighost $C_\mu^*$, while Subset 2 contains the fermionic one $\xi_\mu^*$. In Subset 1 we have

$$a_2 = \begin{cases} p = 1: & igC^{*\mu}\bar{\xi}^\alpha\gamma_\mu\xi_\alpha, \\ p = 2: & igC^{*\mu}\bar{\xi}_{\mu\nu}\xi^\nu + \text{h.c.}, \\ p = 3: & igC^{*\mu}\bar{\xi}^{\alpha\beta}\gamma_\mu\xi_{\alpha\beta}. \end{cases} \quad (6.9)$$

It is easy to see that this list is indeed complete. First, it follows from Lorentz invariance that if $p$ is odd (resp. even), the number of $\gamma$ matrices is also odd (resp. even), and the latter can be chosen simply to be 1 (resp. 0). This is because if more $\gamma$ matrices are there, one can anti-commute them past each other using the Clifford algebra to see that only terms with 1 (resp. 0) $\gamma$-matrix survive, while other terms are either killed (because $\not{\xi} = 0$) or made trivial (because $\gamma^\alpha \xi_{\alpha\beta} = \Gamma$-exact).

Note that the $p = 1$ candidate, $igC^{*\mu}\bar{\xi}^\alpha\gamma_\mu\xi_\alpha$, is easily ruled out as inconsistent. To see this, we simply take its $\Delta$ variation and integrate by parts to find $\Delta a_2 \doteq 2igh^{*\mu\nu}\partial_\nu(\bar{\xi}^\alpha\gamma_\mu\xi_\alpha)$, which contains nontrivial elements of H($\Gamma|$d) involving the ghost-curl $\bar{\xi}_{\alpha\nu}$. Therefore, the consistency condition (4.21b) cannot be satisfied.

Next we consider Subset 2, whose $a_2$'s contain the (undifferentiated) fermionic antighost. Again, the $a_2$'s can be classified based on the value $p$ of the number of derivatives in the corresponding vertex $a_0$. The complete list is

$$a_2 = \begin{cases} p=0: & g\bar{\xi}^{*\mu}\gamma^\alpha \xi_\mu C_\alpha + \text{h.c.}, \\ p=1: & g\bar{\xi}^{*\mu}(\xi^\nu \mathfrak{C}_{\mu\nu} + \alpha_1 \xi_{\mu\nu}C^\nu + \alpha_2 \gamma^{\alpha\beta}\xi_\mu \mathfrak{C}_{\alpha\beta}) + \text{h.c.}, \\ p=2: & g\bar{\xi}^{*\mu}\gamma^\alpha \xi^\beta_{\ \mu}\mathfrak{C}_{\alpha\beta} + \text{h.c.}, \end{cases} \tag{6.10}$$

where $\alpha_1$ and $\alpha_2$ are dimensionless constants. Because both $\xi$ and $\xi^*$ vanish, and $\gamma^\alpha \xi_{\alpha\beta} = \Gamma$-exact, any $\gamma$-matrix must be contracted with the bosonic ghost or with its curl. Then one can easily verify that the list (6.10) indeed gives all possible inequivalent Lorentz scalars.

Here it is easy to rule out the $p=0$ candidate, $g\bar{\xi}^{*\mu}\gamma^\alpha \xi_\mu C_\alpha + \text{h.c.}$, as inconsistent. Again, we simply take its $\Delta$ variation and integrate by parts to obtain $\Delta a_2 \doteq -2g\bar{\chi}^{*\mu\nu}\partial_\nu(\gamma^\alpha \xi_\mu C_\alpha) + \text{h.c.}$, which contains nontrivial elements of H($\Gamma$|d) involving the ghost-curls $\xi_{\nu\mu}$ and $C_{\nu\alpha}$. Hence consistency condition (4.21b) cannot be satisfied.

**Absence of Minimal Coupling**

A possible minimal coupling would correspond to a 1-derivative vertex. The most general $a_2$ can be written as (dropping the already-ruled-out candidate containing $C^*_\mu$)

$$a_2 = g\bar{\xi}^{*\mu}(\xi^\nu \mathfrak{C}_{\mu\nu} + \alpha_1 \xi_{\mu\nu}C^\nu + \alpha_2 \gamma^{\rho\sigma}\xi_\mu \mathfrak{C}_{\rho\sigma}) + \text{h.c.}, \tag{6.11}$$

where $\alpha_1$ and $\alpha_2$ are dimensionless constants. Then we have

$$\Delta a_2 \doteq \Gamma\text{-exact} - g\bar{\chi}^{*\mu\alpha}(\xi_\alpha^{\ \nu}\mathfrak{C}_{\mu\nu} + \alpha_1 \xi_{\mu\nu}\mathfrak{C}_\alpha^{\ \nu} + \alpha_2 \gamma^{\rho\sigma}\xi_{\alpha\mu}\mathfrak{C}_{\rho\sigma}) + \text{h.c.}, \tag{6.12}$$

where we recall that $\bar{\chi}^{*\mu\alpha} \equiv \bar{\psi}^{*\mu\alpha} - \frac{1}{D}\bar{\slashed{\psi}}^{*\alpha}\gamma^\mu$. The nontrivial elements of H($\Gamma$|d) appearing on the right-hand side can actually be canceled by the choice $\alpha_1 = -1$ and $\alpha_2 = \frac{1}{4}$. The only subtlety are the terms containing the $\gamma$-trace $\bar{\slashed{\psi}}^*$ of the fermionic antifield, for which one needs to use the identity $\gamma^\mu \gamma^{\rho\sigma} = \gamma^{\rho\sigma}\gamma^\mu + 4\eta^{\mu[\rho}\gamma^{\sigma]}$. With the cocycle condition (4.21b) thus satisfied, the unambiguous piece in $a_1$ reads

$$\hat{a}_1 = -2\left(g\bar{\chi}^{*\mu\rho}\psi_{\mu\nu\|\rho}C^\nu + \text{h.c.}\right) + \mathcal{Y}^{\mu\nu}\mathfrak{C}_{\mu\nu} + \cdots, \tag{6.13}$$

where the ellipses stand for terms involving the fermionic ghost $\xi_\mu$ but not $C_\mu$. This gives

$$\hat{\beta}^\mu \equiv \frac{\delta}{\delta C_\mu}\Delta\hat{a}_1 = \left(2g\Delta\bar{\chi}^*_{\alpha\beta}\psi^{\mu\alpha\|\beta} + \text{h.c.}\right) + 2\Delta\partial_\nu \mathcal{Y}^{[\mu\nu]}. \tag{6.14}$$

Similarly, because the ambiguity $\tilde{a}_1$ belongs to $H(\Gamma)$, we have

$$\tilde{\beta}^\mu \equiv \frac{\delta}{\delta C_\mu} \Delta \tilde{a}_1 = \Gamma\text{-closed}. \tag{6.15}$$

Now the cocycle condition (4.21c) is fulfilled if

$$\Delta \hat{a}_1 + \Delta \tilde{a}_1 \doteq -\Gamma a_0 \doteq 2 C_\mu \partial_\nu T^{\mu\nu} + \cdots \tag{6.16}$$

for some $a_0 \doteq h_{\mu\nu} T^{\mu\nu}$. Taking a functional derivative w.r.t. $C_\mu$ then yields

$$\hat{\beta}^\mu + \tilde{\beta}^\mu = 2 \partial_\nu T^{\mu\nu}. \tag{6.17}$$

Using Eqs. (6.14) and (6.15), and taking a $\Gamma$ variation one is thus lead to the necessary condition

$$\Gamma \hat{\beta}^\mu = \partial^\beta \left( 2g \, \Delta \bar{\chi}^*_{\alpha\beta} \, \xi^{\alpha\mu} + \text{h.c.} \right) + \partial_\nu \left( 2\Gamma \Delta \mathcal{Y}^{[\mu\nu]} \right) = \partial_\nu \left( 2\Gamma T^{\mu\nu} \right). \tag{6.18}$$

In $D \geq 4$, this condition can never be satisfied, since the terms inside the brackets are not $\Gamma$-exact modulo d. Thus we conclude that there is no 1-derivative $2$–$\frac{5}{2}$–$\frac{5}{2}$ vertex, i.e. a massless spin-$\frac{5}{2}$ field cannot have minimal coupling to gravity in flat space, thereby reproducing the result of [42, 43] in an independent manner.

**The 2-Derivatives Vertex**

Having ruled out minimal coupling, we are lead to consider the next possibility — the 2-derivatives vertex, for which the corresponding $a_2$ reads

$$a_2 = \left( ig \, C^{*\mu} \bar{\xi}_{\mu\nu} \xi^\nu + \text{h.c.} \right) + \left( \tilde{g} \, \mathfrak{C}_{\mu\nu} \bar{\xi}^*_\rho \gamma^{\mu\nu\rho\alpha\beta} \xi_{\alpha\beta} + \text{h.c.} \right), \tag{6.19}$$

where the coupling constants $g$ and $\tilde{g}$ are a priori complex, but will soon be required to be real. Notice that, for future convenience, we wrote the term with fermionic antighost with five $\gamma$-matrices, instead of just one, as it appears in (6.10). The equivalence of the two forms, although rather obvious, is made explicit in Appendix D.3.1 for interested readers. To find a possible $a_1$, we take the $\Delta$ variation of the above equation and integrate by parts, thereby obtaining, up to total derivatives:

$$\Delta a_2 \doteq 2 \left[ ig \, h^{*\mu\nu} \partial_\nu \bar{\xi}_{\mu\lambda} \xi^\lambda + \text{h.c.} \right] + 2 \left[ \tilde{g} \, \bar{\chi}^*_{\rho\sigma} \partial^\sigma \left( \mathfrak{C}_{\mu\nu} \gamma^{\mu\nu\rho\alpha\beta} \xi_{\alpha\beta} \right) + \text{h.c.} \right]. \tag{6.20}$$

Now, in view of Eqs. (E.43) and (E.46), the $\Gamma$-exactness of the second piece on the right-hand side is manifest, while in the first piece one can also use Eq. (E.44) to extract $\Gamma$-exact terms. Then, the contributions that are nontrivial in H($\Gamma$) cancel each other only if $g$ is real. Therefore, the cocycle condition (4.21b) is satisfied and we get, up to an ambiguity $\tilde{a}_1$,

$$a_1 = a_{1g} + a_{1\tilde{g}} + \tilde{a}_1, \tag{6.21}$$

where $\Gamma \tilde{a}_1 = 0$ and the other terms are unambiguously determined to be

$$a_{1g} = ig\, h^{*\mu\nu} \left( \bar{\xi}_{\mu\lambda} \psi_\nu{}^\lambda + \bar{\psi}_\nu{}^\lambda \xi_{\mu\lambda} - 2\bar{\xi}^\lambda \psi_{\mu\lambda\|\nu} - 2\bar{\psi}_{\mu\lambda\|\nu} \xi^\lambda \right), \tag{6.22a}$$

$$a_{1\tilde{g}} = 2\tilde{g} \left( \mathfrak{C}_{\mu\nu} \bar{\chi}^*_{\rho\sigma} \gamma^{\mu\nu\rho\alpha\beta} \psi_{\alpha\beta\|}{}^\sigma - \mathfrak{h}_{\mu\nu\|}{}^\sigma \bar{\chi}^*_{\rho\sigma} \gamma^{\mu\nu\rho\alpha\beta} \xi_{\alpha\beta} \right) + \text{h.c.} \tag{6.22b}$$

We will now compute the $\Delta$ variations of the above quantities. From Eq. (6.22a) one finds

$$\Delta a_{1g} \doteq ig\, \bar{\xi}_\lambda \left[ 2G^{\mu\nu} \partial^\lambda \psi_{\mu\nu} - 3\partial_\mu \left( G^{\mu\nu} \psi_\nu{}^\lambda \right) + \partial^\nu \left( G^{\lambda\mu} \psi_{\mu\nu} \right) \right] + \text{h.c.}, \tag{6.23}$$

which does not contain the bosonic ghost $C_\mu$. Note that neither can $\Delta \tilde{a}_1$ give rise to terms containing $C_\mu$, because if the ambiguity $\tilde{a}_1$ contains $C_\mu$ or its curl, then it must also contain the Fronsdal tensors[1] and thus be $\Delta$-exact, so that $\Delta \tilde{a}_1 = 0$. This fact puts restrictions on $\Delta a_{1\tilde{g}}$: it may contain $C_\mu$ only in the form of symmetrized derivatives, $\partial_{(\mu} C_{\nu)}$, up to total-derivative terms. Otherwise, $\Delta a_1$ will have nontrivial pieces belonging to H($\Gamma$|d), and the condition $\Delta a_1 \doteq -\Gamma a_0$ may never be satisfied.

With the above observations in mind, we compute the following useful quantity:

$$\beta^\mu_C \equiv \frac{\delta}{\delta C_\mu} \Delta a_{1\tilde{g}} = -4\tilde{g} \Delta \partial_{[\nu} \bar{\chi}^*_{\rho]\sigma} \gamma^{\mu\nu\rho\alpha\beta} \psi_{\alpha\beta\|}{}^\sigma - 4\tilde{g}^* \bar{\psi}_{\alpha\beta\|}{}^\sigma \gamma^{\mu\nu\rho\alpha\beta} \Delta \partial_{[\nu} \chi^*_{\rho]\sigma}. \tag{6.24}$$

The right-hand side, if non-zero, must be the divergence of a symmetric tensor: $\partial_\nu \mathcal{X}^{\mu\nu}$ with $\mathcal{X}^{\mu\nu} = \mathcal{X}^{\nu\mu}$. As shown in Appendix D.3.1, this is possible only if $\tilde{g}$ is real, and it yields

$$\mathcal{X}^{\mu\nu} = 2i\tilde{g}\, \bar{\psi}_{\rho\sigma\|\lambda}\, \gamma^{\mu\rho\sigma\alpha\beta,\,\nu\lambda\gamma}\, \psi_{\alpha\beta\|\gamma} + (\mu \leftrightarrow \nu). \tag{6.25}$$

Then, the bosonic ghost $C_\mu$ will appear in $\Delta a_{1\tilde{g}}$ only through $\Gamma$-exact pieces. Explicitly,

$$\Delta a_{1\tilde{g}} + \Gamma \left( \tfrac{1}{2} h_{\mu\nu} \mathcal{X}^{\mu\nu} \right) \doteq \tfrac{1}{2} h_{\mu\nu} \Gamma \mathcal{X}^{\mu\nu} - 2\tilde{g} \left( \mathfrak{h}_{\mu\nu\|}{}^\sigma \Delta \bar{\chi}^*_{\rho\sigma} \gamma^{\mu\nu\rho\alpha\beta} \xi_{\alpha\beta} + \text{h.c.} \right). \tag{6.26}$$

---

[1] Also, it cannot contain only curvatures, because then there are too many derivatives in $\Delta \tilde{a}_1$ to possibly correspond to a vertex with $p = 2$.

One can now simplify the right-hand side, which does not contain the bosonic ghost $C_\mu$, but just the fermionic one $\xi_\mu$. The result is (again see Appendix D.3.1)

$$\Delta a_{1\tilde{g}} + \Gamma\left(\tfrac{1}{2} h_{\mu\nu} \mathcal{X}^{\mu\nu}\right) \doteq -i\tilde{g}\,\bar{\xi}_\lambda \left(R_{\mu\nu\rho\sigma} \gamma^{\mu\nu\lambda\alpha\beta,\,\tau\rho\sigma}\,\psi_{\alpha\beta\|\tau}\right) + \text{h.c.} \quad (6.27)$$

It is easy to see that the right-hand side is a nontrivial element of H(Γ|d). Only if it can be written, up to Γ-exact pieces and total derivatives, in terms of $\Delta\tilde{a}_1$ plus possibly $\Delta a_{1g}$, for some choice of $\tilde{g}$, can one fulfill the condition $\Delta a_1 \doteq -\Gamma a_0$ and thus obtain a vertex. After a tedious but straightforward calculation, shown in Appendix D.3.1, one can write

$$\Delta a_{1\tilde{g}} + \Gamma\left(\tfrac{1}{2} h_{\mu\nu}\mathcal{X}^{\mu\nu}\right) \doteq -8i\tilde{g}\,\Gamma\left(\bar{\psi}_{\mu\alpha} R^{+\mu\nu\alpha\beta}\psi_{\nu\beta} + \tfrac{1}{2}\bar{\psi}_\mu \slashed{R}^{\mu\nu}\slashed{\psi}_\nu\right) - \Delta\mathfrak{a}, \quad (6.28)$$

where $R^{+\mu\nu\alpha\beta} \equiv R^{\mu\nu\alpha\beta} + \tfrac{1}{2}\gamma^{\mu\nu\rho\sigma}R_{\rho\sigma}{}^{\alpha\beta}$ and $\Delta\mathfrak{a}$ is given below. The next step is to relate the latter quantity with $\Delta\tilde{a}_1$ and $\Delta a_{1g}$ up to total derivatives, and as we prove in Appendix D.3.1, this can be achieved. We find

$$\Delta\mathfrak{a} \doteq 8\Big(\frac{\tilde{g}}{g}\Big)\Delta a_{1g} + \Delta\tilde{a}_1, \quad (6.29)$$

for some ambiguity $\tilde{a}_1$ spelled out in Eq. (D.89). Then one can choose

$$\tilde{g} = \tfrac{1}{8}g \quad (6.30)$$

in order to fulfill the cocycle condition (4.21c). That is, Eq. (6.28) takes the form

$$\Delta a_{1g} + \Delta a_{1\tilde{g}} + \Delta\tilde{a}_1 \doteq -\Gamma a_0, \quad (6.31)$$

where the vertex $a_0$ is given by

$$a_0 = ig\left(\bar{\psi}_{\mu\alpha} R^{+\mu\nu\alpha\beta}\psi_{\nu\beta} + \tfrac{1}{2}\bar{\psi}_\mu \slashed{R}^{\mu\nu}\slashed{\psi}_\nu + \tfrac{1}{4} h_{\mu\nu}\bar{\psi}_{\rho\sigma\|\lambda}\,\gamma^{\mu\rho\sigma\alpha\beta,\,\nu\lambda\gamma}\,\psi_{\alpha\beta\|\gamma}\right). \quad (6.32)$$

We emphasize that it is a unique linear combination in Eq. (6.19), with $\tilde{g} = \tfrac{1}{8}g$ being real valued, for which the $a_2$ gets lifted to a vertex $a_0$ through the consistency cascade. The 2-derivatives vertex is therefore unique. While it simplifies in dimension 4, for the last term in the above expression then vanishes, the vertex is non-zero in any $D \geq 4$. We also point out the role played by the combination $R^+_{\mu\nu\alpha\beta}$, and its resemblance with the spin-1 analogous quantity $F^+_{\mu\nu}$.

**The 3-Derivatives Vertex**

In this case, as we see from (6.9) and (6.10), there is just one candidate for $a_2$, namely
$$a_2 = -ig C^*_\lambda \, \bar{\xi}_{\mu\nu} \gamma^{\lambda\mu\nu\alpha\beta} \xi_{\alpha\beta}. \tag{6.33}$$
Again, for future convenience, we have written it with five $\gamma$-matrices, instead of just one as it appears in (6.9), and the equivalence of the two forms is again made explicit in Appendix D.3.2. Acting with $\Delta$ on Eq. (6.33) and integrating by parts one evidently produces only $\Gamma$-exact terms, thanks to the relations (E.46). The corresponding $a_1$ is thus easily seen to be
$$a_1 = -2ig h^{*\sigma}_\lambda \left( \bar{\xi}_{\mu\nu} \gamma^{\lambda\mu\nu\alpha\beta} \psi_{\alpha\beta\|\sigma} - \text{h.c.} \right) + \tilde{a}_1, \tag{6.34}$$
for some ambiguity $\tilde{a}_1$ such that $\Gamma \tilde{a}_1 = 0$.

Now we address the problem of finding the lift to $a_0$. Acting on the above expression with $\Delta$, one again obtains the Einstein tensor, which can be written as $G^\sigma_\lambda = 2\partial_{[\rho}\mathfrak{h}^{\rho\sigma\|}{}_{\lambda]} - \tfrac{1}{2}\delta^\sigma_\lambda R$. Thus one ends up having
$$\Delta a_1 = -2ig \bigl(\partial_\rho \mathfrak{h}^{\rho\sigma\|}{}_\lambda - \partial_\lambda \mathfrak{h}^{\rho\sigma\|}{}_\rho - \tfrac{1}{2}\delta^\sigma_\lambda R\bigr) \left( \bar{\xi}_{\mu\nu} \gamma^{\lambda\mu\nu\alpha\beta} \psi_{\alpha\beta\|\sigma} - \text{h.c.} \right) + \Delta \tilde{a}_1. \tag{6.35}$$
The term proportional to the Ricci scalar is simply zero because of the Bianchi identity $\psi_{[\alpha\beta\|\sigma]} = 0$, while the term containing $\partial_\lambda$ is a total derivative, thanks again to the Bianchi identities $\partial_{[\lambda} \bar{\xi}_{\mu\nu]} = 0$ and $\partial_{[\lambda} \bar{\psi}_{\alpha\beta]\|\sigma} = 0$, enforced by the presence of the antisymmetric 5-$\gamma$ product. Finally, the term containing $\partial_\rho$ can be integrated by parts to give, up to the hermitian conjugates,
$$\Delta a_1 \doteq 2ig \mathfrak{h}^{\rho\sigma\|}{}_\lambda \left( \partial_\rho \bar{\xi}_{\mu\nu} \, \gamma^{\lambda\mu\nu\alpha\beta} \, \psi_{\alpha\beta\|\sigma} + \tfrac{1}{2} \bar{\xi}_{\mu\nu} \, \gamma^{\lambda\mu\nu\alpha\beta} \, \Psi_{\alpha\beta|\rho\sigma} - \text{h.c.} \right) + \Delta \tilde{a}_1. \tag{6.36}$$
The first term in the parentheses and its hermitian conjugate combine into a $\Gamma$-exact term modulo d, since the $\Gamma$ variation of the graviton curl is zero up to a total derivative, again by the Bianchi identities $\partial_{[\lambda}\bar{\xi}_{\mu\nu]} = 0$ and $\partial_{[\lambda}\psi_{\alpha\beta]\|\sigma} = 0$. In the second term, on the other hand, one can pull a derivative out of the ghost-curl and integrate by parts to obtain
$$\Delta a_1 + 2ig\,\Gamma\bigl(\mathfrak{h}^{\rho\sigma\|}{}_\lambda \bar{\psi}_{\mu\nu\|\rho} \, \gamma^{\lambda\mu\nu\alpha\beta} \, \psi_{\alpha\beta\|\sigma}\bigr) \doteq ig R_{\mu\nu\rho\sigma} \left( \bar{\xi}_\lambda \, \gamma^{\lambda\mu\nu\alpha\beta} \, \Psi_{\alpha\beta|}{}^{\rho\sigma} \right) \\ + \tfrac{1}{2}\Delta \tilde{a}_1 + \text{h.c.}. \tag{6.37}$$
Now, as shown in Appendix D.3.2, the right-hand side can be rendered precisely $\Gamma$-exact modulo d for some choice of the ambiguity, given by (D.98), and for that right-hand side we thus get
$$-ig\,\Gamma\bigl(\mathfrak{h}^{\rho\sigma\|\lambda} \bar{\psi}_{\mu\nu\|\gamma} \, \gamma^{\lambda\mu\nu\alpha\beta,\,\rho\sigma\gamma\delta} \, \psi_{\alpha\beta\|\delta}\bigr). \tag{6.38}$$

The two $\Gamma$-exact pieces then combine to fulfill the condition $\Delta a_1 + \Gamma a_0 \doteq 0$ for
$$a_0 = ig\,\bar{\psi}_{\mu\nu\|}{}^\rho \big(\mathfrak{h}^+_{\rho\sigma\|\lambda}\,\gamma^{\lambda\mu\nu\alpha\beta} + \gamma^{\lambda\mu\nu\alpha\beta}\,\mathfrak{h}^+_{\rho\sigma\|\lambda}\big)\psi_{\alpha\beta\|}{}^\sigma, \tag{6.39}$$
where
$$\mathfrak{h}^{+\rho\sigma\|\lambda} \equiv \mathfrak{h}^{\rho\sigma\|\lambda} + \tfrac{1}{2}\gamma^{\rho\sigma\alpha\beta}\mathfrak{h}_{\alpha\beta\|}{}^\lambda. \tag{6.40}$$

The above 3-derivatives vertex vanishes in $D=4$, and this fact is manifest from the presence of the antisymmetrized product of five $\gamma$-matrices. Let us also note the appearance of the above combination of the graviton field, which is some version of $R^+_{\mu\nu\rho\lambda}$ with one curl less.

### 6.1.2 Abelian Vertices

Having exhausted all the nontrivial $a_2$'s, we are only left to consider vertices with trivial $a_2$. In this case, as we show in Subsection 6.2.2 for generic spin, one can always choose to write a vertex as the graviton field $h_{\mu\nu}$ contracted with a gauge-invariant[2] current $T^{\mu\nu}$,
$$a_0 = T^{\mu\nu}h_{\mu\nu}, \qquad \Gamma T^{\mu\nu} = 0, \tag{6.41}$$
where the divergence of the current is the $\Delta$ variation of a $\Gamma$-closed object:
$$\partial_\nu T^{\mu\nu} = \Delta M^\mu, \qquad \Gamma M^\mu = 0. \tag{6.42}$$

The gauge-invariant current $T^{\mu\nu}$ is a bilinear in the fermion fields, which cannot be $\Delta$-exact since otherwise the vertex (6.41) would be trivial. This leaves us with considering only bilinears in the curvature $\Psi_{\mu\nu|\rho\sigma}$. Schematically, the current is of the form
$$T^{\mu\nu} = \bar{\Psi}^M \hat{\mathcal{O}}^{\mu\nu}{}_{MN}\Psi^N, \tag{6.43}$$
where $M,N$ are compound indices and $\hat{\mathcal{O}}$ is an operator built out of derivatives, $\gamma$-matrices and the metric tensor. This immediately implies that an abelian vertex will contain at least four derivatives — two from both curvatures with $\hat{\mathcal{O}}$ containing no derivative.

To find the possible tensor structure of $\hat{\mathcal{O}}$, let us first note that we can forgo contractions of any pair of indices in the same curvature tensor since the result is always $\Delta$-exact, if not zero. Moreover, it is sufficient to consider in $\hat{\mathcal{O}}$ no more than one $\gamma$-matrix, which must carry either the $\mu$ index or $\nu$. To see this, notice that if a $\gamma$-matrix carries one of the indices of the curvatures — any from the sets $M$ and $N$ — one can use the Clifford algebra to anticommute it past other possible $\gamma$-matrices, ending up producing a $\gamma$-trace of the curvature, which is $\Delta$-exact. This leaves us with only $\gamma^\mu$ and $\gamma^\nu$ which, however, cannot appear simultaneously because their symmetrization would eliminate them both. A similar reasoning rules out the appearance of the operator $\slashed{\partial}$, and therefore of $\Box$, in $\hat{\mathcal{O}}$.

---

[2] Gauge invariance of $T^{\mu\nu}$ is the whole point here: one can always write a vertex as $a_0 \approx T^{\mu\nu}h_{\mu\nu}$, but in general $T^{\mu\nu}$ will not be strictly gauge invariant.

How many derivatives may $\hat{\mathcal{O}}$ contain? If it contains one derivative, there will be one $\gamma$-matrix carrying either the index $\mu$ or $\nu$, say $\gamma^\mu$. One can always choose the other index $\nu$ to appear on the derivative under consideration. In the only other nontrivial possibility, the latter index is contracted with, and therefore appears on, a curvature on which the derivative must act. Then one can pull out the derivative $\partial^\nu$ by using the second Bianchi identity and symmetry properties of the curvatures. Similarly, when $\hat{\mathcal{O}}$ contains more derivatives, one can forgo the appearance of the indices $\mu$ and $\nu$ on the curvatures. However, the number of derivatives cannot exceed two. To see this, let us consider the possibility of having three derivatives or more:

$$T^{\mu\nu} = \bar{\Psi}^M \overleftarrow{\partial}_\rho \hat{\mathcal{P}}^{\mu\nu}{}_{MN} \vec{\partial}^\rho \Psi^N,$$

where $\hat{\mathcal{P}}$ is a 1- or higher-derivative operator. Then one can use the so-called 3-box rule: $2\partial_\rho X \partial^\rho Y = \Box(XY) - X\Box Y - Y\Box X$, integrate by parts, and drop $\Delta$-exact terms to write

$$a_0 \approx \Box h_{\mu\nu} \left(\tfrac{1}{2} \bar{\Psi}^M \hat{\mathcal{P}}^{\mu\nu}{}_{MN} \Psi^N\right) \approx \left(\tfrac{1}{2} \partial_\mu h' - \partial \cdot h_\mu\right) \partial_\nu \left(\bar{\Psi}^M \hat{\mathcal{P}}^{\mu\nu}{}_{MN} \Psi^N\right),$$

where the last equivalence comes from $R_{\mu\nu} \equiv \Box h_{\mu\nu} - 2\partial_{(\mu}\partial \cdot h_{\nu)} + \partial_\mu \partial_\nu h'$ being a $\Delta$-exact quantity. Therefore, the vertex is trivial since the divergence of the fermion bilinear is $\Delta$-exact. The latter fact originates from $\partial_\nu T^{\mu\nu} = \Delta M^\mu$, and from that the divergence is blind to the presence of the extra derivatives $\overleftarrow{\partial}_\rho \vec{\partial}^\rho$ in $T^{\mu\nu}$. On the other hand, if the extra derivatives carry any indices belonging to the sets $M$ and $N$, one must keep in mind that a divergence of the curvature is $\Delta$-exact. Given the hermiticity of $T^{\mu\nu}$, the commutativity of covariant derivatives, the antisymmetry of paired indices and the Bianchi identities obeyed by the curvature it is then easy to convince ourselves that this vertex is always equivalent to the previous one, which we already ruled out. This proves our claim that $\hat{\mathcal{O}}$ may contain at most two derivatives, and hence sets an upper bound of 6 on the number of derivatives in $T^{\mu\nu}$, and hence also in $a_0$.

**The 4-Derivatives Vertex**

When the operator $\hat{\mathcal{O}}$ in Eq. (6.43) does not contain any derivatives, the corresponding vertex (6.41) is a 4-derivatives one. The generic form of the

current is

$$T^{\mu\nu} = ig\big(\bar{\Psi}^{(\mu}{}_{\lambda|\alpha\beta}\Psi^{\nu)\lambda|\alpha\beta} + \alpha\eta^{\mu\nu}\bar{\Psi}_{\rho\sigma|\alpha\beta}\Psi^{\rho\sigma|\alpha\beta}\big), \qquad (6.44)$$

where the parameter $\alpha$ is to be fixed by requiring that $\partial_\nu T^{\mu\nu}$ be $\Delta$-exact. Now the divergence of the above equation contains nontrivial pieces in $H(\Delta)$, which are given by

$$\partial_\nu T^{\mu\nu} = \Delta M^\mu + ig\big(\tfrac{1}{2}\bar{\Psi}^{\nu\lambda|\alpha\beta}\partial_\nu\Psi^\mu{}_{\lambda|\alpha\beta} + \alpha\bar{\Psi}^{\rho\sigma|\alpha\beta}\partial^\mu\Psi_{\rho\sigma|\alpha\beta} - \text{h.c.}\big). \quad (6.45)$$

By using the Bianchi identity $\partial_{[\nu}\Psi^\mu{}_{\lambda]|\alpha\beta} = \tfrac{1}{2}\partial^\mu\Psi_{\nu\lambda|\alpha\beta}$, the first term in the parentheses is rendered the same as the second one. These terms cancel each other if we set $\alpha = -\tfrac{1}{4}$. Thus, there is just one 4-derivatives vertex, given by

$$\begin{aligned}a_0 &= ig\left(h_{\mu\nu} - \tfrac{1}{4}\eta_{\mu\nu}h'\right)\bar{\Psi}^\mu{}_{\lambda|\alpha\beta}\Psi^{\nu\lambda|\alpha\beta}\\ &\approx -\tfrac{i}{2}g\left(h_{\mu\nu} - \tfrac{1}{4}\eta_{\mu\nu}h'\right)\bar{\Psi}^\mu{}_{\lambda|\rho\sigma}\gamma^{\rho\sigma\alpha\beta}\Psi^{\nu\lambda}{}_{\alpha\beta},\end{aligned} \qquad (6.46)$$

where the last equivalent form owes its existence to the identity (D.61), which can be rewritten as $\eta^{\rho\sigma|\alpha\beta} = -\tfrac{1}{2}\gamma^{\rho\sigma\alpha\beta} + \tfrac{1}{2}\gamma^{\rho\sigma}\gamma^{\alpha\beta} - 2\gamma^{[\rho}\eta^{\sigma][\alpha}\gamma^{\beta]}$, and to the EoMs (E.19) and (E.20).

Now let us compute the quantity $\Delta M^\mu = \partial_\nu T^{\mu\nu}$ from Eq. (6.46). One gets

$$\Delta M^\mu = -\tfrac{i}{4}g\,\bar{\Psi}^\mu{}_{\lambda|\rho\sigma}\gamma^{\rho\sigma\alpha\beta}\partial_\nu\Psi^{\nu\lambda}{}_{\alpha\beta} + \text{h.c.} \qquad (6.47)$$

One can then use the identity (E.22) for the divergence of the curvature, thereby obtaining

$$\Delta M^\mu = -\tfrac{1}{2}g\,\bar{\Psi}^{\mu\lambda|}{}_{\rho\sigma}\gamma^{\rho\sigma\alpha\beta}\slashed{\partial}\,\partial_{[\alpha}\mathcal{S}_{\beta]\lambda} + \tfrac{1}{4}g\,\bar{\Psi}^{\mu\lambda|}{}_{\rho\sigma}\gamma^{\rho\sigma\alpha\beta}\partial_\lambda\partial_{[\alpha}\slashed{\mathcal{S}}_{\beta]} + \text{h.c.} \qquad (6.48)$$

In the first term on the right-hand side, one can make use of the identity $\gamma^{\rho\sigma\alpha\beta}\slashed{\partial} = (2\gamma^{\rho\sigma\alpha\beta\tau} - \gamma^\tau\gamma^{\rho\sigma\alpha\beta})\partial_\tau$ and then integrate by parts w.r.t. $\partial_\tau$, noticing that the 5-$\gamma$ piece is killed by a Bianchi identity. In the second term, on the other hand, one can integrate by parts w.r.t. $\partial_\lambda$, which results in

$$\begin{aligned}\Delta M^\mu &= \tfrac{1}{2}g\,\partial_\tau\big(\bar{\Psi}^{\mu\lambda|}{}_{\rho\sigma}\gamma^\tau\gamma^{\rho\sigma\alpha\beta}\partial_{[\alpha}\mathcal{S}_{\beta]\lambda}\big) + \tfrac{1}{4}g\,\partial_\lambda\big(\bar{\Psi}^{\mu\lambda|}{}_{\rho\sigma}\gamma^{\rho\sigma\alpha\beta}\partial_{[\alpha}\slashed{\mathcal{S}}_{\beta]}\big)\\ &\quad - \tfrac{1}{2}g\,\bar{\Psi}^{\mu\lambda|}{}_{\rho\sigma}\overleftarrow{\slashed{\partial}}\gamma^{\rho\sigma\alpha\beta}\partial_{[\alpha}\mathcal{S}_{\beta]\lambda} - \tfrac{1}{4}g\,\bar{\Psi}^{\mu\lambda|}{}_{\rho\sigma}\overleftarrow{\partial}_\lambda\gamma^{\rho\sigma\alpha\beta}\partial_{[\alpha}\slashed{\mathcal{S}}_{\beta]} + \text{h.c.}\end{aligned} \qquad (6.49)$$

The first line on the right-hand side is a double divergence, because one can pull out the $\partial_\alpha$ from the Fronsdal tensor, and make it a total derivative by using the Bianchi identities. That is, the first line plus its hermitian conjugate reduces to the form $\partial_\alpha \partial_\tau \mathcal{Y}_1^{\mu\alpha\tau}$, where

$$\mathcal{Y}_1^{\mu\alpha\tau} = \tfrac{1}{2} g \big( \bar{\Psi}^{\mu\lambda|}{}_{\rho\sigma} \gamma^\tau \gamma^{\rho\sigma\alpha\beta} \mathcal{S}_{\beta\lambda} + \tfrac{1}{2} \bar{\Psi}^{\mu\tau|}{}_{\rho\sigma} \gamma^{\rho\sigma\alpha\beta} \mathcal{\slashed{S}}_\beta + \text{h.c.} \big), \qquad (6.50)$$

which is both $\Gamma$-closed and $\Delta$-exact. On the other hand, the second line of Eq. (6.49) contains bilinears in the Fronsdal tensor by virtue of the EoMs (E.21) and (E.22). The first piece contains the double curl, $\partial^{[\mu} \partial_{[\rho} \bar{\mathcal{S}}_{\sigma]}{}^{\lambda]}$, while the second one includes $\partial_{[\rho} \bar{\mathcal{S}}_{\sigma]}{}^\mu \overleftarrow{\slashed{\partial}}$. In the former of these, one pulls out $\partial_\mu$ to integrate by parts, while in the latter one uses $\overleftarrow{\slashed{\partial}} \gamma^{\rho\sigma\alpha\beta} = \overleftarrow{\partial}_\tau (2\gamma^{\rho\sigma\alpha\beta\tau} - \gamma^{\rho\sigma\alpha\beta} \gamma^\tau)$ and then integrates by parts w.r.t. $\partial_\tau$. The last step produces $\slashed{\partial} \partial_{[\alpha} \mathcal{\slashed{S}}_{\beta]}$, which then can be replaced, thanks to Identity (E.23), by $2\partial^\lambda \partial_{[\alpha} \mathcal{S}_{\beta]\lambda}$. The same step also gives a total derivative: $\partial_\tau (\partial_{[\rho} \bar{\mathcal{S}}_{\sigma]}{}^\mu \gamma^{\rho\sigma\alpha\beta} \gamma^\tau \partial_{[\alpha} \mathcal{\slashed{S}}_{\beta]})$, which can be turned into a double divergence by pulling out $\partial_\alpha$ and integrating by parts. When hermitian conjugates are taken into account, the end result is that the second line of Eq. (6.49) reduces to the form $\partial_\nu \mathcal{X}^{(\mu\nu)} + \partial_\alpha \partial_\tau \mathcal{Y}_2^{\mu\alpha\tau}$, where $\mathcal{X}$ and $\mathcal{Y}_2$ are both $\Gamma$-closed and $\Delta$-exact:

$$\mathcal{Y}_2^{\mu\alpha\tau} = -\tfrac{i}{2} g \left( \partial_{[\rho} \bar{\mathcal{S}}_{\sigma]}{}^\mu \gamma^{\rho\sigma\alpha\beta} \gamma^\tau \mathcal{\slashed{S}}_\beta - \text{h.c.} \right), \qquad (6.51a)$$

$$\mathcal{X}^{(\mu\nu)} = -\tfrac{i}{4} g \left( \partial_{[\rho} \bar{\mathcal{S}}_{\sigma]}{}^\mu \gamma^{\rho\sigma\alpha\beta} \partial_{[\alpha} \mathcal{S}_{\beta]}{}^\nu + \partial_{[\rho} \bar{\mathcal{S}}_{\sigma]}{}^\nu \gamma^{\rho\sigma\alpha\beta} \partial_{[\alpha} \mathcal{S}_{\beta]}{}^\mu \right) \qquad (6.51b)$$
$$+ \tfrac{i}{4} g \eta^{\mu\nu} \left( \partial_{[\rho} \bar{\mathcal{S}}_{\sigma]}{}^\lambda \gamma^{\rho\sigma\alpha\beta} \partial_{[\alpha} \mathcal{S}_{\beta]\lambda} + \tfrac{1}{4} \partial_{[\rho} \bar{\mathcal{\slashed{S}}}_{\sigma]} \gamma^{\rho\sigma\alpha\beta} \partial_{[\alpha} \mathcal{\slashed{S}}_{\beta]} \right).$$

Thus, we have shown that $\Delta M^\mu$ can be rewritten as

$$\Delta M^\mu = \partial_\nu \mathcal{X}^{(\mu\nu)} + \partial_\alpha \partial_\tau \left( \mathcal{Y}_1^{\mu\alpha\tau} + \mathcal{Y}_2^{\mu\alpha\tau} \right). \qquad (6.52)$$

This, along with Eqs. (6.50)–(6.51), fulfills the sufficient condition (D.112) for the triviality of $a_1$. That is, the vertex does not actually deform the gauge transformations: one can make it strictly gauge-invariant modulo d, by adding $\Delta$-exact terms, which are spelled out in (D.115).

Although not manifest, this vertex actually vanishes in $D = 4$. In fact, one can find the following form for the vertex:

$$a_0 \approx -\tfrac{i}{8} g \, h_{\mu\nu} \bar{\Psi}_{\rho\sigma|\tau\lambda} \, \gamma^{\mu\rho\sigma\alpha\beta, \, \nu\tau\gamma} \, \Psi_{\alpha\beta|\gamma}{}^\lambda, \qquad (6.53)$$

which makes the triviality in dimension 4 manifest. To see that this is indeed equivalent to the vertex (6.46), let us use the $\gamma$-matrix identity (D.76) in the vertex (6.53) in order to break it into terms containing only antisymmetric products of six $\gamma$-matrices or two. The former kind of terms all vanish because of either the Bianchi identities or the symmetry in the indices carried by the graviton. On the other hand, the terms containing two $\gamma$-matrices are actually equivalent to terms containing none. This is because the symmetry in the graviton indices requires that at least one $\gamma$-matrix be contracted with a spin-$\frac{5}{2}$ curvature; then the Clifford algebra gives a $\gamma$-trace of the curvature, which is $\Delta$-exact. Thus we get

$$a_0 \approx -\tfrac{i}{8} g h_\mu{}^\nu \, \bar{\Psi}_{\rho\sigma|\lambda}{}^\tau \left( 12\, \delta^{\beta\rho\sigma}_{\nu\tau\gamma}\, \eta^{\mu\alpha} + 24\, \delta^{\mu\rho\beta}_{\nu\tau\gamma}\, \eta^{\sigma\alpha} - 12\, \delta^{\sigma\alpha\beta}_{\nu\tau\gamma}\, \eta^{\mu\rho} \right) \Psi_{\alpha\beta|}{}^{\lambda\gamma}.$$

Having gotten rid of $\gamma$-matrices, it is now straightforward to carry out the computation; the number of possible terms are greatly reduced by the symmetry properties of the associated fields and curvatures, and one can also drop traces of the curvatures, since they are $\Delta$-exact. Thus, one ends up with the first form of the vertex presented in (6.46).

**The 5-Derivatives Vertex**

When the vertex contains five derivatives, the operator $\hat{\mathcal{O}}$ in Eq. (6.43) includes one. As we discussed already, the form of $\hat{\mathcal{O}}$ is much restricted. Indeed, we have just one possibility:

$$T^{\mu\nu} = ig\, \bar{\Psi}^{\rho\sigma|\alpha\beta} \gamma^{(\mu} \overleftrightarrow{\partial}^{\nu)} \Psi_{\rho\sigma|\alpha\beta}, \tag{6.54}$$

where the operator $\overleftrightarrow{\partial}_\mu \equiv \vec{\partial}_\mu - \overleftarrow{\partial}_\mu$ plays a crucial role in eliminating from $\partial_\nu T^{\mu\nu}$ terms that are not $\Delta$-exact. The vertex is given, by virtue of Eq. (6.41), as

$$a_0 = igh_{\mu\nu} \bar{\Psi}^{\rho\sigma|\alpha\beta} \gamma^{(\mu} \overleftrightarrow{\partial}^{\nu)} \Psi_{\rho\sigma|\alpha\beta} \approx -ig\, \mathfrak{h}_{\mu\nu\|\lambda} \bar{\Psi}^\mu{}_{\tau|\rho\sigma}\, \gamma^\lambda\, \Psi^{\nu\tau|\rho\sigma}. \tag{6.55}$$

To see the equivalence with the second form, let us remove therein any derivatives on the graviton field by partial integration. This gives a derivative of the spin-$\frac{5}{2}$ curvatures: the divergence is $\Delta$-exact, while in the gradient one can use the second Bianchi identity and the symmetry properties of the curvatures to pull out a derivative with an index of the graviton field. The equivalence of the vertices then follows immediately.

We can write the second equivalent form as

$$\tfrac{1}{2} \mathfrak{h}_{\mu\nu\|\lambda} \bar{\Psi}^{\mu\tau|}{}_{\rho\sigma} \left( \eta^{\rho\sigma|\alpha\beta} \gamma^\lambda + \gamma^\lambda \eta^{\rho\sigma|\alpha\beta} \right) \Psi^\nu{}_{\tau|\alpha\beta}. \tag{6.56}$$

Then the identity $\eta^{\rho\sigma|\alpha\beta} = -\tfrac{1}{2}\gamma^{\rho\sigma\alpha\beta} + \tfrac{1}{2}\gamma^{\rho\sigma}\gamma^{\alpha\beta} - 2\gamma^{[\rho}\eta^{\sigma][\alpha}\gamma^{\beta]}$ is again helpful, for it helps us drop some $\Delta$-exact pieces, thanks to Eqs. (E.19)–(E.20), so to be left with $\tfrac{1}{2}(\gamma^{\rho\sigma\alpha\beta}\gamma^\lambda + \gamma^\lambda \gamma^{\rho\sigma\alpha\beta}) = \gamma^{\lambda\rho\sigma\alpha\beta}$. Therefore, we have another equivalent form of the vertex:

$$a_0 \approx \tfrac{i}{2} g\, \mathfrak{h}_{\mu\nu\|\lambda} \bar{\Psi}^{\mu\tau|}{}_{\rho\sigma}\, \gamma^{\lambda\rho\sigma\alpha\beta}\, \Psi^\nu{}_{\tau|\alpha\beta}. \tag{6.57}$$

The virtue of this latter form is twofold. First, the presence of an antisymmetric product of five $\gamma$-matrices *manifestly* renders this vertex trivial in $D = 4$. Second, because of Bianchi identities, the gauge variation of the vertex is just a total derivative, which means that it does not deform the gauge transformations, and that feature is also quite manifest in the above expression.

**The 6-Derivatives Vertex**

There is a unique 6-derivatives hermitian current whose divergence is $\Delta$-exact. It reads

$$T^{\mu\nu} = ig\bar{\Psi}^{\rho\sigma|\alpha\beta}\big(\vec{\partial}^\mu\vec{\partial}^\nu + \overleftarrow{\partial}^\mu\overleftarrow{\partial}^\nu - \eta^{\mu\nu}\overleftarrow{\partial}^\lambda\vec{\partial}_\lambda\big)\Psi_{\rho\sigma|\alpha\beta}. \qquad (6.58)$$

While the vertex is simply given by $T^{\mu\nu}h_{\mu\nu}$, one can also cast it into a more 'geometrical' form that involves the product of all three curvatures:

$$a_0 \approx igR_{\mu\nu\rho\sigma}\bar{\Psi}^{\rho\sigma|\alpha\beta}\Psi_{\alpha\beta}{}^{\mu\nu}. \qquad (6.59)$$

This form is strictly gauge invariant in a manifest way, and the vertex exists in all $D \geq 4$. To see the equivalence with the two forms of the vertex, let us remove in the vertex above all the derivatives from the graviton field, by integrations by parts. Dropping divergences of the spin-$\frac{5}{2}$ curvature, that are $\Delta$-exact, we thus arrive at

$$\begin{aligned}a_0 &\approx 4igh_{\mu\nu}\bar{\Psi}^{\mu\alpha|\rho\sigma}\overleftarrow{\partial}^\beta\vec{\partial}_\alpha\Psi^\nu{}_{\beta|\rho\sigma} \\ &\approx 4igh_{\mu\nu}\big(-\Psi^{\alpha\beta|\rho\sigma}\overleftarrow{\partial}^\mu + \bar{\Psi}^{\mu\beta|\rho\sigma}\overleftarrow{\partial}^\alpha\big)\vec{\partial}_\alpha\Psi^\nu{}_{\beta|\rho\sigma},\end{aligned} \qquad (6.60)$$

where the second equivalence results from the Bianchi identity. The first term in the above parentheses imposes the useful Bianchi identity $\partial_{[\alpha}\Psi^\nu{}_{\beta]|\rho\sigma} = \frac{1}{2}\partial^\nu\Psi_{\alpha\beta|\rho\sigma}$, whereas the second term enables us to use the 3-box rule, so that we can drop $\Delta$-exact terms, like $\Box\Psi_{\mu\beta|\rho\sigma}$, and integrate by parts to obtain

$$a_0 \approx -2igh_{\mu\nu}\bar{\Psi}^{\alpha\beta|\rho\sigma}\overleftarrow{\partial}^\mu\vec{\partial}^\nu\Psi^{\alpha\beta|\rho\sigma} + 2ig\,\Box h_{\mu\nu}\bar{\Psi}^{\mu\beta|\rho\sigma}\Psi^\nu{}_{\beta|\rho\sigma}. \qquad (6.61)$$

Now, let us replace $\Box h_{\mu\nu}$ by $2\partial_{(\mu}\partial\cdot h_{\nu)} - \partial_\mu\partial_\nu h'$, since their difference is $R_{\mu\nu} = \Delta$-exact. In the resulting expression, let us remove all the derivatives from the graviton field, which yields

$$\begin{aligned}a_0 \approx &-2igh_{\mu\nu}\bar{\Psi}^{\alpha\beta|\rho\sigma}\overleftarrow{\partial}^\mu\vec{\partial}^\nu\Psi^{\alpha\beta|\rho\sigma} \\ &+ 2ig\big(h_{\mu\nu} - \tfrac{1}{2}\eta_{\mu\nu}h'\big)\partial^\nu\big(\bar{\Psi}^{\lambda\beta|\rho\sigma}\vec{\partial}_\lambda\Psi^\mu{}_{\beta|\rho\sigma} - \text{h.c.}\big).\end{aligned} \qquad (6.62)$$

In the second term, we can again use $\partial_{[\lambda}\Psi^{\mu}{}_{\beta]|\rho\sigma} = \frac{1}{2}\partial^{\mu}\Psi_{\lambda\beta|\rho\sigma}$ to find that some of the resulting pieces cancel the first term. The remaining part adds to the form $T^{\mu\nu}h_{\mu\nu}$, with $T^{\mu\nu}$ given precisely by (6.58). Hence the vertices are equivalent. The summary of our couplings for this setup is given in Table 7

## 6.2 Arbitrary-Spin Couplings

The sets of fields and antifields for the arbitrary-spin case are given by

$$\Phi^A = \{h_{\mu\nu}, C_\mu, \psi_{\mu_1\ldots\mu_n}, \xi_{\mu_1\ldots\mu_{n-1}}\}, \tag{6.63a}$$

$$\Phi^*_A = \{h^{*\mu\nu}, C^{*\mu}, \bar{\psi}^{*\mu_1\ldots\mu_n}, \bar{\xi}^{*\mu_1\ldots\mu_{n-1}}\}. \tag{6.63b}$$

For $n > 2$, there is a triple $\gamma$-trace constraint on the field and antifield, i.e.

$$\slashed{\psi}'_{\mu_1\ldots\mu_{n-3}} = 0, \qquad \bar{\slashed{\psi}}^{*\prime}_{\mu_1\ldots\mu_{n-3}} = 0. \tag{6.64}$$

The rank-$(n-1)$ fermionic ghost and its antighost are $\gamma$-traceless as usual:

$$\slashed{\xi}_{\mu_1\ldots\mu_{n-2}} = 0, \qquad \bar{\slashed{\xi}}^{*}_{\mu_1\ldots\mu_{n-2}} = 0, \tag{6.65}$$

and the spin-$s$ Lagrangian EoMs are given by the rank-$n$ tensor-spinor $\mathcal{R}_{\mu_1\ldots\mu_n}$, which we have already introduced in Chapter 5 and we recall its expression:

$$\mathcal{R}_{\mu_1\ldots\mu_n} = \mathcal{S}_{\mu_1\ldots\mu_n} - \tfrac{1}{2}n\,\gamma_{(\mu_1}\,\slashed{\mathcal{S}}_{\mu_2\ldots\mu_n)} - \tfrac{1}{4}n(n-1)\,\eta_{(\mu_1\mu_2}\mathcal{S}'_{\mu_3\ldots\mu_n)}. \tag{6.66}$$

Let us recall once again that an account of the cohomology of $\Gamma$ is given in Appendix E, and in Table 6.2 below we spell out some important properties of the various fields and antifields.
Again, the antifield $\bar{\chi}^{*\mu_1\cdots\mu_n}$ is given by Eqs. (E.34)–(E.35), while the BRST-closed free master action for the arbitrary-spin case now reads:

$$S_0 = \int d^D x \left[ G^{\mu\nu}h_{\mu\nu} + \tfrac{1}{2}\left(\bar{\mathcal{R}}^{\mu_1\ldots\mu_n}\psi_{\mu_1\ldots\mu_n} - \bar{\psi}_{\mu_1\ldots\mu_n}\mathcal{R}^{\mu_1\ldots\mu_n}\right)\right] \tag{6.67}$$

$$+ \int d^D x \left[-2h^{*\mu\nu}\partial_\mu C_\nu + \tfrac{1}{2}n\left(\bar{\psi}^{*\mu_1\ldots\mu_n}\partial_{\mu_1}\xi_{\mu_2\ldots\mu_n} - \partial_{\mu_1}\bar{\xi}_{\mu_2\ldots\mu_n}\psi^{*\mu_1\ldots\mu_n}\right)\right].$$

Now we are ready to construct the $2$–$s$–$s$ cubic vertices. Having worked out the spin-$\frac{5}{2}$ case as a prototypical example, our job has become easy, since many of the statements made for spin $\frac{5}{2}$ will translate verbatim to arbitrary spin.

Table 6.2: Properties of the Various Fields & Antifields ($\forall\, n$)

| $Z$ | $\Gamma(Z)$ | $\Delta(Z)$ | pgh($Z$) | agh($Z$) | gh($Z$) | $\epsilon(Z)$ |
|---|---|---|---|---|---|---|
| $h_{\mu\nu}$ | $2\partial_{(\mu}C_{\nu)}$ | 0 | 0 | 0 | 0 | 0 |
| $C_\mu$ | 0 | 0 | 1 | 0 | 1 | 1 |
| $h^{*\mu\nu}$ | 0 | $G^{\mu\nu}$ | 0 | 1 | $-1$ | 1 |
| $C^{*\mu}$ | 0 | $-2\partial_\nu h^{*\mu\nu}$ | 0 | 2 | $-2$ | 0 |
| $\psi_{\mu_1\ldots\mu_n}$ | $n\partial_{(\mu_1}\xi_{\mu_2\ldots\mu_n)}$ | 0 | 0 | 0 | 0 | 1 |
| $\xi_{\mu_1\ldots\mu_{n-1}}$ | 0 | 0 | 1 | 0 | 1 | 0 |
| $\psi^{*\mu_1\ldots\mu_n}$ | 0 | $\bar{\mathcal{R}}^{\mu_1\ldots\mu_n}$ | 0 | 1 | $-1$ | 0 |
| $\bar\xi^{*\mu_1\ldots\mu_{n-1}}$ | 0 | $2\partial_{\mu_n}\bar\chi^{*\mu_1\ldots\mu_n}$ | 0 | 2 | $-2$ | 1 |

### 6.2.1 Non-Abelian Vertices

Let us recall that any $a_2$ consists of two ghost fields and a single antighost, and that the latter can be chosen to be undifferentiated without loss of generality. As explained in Appendix E, a single derivative acting on the ghost $C_\mu$ can be realized as a 1-curl $\mathfrak{C}_{\mu\nu}$ modulo irrelevant $\Gamma$-exact terms, while two or more derivatives are never nontrivial. For the fermionic ghost $\xi_{\mu_1\ldots\mu_{n-1}}$, on the other hand, one can choose any $m$-curl $\xi_{\mu_1\nu_1|\ldots|\mu_m\nu_m\|\mu_{m+1}\ldots\mu_{n-1}}$ with $m = 1, 2, \ldots, n-1$, and more than $n-1$ derivatives give $\Gamma$-exact terms. Clearly, a nontrivial $a_2$ cannot contain more than $2n-2$ derivatives. This sets an upper bound of $2n-1$ on the number of derivatives in a non-abelian vertex given the actions of $\Gamma$ and $\Delta$ on various (anti)fields and the consistency cascade (4.21).

Again, all nontrivial $a_2$'s fall into two subsets: Subset 1 contains the bosonic antighost $C^{*\mu}$, and subset 2 the fermionic one $\xi^{*\mu_1\ldots\mu_{n-1}}$. Subset 1 has the form $a_2 = C^{*\mu}X_\mu$, where $X_\mu$ is some bilinear in the fermionic ghost-curls. Then we have: $\Delta a_2 \doteq 2h^{*\mu\nu}\partial_{(\mu}X_{\nu)}$, which must be $\Gamma$-exact modulo d if the cocycle condition (4.21b) is to be satisfied. Then, because $\Gamma$ does not act on the antifields, a functional derivative w.r.t. $h^{*\mu\nu}$ gives

$$\partial_{(\mu}X_{\nu)} = \Gamma\text{-exact}. \tag{6.68}$$

Now, the symmetrized derivative of $X_\mu$ can be schematically written as

$$\partial X \sim \partial\big(\bar\xi^{(m_1)}\xi^{(m_2)} \pm \bar\xi^{(m_2)}\xi^{(m_1)}\big) \tag{6.69}$$
$$\sim \Gamma(\ldots) + \bar\xi^{(m_1+1)}\xi^{(m_2)} + \bar\xi^{(m_1)}\xi^{(m_2+1)} \pm (m_1 \leftrightarrow m_2).$$

When $m_1$ and $m_2$ are equal, we have the plus sign for a nonzero $X$, and nontrivial elements of H($\Gamma$) are absent only when $m_1 = m_2 = n-1$.

When they are unequal, let us take $m_1 > m_2$, and then $\partial X$ is $\Gamma$-exact with the minus sign if $m_1 = m_2 + 1 = n - 1$. The only $a_2$'s that pass the condition (4.21b) thus contain $2n - 3$ and $2n - 2$ derivatives. More explicitly,

$$a_2 = \begin{cases} p = 2n-2: & igC^{*\mu}(\bar{\xi}^{(n-1)}_{\mu\ldots}\xi^{(n-2)\cdots} - \bar{\xi}^{(n-2)\cdots}\xi^{(n-1)}_{\mu\ldots}), \\ p = 2n-1: & igC^{*\mu}\bar{\xi}^{(n-1)\cdots}\gamma_\mu\xi^{(n-1)}_{\ldots}, \end{cases} \quad (6.70)$$

where the ellipses mean contracted indices. The similarity with the spin-$\frac{5}{2}$ case is manifest.

Subset 2, on the other hand, involves the (undifferentiated) fermionic antighost. In this case, the $a_2$ has the form $a_2 = \bar{\xi}^{*\mu_1\ldots\mu_{n-1}}Y_{\mu_1\ldots\mu_{n-1}} + \text{h.c.}$ Symmetry is imposed in the indices of $Y$, which comprises both of the ghosts and curls thereof. Following the same logic as presented for spin $\frac{5}{2}$, it is clear that $a_2$ can contain at most two derivatives: one in $\mathfrak{C}_{\mu\nu}$ and the other in the 1-curl $\xi_{\mu_1\nu_1\|\nu_2\ldots\nu_{n-1}}$, as higher-curls of the latter are incompatible with the symmetry of the indices. At this point the possibilities (all to be ruled out) are thus:

$$a_2 = \begin{cases} p = 0: & g\bar{\xi}^{*\mu\cdots}\gamma^\alpha\xi_{\mu\ldots}C_\alpha + \text{h.c.}, \\ p = 1: & g\bar{\xi}^{*\mu\cdots}(\xi^\nu_{\ldots}\mathfrak{C}_{\mu\nu} + \alpha_1\xi^{(1)}_{\mu\nu\|\ldots}C^\nu + \alpha_2\gamma^{\alpha\beta}\xi_{\mu\ldots}\mathfrak{C}_{\alpha\beta}) + \text{h.c.}, \\ p = 2: & g\bar{\xi}^{*\mu\cdots}\gamma^\alpha\xi^\beta{}_{\mu\|\ldots}\mathfrak{C}_{\alpha\beta} + \text{h.c.} \end{cases} \quad (6.71)$$

However, one can derive quite similarly a counterpart of condition (6.68), namely

$$\partial_{(\mu_1}Y_{\mu_2\ldots\mu_n)} = \Gamma\text{-exact}, \quad (6.72)$$

and when $n > 2$ it is impossible for any element in the list (6.71) to fulfill this condition, because $\partial Y$ will always contain nontrivial elements of $H(\Gamma)$. This rules out all of them.

**The ($2n$-2)-Derivative Vertex**

In this case, one can proceed along the same lines as for the 2-derivatives spin-$\frac{5}{2}$ vertex, and to make the steps go verbatim we add a trivial term to the first element of (6.70), thus writing

$$\begin{aligned} a_2 &= igC^{*\mu}\big(\bar{\xi}^{(n-1)}_{\mu\ldots}\xi^{(n-2)\cdots} - \bar{\xi}^{(n-2)\cdots}\xi^{(n-1)}_{\mu\ldots}\big) \\ &\quad + \tfrac{1}{8}g\mathfrak{C}_{\mu\nu}\big(\bar{\xi}^{*(n-2)}_{\ldots\|\rho}\gamma^{\mu\nu\rho\alpha\beta}\xi^{(n-1)\cdots}\|_{\alpha\beta} - \text{h.c.}\big), \end{aligned} \quad (6.73)$$

which looks quite similar to the spin-$\frac{5}{2}$ counterpart (6.19), given the relation (6.30). To obtain the vertex, one can now simply redo the steps of Subsection 6.1.1, and consequently find

$$a_0 = ig\big(\bar{\psi}^{(n-2)}_{\cdots\|\mu\alpha} R^{+\mu\nu\alpha\beta} \psi^{(n-2)\cdots\|}{}_{\nu\beta} + \tfrac{1}{2}\bar{\not{\psi}}^{(n-2)}_{\cdots\|\mu} \not{R}^{\mu\nu} \not{\psi}^{(n-2)\cdots\|}{}_{\nu}\big) \\ + \tfrac{i}{4} g\, h_{\mu\nu}\, \bar{\psi}^{(n-1)}_{\cdots\rho\sigma\|\lambda} \gamma^{\mu\rho\sigma\alpha\beta,\,\nu\lambda\gamma} \psi^{(n-1)\cdots}{}_{\alpha\beta\|\gamma} \tag{6.74}$$

as the searched-for non-abelian $2-s-s$ vertex containing $2n-2$ derivatives. Again, let us notice the striking similarity with its spin-$\frac{5}{2}$ counterpart, given in (6.32), including the appearance of the $R^+_{\mu\nu\alpha\beta}$ combination.

**The ($2n$-1)-Derivative Vertex**

Here one starts with the second element of (6.70). In order to have a direct generalization of (6.33) we use five $\gamma$-matrices instead of one, thus writing

$$a_2 = -ig\, C^*_\lambda\, \bar{\xi}^{(n-1)}_{\cdots|\,\mu\nu}\, \gamma^{\lambda\mu\nu\alpha\beta}\, \xi^{(n-1)\cdots|}{}_{\alpha\beta}. \tag{6.75}$$

One can then proceed in the same way as in Subsection 6.1.1 to find:

$$a_0 = ig\bar{\psi}^{(n-1)}_{\cdots\mu\nu\|}{}^\rho \big(\mathfrak{h}^+_{\rho\sigma\|\lambda}\gamma^{\lambda\mu\nu\alpha\beta} + \gamma^{\lambda\mu\nu\alpha\beta}\, \mathfrak{h}^+_{\rho\sigma\|\lambda}\big)\psi^{(n-1)\cdots}{}_{\alpha\beta\|}{}^\sigma, \tag{6.76}$$

which is our non-abelian $2-s-s$ vertex with $2n-1$ derivatives. Again, the comparison with the spin-$\frac{5}{2}$ counterpart (6.39) reveals that they are very similar, and in particular the '$+$' operator again plays a role.

### 6.2.2 Abelian Vertices

Abelian vertices are those that do not deform the gauge algebra. Such a vertex corresponds to a trivial $a_2$, and therefore to an $a_1$ which can always be chosen to be $\Gamma$-closed [177, 178],

$$\Gamma a_1 = 0, \tag{6.77}$$

and which is related to the vertex $a_0$ through the cocycle condition (4.21c), that is:

$$\Delta a_1 + \Gamma a_0 \doteq 0. \tag{6.78}$$

Now, in full analogy with the electromagnetic case, one can again prove in this instance that abelian vertices do not deform the gauge algebra. The gravitational-case proof is analogous to its electromagnetic counterpart, so that the former is to be found in Appendix D, and we shall thus hereafter assume $a_1$ to be trivial.

The arguments presented in the beginning of Subsection 6.1.2 then go verbatim for arbitrary spin, except that now the number of derivatives in the abelian vertex can take the values $2n, 2n+1$ and $2n+2$, since the spin-$s$ curvature tensor contains $n$ derivatives ($s = n + \frac{1}{2}$). The corresponding currents can be written as direct generalizations of those for spin $\frac{5}{2}$, given respectively by Eq. (6.44) with $\alpha = \frac{1}{4}$ and by Eqs. (6.54) and (6.58). Explicitly, the vertices are:

$$p = 2n: \qquad a_0 = ig(h_{\mu\nu} - \tfrac{1}{4}\eta_{\mu\nu}h')\bar{\Psi}^{\mu\cdots}\Psi^{\nu\cdots}, \tag{6.79a}$$

$$p = 2n+1: \qquad a_0 = igh_{\mu\nu}\bar{\Psi}^{\cdots}\gamma^{(\mu}\vec{\partial}^{\nu)}\Psi_{\cdots}, \tag{6.79b}$$

$$p = 2n+2: \qquad a_0 = igh_{\mu\nu}\bar{\Psi}^{\cdots}\big(\vec{\partial}^{\mu}\vec{\partial}^{\nu} + \cev{\partial}^{\mu}\cev{\partial}^{\nu} - \eta^{\mu\nu}\cev{\partial}^{\lambda}\vec{\partial}_{\lambda}\big)\Psi_{\cdots}. \tag{6.79c}$$

None of these vertices deform the gauge transformations. The $2n$-derivatives vertex can be shown to fulfill the sufficient condition (D.112) in order for its $a_1$ to be trivial, and the proof follows exactly the same steps as in the spin-$\frac{5}{2}$ case. On the other hand, one can render the $(2n+1)$-derivatives vertex manifestly $\Gamma$-closed modulo d by casting it into a generalization of the expression (6.57), while the $(2n+2)$-derivatives one takes the 3-curvatures form akin to that of (6.59). These proofs are also straightforward generalizations of the spin-$\frac{5}{2}$ case.

Finally, direct generalizations of the prototypical spin-$\frac{5}{2}$ example also show that the $2n$- and $(2n+1)$-derivatives vertices are trivial in $D = 4$, while the $(2n+2)$-derivatives, 3-curvatures vertex exits in all $D \geq 4$. More comments are found in the following chapter, where we also comment on quartic consistency. In the case of gravitational coupling, the proof that our non-abelian cubic vertices are obstructed at higher-order is relegated to Appendix D.5. Again, a summary of our results is given in Table 7 of the following chapter.

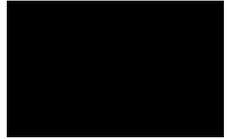

CHAPTER 7

# Conclusions

In this last chapter, after summarizing our results, we comment on them and discuss their relation to other works.

**Summary of Results**

Making use of the BRST-Antifield reformulation of the deformation problem we have obtained the exhaustive list of $1-s-s$ and $2-s-s$ consistent couplings in Minkowski spacetime of dimension $D \geq 4$, for all gauge fermions of spin $s = n + \frac{1}{2}$. The assumptions have been locality, Lorentz invariance and parity invariance. The vertices have been obtained in their full off-shell form, and no gauge fixing or on-shell condition has been imposed. Moreover, for those vertices that are non-abelian, the corresponding gauge-algebra and gauge-transformation deformations have been given explicitly. For spin $s = n + \frac{1}{2}$, we find that the possible number of derivatives in a cubic $1-s-s$ vertex is restricted to only three values: $2n-1$, $2n$ and $2n+1$, while for the $2-s-s$ vertices they are five: $2n-2, 2n-1, 2n, 2n+1$ and $2n+2$, with only one inequivalent vertex for each value, displayed hereafter.

Our notation is as follows: by 'NA' we mean that the vertex is non-abelian; 'CS', which stands for Chern–Simons, means that the vertex is displayed in a form where it is manifestly gauge invariant up to total derivatives; and by 'BI', which stands for Born–Infeld we mean the vertex is strictly gauge invariant, without use of partial integration (and is displayed in that form). The number of derivatives is noted $p$.

The electromagnetic couplings of the Rarita–Schwinger read as follows.

Table 7.1: Summary of $1-\frac{3}{2}-\frac{3}{2}$ Vertices with $p$ Derivatives

| $p$ | Vertex | Nature |
|---|---|---|
| 1 | $\bar{\psi}_\mu F^{+\mu\nu}\psi_\nu$ | $D \geq 4$, NA |
| 2 | $\left(\bar{\Psi}_{\mu\nu}\,\gamma^{\mu\nu\alpha\beta\lambda}\,\Psi_{\alpha\beta}\right)A_\lambda$ | $D \geq 5$, CS |
| 3 | $\bar{\Psi}_{\mu\alpha}\Psi^\alpha{}_\nu F^{\mu\nu}$ | $D \geq 4$, BI |

We note the appearance of the 'geometric' quantity

$$F^{+\mu\nu} \equiv F^{\mu\nu} + \tfrac{1}{2}\gamma^{\mu\nu\rho\lambda}F_{\rho\lambda} \qquad (7.1)$$

in the non-abelian vertex with one derivative. Let us note that the deformation with two derivatives has been put in a form $A_\mu J^\mu$, where $J^\mu$ is the gauge-invariant current, bilinear in the fermion curvatures and containing the maximum possible number of $\gamma$-matrices. Also, we point out that the strictly gauge-invariant vertex with three derivatives is simply a product of curvatures — the only one that can be written down, actually.[1] One also notices that the 'Chern–Simons' vertex vanishes in $D=4$, as is made manifest in the given expression. Finally, it is instructive to compare the above to the spin-$\tfrac{1}{2}$ couplings, given in Table 4.3: one observes that for $s=\tfrac{1}{2}$ ($n=0$), the bounds $2n-1$, $2n$, $2n+1$ on $p$ would formally give rise to vertices with 1, 0 and $-1$ derivatives. This means that the non-abelian coupling thereof does not exist, as Table 4.3 indeed shows, whereas there *is*, in that case, a deformation without derivatives, which is the minimal coupling — absent for $s \geq \tfrac{3}{2}$.

Note that minimal coupling is usually defined as the cubic term produced by the covariantization of derivatives in the corresponding kinetic term. In our context, where we deal with fermions, the minimal coupling would have zero derivatives, because the free EoMs are of order one in the derivatives. As aforesaid, we have demonstrated that such vertices do not exist for higher-spin gauge fermions.

---

[1] Such higher-derivative vertices for higher spins were noticed to exist at an early stage already, e.g. in [193].

For the spin-$\frac{5}{2}$ field, the couplings with the photon are given below.

Table 7.2: Summary of $1-\frac{5}{2}-\frac{5}{2}$ Vertices with $p$ Derivatives

| $p$ | Vertex | Nature |
|---|---|---|
| 3 | $\bar{\psi}^{(1)}_{\alpha\beta\|\mu} F^{+\mu\nu} \psi^{(1)\alpha\beta\|}{}_{\nu}$ | $D \geq 4$, NA |
| 4 | $\left(\bar{\Psi}_{\mu\nu\|\rho\sigma}\, \gamma^{\mu\nu\alpha\beta\lambda}\, \Psi_{\alpha\beta\|}{}^{\rho\sigma}\right) A_\lambda$ | $D \geq 5$, CS |
| 5 | $\bar{\Psi}_{\alpha\beta\|\mu\rho}\Psi^{\alpha\beta\|\rho}{}_{\nu} F^{\mu\nu}$ | $D \geq 4$, BI |

Interestingly, the expressions are really akin to those for the Rarita-Schwinger, and one simply adds the necessary indices by taking curls wherever the fermion appears and contracting the additional indices in the only possible way. The structure of gauge invariance, together with the dependence on the spacetime dimension, is the same as for $s = \frac{3}{2}$, and so is the appearance of the various curvatures.

Having in mind the above cases, it is no surprise that the generic couplings of a spin-$s$ gauge fermion with a vector field are as follows.

Table 7.3: Summary of $1-s-s$ Vertices with $p$ Derivatives

| $p$ | Vertex | Nature |
|---|---|---|
| $2n-1$ | $\bar{\psi}^{(n-1)}_{\mu_1\nu_1\|\dots\|\mu_{n-1}\nu_{n-1}\|\mu_n} F^{+\mu_n}{}_{\nu_n} \psi^{(n-1)\mu_1\nu_1\|\dots\|\mu_{n-1}\nu_{n-1}\|\nu_n}$ | $D \geq 4$, NA |
| $2n$ | $\left(\bar{\Psi}_{\mu_1\nu_1\|\mu_2\nu_2\|\dots\|\mu_n\nu_n}\gamma^{\mu_1\nu_1\alpha_1\beta_1\lambda}\Psi_{\alpha_1\beta_1\|}{}^{\mu_2\nu_2\|\dots\|\mu_n\nu_n}\right)A_\lambda$ | $D \geq 5$, CS |
| $2n+1$ | $\bar{\Psi}_{\mu_1\nu_1\|\mu_2\nu_2\|\dots\|\mu_n\alpha}\Psi^{\mu_1\nu_1\|\mu_2\nu_2\|\dots\|\alpha\nu_n} F^{\mu_n}{}_{\nu_n}$ | $D \geq 4$, BI |

Again, the pattern very much resembles that of the simplest non-trivial case, namely the spin-$\frac{3}{2}$ one. However, we see another interesting feature emerge: the $s = n + \frac{1}{2}$ non-abelian vertex contains the $(n-1)$-curls of the fermion field, whereas the tensor $F^{+\mu\nu}$ still appears in the same way.

We now move on to recalling our results which concern the gravitational couplings. Unlike in the electromagnetic case, a higher-spin gauge fermion can couple to a graviton in five different ways, not three.

The simplest gravitational higher-spin vertices are given hereafter.

Table 7.4: Summary of $2-\frac{5}{2}-\frac{5}{2}$ Vertices with $p$ Derivatives

| $p$ | Vertex | Nature |
|---|---|---|
| 2 | $i\bar{\psi}_{\mu\alpha}R^{+\mu\nu\alpha\beta}\psi_{\nu\beta} + \frac{i}{4}h_{\mu\nu}\bar{\psi}_{\rho\sigma\|\lambda}\,\gamma^{\mu\rho\sigma\alpha\beta,\nu\lambda\gamma}\,\psi_{\alpha\beta\|\gamma} + \text{TT}$ | $D \geq 4$, NA |
| 3 | $i\bar{\psi}_{\mu\nu\|}{}^{\rho}\bigl(\mathfrak{h}^{+}_{\rho\sigma\|\lambda}\,\gamma^{\lambda\mu\nu\alpha\beta} + \gamma^{\lambda\mu\nu\alpha\beta}\,\mathfrak{h}^{+}_{\rho\sigma\|\lambda}\bigr)\psi_{\alpha\beta\|}{}^{\sigma}$ | $D \geq 5$, NA |
| 4 | $ih_{\mu\nu}\bar{\Psi}_{\rho\sigma\|\tau\lambda}\,\gamma^{\mu\rho\sigma\alpha\beta,\nu\tau\gamma}\,\Psi_{\alpha\beta\|\gamma}{}^{\lambda}$ | $D \geq 5$, A |
| 5 | $i\mathfrak{h}_{\mu\nu\|\lambda}\bar{\Psi}^{\mu\tau\|}{}_{\rho\sigma}\,\gamma^{\lambda\rho\sigma\alpha\beta}\,\Psi^{\nu}{}_{\tau\|\alpha\beta}$ | $D \geq 5$, CS |
| 6 | $iR_{\mu\nu\rho\sigma}\bar{\Psi}^{\rho\sigma\|\alpha\beta}\Psi_{\alpha\beta}{}^{\mu\nu}$ | $D \geq 4$, BI |

The transverse-traceless terms in the 2-derivatives vertex, denoted 'TT', are given by

$$\tfrac{i}{2}\,\bar{\slashed{\psi}}_{\mu}\slashed{R}^{\mu\nu}\slashed{\psi}_{\nu}. \qquad (7.2)$$

The vertex with the highest number of derivatives is again a product of curvatures, whereas the $p = 5$ and $p = 4$ ones only contain the fermion curvatures. Moving down in number of derivatives we face one of the two non-abelian vertices, where we start seeing the appearance of the '+' operator, which turns a tensor with two indices into the equivalent of (7.1), where the tensor is $F$. Finally, the deformation with two derivatives is seen to involve a term with the tensor $R^{+}_{\mu\nu\alpha\beta}$, which is really the spin-2 analogue of $F^{+}_{\mu\nu}$ for the photon field, and that term is seen to be a straightforward generalization of the non-abelian electromagnetic dipole term. However, in the gravitational case this term is dressed by another one, which can be seen to vanish in the traceless, transverse gauge and which we give hereabove with the correct normalization. Again, we see that only the vertices with the highest and the lowest number of derivatives survive in dimension 4, just like for the electromagnetic case. Let us also point out that the four-derivatives vertex has been given in a form where it is not gauge invariant (up to total derivatives) off-shell, and one needs to use the equations of motion.[2]

---

[2] Of course, since we know it does not deform the gauge transformations (its associated $a_1$ is trivial), it is possible to perform field redefinitions on it so that it will be strictly gauge invariant up to partial integration only, but we chose not do so.

Last of all, we give the gravitational couplings for arbitrary spin $s$.

Table 7.5: Summary $2-s-s$ Vertices with $p$ Derivatives

| $p$ | Vertex | Nature |
|---|---|---|
| $2n-2$ | $ig\,\bar{\psi}^{(n-2)}_{\cdots\|\mu\alpha}R^{+\mu\nu\alpha\beta}\psi^{(n-2)\cdots\|}{}_{\nu\beta}$ (TT, $D\geq 5$) | $D\geq 4$, NA |
| $2n-1$ | $ig\,\bar{\psi}^{(n-1)}{}_{\cdots\,\mu\nu\|}{}^{\rho}\big(\mathfrak{h}^{+}_{\rho\sigma\|\lambda}\gamma^{\lambda\mu\nu\alpha\beta}+\gamma^{\lambda\mu\nu\alpha\beta}\mathfrak{h}^{+}_{\rho\sigma\|\lambda}\big)\psi^{(n-1)\cdots}{}_{\alpha\beta\|}{}^{\sigma}$ | $D\geq 5$, NA |
| $2n$ | $ig\left(h_{\mu\nu}-\tfrac{1}{4}\eta_{\mu\nu}h'\right)\bar{\Psi}^{\mu\cdots}\Psi^{\nu\cdots}$ | $D\geq 5$, A |
| $2n+1$ | $ig\,h_{\mu\nu}\bar{\Psi}^{\cdots}\gamma^{(\mu}\overrightarrow{\partial}{}^{\nu)}\Psi_{\cdots}$ | $D\geq 5$, CS |
| $2n+2$ | $ig\,h_{\mu\nu}\bar{\Psi}^{\cdots}\big(\overleftarrow{\partial}{}^{\mu}\overrightarrow{\partial}{}^{\nu}+\overleftarrow{\partial}{}^{\mu}\overrightarrow{\partial}{}^{\nu}-\eta^{\mu\nu}\overleftarrow{\partial}{}^{\lambda}\overrightarrow{\partial}_{\lambda}\big)\Psi_{\cdots}$ | $D\geq 4$, BI |

Just as for the vector-field couplings we observe that, in the given form, the generalization to generic spin proceeds in a trivial manner, by simply adding indices in the only way compatible with the symmetry of the tensors — and for the non-abelian vertices those extra indices have been hidden hereabove. We note that the non-abelian vertex with $2n-2$ derivatives is given here in the traceless, transverse gauge in four dimensions, while the full expression would be obtained by adding to the latter the terms

$$\tfrac{i}{2}g\bar{\slashed{\psi}}^{(n-2)}_{\cdots\|\mu}\slashed{R}^{\mu\nu}\slashed{\psi}^{(n-2)\cdots\|}{}_{\nu}+\tfrac{i}{4}g\,h_{\mu\nu}\,\bar{\psi}^{(n-1)}_{\cdots\rho\sigma\|\lambda}\gamma^{\mu\rho\sigma\alpha\beta,\nu\lambda\gamma}\psi^{(n-1)\cdots}{}_{\alpha\beta\|\gamma}. \quad (7.3)$$

Our results also include a cohomological proof of the following facts.

**Minimal coupling:** the well-known fact that in flat space a massless spin-$\tfrac{3}{2}$ (resp. spin-$\tfrac{5}{2}$) field cannot have minimal coupling to Electromagnetism (resp. Gravity) has been demonstrated.[3]

**Abelianity:** we have given a generic proof that the abelian vertices preserve the gauge symmetries (and not only the gauge algebra), which we have checked explicitly on the above expressions.

**Second-order:** our analysis of second-order consistency reveals that in a local theory, without additional degrees of freedom, our vertices cannot be made fully consistent by the addition of quartic terms.

Let us also note that our results heavily rely, among other things, on the use of $\gamma$-matrix identities, for which we have sometimes used the most useful Mathematica package 'GAMMA' [194].

---

[3] This result generalizes easily to higher-spin fermions of any spin.

**Light-Cone Approach**

A larger class of couplings, involving the $1-s-s$ and $2-s-s$ fermionic cases studied here, have been investigated in [62] by means of the Light-Cone formulation, where it was found that there is one and only one cubic $s'-s-s$ vertex (up to equivalency) for each of the following number of derivatives it contains: $2n - s'$, $2n - s' + 1$, ..., $2n + s' - 1$ and $2n + s'$, where $s'$ is the spin of the bosonic gauge field and $s = n + \frac{1}{2}$ is that of the fermionic tensor-spinor, such that $s' \leq n$. These restrictions evidently coincide with those derived in the present work and reported hereabove. Moreover, the vanishing in $D = 4$ of the vertices containing $2n-s'$, $2n-s'+1$ to $2n+s'-1$ derivatives, also stated in [62], is in full agreement with our expressions.

As the Light-Cone approach is a complete gauge fixing, it can be said that it is somewhat the opposite of our procedure, which is not only covariant but also fully off-shell, and the full agreement between the outputs of both techniques gives further confidence in the validity of the result. On the other hand, different techniques allow for different features to be within reach. For example, in the Light-Cone approach of [62] the gauge-algebra and gauge-symmetry deformations are not discussed, but the analysis is extended beyond the gravitational coupling and is more generic in that sense. With our methods we only addressed the electromagnetic and gravitational cases, and although one may think of considering e.g. $3-s-s$ couplings, at the present moment we have not yet found a way to straightforwardly generalize our results to $s'-s-s$ vertices for arbitrary $s'$.

**String Theory Lessons**

The couplings we have obtained in this work were obtained before in [25], and we now comment on the comparison. The analysis performed in that reference is remarkable: previously, all attempts at extracting information from some tensionless limit of String Theory mainly resulted in obtaining information about free higher-spin gauge fields. Indeed, if one is too naive in taking the tensionless limit $\alpha' \to \infty$, the typical situation is that the interactions among the massive (higher-spin) modes, made massless, are lost. However, the authors of [25] found a way of taking the tensionless limit of tree-level amplitudes for the open, bosonic string in such a way as to allow for consistent gauge cubic couplings to be extracted therefrom. The expressions are naturally extracted in an on-shell form, and then completed to form a cubic off-shell interaction. The results, of course, agree with the number-of-derivatives bounds found in [62] and with the results of [184].

Given that our cubic couplings also agree with the restrictions imposed by [62] on the number of derivatives contained in the vertex, it should come as no surprise that we find an agreement between our expressions and those of [25]. The match between both versions of the results is worked out in Appendix D.2, where for simplicity we address the electromagnetic couplings only. In fact, our proof that both expressions coincide is off-shell only for the spin-$\frac{3}{2}$ electromagnetic couplings, and for the other $1-s-s$ interactions we give a proof of agreement in the transverse-traceless gauge up to EoMs. The zealous reader shall easily extend such demonstrations to the off-shell case, and also work out the equivalent proof for the gravitational couplings. Let us point out that the fact that the work of [25] gives *all* the possible cubic couplings is remarkable. Indeed, although it was to be expected that the tensionless limit of String Theory contains some of these couplings, there is a priori no reason why it should contain them all (which our work has again made certain in the fermionic sector).

Another related and interesting observation is the following: in [25], the various couplings that we have obtained appear with related coupling constants, that is, the prefactors to all the couplings depend on one parameter only. This observation was made in the said works already, where it is noticed that String Theory makes use of the exponential of some differential operator in order to act on the generating function of all the couplings. It is the presence of the exponential, then, which relates all the coupling constants, and the authors of [25] pointed out that there was a priori no reason why String Theory should use the exponential and not another function of the differential operator they write down. In our work, we see that cubic-order consistency does not relate the coupling constants (as usual). However, it is well known that higher-order consistency may require them to be related to one another [103], and as the consistency of String Theory is not restricted to the first order the observation made in [25] is not surprising. It thus seems as if String Theory needs all the couplings for consistency. At any rate, String Theory may not be the unique consistent theory of higher-spin fields. If this is true, other possible choices of the cubic couplings would pertain to other consistent theories, which from the standpoint of [25] would amount to using another functional of their differential operator and not the exponential.

Once again, we see that different approaches to the same problem have different advantages. For example, the work of Sagnotti and Taronna is impressive in that a generating functional is given from which one can easily extract all the $s-s'-s''$ couplings, both bosonic and fermionic, whereas our methods did not allow for such a general treatment so far. On the other hand, when one reads off the couplings extracted from the generating functional given in [25] in the most naive way, they are seen to contain many terms, and a 'geometric' repackaging thereof is far from simple. Moreover, features such as the vanishing of some interactions in $D=4$ are hardly visible in that approach. With our methods, for the cases dealt with, we have obtained simple and rather appealing forms for all our couplings, and all their properties are made manifest. Finally, let us highlight other references where the relationship between String Theory and flat-space higher spins is investigated, such as [195–198].

**Second-Order Consistency**

Our non-abelian vertices have been found to be inconsistent beyond the cubic order. The precise meaning of this statement is that there is no quartic term, built in terms of our original fields, that one can add to the theory so to make it fully consistent. However, as is well known, an enlargement of the spectrum sometimes cures this obstruction. A rather famous example is that of the Pauli term in $\mathcal{N}=2$ Supergravity [199, 200],[4] which is precisely our non-abelian $1-\frac{3}{2}-\frac{3}{2}$ vertex. Needless to say, $\mathcal{N}=2$ Supergravity is a fully consistent theory, and one thus wonders how to reconcile that fact with the quartic obstruction we have unveiled here. The answer, of course, lies in the presence of the graviton[5] in $\mathcal{N}=2$ Supergravity, and indeed one can observe that the gauge invariance of the full Lagrangian thereof involves cancellations between the (first-order) deformed gauge variation of the Pauli term and the (zeroth-order) gauge variation of quartic terms which involve the graviton.

Let us point out that the chosen example hereabove is rather peculiar, and that a quartic completion (by an enlargement of the spectrum) is possible in that case because the spin-$\frac{3}{2}$ field is somewhat in between pertaining to standard Field Theory and Higher-Spin Theory. In general, however, in the field of higher-spin couplings, the typical situation is that no such cure is available, and sometimes even the inclusion of an infinite number of additional degrees of freedom is vain [201]. In that case, a fully consistent theory including those cubic vertices is necessarily non-local. Indeed, the underlying but crucial assumption in all our studies is that of locality. An interesting perspective on the matter, again based on the example of $\mathcal{N} = 2$ Supergravity, is to try to recover from the latter a consistent theory of the Pauli term without involving the graviton. This is impossible if one insists on locality: if one decouples Gravity therefrom by taking $M_P \to \infty$, the Pauli term therein vanishes because the dimensionful coupling constant is nothing but $1/M_P$, the inverse of the Planck mass. Alternatively, one could integrate out the massless graviton to obtain a system of spin-$\frac{3}{2}$ and spin-1 fields only: this time the resulting theory surely contains the Pauli term, but it is necessarily non-local, because one has integrated out massless degrees of freedom.

---

[4] Recall that $\mathcal{N}=2$ SUGRA allows massless gravitini to have dipole and higher-derivative couplings, but forbids a non-zero U(1) charge in flat space.

[5] Recall the spectrum of $\mathcal{N}=2$ Supergravity is that of a photon, a graviton, and a complex Rarita–Schwinger tensor-spinor.

In fact, one can even show that, if locality is not insisted on, then *any* consitent cubic vertex admits a quartic completion to a fully consistent theory [176]. For higher spins in flat space of dimension four or greater, it might thus be that non-locality is a crucial ingredient, and in fact one can see signs of it already at the free level, when formulated in its 'geometric' form [202–206]. Moreover, other investigations further confirm that non-locality necessarily creeps in beyond the cubic order in (the tensionless limit of) String Theory [171]. If one has to give up locality, one way of continuing the study of higher-spin interactions is a formulation that does not require locality as an input. In that line of thought, it seems as if modern approaches that do not assume a Lagrangian formulation may have lessons in store concerning the systematic search of consistent interactions of massless higher-spin particles in Minkowski space of dimension four [207–209]. On the other hand, non-local Lagrangians are still a largely unexplored topic in field theory. The reason behind the usual lack of interest in the field is the standard lore according to which a non-local Lagrangian would yield non-unitary or causality-violating amplitudes. However, as pointed out in [171], this is not necessarily true in all cases and, in particular, having an infinite number of degrees of freedom to accommodate for might change that paradigm. It is the author's belief that Higher-Spin Theory, and more generally Field Theory may benefit from a more systematic study of non-local Lagrangians. Moreover, let us point out that quartic couplings and their consistency has also been studied, for example in [171, 210, 211].

**Bosonic Counterparts**

Let us briefly comment on the comparison between our results and their bosonic analogues, obtained in [24, 184] via the same cohomological methods[6] and which also obey the corresponding restrictions on the number of derivatives [62, 215]. For arbitrary spin, we should compare a fermion of spin $s = n + \frac{1}{2}$ with a boson of spin $s = n$, for those would have the same number of (symmetric) spacetime indices. The number-of-derivatives bounds on the electromagnetic couplings of a spin-$s$ boson are $2s - 1$ and $2s + 1$, which is a different structure from that which is found in the fermionic case. However, in the case of bosons, the structure changes as one moves to the gravitational interactions, for which we have the possibilities $2s - 2$, $2s$ and $2s + 2$. Interestingly, our gravitational deformations for fermions are also bounded by $2n - 2$ and $2n + 2$, but the jump in $p$ is by one unit and not two, because of the freedom to use $\gamma$-matrices.

---

[6] Note that these bosonic flat-space couplings, originally obtained in [39, 212] in the Light-Cone gauge, were also found via the Noether procedure in [213, 214].

Also, the nature of bosonic vertices is very similar to that of fermionic ones: the $(2s+2)$-derivatives one is of the Born–Infeld type and thus exists also in $D = 4$: the $p = 2s - 2$ is the only non-abelian one and survives in dimension 4 as well, whereas the one with $2s$ derivatives is zero in four dimensions and is abelian but not strictly gauge invariant. In dimension four we thus have the same number of vertices for a boson and a fermion; one of them is abelian and the other is simply the product of curvatures. In $D \geq 5$ the fermion exceeds the boson by two vertices: an abelian one and a non-abelian one.

In a similar fashion, our electromagnetic deformations resemble their bosonic counterparts: the number of derivatives they may contain are bounded from above and below by the same numbers, $2n-1$ and $2n+1$ (or $2s - 1$ and $2s + 1$), but the fermion has an additional $p = 2n$ deformation, which requires making use of at least one $\gamma$-matrix.

**Anti-de Sitter Setups**

The relation between flat-space vertices and those existing in AdS is rather insightful. In anti-de Sitter spacetimes, for the bosonic couplings of the graviton, the Fradkin–Vasiliev construction [19, 20] yields a cubic vertex with derivatives ranging from 1 to $2s - 2$. In Minkowski spacetimes, as we have just seen, $2s - 2$ is the minimum number of derivatives a gravitational interaction with a spin-$s$ field can contain, whereas minimal coupling never exists. In [184], it was proved that a suitable flat limit can be taken where the cosmological constant $\Lambda \to 0$ and where only the piece with $2s - 2$ derivatives survives, thus yielding the non-abelian coupling of the spin-$s$ boson in flat space.

The 'fermionic analogue' of the work [184] does not exist yet, but it is expected that the behavior of fermions is alike to that of bosons. In fact, the study of gravitational interaction vertices of a massive spin-$\frac{5}{2}$ field in AdS was actually carried out in [216], where it was noticed that what survives in the massless flat limit is only a 2-derivatives vertex when $D = 4$, or a 3-derivatives one when $D > 4$. These *must* precisely be our flat-space highest-number-of-derivatives non-abelian cubic vertices in the respective dimensions.

It would be interesting to extend our systematic analysis to (A)dS spaces. There are certain technical difficulties, though, in extending the applicability of the BRST deformation scheme to spaces of constant curvature. One may use the ambient-space formulation [217–219] for AdS space, in particular, to avoid these issues. Then one could construct covariant higher-spin vertices in AdS, and the results could be compared with those obtained recently e.g. in Refs. [21, 220, 221] for symmetric fields. This would help us understand better the rather intricate structure of the Vasiliev higher-spin systems [50, 104, 105], possibly by leading us a step closer to a yet-to-be-found standard action.

**Massive Fields**

What connection may our vertices have with the massive theories? For a massive spin-$\frac{5}{2}$ field, coupled to gravity in flat space, it was noticed in [222] that suitable non-minimal couplings improve the high-energy behavior of the theory by pushing higher the scale at which tree-level unitarity is violated. The simplest of these terms has two derivatives, and in dimension four it reads $\bar{\psi}_{\mu\alpha} R^{+\mu\nu\alpha\beta} \psi_{\nu\beta}$ up to on-shell terms. Interestingly, this is nothing but the first piece in our 2-derivatives vertex (6.32) — the part surviving in the transverse-traceless gauge. This may not come as a surprise: after all, consistent massive theories are expected to originate from massless ones. In fact, a similar thing happens for the spin-$\frac{3}{2}$ electromagnetic coupling: the gauge-invariant Pauli term improves the tree-level unitarity of the aforementioned massive theory [223] and shows up in the consistent $\mathcal{N} = 2$ broken Supergravity theory [224–227].

Let us also discuss the relation with electromagnetic, massive higher-spin couplings in flat space, which have been studied a little more, for example in [228–239] and in references therein. In a nutshell, if Lorentz, parity and time-reversal symmetries hold good, a massive spin-$s$ particle will have $2s+1$ electromagnetic couplings [239], and this immediately sets for the possible number of derivatives in a $1-s-s$ vertex an upper bound, which remains the same in the massless limit. Then, the assumption of Light-Cone helicity conservation in $D = 4$ uniquely determines all the couplings. However, only the highest-$p$ vertex survives in an appropriate massless, chargeless scaling limit. This observation is in harmony with our results, since all of our low-$p$ vertices either vanish in $D = 4$ or are not consistent by themselves in a local theory (because of their non-abelian nature).

**Outro: Back to Dimension 3**

As a final word, we comment on the following, legitimate question: in Part I, three-dimensional higher-spin theories are built which interact, and we have pointed out that they reproduce the minimal coupling of Gravity with higher-spin fields. Such minimal coupling, as we have demonstrated in the present part, does not exist in dimension four or greater. However, as can be checked, in dimension three the proof given in Subsection 6.1.1 that minimal coupling is absent is no longer valid, because of the many dimension-dependent identities (e.g. among $\gamma$-matrices) which hold good in dimension three.

# Appendices

# Notation Conventions

We summarize our conventions and notations so that they are easily found.

## A.1 Dimension 3

We denote AdS$_3$ radius as $\ell$. We adopt the global coordinates of AdS$_3$:

$$(x) = (x^0, x^1, x^2) = (t, \ell\theta, r), \tag{A.1}$$

n the former system, the AdS$_3$ metric reads

$$ds^2 = -\left(1 + \left(\frac{x^2}{\ell}\right)^2\right)(dx^0)^2 + \left(1 + \left(\frac{x^2}{\ell}\right)^2\right)^{-1}(dx^2)^2 + \left(\frac{x^2}{\ell}\right)^2 (dx^1)^2. \tag{A.2}$$

To leading order at infinity, the '1' is negligible and one can replace asymptotically the metric by that of the zero mass black hole [240, 241],

$$ds^2 = -\left(\frac{x^2}{\ell}\right)^2 (dx^0)^2 + \left(\frac{x^2}{\ell}\right)^{-2} (dx^2)^2 + \left(\frac{x^2}{\ell}\right)^2 (dx^1)^2. \tag{A.3}$$

The Light-Cone coordinates are defined by

$$(x) = (x^\pm, x^2) = (t \pm \ell\theta, r). \tag{A.4}$$

Unless otherwise specified we always work with AdS radius $\ell = 1$. We also recall our notation $\partial$ means $\partial/\partial x^+$ and $\cdot'$ means the derivative with respect to the argument, whereas 'd' is the exterior spacetime derivative. Spinor indices are raised and lowered with a spinor metric

$$(\epsilon^{\alpha\beta}) \equiv (\epsilon_{\alpha\beta}) \equiv \begin{pmatrix} 0 & 1 \\ -1 & 0 \end{pmatrix}, \qquad \alpha, \beta \in \{1,2\}, \tag{A.5}$$

and we use so-called 'North-West/South-East' conventions such that $q_\alpha = q^\beta \epsilon_{\beta\alpha}$. Our spinor indices are denoted by Greek letters from the beginning of the Greek alphabet and take values in $\{1,2\}$, whereas spacetime indices are denoted by Greek letters such as $\mu, \nu,$ etc. and take values in $\{0,1,2\}$.

The conventions and notations having to do with the various algebras which are dealt with in the main text are found in Appendix B below.

## A.2 Dimension 4 and Higher

### Clifford Algebra and Symmetrization

The flat metric $\eta_{\mu\nu}$ on the Minkowski spacetime of dimension $D$ is taken to be of 'mostly-positive' signature and our spacetime indices run from 0 to $D-1$. The Clifford algebra is

$$\{\gamma^\mu, \gamma^\nu\} = +2\eta^{\mu\nu}, \tag{A.6}$$

where $\{\cdot,\cdot\}$ is the anticommutator and $\gamma^\mu$ generically denotes the $\gamma$-matrices of Dirac, whose hermitian conjugate is taken to be $\gamma^{\mu\,\dagger} \equiv \eta^{\mu\mu}\gamma^\mu$ (without summation on the repeated indices). Furthermore, the Dirac adjoint is defined as $\bar\psi_\mu = \psi_\mu^\dagger \gamma^0$. The $D$-dimensional Levi-Civita tensor, $\epsilon_{\mu_1\mu_2...\mu_D}$, is normalized as $\epsilon_{01...(D-1)} = +1$. We define $\gamma^{\mu_1....\mu_n} \equiv \gamma^{[\mu_1}\gamma^{\mu_2}...\gamma^{\mu_n]}$, where the notation $[i_1...i_n]$ means totally antisymmetric expression in all the indices $i_1,...,i_n$ with the normalization factor $\frac{1}{n!}$, and the totally symmetric expression $(i_1...i_n)$ has the same normalization. In particular,

$$\gamma^\mu\gamma^\nu = \gamma^{\mu\nu} + \eta^{\mu\nu}. \tag{A.7}$$

We also use the slash notation $\gamma^\mu Q_\mu \equiv \slashed{Q}$, which applied to a partial derivative gives the Dirac operator $\slashed{\partial}$, the square of which is the Klein–Gordon operator $\Box \equiv \partial^\alpha \partial_\alpha$. For antisymmetric tensors of rank 2, $Q_{\mu\nu}$, we also sometimes use the double-slash notation, $\gamma^{\mu\nu}Q_{\mu\nu} \equiv \slashed{\slashed{Q}}$. Another operation on such tensors is that which yields the following expression:

$$T^+_{\mu\nu} \equiv T_{\mu\nu} + \tfrac{1}{2}\gamma_{\mu\nu\alpha\beta}T^{\alpha\beta}, \tag{A.8}$$

*APPENDIX A. NOTATION CONVENTIONS* 185which we often use, together with the double-slash, on the linearized Riemann tensor as well as on the Faraday field strength.[1] When a rank-2 tensor is symmetric the double-slash is in fact the trace, which we simply denote with the original letter but foregoing the indices, or alternatively by a prime, that is $\eta^{\mu\nu} S_{\mu\nu} \equiv S \equiv S'$, and for higher-rank tensors multiple primes if multiple traces are taken. The 'anticommutator' of two antisymmetric products of $\gamma$-matrices is denoted as $\gamma^{\mu_1\ldots\mu_m,\,\nu_1\ldots\nu_n} \equiv \frac{1}{2}\{\gamma^{\mu_1\ldots\mu_m}, \gamma^{\nu_1\ldots\nu_n}\}$, and we also use the following symbol:

$$\eta^{\mu\nu|\rho\sigma} \equiv \eta^{[\mu\nu]|[\rho\sigma]} = \eta^{\rho\sigma|\mu\nu} \equiv \tfrac{1}{2}\left(\eta^{\mu\rho}\eta^{\nu\sigma} - \eta^{\mu\sigma}\eta^{\nu\rho}\right). \tag{A.9}$$

**Curvatures and Curls**

For any totally symmetric tensor (or tensor-spinor) of rank $n$, $T_{\mu_1\ldots\mu_n}$, its *curvature* is defined as the rank-$2n$ tensor of mixed symmetry type obtained by taking $n$ successive antisymmetrized gradients of $T$ without normalization factor, i.e.

$$T_{\mu_1\nu_1|\ldots|\mu_n\nu_n} \equiv [\ldots [\, [\partial_{\mu_1}\ldots\partial_{\mu_n} T_{\nu_1\ldots\nu_n} - (\mu_1 \leftrightarrow \nu_1)] - (\mu_2 \leftrightarrow \nu_2)]\ldots]$$
$$- (\mu_n \leftrightarrow \nu_n). \tag{A.10}$$

Similarly, we also define the $m$-th *curl* of $T$ as the rank-$(m+n)$ tensor of mixed symmetry type obtained by taking only $m$ successive antisymmetrized derivatives of $T$, that is,

$$T^{(m)}_{\mu_1\nu_1|\ldots|\mu_m\nu_m\|\nu_{m+1}\ldots\nu_n} \equiv [\ldots [\, [\partial_{\mu_1}\ldots\partial_{\mu_m} T_{\nu_1\ldots\nu_n} - (\mu_1 \leftrightarrow \nu_1)] - (\mu_2 \leftrightarrow \nu_2)]]$$
$$- (\mu_m \leftrightarrow \nu_m). \tag{A.11}$$

Evidently, for a rank-$n$ tensor the curvature is the $n$-curl, while the zeroth curl is the original tensor itself. For the fields that are most used the notation shall sometimes be simplified by not displaying the order of the curl explicitly, and instead changing font (see below). We further note that any curl (including the curvature) is antisymmetric under the exchange of two paired indices (such as $\mu_1$ and $\nu_1$), and symmetric under the exchange of two pairs — the inexperienced reader shall have in mind some generalization of the symmetries of the Riemann tensor. Let us also point out the two following Bianchi identities:

$$\partial_{[\rho} T^{(m)}_{\mu_1\nu_1]|\ldots|\mu_m\nu_m\|\nu_{m+1}\ldots\nu_n} = 0, \tag{A.12a}$$

$$T^{(m)}_{\mu_1\nu_1|\ldots|[\mu_m\nu_m\|\nu_{m+1}]\nu_{m+2}\ldots\nu_n} = 0, \tag{A.12b}$$

---

[1] The Riemann tensor is not of rank 2, and has two such blocks of indices, but it is symmetric under the exchange of its two blocks, so there is no notation ambiguity.

which holds for any $m = 1, \ldots, n$. Another relation is the simple identity

$$\partial_{[\rho} T^{(m)}_{\mu_1]\nu_1|\ldots|\mu_m\nu_m\|\nu_{m+1}\ldots\nu_n} = -\tfrac{1}{2}\partial_{\nu_1} T^{(m)}_{\rho\mu_1|\ldots|\mu_m\nu_m\|\nu_{m+1}\ldots\nu_n}, \qquad (A.13)$$

which is ubiquitous in our computations.

The curvature for the spin-1 field is its 1-curl,

$$A_{\mu\nu} \equiv \partial_\mu A_\nu - \partial_\nu A_\mu \equiv F_{\mu\nu}, \qquad (A.14)$$

which is just the electromagnetic field strength. For the spin-2 field, which has one more spacetime index, the curvature is given by the 2-curl,

$$h_{\mu\nu|}{}^{\rho\sigma} \equiv 4\partial_{[\mu}\partial^{[\rho} h_{\nu]}{}^{\sigma]} \equiv R_{\mu\nu}{}^{\rho\sigma}, \qquad (A.15)$$

which we know as the linearized Riemann tensor. Because of the extra index, there is an 'intermediate' curl for the graviton field, namely its 1-curl $h^{(1)}_{\mu\nu\|\rho} \equiv 2\partial_{[\mu} h_{\nu]\rho} \equiv \mathfrak{h}_{\mu\nu\|\rho}$.

As for the fermion fields, the curvatures are denoted with the letter $\Psi$. Given that only the spacetime indices are concerned by such operations as those that we have just recalled, the structure for the spin-$\frac{3}{2}$ (resp. spin-$\frac{5}{2}$) field is the same as that of the photon (resp. the graviton). In general, for arbitrary spin $s \equiv n + \frac{1}{2}$, the so-called Weinberg curvature [242–244] for the tensor-spinors are thus given by

$$\Psi_{\mu_1\nu_1|\ldots|\mu_n\nu_n} \equiv [\ldots [[\partial_{\mu_1}\ldots\partial_{\mu_n}\psi_{\nu_1\ldots\nu_n} - (\mu_1 \leftrightarrow \nu_1)] - (\mu_2 \leftrightarrow \nu_2)]\ldots]$$
$$- (\mu_n \leftrightarrow \nu_n), \qquad (A.16)$$

and when $n \geq 2$ one can again consider intermediate $m$-curls $\psi^{(m)}_{\mu_1\nu_1|\ldots|\mu_m\nu_m\|\nu_{m+1}\ldots\nu_n} \equiv \Psi_{\mu_1\nu_1|\ldots|\mu_m\nu_m\|\nu_{m+1}\ldots\nu_n}$.

The relation between the equations of motion and the various curvatures and curls for the fermion fields as well as for the spin-1 and spin-2 tensors are discussed in Appendix E, where the gauge invariance of the quantities defined hereabove is also discussed. Here we simply point out that the fermionic curvatures (but not the other curls) are gauge invariant under the unconstrained gauge parameters.[2] However, other quantities built out of the original tensor-spinors are gauge invariant under the constrained ($\gamma$-traceless) gauge parameter only, and those are all functions of the *Fronsdal tensor* [34]

$$\mathcal{S}_{\mu_1\ldots\mu_n} \equiv i\bigl(\slashed{\partial}\psi_{\mu_1\ldots\mu_n} - n\partial_{(\mu_1}\slashed{\psi}_{\mu_2\ldots\mu_n)}\bigr). \tag{A.17}$$

---

[2] This is true for the bosonic ones as well, but the (trace) constraints on the bosonic gauge parameters appear when considering the spin-3 field only, simply because the graviton gauge parameter has only one index.

In particular, there is one combination of $\mathcal{S}$ which gives the original equations of motion (those derived from varying the Lagrangian):

$$\mathcal{R}_{\mu_1\ldots\mu_n} \equiv \mathcal{S}_{\mu_1\ldots\mu_n} - \tfrac{1}{2}n\,\gamma_{(\mu_1}\,\slashed{\mathcal{S}}_{\mu_2\ldots\mu_n)} - \tfrac{1}{4}n(n-1)\,\eta_{(\mu_1\mu_2}\mathcal{S}'_{\mu_3\ldots\mu_n)}. \tag{A.18}$$

More details can be found in Appendix E.

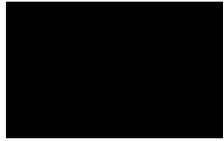

APPENDIX B

# Algebras in Dimension 3

Let us recall the conventions and the properties pertaining to the various Lie algebras we make use of in this work. In section B.1 we start by recalling the properties of sl(2|$\mathbb{R}$), which is (half of) the algebra of pure AdS$_3$ Gravity. Then, in section B.2 we move on to recapitulating the relations defining the osp($N, 2|\mathbb{R}$) superalgebra, which underlies the type of Supergravity which we focus on in the text, and end the 'low-spin' reminders with the list of all extended AdS$_3$ Supergravity superalgebras, found in Subsection B.2.2. Finally, Section B.3 first details the structure of the non-extended shs(1, 1) higher-spin superalgebra, which characterizes the higher-spin theory which we explicitly deal with in the main text, and then ends this appendix with quite an in-depth treatment of its extended version shs($N, 2|\mathbb{R}$), which is only briefly touched upon in the bulk of this thesis.

## B.1 Pure Gravity

Starting from the basic isomorphism so(2, 2) $\simeq$ sl(2|$\mathbb{R}$)$\oplus$sl(2|$\mathbb{R}$) we give hereafter different sets of generators for sl(2|$\mathbb{R}$) $\simeq$ sp(2|$\mathbb{R}$) $\simeq$ so(1, 2) $\simeq$ su(1, 1).

The real algebra sl(2|$\mathbb{R}$) can be realized as the real vector space of $2 \times 2$ traceless real matrices equipped with the usual Lie bracket

$$[M, M'] \equiv MM' - M'M, \tag{B.1}$$

where the multiplication is the matrix multiplication. Such matrices have the form

$$\begin{pmatrix} a & b \\ c & -a \end{pmatrix} \tag{B.2}$$

with $a$, $b$, $c$ real. For the standard (Chevalley–Serre) generators

$$H \equiv \begin{pmatrix} 1 & 0 \\ 0 & -1 \end{pmatrix}, \qquad E \equiv \begin{pmatrix} 0 & 1 \\ 0 & 0 \end{pmatrix}, \qquad F \equiv \begin{pmatrix} 0 & 0 \\ 1 & 0 \end{pmatrix}, \tag{B.3}$$

we find the commutators

$$[H, E] = 2E, \qquad [H, F] = -2F, \qquad [E, F] = H, \tag{B.4}$$

with the usual matrix trace defining a scalar product

$$(M, M') \equiv \operatorname{Tr}(MM'), \tag{B.5}$$

which yields the non-zero projections

$$(H, H) = 2, \qquad (E, F) = (F, E) = 1. \tag{B.6}$$

Another quite useful basis is obtained by performing the redefinitions

$$E \equiv T_1 + T_2, \qquad F \equiv T_1 - T_2, \qquad H \equiv 2T_3, \tag{B.7}$$

which yield the relations

$$[T_a, T_b] = \epsilon_{abc} T^c, \tag{B.8}$$

where it should be noted that Latin indices are raised and lowered with the Minkowski metric $\eta_{ab}$ and its inverse, with $a, b, c = 1, 2, 3$. The trace defined above now reads

$$(J_a, J_b) = \frac{1}{2} \eta_{ab}. \tag{B.9}$$

Finally, we also sometimes use a slightly different basis than $\{E, F, H\}$, noted

$$\sigma^+ = T_1 + T_2, \quad \sigma^- = T_1 - T_2, \quad \sigma^3 = T_3, \tag{B.10}$$

which implies

$$[\sigma^3, \sigma^\pm] = \pm\sigma^\pm, \quad [\sigma^+, \sigma^-] = 2\sigma^3, \quad (\sigma^3, \sigma^3) = \frac{1}{2}, \quad (\sigma^+, \sigma^-) = 1. \tag{B.11}$$

Moreover, a matrix representation of the latter generators is given by

$$\sigma^3 \equiv \begin{pmatrix} \frac{1}{2} & 0 \\ 0 & -\frac{1}{2} \end{pmatrix}, \qquad \sigma^+ \equiv \begin{pmatrix} 0 & 1 \\ 0 & 0 \end{pmatrix}, \qquad \sigma^- \equiv \begin{pmatrix} 0 & 0 \\ 1 & 0 \end{pmatrix}, \tag{B.12}$$

which we also make use of.

## B.2 Supergravity

### B.2.1 Orthosymplectic Supergravity

**The Non-Extended Case**

The orthosymplectic $\mathrm{osp}(1,2|\mathbb{R})$ superalgebra can be realized in the following way: the real vector space of even (grading-preserving) $3 \times 3$ supermatrices acting on 1 commuting real Grassmann-even variable $x$ and 2 anticommuting real Grassmann-odd variables $\theta^1$ and $\theta^2$ and which preserve the quadratic form

$$x^2 + 2i\theta^1\theta^2 = x^2 + i\epsilon_{\alpha\beta}\theta^\alpha\theta^\beta \tag{B.13}$$

as well as the real character of the coordinates — equipped with the usual Lie bracket

$$[\Gamma, \Gamma'] \equiv \Gamma\Gamma' - \Gamma'\Gamma, \tag{B.14}$$

where the multiplication is the matrix multiplication. Such supermatrices have the form

$$\begin{pmatrix} 0 & i\mu & -i\lambda \\ \lambda & a & b \\ \mu & c & -a \end{pmatrix}, \tag{B.15}$$

with $a$, $b$, $c$ real and commuting, while $\lambda$ and $\mu$ are real and anticommuting. We identify the generators:

$$H \equiv \begin{pmatrix} 0 & 0 & 0 \\ 0 & 1 & 0 \\ 0 & 0 & -1 \end{pmatrix}, \qquad E \equiv \begin{pmatrix} 0 & 0 & 0 \\ 0 & 0 & 1 \\ 0 & 0 & 0 \end{pmatrix}, \qquad F \equiv \begin{pmatrix} 0 & 0 & 0 \\ 0 & 0 & 0 \\ 0 & 1 & 0 \end{pmatrix},$$

$$R^- \equiv \begin{pmatrix} 0 & i & 0 \\ 0 & 0 & 0 \\ 1 & 0 & 0 \end{pmatrix}, \qquad R^+ \equiv \begin{pmatrix} 0 & 0 & -i \\ 1 & 0 & 0 \\ 0 & 0 & 0 \end{pmatrix}, \tag{B.16}$$

in terms of which we find the supercommutators

$$\begin{aligned}
[H, E] &= 2E, & [H, F] &= -2F, & [E, F] &= H, \\
[H, R^+] &= R^+, & [E, R^+] &= 0, & [F, R^+] &= R^-, \\
[H, R^-] &= -R^-, & [E, R^-] &= R^+, & [F, R^-] &= 0, \\
\{R^+, R^+\} &= -2iE, & \{R^-, R^-\} &= 2iF, & \{R^+, R^-\} &= iH,
\end{aligned} \tag{B.17}$$

where the anticommutator is defined as always by $\{\Gamma, \Gamma'\} \equiv \Gamma\Gamma' + \Gamma'\Gamma$.

The supertrace and scalar product are defined as

$$\text{STr}(\Gamma) \equiv \Gamma_{11} - \text{Tr}\left(\Gamma_{\text{sp}(2)}\right) = \Gamma_{11} - \Gamma_{22} - \Gamma_{33} = -\Gamma_{22} - \Gamma_{33}, \quad \text{(B.18a)}$$
$$(\Gamma, \Gamma') \equiv \text{STr}(\Gamma\Gamma'), \quad \text{(B.18b)}$$

where $\Gamma_{\text{sp}(2)}$ is the submatrix generated by $E$, $F$ and $H$ ('spacetime' algebra), and there is no internal algebra because $N = 1$. In our representation, the fermionic sector is thus encoded in the $\Gamma_{1a}$ and $\Gamma_{a1}$ components of the matrices and the sp(2) subalgebra of osp(1, 2|$\mathbb{R}$) thus lies in the $\Gamma_{ab}$ components, with $a, b = 2, 3$.

**The Extended Case**

The orthosymplectic osp($N$, 2|$\mathbb{R}$) superalgebra can be realized as the real vector space of even (grading-preserving) $(N+2) \times (N+2)$ supermatrices acting on $N$ commuting real Grassmann-even variables $x^i$ and 2 anticommuting real Grassmann-odd variables $\theta^1$ and $\theta^2$ and which preserve the quadratic form

$$\sum_{i=1}^{N}(x^i)^2 + 2i\theta^1\theta^2 = \delta_{ij}x^i x^j + i\epsilon_{\alpha\beta}\theta^\alpha\theta^\beta \quad \text{(B.19)}$$

as well as the real character of the coordinates — equipped with the usual Lie bracket

$$[\Gamma, \Gamma'] \equiv \Gamma\Gamma' - \Gamma'\Gamma, \quad \text{(B.20)}$$

where the multiplication is the matrix multiplication. Such supermatrices have the form

$$\begin{pmatrix} & & i\mu_1 & -i\lambda_1 \\ O_{ij} & & \vdots & \vdots \\ & & i\mu_N & -i\lambda_N \\ \lambda_1 & \cdots & \lambda_N & a & b \\ \mu_1 & \cdots & \mu_N & c & -a \end{pmatrix}, \quad \text{(B.21)}$$

with $O_{ij} = -O_{ji}$, $a$, $b$, $c$ real and commuting, and $\lambda_i$, $\mu_i$ real and anticommuting. We identify the generators:

$$H \equiv \begin{pmatrix} & & & 0 & 0 \\ & 0 & & \vdots & \vdots \\ & & & 0 & 0 \\ 0 & \cdots & 0 & 1 & 0 \\ 0 & \cdots & 0 & 0 & \bar{1} \end{pmatrix}, \quad E \equiv \begin{pmatrix} & & & 0 & 0 \\ & 0 & & \vdots & \vdots \\ & & & 0 & 0 \\ 0 & \cdots & 0 & 0 & 1 \\ 0 & \cdots & 0 & 0 & 0 \end{pmatrix}, \quad F \equiv \begin{pmatrix} & & & 0 & 0 \\ & 0 & & \vdots & \vdots \\ & & & 0 & 0 \\ 0 & \cdots & 0 & 0 & 0 \\ 0 & \cdots & 0 & 1 & 0 \end{pmatrix},$$

$$J_{ij} \equiv \begin{pmatrix} 0 & & 1 & 0 & 0 \\ & \ddots & & \vdots & \vdots \\ \bar{1} & & 0 & 0 & 0 \\ 0 & \cdots & 0 & 0 & 0 \\ 0 & \cdots & 0 & 0 & 0 \end{pmatrix}, \quad (B.22)$$

$$R_i^- \equiv \begin{pmatrix} & & & & 0 & 0 \\ & & & & \vdots & \vdots \\ & 0 & & 0 & \bar{i} \\ & & & & \vdots & \vdots \\ & & & & 0 & 0 \\ 0 & \cdots & 1 & \cdots & 0 & 0 & 0 \\ 0 & \cdots & 0 & \cdots & 0 & 0 & 0 \end{pmatrix}, \quad R_i^+ \equiv \begin{pmatrix} & & & & 0 & 0 \\ & & & & \vdots & \vdots \\ & 0 & & i & 0 \\ & & & & \vdots & \vdots \\ & & & & 0 & 0 \\ 0 & \cdots & 0 & \cdots & 0 & 0 & 0 \\ 0 & \cdots & 1 & \cdots & 0 & 0 & 0 \end{pmatrix},$$

where in $R_i^+$ and $R_i^-$ (odd generators) the $i$ factors sit in the $i$-th line and the 1 factors in the $i$-th column, and in $J_{ij}$ the 1 (resp. $-1$) factors sit in the position $(i,j)$ (resp. $(j,i)$). We find the supercommutators

$$\begin{aligned} &[H, E] = 2E, & &[H, F] = -2F, & &[E, F] = H, \\ &[H, R_i^+] = R_i^+, & &[E, R_i^+] = 0, & &[F, R_i^+] = R_i^-, \\ &[H, R_i^-] = -R_i^-, & &[E, R_i^-] = R_i^+, & &[F, R_i^-] = 0, \\ &i\{R_i^+, R_j^+\} = 2\delta_{ij} E, & &i\{R_i^-, R_j^-\} = -2\delta_{ij} F, & &i\{R_i^+, R_j^-\} = J_{ij} - \delta_{ij} H, \\ &[J_{ij}, E] = 0, & &[J_{ij}, F] = 0, & &[J_{ij}, H] = 0, \quad (B.23) \\ &[J_{ij}, R_k^+] = \delta_{jk} R_i^+ - \delta_{ik} R_j^+, & &[J_{ij}, R_k^-] = \delta_{jk} R_i^- - \delta_{ik} R_j^-, \\ &[J_{ij}, J_{kl}] = \delta_{jk} J_{il} + \delta_{il} J_{jk} - \delta_{ik} J_{jl} - \delta_{jl} J_{ik}, \end{aligned}$$

with the supercommutator again defined in the usual way (see non-extended case).

The supertrace and scalar product are defined as

$$\operatorname{STr}(\Gamma) \equiv \operatorname{Tr}(\Gamma_{\operatorname{so}(N)}) - \operatorname{Tr}(\Gamma_{\operatorname{sp}(2)}), \tag{B.24a}$$

$$(\Gamma, \Gamma') \equiv \operatorname{STr}(\Gamma\Gamma'), \tag{B.24b}$$

where $\Gamma_{\operatorname{so}(N)}$ is the submatrix of $\Gamma$ generated by the $J_{ij}$ basis elements (internal algebra) and $\Gamma_{\operatorname{sp}(2)}$ is the submatrix generated by $E$, $F$ and $H$ ('spacetime' algebra). In our representation, the fermionic sector is thus encoded in the $\Gamma_{ia}$ and $\Gamma_{ai}$ components of the matrices, the sp(2) subalgebra of osp($N, 2|\mathbb{R}$) thus lies in the $\Gamma_{ab}$ components while the internal so($N$) algebra is formed by the $\Gamma_{ij}$ components, with $i, j = 1, \ldots, N$ and $a, b = N+1, N+2$.

### B.2.2 Other Extended Supergravities

A few, very basic requirements very much constrain the gauge superalgebras one can use to build extended supergravities on AdS$_3$ — recall that the latter can be reformulated in terms of a Chern–Simons theory with a connection one-form valued in some gauge superalgebra [58], so that all the local (bulk) information about the theory is really encoded in that algebra — and in [151, 245–247] it was proved that only seven (families of) candidates satisfy those requirements. We give them in Table B.2.2 below, where $\mathcal{A}$ is the (extended) superalgebra, $\mathcal{G}$ is the corresponding internal subalgebra, $\rho$ is the representation of $\mathcal{G}$ (denoted in boldface by its dimension) in which the spinors transform and $D$ is the dimension of $\mathcal{G}$. Note that, evidently, one is to consider two chiral copies of each of the candidates below to construct the corresponding supergravity gauge superalgebra. The first four superalgebras belong to the osp($m, 2n$) and spl($m, n$) infinite families while the last three are 'exceptional' Lie superalgebras.

Table B.1: Superalgebras of Extended AdS Supegravities in Dimension 3.

| $\mathcal{A}$ | $\mathcal{G}$ | $\rho$ | $D$ |
|---|---|---|---|
| osp($N|2, \mathbb{R}$) | so($N$) | $\mathbf{N}$ | $N(N-1)/2$ |
| su($1,1|N$), $N > 2$ | su($N$) $\oplus$ u(1) | $\mathbf{N} + \overline{\mathbf{N}}$ | $N^2$ |
| su($1,1|2$) / u(1) | su(2) | $\mathbf{2} + \overline{\mathbf{2}}$ | 3 |
| osp($4^*|2N$) | su(2) $\oplus$ usp($2N$) | $(\mathbf{2}, \mathbf{2N})$ | $N(2N+1)+3$ |
| D$^1(2,1|\alpha)$ | su(2) $\oplus$ su(2) | $(\mathbf{2}, \mathbf{2})$ | 6 |
| G(3) | G$_2$ | $\mathbf{7}$ | 14 |
| F(4) | spin(7) | $\mathbf{8}_s$ | 21 |

In fact, each one of these (families of) superalgebras can be associated with a (family of) superconformal algebra(s) in dimension 2 with quadratic non-linearities, which is not an accident at all, as was unveiled in [60], where the said superconformal algebras were found to be precisely the asymptotic symmetry algebras of supergravity models based on the above superalgebras.

## B.3 Higher-Spin Algebras

Let us detail various properties of some higher-spin infinite-dimensional algebras in dimension 3, which we present in their oscillator realization. For the sake of simplicity, and also because such is not the focus of the present work, we shall keep the oscillators undeformed, that is, we shall proceed with $\lambda = \frac{1}{2}$, and we refer to [119, 120] for the treatment of the deformed cases. The bosonic higher-spin algebra $hs(1,1)$ has been somewhat introduced in Subsection 1.2.2 of the main text, and here we mostly comment on its supersymmetric, undeformed versions $shs(1,1)$ and $shs(N, 2|\mathbb{R})$.

### B.3.1 The Algebra $shs(1,1)$: the Non-Extended Case

In this section we give a detailed account of the structure of the superalgebra $shs(1,1) \simeq shs(1,2|\mathbb{R})$ using its oscillator realization. In [56], the superalgebra $shs(1,2|\mathbb{R})$ was not directly recognized as the relevant one for the three-dimensional higher-spin supergravity problem. Rather, it was first showed that the relevant superalgebra should be some real form of the complex superalgebra $shs(1,2|\mathbb{C})$ of Vasiliev [130], after which a real form of the later is chosen. Thus, we shall first exhibit the construction of the realization of $shs(1,2|\mathbb{C})$ in terms of oscillators and then recall how to chose a real form of the later.

**Supercommutator on Polynomials**

A handy realization of $shs(1,2|\mathbb{C})$ has been given in [130] in terms of polynomials of all degrees but zero in two real commuting spinors (the so-called oscillators) with coefficients in $\mathcal{G}$, the 'physical' Grassmann algebra — see below for conventions — , and with a commutator built out of the $\star$-product on these polynomial functions (also called Moyal product). This is the so-called oscillator realization we work with. More precisely, our polynomials are functions on which the $\star$-bracket is defined as

$$[f,g]_\star \equiv \frac{1}{2i}\left(f \star g - g \star f\right), \tag{B.25}$$

where '$\star$' is the star-product of two polynomials $f$ and $g$, defined by

$$(f \star g)(q'') \equiv \exp\left(i\,\epsilon_{\alpha\beta}\frac{\partial}{\partial q_\alpha}\frac{\partial}{\partial q'_\beta}\right) f(q)g(q')\,|_{q=q'=q''}, \tag{B.26}$$

where $f(q) \equiv f(q_1, q_2)$ and so on. The $1/2i$ factor will be explained below. Let us make it clear that in the $f(q)g(q')$ term of the right-hand side of the above equation the Grassmann product in $\mathcal{G}$ of the coefficients of the polynomials $f$ and $g$ is implicit. Also observe that in the above expression the real commuting spinors $q_1$ and $q_2$ are packaged into $q_\alpha$, these indices being understood as raised and lowered with spinor metric

$$(\epsilon^{\alpha\beta}) \equiv (\epsilon_{\alpha\beta}) \equiv \begin{pmatrix} 0 & 1 \\ -1 & 0 \end{pmatrix}, \qquad \alpha,\,\beta \in \{1,2\}, \tag{B.27}$$

using the so-called 'North-West/South-East' convention (see Appendix A). Note that (1.86) is the usual definition of a Lie bracket on an associative algebra with some product $\star$ (up to a normalization explained below). Furthermore, our coefficients are not arbitrary in $\mathcal{G}$. Rather, they are restricted to be Grassmann even (resp. odd) in $\mathcal{G}$ when they multiply monomials of even (resp. odd) degree in the $q_\alpha$'s. Our commutator thus yields

$$[f,g]_\star = \sin\left(\epsilon_{\alpha\beta}\frac{\partial}{\partial q_\alpha}\frac{\partial}{\partial q'_\beta}\right) f(q)g(q')\,|_{q=q'=q''} \qquad \text{when } \pi_f\pi_g = 0, \tag{B.28a}$$

$$[f,g]_\star = -i\cos\left(\epsilon_{\alpha\beta}\frac{\partial}{\partial q_\alpha}\frac{\partial}{\partial q'_\beta}\right) f(q)g(q')\,|_{q=q'=q''} \quad \text{when } \pi_f\pi_g = 1, \tag{B.28b}$$

where our parities $\pi_f$ take values $0, 1$ (0 being the bosonic case), and are defined to be the degree of $f$ modulo 2.

In the 'supercommutator description' of this superalgebra, the coefficients of the polynomials take values in $\mathbb{C}$ and the Lie bracket (1.86) becomes a Lie superbracket[1]

$$[f,g]_\star \equiv \frac{1}{2i}\left(f \star g - (-)^{\pi_f \pi_g} g \star f\right), \tag{B.29}$$

---

[1] The definition (B.29) holds for polynomials $f$ and $g$ with definite parity only, but as always this is straightforwardly extended to the whole vector space of polynomials.

satisfying the graded Jacobi identity. As always, in this description we find the Lie superbracket to yield the same expression as the Lie bracket acting on Grassmann-like objects, that is,

$$[f,g\}_\star = [f,g]_\star = \frac{1}{2}(f \star g - g \star f) \tag{B.30a}$$

$$= \sin\left(\epsilon_{\alpha\beta}\frac{\partial}{\partial q_\alpha}\frac{\partial}{\partial q'_\beta}\right)f(q)g(q')\,|_{q=q'=q''} \qquad \text{when } \pi_f\pi_g = 0,$$

$$[f,g\}_\star = \{f,g\}_\star = \frac{1}{2}(f \star g + g \star f) \tag{B.30b}$$

$$= -i\cos\left(\epsilon_{\alpha\beta}\frac{\partial}{\partial q_\alpha}\frac{\partial}{\partial q'_\beta}\right)f(q)g(q')\,|_{q=q'=q''} \qquad \text{when } \pi_f\pi_g = 1.$$

Let us further point out that our Lie bracket does not produce constant polynomials, so that our restriction to polynomials of all degrees but zero is consistent with the Lie structure.

**Basis**

We will work with the following basis of our realization:

$$X_{(p,q)} \equiv X_{\underbrace{1\ldots1}_{p}\underbrace{2\ldots2}_{q}} \equiv \frac{1}{p!q!}(q_1)^p(q_2)^q, \qquad p+q \in \mathbb{N}_0, \tag{B.31}$$

where $q_1, q_2$ are the two commuting spinor variables used above. Let us stress again that, as the above definition states, we do not consider the (unique) generator with no indices (zero degree polynomials). The elements of our superalgebra $\text{shs}(1,2|\mathbb{C})$ are thus the linear combinations of the above generators, with coefficients in $\mathcal{G}$ (resp. $\mathbb{C}$) in the description in which the Lie bracket (1.86) (resp. Lie superbracket (B.29)) is used. We also recall that, when in $\mathcal{G}$, the coefficients also have to satisfy certain parity conditions, namely they are even (resp. odd) when they multiply basis elements of even (resp. odd) degree in the $q$'s.

The generators (basis elements) with $n = p+q$ indices above are said to carry spin $n/2 + 1$ (conformal spin $n/2$). The generators with two and one indices are thus carrying spin 2 and 3/2 (conformal spin 1 and 1/2), which is indeed consistent with the fact that upon truncation to $p+q \leq 2$, $\text{shs}(1,2|\mathbb{R})$ is seen to reduce to $\text{osp}(1,2|\mathbb{R})$ (see below), which we know is (half of) the superalgebra describing $\mathcal{N} = (1,1)$ three-dimensional supergravity [58], which in turn is known to contain the graviton and the gravitino: fields of spin 2 and 3/2 respectively (conformal spin 1 and 1/2).

**Real Form of shs$(1,2|\mathbb{C})$**

As always, the number of anticommuting generators of our Grassmann algebra $\mathcal{G}$ is left arbitrary [103] and these are assumed to be real with respect to a certain conjugation operation on $\mathcal{G}$ that we denote by '$*$'. We shall use conventions according to which our Grassmann product satisfies $(ab)^* = b^*a^*$ and $(a^*)^* = a$ for any two elements $a, b \in \mathcal{G}$, so that the subspace $\mathcal{G}^{\mathbb{R}}$ of real elements of $\mathcal{G}$ is not a subalgebra for the Grassmann product, as the product of two real anticommuting elements is imaginary.

We now want to extract a real form shs$(1,2|\mathbb{R})$ of shs$(1,2|\mathbb{C})$, i.e. the subsuperalgebra of fixed points under some conjugation operation on shs$(1,2|\mathbb{C})$ (we now think in terms of Grassmann-like objects). One rather natural candidate is the operation which acts semilinearly on our polynomials and on the coefficients acts as $*$ on $\mathcal{G}$ (and on the $q_\alpha$ variables does nothing). This operation we denote by $\circledast$, and we see that its fixed points are simply the polynomials (of all degree but zero in the $q_\alpha$'s) with coefficients in $\mathcal{G}^{\mathbb{R}}$.

We have not yet checked that $\circledast$ is a proper conjugation operation, and we now do so. From its definition and from the properties of $*$, it is evidently a semilinear and involutive operation, so that it only remains to be checked that the Lie bracket preserves the reality condition, which is easily seen to be true by inspection of (B.30) if we recall that our coefficients are even or odd in $\mathcal{G}$ according to the degree of the monomial they multiply, and that with our conventions the Grassmann product of two real anticommuting numbers is imaginary (so the $i$ factor in the second line of (B.30) actually makes the anticommutator of two real elements a real element of the algebra).

In the superbracket description of shs$(1,2|\mathbb{R})$ the coefficients are real numbers, in $\mathbb{R}$, and we thus see that the anticommutator (of two real elements) is imaginary. As always, this is consistent with our reality conditions, because in physical computations the anticommutator is always multiplied by a product of two real odd elements of $\mathcal{G}$, so that physical quantities are real (because a product of two real odd elements of $\mathcal{G}$ is imaginary).

**Supertrace**

For the chosen realization, the construction of a (real) non-degenerate, supersymmetric and invariant bilinear application turns out to be quite easy. Indeed, defining the supertrace as

$$\text{STr}(f) \equiv 2f(0), \tag{B.32}$$

we easily check that the following bilinear form on $\mathrm{shs}(1,2|\mathbb{R})$ satisfies all the asked-for properties:

$$(f, g) \equiv \mathrm{STr}(f \star g). \tag{B.33}$$

It is easily verified that this bilinear form is also consistent, so that the above expression provides us with a non-degenerate inner product on $\mathrm{shs}(1,2|\mathbb{R})$.[2]

**Low-Spin and Bosonic Truncation**

By 'low-spin truncation' we mean retaining only the spin-2 and spin-3/2 generators (conformal spin 1 and 1/2), and one can easily convince oneself that this truncation yields a subsuperalgebra isomorphic to $\mathrm{osp}(1,2|\mathbb{R})$. Indeed, in this sector one finds the non-zero Lie superbrackets

$$\begin{aligned}
[X_{11}, X_{12}] &= 2X_{11}, & [X_{11}, X_{22}] &= X_{12}, & [X_{12}, X_{22}] &= 2X_{22}, \\
\{X_1, X_1\} &= -2iX_{11}, & \{X_2, X_2\} &= -2iX_{22}, & \{X_1, X_2\} &= -iX_{12}, \\
[X_1, X_{22}] &= X_2, & [X_2, X_{12}] &= -X_2, & [X_1, X_{12}] &= X_1, \\
& & [X_2, X_{11}] &= -X_1,
\end{aligned} \tag{B.34}$$

which can be seen to agree with the commutation relations of $\mathrm{osp}(1,2|\mathbb{R})$. In fact, upon setting $X_{11} \equiv E$, $X_{22} \equiv -F$, $X_{12} \equiv -H$, $X_1 \equiv R_+$ and $X_2 \equiv R_-$ the above supercommutation relations can be seen to match the supercommutation relations given in Appendix B.3.2. Note that we have dropped the subscript $\star$ for the supercommutators, as will be also sometimes done in the sequel.

Also note that the bosonic algebra $\mathrm{hs}(1,1)$ can be seen to be a subalgebra of $\mathrm{shs}(1,1)$. Indeed, upon retaining only the generators with an even number of indices the commutator (B.29) never produces odd-degree generators, and the commutation relations thus close on the bosonic subsector.

**Scalar Products and Higher-Spin Commutators**

It is easily checked that the expression (B.33) for the scalar product implies that elements with different number of indices (different spin) are orthogonal, so that in particular it is indeed consistent, as aforesaid. Also, it is rather straightforward to show that the $X_{(n,0)}$ generators have non-zero scalar products only with the $X_{(0,n)}$ generators (and conversely) and that these are given by

$$(X_{(n,0)}, X_{(0,n)}) = 2\frac{i^n}{n!}, \tag{B.35}$$

which at the first levels yield the non-zero projections

$$\begin{aligned}
(X_1, X_2) &= 2i, \\
(X_{11}, X_{22}) &= -1, & (X_{12}, X_{12}) &= 2, \\
(X_{111}, X_{222}) &= -i/3, & (X_{112}, X_{122}) &= i, \\
(X_{1111}, X_{2222}) &= 1/12, & (X_{1112}, X_{1222}) &= -1/3, & (X_{1122}, X_{1122}) &= 1/2.
\end{aligned} \tag{B.36}$$

---

[2] We call inner product any (real) consistent, invariant and supersymmetric bilinear form on a superalgebra.

The $shs(1,2|\mathbb{R})$ superalgebra can thus really be thought of as being composed of 'layers' of definite spin, all orthogonal to the others, with generators $X_{(n,0)}$ and $X_{(0,n)}$ spanning an orthogonal subspace (but not a subalgebra) for each $n$. We point out that, as always, because of the *super*symmetric invariance of the scalar product, the norm (scalar product with itself) of all fermionic generators is zero. Also observe that the scalar product of two odd generators is an imaginary number, which is consistent with the fact that the scalar product of two elements of the algebra belongs to $\mathcal{G}_0^{\mathbb{R}}$, for such 'odd-odd' scalar products are multiplied by two real anticommuting elements of $\mathcal{G}$.

In the above discussion we say that $shs(1,2|\mathbb{R})$ can be viewed as composed of layers, or levels of different spin. However, this picture is only valid as far as the scalar product is concerned, and the supercommutator evidently mixes different layers, which can be seen for example by the following supercommutation relations, of which not all are useful for the derivation of the asymptotic symmetries described in the main text but which are also given to provide further insight into the structure (recall we don't consider the spin-1 generator, i.e. the one with no indices at all).

$$\{X_{(n,0)}, X_{(0,m)}\}_\star = -i\sum_{j=0}^{\frac{r-1}{2}} \frac{(-)^j}{(2j)!} X_{(n-2j, m-2j)} \quad (n, m \text{ both odd}), \tag{B.37a}$$

$$[X_{(n,0)}, X_{(0,m)}]_\star = \sum_{j=0}^{f(r)} \frac{(-)^j}{(2j+1)!} X_{(n-2j-1, m-2j-1)} \quad (n \text{ and/or } m \text{ even}), \tag{B.37b}$$

where $r \equiv \min(n,m)$ and

$$f(r) \equiv \frac{r - 2^{1-\pi_r}}{2}. \tag{B.38}$$

Note that the $X_{(i,0)}$ generators are produced in the commutator above when $r = m$ is odd ($n$ even) only, and the $X_{(0,j)}$ are produced when $r = n$ is odd ($m$ even) only. This is actually of much use in the computation of the corresponding asymptotic symmetries.

### B.3.2 The Algebra shs$(N, 2|\mathbb{R})$: the Extended Case

Constructing the oscillator realization for the extended version of the superalgebra shs$(1,1)$ explored above involves considering Grassmann-odd oscillators, which arguably adds a qualitative technical complication to it [130]. Thus, although quite similar in spirit to its non-extended version we proceed hereafter with the construction, again from scratch, of the shs$(N, 2|\mathbb{R})$ superalgebra in terms of oscillators. Much like in the previous section, we begin by constructing the complex version of the latter and then take a real form thereof.

**Supercommutator on Polynomials**

Consider the following $N + 2$ Grassmann variables: two commuting ones, $q_\alpha$ ($\alpha = 1, 2$), together with $N$ anticommuting ones, $\psi_i$ ($i = 1, \ldots, N$). Adapting to the terminology used in the literature, we refer to the index $i$ as the 'color' or 'internal' index (sometimes also as the 'Latin' index). As such,

$$q_\alpha q_\beta = q_\beta q_\alpha \qquad \forall\, \alpha, \beta = 1, 2, \tag{B.39a}$$
$$\psi_i \psi_j = -\psi_j \psi_i \qquad \forall\, i, j = 1, \ldots, N, \tag{B.39b}$$
$$q_\alpha \psi_i = \psi_i q_\alpha \qquad \forall\, \alpha = 1, 2 \ \ \&\ \ i = 1, \ldots, N. \tag{B.39c}$$

These variables are all taken to be real-valued, $q_\alpha^* = q_\alpha$, $\psi_i^* = \psi_i$. We construct polynomials in these $N + 2$ variables, with coefficients which can be themselves commuting or anticommuting, i.e. which belong also to a different Grassmann algebra $\mathcal{G}$ (the 'physical' Grassmann algebra). Thus, we formally consider the (graded) tensor product $\mathcal{A} = \mathcal{G} \otimes \mathcal{P}$ of the polynomial algebra $\mathcal{P}$ in $q_\alpha$, $\psi_i$ with the Grassmann algebra $\mathcal{G}$. The sign in the commutation relations for the multiplication of elements in the graded tensor product is dictated by the total grading, so that odd elements of $\mathcal{G}$ and $\mathcal{P}$ anticommute. The Grassmann parity used below will always be the total grading. A complex conjugation is assumed to be defined in $\mathcal{G}$, and can be extended to $\mathcal{A}$ taking into account that $q_\alpha$ and $\psi_i$ are real, and we systematically use the convention $(ab)^* = b^* a^*$.

Let $\mathcal{A}^\mathrm{E}$ be the subalgebra of Grassmann-even polynomials in $q_\alpha$, $\psi_i$ containing only monomials of even degree and no constant term. A general element of $\mathcal{A}^\mathrm{E}$ thus reads

$$\begin{aligned}
f &= f^{\alpha\beta} q_\alpha q_\beta + f^{\alpha,i} q_\alpha \psi_i + f^{ij} \psi_i \psi_j \\
&\quad + f^{\alpha\beta\gamma\delta} q_\alpha q_\beta q_\gamma q_\delta + f^{\alpha\beta\gamma,i} q_\alpha q_\beta q_\gamma \psi_i + f^{\alpha\beta,ij} q_\alpha q_\beta \psi_i \psi_j + \cdots \\
&\quad + f^{\alpha\beta\gamma\delta\epsilon\eta} q_\alpha q_\beta q_\gamma q_\delta q_\epsilon q_\eta + \cdots \\
&\quad + \cdots,
\end{aligned} \quad (\mathrm{B.40})$$

where terms of arbitrarily high power are allowed. The coefficients in this expansion are completely symmetric (respectively antisymmetric) in the Greek (respectively Latin) indices. They are commuting (respectively anticommuting) whenever they multiply an even (respectively odd) number of $\psi$'s.

A $\star$-product is defined on $\mathcal{A}$ as follows:

$$(f \star g)(z'') \equiv \exp\left( i\, \epsilon_{\alpha\beta} \frac{\partial}{\partial q_\alpha} \frac{\partial}{\partial q'_\beta} + \delta_{ij} \frac{\overleftarrow{\partial}}{\partial \psi_i} \frac{\vec{\partial}}{\partial \psi'_j} \right) f(z) g(z') \,\big|_{z=z'=z''}, \quad (\mathrm{B.41})$$

here $f(z) \equiv f(q_\alpha, \psi_i)$ and so on. In this expression, $f(z) g(z')$ is the standard Grassmann product, and left and right derivatives with respect to the anticommuting variables are defined by

$$\delta f = \delta\psi_i \frac{\vec{\partial} f}{\partial \psi_i}, \quad (\mathrm{B.42a})$$

$$\delta f = \frac{\overleftarrow{\partial} f}{\partial \psi_i} \delta\psi_i, \quad (\mathrm{B.42b})$$

and the conventions for the use of the epsilon symbol are again the same (see Appendix A).

The above $\star$-product is well known to be associative. However, it does not preserve the reality condition, in the sense that $f \star g$ is not real even if $f$ and $g$ are so. On the other hand, one can check that if $f$ and $g$ are both real elements of $\mathcal{A}^\mathrm{E}$, or both purely imaginary elements of $\mathcal{A}^\mathrm{E}$, of respective order $2n$ and $2m$, then the homogenous polynomials appearing in the expansion of $f \star g$,

$$f \star g = \sum_{j=0}^{m+n} h_{2(m+n-j)}, \qquad (B.43)$$

are alternatively real and imaginary. More precisely, the homogeneous polynomial $h_{2(m+n-j)}$ of degree $2(m+n-j)$ in $q_\alpha$ and $\psi_i$ is:

- real and symmetric for the exchange of $f$ and $g$ when $j$ is even,
- imaginary and antisymmetric under exchange of $f$ and $g$ for $j$ odd.

We then define the $\star$-commutator as

$$[f, g]_\star \equiv f \star g - g \star f, \qquad (B.44)$$

which fulfills the Jacobi identity since the $\star$-product is associative. From what we have just seen, $[f, g]_\star$ is purely imaginary whenever $f$ and $g$ are both real or both pure imaginary.

The Lie superalgebra $\mathrm{shs}^{\mathrm{E}}(N, 2|\mathbb{R}) \simeq \mathrm{shs}(N, 2|\mathbb{R})$ is the real subspace of pure imaginary elements of $\mathcal{A}^{\mathrm{E}}$ equipped with the $\star$-bracket[3]. A general element of $\mathrm{shs}(N, 2|\mathbb{R})$ is thus of the form

$$\begin{aligned} f =& f^{\alpha\beta} q_\alpha q_\beta + f^{\alpha,i} q_\alpha \psi_i + f^{ij} \psi_i \psi_j \\ &+ f^{\alpha\beta\gamma\delta} q_\alpha q_\beta q_\gamma q_\delta + f^{\alpha\beta\gamma,i} q_\alpha q_\beta q_\gamma \psi_i + f^{\alpha\beta,ij} q_\alpha q_\beta \psi_i \psi_j + \cdots \\ &+ f^{\alpha\beta\gamma\delta\epsilon\eta} q_\alpha q_\beta q_\gamma q_\delta q_\epsilon q_\eta + \cdots \\ &+ \cdots , \end{aligned} \qquad (B.45)$$

but the coefficients are further restricted so as to make $f$ imaginary. For instance, the coefficient $f^{\alpha\beta}$ is imaginary while $f^{\alpha,i}$ and $f^{ij}$ are real. Also, one can alternatively rewrite (B.44) as

$$\begin{aligned} [f, g]_\star(z'') =& \left[ 2i \sin\left(\epsilon_{\alpha\beta} \frac{\partial}{\partial q_\alpha} \frac{\partial}{\partial q'_\beta}\right) \cosh\left(\delta_{ij} \frac{\overleftarrow{\partial}}{\partial \psi_i} \frac{\overrightarrow{\partial}}{\partial \psi'_j}\right) \right. \\ &\left. + 2\cos\left(\epsilon_{\alpha\beta} \frac{\partial}{\partial q_\alpha} \frac{\partial}{\partial q'_\beta}\right) \sinh\left(\delta_{ij} \frac{\overleftarrow{\partial}}{\partial \psi_i} \frac{\overrightarrow{\partial}}{\partial \psi'_j}\right) \right] f(z) g(z') \big|_{z=z'=z''}. \end{aligned} \qquad (B.46)$$

---

[3] One could equivalently insert a factor of $i$ in the definition of the $\star$-bracket, which would no longer coincide with the star commutator, and define $\mathrm{shs}(N, 2|\mathbb{R})$ as the subspace of real polynomials equipped with that alternative bracket. Either convention has its own advantages.

It should be stressed that the polynomial $[f,g]_\star$ starts at highest polynomial degree $2(n+m-1)$. Note also that the lowest polynomial degree term in the expansion (B.43) is $h_{2(|m-n|)}$ so that there is a term of degree zero in (B.43) only if $n = m$, in which case $j = 2m$ is even. This implies that the term of degree zero (when present) is symmetric for the exchange of $f$ with $g$ and in particular that the constant term (when present in $f \star g$) drops from the $\star$-commutator so that $[f,g]_\star$ has indeed no constant term and belongs to $\text{shs}(N,2|\mathbb{R})$.

**Basis**

A basis of $\text{shs}(N,2|\mathbb{R})$ is given by the monomials

$$X_{p,q;\,i_1,i_2,\cdots,i_N} \equiv \frac{i^{\lfloor \frac{K+1}{2} \rfloor}}{2\,i\,p!\,q!} q_1^p q_2^q \psi_1^{i_1} \ldots \psi_N^{i_N}, \tag{B.47}$$

where $p,q \in \mathbb{N}$ and $i_k \in \{0,1\}$. The degree of $X_{p,q;\,i_1,i_2,\cdots,i_N}$, which is $p + q + K$, must be even and positive, where $K = \Sigma_{k=1}^N i_k$ is the degree in the $\psi$'s. The power of $i$ has been inserted in such a way that the elements of even Grassmann parity are imaginary, while those of odd Grassmann parity are real.

With this choice, a general element of $\text{shs}^{\text{E}}(N,2|\mathbb{R})$ is of the form

$$\sum \mu^{p,q;\,i_1,i_2,\cdots,i_N} X_{p,q;\,i_1,i_2,\cdots,i_N}, \tag{B.48}$$

where $\mu^{p,q;\,i_1,i_2,\cdots,i_N}$ is real and of Grassmann parity $(-1)^K = (-1)^{p+q}$. Note that the above generators are antisymmetric under the exchange of any two indices $i_a$, $i_b$. Also observe that the $X_{(p,q)}$ ($N=0$) generators above are the $X_{(p,q)}$ generators of the non-extended case (and when $N=0$ they will always we written like that). We also point out that, as the $\psi$'s are anticommuting, all the $i_a$ indices in one of the above generators have to be different for the corresponding generator to be non-zero.

**Supertrace and Scalar Products**

The supertrace of a polynomial in the $q$'s and the $\psi$'s is defined by its component of degree zero:

$$\text{STr}f(q,\psi) = 8f(0). \tag{B.49}$$

Although $\text{STr}f = 0\; \forall f \in \text{shs}(N,2|\mathbb{R})$, it turns out that $\text{STr}(f \star g)$ may differ from zero even if $f,g \in \text{shs}(N,2|\mathbb{R})$. One thus defines a scalar product on $\text{shs}(N,2|\mathbb{R})$ by

$$(f,g) \equiv \text{STr}(f \star g). \tag{B.50}$$

The scalar product is evidently bilinear, real and symmetric (given our discussions in the previous subsection). Using the symmetry together with the associativity of the $\star$-product, we further conclude that it is also invariant:

$$([f,g]_\star, h) = (f, [g,h]_\star). \tag{B.51}$$

In addition, it is non-degenerate. It is non-zero only when $f$ and $g$ have same degree in both the $\psi_i$'s and the $q_\alpha$'s. It is this scalar product that is used to define a Chern–Simons action term.

**Low-Spin Truncation and Internal Subalgebra**

The subspace of quadratic polynomials is a subalgebra isomorphic to $\mathrm{osp}(N,2|\mathbb{R})$, as is known from the familiar oscillator realization of $\mathrm{osp}(N,2|\mathbb{R})$ [151]. Renormalizing and relabeling[4] the quadratic basis elements as

$$Y_{\alpha\beta} = -\tfrac{i}{2} q_\alpha q_\beta, \quad X_{\alpha i} = \tfrac{1}{2} q_\alpha \psi_i, \quad X_{ij} = \tfrac{1}{2} \psi_i \psi_j, \tag{B.52}$$

one finds that the non-zero Lie superbrackets read explicitly

$$[Y_{\alpha\beta}, Y_{\gamma\delta}]_\star = \epsilon_{\alpha\gamma} Y_{\beta\delta} + \epsilon_{\alpha\delta} Y_{\beta\gamma} + \epsilon_{\beta\gamma} Y_{\alpha\delta} + \epsilon_{\beta\delta} Y_{\alpha\gamma}, \tag{B.53a}$$

$$[X_{\alpha i}, Y_{\beta\gamma}]_\star = \epsilon_{\alpha\beta} X_{\gamma i} + \epsilon_{\alpha\gamma} X_{\beta i}, \tag{B.53b}$$

$$i\{X_{\alpha i}, X_{\beta j}\}_\star = \delta_{ij} Y_{\alpha\beta} + \epsilon_{\alpha\beta} X_{ij}, \tag{B.53c}$$

$$[X_{ij}, X_{\alpha k}]_\star = \delta_{jk} X_{\alpha i} - \delta_{ik} X_{\alpha j}, \tag{B.53d}$$

$$[X_{ij}, X_{kl}]_\star = \delta_{il} X_{jk} + \delta_{jk} X_{il} - \delta_{ik} X_{jl} - \delta_{jl} X_{ik}. \tag{B.53e}$$

which, upon performing the redefinitions

$$\begin{aligned} E &\equiv \tfrac{1}{2} X_{11}, & F &\equiv -\tfrac{1}{2} X_{22}, & H &\equiv -X_{12}, \\ R_i^+ &\equiv X_{1i}, & R_i^- &\equiv X_{2i}, & J_{ij} &\equiv X_{ij} \end{aligned} \tag{B.54}$$

are seen to match the relations given in appendix B.2.1. Hence, one goes from $\mathrm{shs}(N,2|\mathbb{R})$ to $\mathrm{osp}(N,2|\mathbb{R})$ by restricting the $\star$-algebra of polynomials of even degree in the $q$'s and the $\psi$'s to the $\star$-subalgebra of polynomials of degree two. Conversely, one goes from $\mathrm{osp}(N,2|\mathbb{R})$ to $\mathrm{shs}(N,2|\mathbb{R})$ by relaxing the condition that the polynomials should be quadratic, i.e. by allowing arbitrary (pure imaginary) polynomials of even degree modulo a zero-degree term.

---

[4] Note that we have changed the letter $X$ to $Y$ for the generators with no $\psi$'s since these differ from the corresponding $X$'s by a factor.

The finite-dimensional subalgebra of polynomials involving only $\psi$'s and no $q$'s is called the internal subalgebra U. The internal subalgebra U contains so($N$) as the subalgebra generated by the quadratic monomials $X_{ij}$, which we know is the internal subalgebra of osp($N, 2|\mathbb{R}$). To identify U completely, we recall that the $\psi_i$'s are the generators of a Clifford algebra, which implies

$$\text{U} = \text{su}(2^{\frac{N-2}{2}}) \oplus \text{su}(2^{\frac{N-2}{2}}) \oplus \text{u}(1) \qquad (N \text{ even}), \tag{B.55a}$$

$$\text{U} = \text{su}(2^{\frac{N-1}{2}}) \qquad\qquad\qquad\qquad\quad (N \text{ odd}), \tag{B.55b}$$

and one can indeed check that so($N$) is indeed a subalgebra thereof for all values of $N$.

**Non-Extended Case: Alternative Description**

For $N = 1$ supersymmetry, an alternative description of the superalgebra is available. Since there is only one $\psi$, any element of shs($N, 2|\mathbb{R}$) can be decomposed as

$$f = P_0 + p_1, \tag{B.56}$$

where $P_0$ is a Grassmann-even polynomial in the $q$'s containing no $\psi$ while $p_1$ is linear in $\psi$ and reads

$$p_1 = iP_1\psi. \tag{B.57}$$

Here, $P_1$ is a Grassmann-odd polynomial in the $q$'s. Furthermore, $P_0$ contains only terms of even degrees in the $q$'s while $P_1$ contains only terms of odd degrees. We can associate to $f$ a polynomial $F$ in the $q$'s with no constant term as follows:

$$f = P_0 + p_1 \mapsto F = P_0 + P_1. \tag{B.58}$$

Here, $F$ is pure imaginary and contains both even ($P_0$) and odd ($P_1$) powers in the $q$'s. The even part $P_0$ is also Grassmann-even, while the odd part $P_1$ is Grassmann-odd. In terms of this new representation, the $\star$-product reads

$$(F \star G)(q'') \equiv \exp\left(i\,\epsilon_{\alpha\beta}\frac{\partial}{\partial q_\alpha}\frac{\partial}{\partial q'_\beta}\right) F(q)G(q')\,|_{q=q'=q''}, \tag{B.59}$$

and the $\star$-bracket becomes

$$[F, G]_\star = 2i\sin\left(\epsilon_{\alpha\beta}\frac{\partial}{\partial q_\alpha}\frac{\partial}{\partial q'_\beta}\right)(F_0(q)G_0(q') + F_1(q)G_0(q') + F_0(q)G_1(q'))\,|_{\ldots}$$

$$+ 2\cos\left(\epsilon_{\alpha\beta}\frac{\partial}{\partial q_\alpha}\frac{\partial}{\partial q'_\beta}\right)(F_1(q)G_1(q'))\,|_{q=q'=q''}. \tag{B.60}$$

The $N = 1$ super-algebra $\text{she}(1, 2|\mathbb{R})$ is thus isomorphic to the super-algebra $\text{shs}(2|\mathbb{R}) \simeq \text{shs}(1, 1)$, defined to be the superalgebra of imaginary polynomials in the $q$'s with no constant term but with both even and odd powers (the coefficients of the even — respectively odd — powers being Grassmann-even — respectively Grassmann-odd), equipped with the $\star$-bracket (B.60). The above basis (B.47) becomes in this alternative description

$$X_{(p,q)} \equiv X_{\underbrace{1\ldots1}_{p}\underbrace{2\ldots2}_{q}} \equiv \frac{1}{2\,i\,p!\,q!}(q_1)^p(q_2)^q, \qquad p + q \in \mathbb{N}_0. \quad \text{(B.61)}$$

**Spin Structure**

For any $N$-extended supersymmetry, the superalgebra $\text{shs}(N, 2|\mathbb{R})$ contains the spacetime algebra $\text{sl}(2, \mathbb{R}) \simeq \text{su}(1, 1)$ under which it decomposes as a direct sum of irreducible representations. To exhibit this decomposition, it is convenient to write

$$\text{shs}(N, 2|\mathbb{R}) = \oplus_{j \geq 0} V_j, \quad \text{(B.62)}$$

where $j$ is a non-negative integer or half-integer, and $V_j$ is the vector subspace containing the polynomials in the $q$'s of degree $2j$ (with no restriction on the degree in the $\psi$'s, which are spacetime scalars). The subspaces $V_j$ are invariant under the action of $\text{sl}(2, \mathbb{R})$ and are reducible for $N > 1$. More precisely,

$$V_0 = D_0 \otimes \mathcal{E}', \quad \text{(B.63a)}$$
$$V_j = D_j \otimes \mathcal{O} \quad (j \text{ half-integer} \geq \tfrac{1}{2}), \quad \text{(B.63b)}$$
$$V_j = D_j \otimes \mathcal{E} \quad (j \text{ integer} \geq 1), \quad \text{(B.63c)}$$

where $D_j$ is the $(2j+1)$-dimensional space of the $\text{sl}(2, \mathbb{R})$-spin $j$ irreducible representation. Furthermore, $\mathcal{E}$ is the space of polynomials of even degree in $\psi_i$, $\mathcal{E}'$ is the space of polynomials of even degree in $\psi_i$ with no constant term, and $\mathcal{O}$ is the space of polynomials of odd degree in $\psi_i$. The subalgebra $\text{sl}(2, \mathbb{R})$ appears in $V_1$, as $D_1$ times the constants. The subspaces $\mathcal{E}$, $\mathcal{E}'$ and $\mathcal{O}$ have respective dimensions displayed in Table B.3.2 below. For $N \leq 1$, the space $\mathcal{E}'$, and hence also the space $V_0$, is trivial. Hence, for $N \leq 1$, the spin-0 representation does not occur. Furthermore, the spaces $\mathcal{O}$ and $\mathcal{E}$ are then one-dimensional, so that the subspaces $V_j$ are irreducible and each value of the spin appearing in the theory is non-degenerate. Neither of these features holds for $N > 1$.

Table B.2: Dimensions of the Internal Subspaces of $\text{shs}(N, 2|\mathbb{R})$

|  | dim $(N = 0)$ | dim $(N > 0)$ |
|---|---|---|
| $\mathcal{E}$ | 1 | $2^{N-1}$ |
| $\mathcal{E}'$ | 0 | $2^{N-1} - 1$ |
| $\mathcal{O}$ | 0 | $2^{N-1}$ |

To summarize, one encounters the following higher spin superalgebras as one increases the number $N$ of supersymmetries:

- $\text{shs}(0, 2|\mathbb{R}) \simeq \text{hs}(2|\mathbb{R}) \simeq \text{hs}(1, 1)$ is the bosonic higher spin algebra involving only integer spins $\geq 1$ (no Supersymmetry). Each value of the spin is non-degenerate.

- $\text{shs}(1, 2|\mathbb{R}) \simeq \text{shs}(2|\mathbb{R}) \simeq \text{shs}(1, 1)$ is the higher spin superalgebra for simple Supergravity. It contains $\text{osp}(1, 2|\mathbb{R})$, which in turn contains $\text{sl}(2|\mathbb{R})$, and has no non-trivial internal subalgebra. It involves both half-integer and integer spins $\geq \frac{1}{2}$. Each value of the spin is again non-degenerate.

- $\text{shs}(N, 2|\mathbb{R})$ is relevant for the extended models. It involves both half-integer and integer spins $\geq \frac{3}{2}$. Spin 0 is degenerate $2^{N-1} - 1$ times, while spins $\geq \frac{1}{2}$ are degenerate $2^{N-1}$ times.

At infinity, by the boundary conditions discussed in the text, each $\text{sl}(2|\mathbb{R})$-representation $D_j$ yields a generator of conformal dimension $j + 1$.

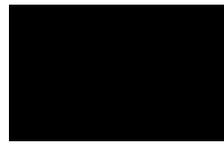

APPENDIX C

# Details in Dimension 3

In this appendix we detail some computations supporting some of the claims made in the text. In Section C.1 we start by the simple but perhaps useful presentation, in detail, of the matching of the gauge symmetries between the frame formalism and the Chern–Simons gauge picture for pure Gravity on flat spacetime. When higher spins are included, those symmetries are extended to some higher-spin algebra and in Section C.2 we study how the supersymmetric version thereof is embedded into the corresponding asymptotic symmetry algebra. The computation of the latter has been performed explicitly in the text for the non-extended, supersymmetric case and in Section C.3 we sketch extended version of the computation.

## C.1 Minkowskian Gravity in the Gauge Picture

There is a last non-trivial check to do before one can safely claim the frame formalism and the Chern–Simons formalism to be equivalent for Gravity; namely, we need verify the gauge transformations on both sides to be the same. Indeed, while both the first-order formulation and the Chern–Simons action are manifestly invariant under diffeomorphisms, in the first-order formulation we also have the local Lorentz transformations as gauge symmetries, whereas in the Chern–Simons picture we have the full $iso(2,1)$ gauge symmetries. As we shall now show, the homogeneous part of the $iso(2,1)$ gauge symmetries are easily seen to correspond to the LLTs in the first-order formalism but, as for the infinitesimal gauge translations of $iso(2,1)$, one has to show that they are not extra gauge symmetries (which would be bad for our rewriting of the action would then eliminate degrees of freedom in some sense) but, rather, that they correspond to some combination of the symmetries of the first-order formalism action.

The gauge transformations in the Chern–Simons picture are parametrized by a zero-form gauge parameter taking values in the gauge algebra,
$$u \equiv \rho^a P_a + \tau^a J_a, \tag{C.1}$$
with $\rho^a$ and $\tau^a$ being infinitesimal parameters, and the transformation law for the gauge connection (sitting in the adjoint representation of the gauge algebra) is $A \to A + \delta A$ with
$$\delta A_\mu = \partial_\mu u + [A_\mu, u]. \tag{C.2}$$
Upon now plugging the expression for $u$ and the decomposition of $A$ in terms of the dreibein and spin-connection in the above equation we can read off the variations of $e$ and $\omega$, which are given by
$$\delta e^a_\mu = \partial_\mu \rho^a + \epsilon^{abc} e_{b\mu} \tau_c + \epsilon^{abc} \omega_{b\mu} \rho_c, \tag{C.3a}$$
$$\delta \omega^a_\mu = \partial_\mu \tau^a + \epsilon^{abc} \omega_{b\mu} \tau_c. \tag{C.3b}$$
These are all the (infinitesimal) local symmetries of the action in the gauge (Chern–Simons) picture and they can be decomposed into those generated by $\rho^a$,
$$\delta e^a_\mu = \partial_\mu \rho^a + \epsilon^{abc} \omega_{b\mu} \rho_c, \tag{C.4a}$$
$$\delta \omega^a_\mu = 0, \tag{C.4b}$$
and those generated by $\tau^a$,
$$\delta e^a_\mu = \epsilon^{abc} e_{b\mu} \tau_c, \tag{C.5a}$$
$$\delta \omega^a_\mu = \partial_\mu \tau^a + \epsilon^{abc} \omega_{b\mu} \tau_c. \tag{C.5b}$$
Moreover, as we already explained, the Chern–Simons term is also manifestly invariant under diffeomorphisms because it is written in terms of forms. The diffeomorphisms act by the well-known formula
$$\delta A_\mu = \xi^\nu \partial_\nu A_\mu + A_\nu \partial_\mu \xi^\nu, \tag{C.6}$$
where $\xi^\nu$ is some infinitesimal spacetime vector. We may again use the identification (1.37) to now simply find
$$\delta e^a_\mu = \xi^\nu \partial_\nu e^a_\mu + e^a_\nu \partial_\mu \xi^\nu = \xi^\nu (\partial_\nu e^a_\mu - \partial_\mu e^a_\nu) + \partial_\mu (e^a_\nu \xi^\nu), \tag{C.7a}$$
$$\delta \omega^a_\mu = \xi^\nu \partial_\nu \omega^a_\mu + \omega^a_\nu \partial_\mu \xi^\nu = \xi^\nu (\partial_\nu \omega^a_\mu - \partial_\mu \omega^a_\nu) + \partial_\mu (\omega^a_\nu \xi^\nu), \tag{C.7b}$$

which are the usual diffeomorphism transformations in the frame picture. Then, when dealing with the first-order action we also have the local Lorentz transformations, which act on $e$ and $\omega$ as in (1.5) and (1.8) respectively, the infinitesimal version of which is

$$\delta e_\mu^a = -\alpha^a{}_b e_\mu^b, \tag{C.8a}$$
$$\delta \omega_\mu^a = -\alpha^a{}_b \omega_\mu^b + \tfrac{1}{2}\epsilon^{abc}\partial_\mu \alpha_{bc}, \tag{C.8b}$$

as is easily derived taking $\alpha^a{}_b$ to be an element of the Lorentz algebra with coefficients depending on spacetime coordinates. These are the local symmetries of the frame formulation.

Now, the later LLTs are quite easily seen to be in one-to-one correspondence with the gauge transformations generated by the parameters $\tau^a$ on the Chern–Simons side. Indeed, upon setting

$$\alpha^{ab} = -\epsilon^{abc}\tau_c \quad \Leftrightarrow \quad \tau^a = \tfrac{1}{2}\epsilon^{abc}\alpha_{bc}, \tag{C.9}$$

one sees that (C.8) and (C.5) agree with one another. As for the infinitesimal gauge transformations of the gauge picture generated by the $\rho^a$'s the story is a little more subtle, and indeed at first sight one wonders what they could correspond to in the frame formulation. Actually, we shall need use both the equations of motion and the invariance under local Lorentz transformations to make them match with the infinitesimal diffeomorphisms or, differently put, we will show that the gauge transformations generated by the $\rho^a$'s are somehow not extra gauge transformations but, rather, on-shell they are simply some combination of diffeomorphisms and LLTs. Let us then try to relate the infinitesimal parameter $\xi^\mu$ to $\rho^a$. We are tempted to try

$$\rho^a = \xi^\mu e_\mu^a, \tag{C.10}$$

which yields

$$(\delta_\xi - \delta_\rho)e_\mu^a = \xi^\nu(\partial_\nu e_\mu^a - \partial_\mu e_\nu^a) + \epsilon^{abc}\xi^\nu e_{b\nu}\omega_{c\mu}, \tag{C.11a}$$
$$(\delta_\xi - \delta_\rho)\omega_\mu^a = \xi^\nu(\partial_\nu \omega_\mu^a - \partial_\mu \omega_\nu^a) + \partial_\mu(\omega_\nu^a \xi^\nu). \tag{C.11b}$$

Now, the first terms in the right-hand sides of the two above equations are seen to be the 'abelian' part of the equations of motion (1.40), so we try making these terms exactly the whole equations of motion, which yields

$$(\delta_\xi - \delta_\rho)e_\mu^a = \xi^\nu(D_\nu e_\mu^a - D_\mu e_\nu^a) + \epsilon^{abc}\xi^\nu e_{b\mu}\omega_{c\nu}, \tag{C.12a}$$
$$(\delta_\xi - \delta_\rho)\omega_\mu^a = \xi^\nu(\partial_\nu \omega_\mu^a - \partial_\mu \omega_\nu^a + \epsilon^{abc}\omega_{b\nu}\omega_{c\mu}) + \xi^\nu \epsilon^{abc}\omega_{b\mu}\omega_{c\nu} + \partial_\mu(\omega_\nu^a \xi^\nu). \tag{C.12b}$$

We now see that the first terms are proportional to the equations of motion whereas the last terms are local Lorentz transformations with parameters

$$\alpha^{ab} = -\epsilon^{abc}\xi^\nu \omega_{c\nu} \quad \Leftrightarrow \quad \tau^a = \xi^\nu \omega_\nu^a, \tag{C.13}$$

so that gauge transformations generated by $\rho^a$, which are also named infinitesimal gauge translations, indeed correspond to diffeomorphisms in the frame formulation (up to LLTs and EoMs). As already stated, this is well, since the point was to check that there are no extra gauge symmetries. This achieves the proof of the equivalence for the $\lambda = 0 = \Lambda$ case. Three-dimensional gravity is thus a gauge theory for the gauge group iso(2, 1), the Poincaré group (for zero cosmological constant).

In fact, an important point is the following: what is really required for the matching of the above gauge symmetries is not that their difference is proportional to the EoMs (up to LLTs) but, rather, that the latter piece is some antisymmetric combination of the EoMs, which is what ensures that the difference is a so-called 'trivial' gauge transformation [103]. This can be seen to be the case in present setup.

## C.2 Bulk Symmetries and Enhancements

In the text we showed that the asymptotic symmetry algebra provides an enhancement from the naive global gauge symmetry algebra shs$(N, 2|\mathbb{R})$ to s$\mathcal{W}_\infty$. To deepen our understanding of this remarkable feature, we would like to identify the way in which shs$(N, 2|\mathbb{R})$ is embedded in the s$\mathcal{W}_\infty$ algebra. In this appendix we carry out this analysis in detail, and we do so in full generality, i.e. for the generic $N \geq 1$ extended case.

### C.2.1 Exact Symmetries

The exact symmetry algebra of the AdS$_3$ background is the 'rigid' shs$(N, 2|\mathbb{R})$. Indeed, the AdS$_3$ connection is locally gauge equivalent to zero (it is pure gauge), and the zero connection is clearly invariant under gauge transformations that are constant but otherwise arbitrary:

$$0 \to S^{-1} \star dS + S^{-1} \star 0 \star S = 0 \quad \text{iff} \quad dS = 0, \quad \text{viz.} \quad S = S_0, \tag{C.14}$$

with $S_0$ a constant function belonging to the group which would correspond to shs$(N, 2|\mathbb{R})$ (let us think of it as the exponential of the latter). Moreover, the algebra of constant gauge transformations $S_0$ is of course isomorphic to shs$(N, 2|\mathbb{R})$.

If we denote by $\Gamma^{\text{AdS}}$ the AdS$_3$ connection in the standard static-polar reference frame (see Appendix A), we can express it as

$$\Gamma^{\text{AdS}} = U^{-1} \, \mathrm{d}U, \tag{C.15}$$

where $U$ is given by the simple expression

$$U = \exp\left(-\tfrac{1}{2} x^+ (X_{11} + X_{22})\right), \tag{C.16}$$

which contains only generators of the sl$(2,\mathbb{R})$ subalgebra (not higher-spin generators nor generators with color indices). The generator $X_{11} + X_{22}$ is the compact generator $E - F$ in the Chevalley basis (see Appendix B) and generates SO(2). A few comments are now in order:

- $U$ involves also $\exp\left(f(r) X_{12}\right)$ for some definite function $f(r)$ but this gauge transformation is irrelevant for the present considerations, so we drop it. The AdS$_3$ connection is then seen to read

$$\Gamma^{\text{AdS}} = -\tfrac{1}{2} \left(X_{11} + X_{22}\right) \mathrm{d}x^+ . \tag{C.17}$$

- We shall focus on an equal time slice, which we can assume to be $x^0 = 0$, and so we set $x^+ = \theta$ (at $\ell = 1$).

- The transformation $U$ is in the SL$(2|\mathbb{R})$ subgroup generated by the $X_{\alpha\beta}$'s (even in its SO(2) subgroup) and hence is the direct sum of the $2 \times 2$ matrix $R$,

$$R = \begin{pmatrix} \cos\tfrac{\theta}{2} & -\sin\tfrac{\theta}{2} \\ \sin\tfrac{\theta}{2} & \cos\tfrac{\theta}{2} \end{pmatrix} \tag{C.18}$$

and trivial identity terms in the complementary subspaces. Note that $R^{-1}$ is given by

$$R^{-1} = \begin{pmatrix} \cos\tfrac{\theta}{2} & \sin\tfrac{\theta}{2} \\ -\sin\tfrac{\theta}{2} & \cos\tfrac{\theta}{2} \end{pmatrix}, \tag{C.19}$$

and

$$R^{-1} \mathrm{d}R = \begin{pmatrix} 0 & -\tfrac{1}{2} \\ \tfrac{1}{2} & 0 \end{pmatrix} . \tag{C.20}$$

- To match (C.17) with the asymptotic behavior we have postulated in Subsection 2.2.1, a further constant gauge transformation $T$ must actually be performed, with

$$T = \exp\left(-\sqrt{2} X_{12}\right) = \begin{pmatrix} \sqrt{2} & 0 \\ 0 & \tfrac{1}{\sqrt{2}} \end{pmatrix} . \tag{C.21}$$

It then follows that

$$(RT)^{-1}\mathrm{d}(RT) = \begin{pmatrix} 0 & -\frac{1}{4} \\ 1 & 0 \end{pmatrix} = -X_{22} - \frac{1}{4}X_{11}, \qquad \text{(C.22)}$$

which indeed fulfills the asymptotic condition (2.35). The group element $T$ can be combined with $\exp(f(r)X_{12})$ above. Note that the motivation for including $T$ is not only that it makes the coefficient of $-X_{22} \equiv F$ equal to one, but that the connection corresponding to the zero mass black hole is then simply given by

$$\begin{pmatrix} 0 & 0 \\ 1 & 0 \end{pmatrix} = -X_{22} \qquad \text{(C.23)}$$

in that gauge. However, for the analysis of this section, we find it more convenient not to include $T$ so that the group element is in SO(2). The effect of $T$ is to rescale $q_1$ by $\sqrt{2}$ and $q_2$ by $\frac{1}{\sqrt{2}}$, a transformation that does not mix components of different sl(2|$\mathbb{R}$) weights. For that reason, the asymptotic analysis we made in the main text remains unchanged if we do not include $T$.

Now, we can rewrite the constant gauge transformations that leave the zero connection invariant in the representation in which the connection takes the form (C.15). These gauge transformations are just

$$S = U^{-1} S_0 \, U, \qquad \text{(C.24)}$$

where $S_0$ is constant, $\mathrm{d}S_0 = 0$. In infinitesimal form, $S = I + \Lambda^{\text{AdS}}$ with

$$\Lambda^{\text{AdS}} = \sum_{m,n,i_1,\cdots,i_N} \Lambda_0^{m,n;i_1,\cdots,i_N} U^{-1} X_{m,n;i_1,\cdots,i_N} U \,, \qquad \text{(C.25)}$$

where $\Lambda_0^{m,n;i_1,\cdots,i_N}$ are constants. For these gauge transformations, one evidently finds that

$$\delta\Gamma^{\text{AdS}} = \mathrm{d}\Lambda^{\text{AdS}} + [\Gamma^{\text{AdS}}, \Lambda^{\text{AdS}}] = 0, \qquad \text{(C.26)}$$

and it also straightforwardly verified that the algebra $[\Lambda_1^{\text{AdS}}, \Lambda_2^{\text{AdS}}]$ of the exact symmetries of the AdS$_3$ superconnection is shs($N, 2|\mathbb{R}$).

The algebra elements $U^{-1}X_{m,n;i_1,\cdots,i_N}U$ can then be expanded in the basis the $X_{m,n;i_1,\cdots,i_N}$'s. In particular, since we have observed that the lowest sl(2|$\mathbb{R}$)-weight components of the gauge transformations play a

central role, we find it interesting to work out the components of (C.25) along the lowest-weight generators $X_{0,\ell;i_1,\cdots,i_N}$. To that end we observe that, as $U$ belongs to the SO(2) subalgebra of SL(2|$\mathbb{R}$), it does not mix different spins and just acts on the generators $X_{m,n;i_1,\cdots,i_N}$, $m+n=\ell=2s$ of the spin $s$ representation by the symmetrized $\ell$-th tensor power of the rotation matrix $R$, without affecting the internal indices $i_k$. The formulas are thus more transparent if we drop the passive (and hence irrelevant for the present considerations) indices $i_k$ and work in the basis $Y_{m,n}$ with

$$Y_{m,n} \equiv z^m \bar{z}^n, \quad \text{where} \quad z = q_1 + iq_2, \quad \bar{z} = q_1 - iq_2. \tag{C.27}$$

Let us name $\Xi_0^{m,n}$ the coefficients in the new basis. For the spin $s = \frac{\ell}{2}$ subspace,

$$\sum_{m+n=\ell} \Lambda_0^{m,n} X_{m,n} = \sum_{m+n=\ell} \Xi_0^{m,n} Y_{m,n} \,. \tag{C.28}$$

Therefore,

$$Y_{m,n} = i^{m-n}(q_2)^\ell + \text{'higher'}, \quad (\ell = m+n), \tag{C.29}$$

and hence each vector in the basis $\{Y_{m,n}\}$ ($m+n=\ell=2s$) of the spin-$s$ subspace has a non-vanishing component along the lowest weight vector $X_{0,\ell} \propto (q_2)^\ell$. Here, 'higher' stands for the higher-weight terms containing at least one $q_1$.

Under the rotation $R$, the $z$'s transform as

$$z' = e^{\frac{i}{2}\theta} z, \quad \bar{z}' = e^{-\frac{i}{2}\theta} \bar{z}, \tag{C.30}$$

and consequently

$$U^{-1} Y_{m,n} U = e^{\frac{i}{2}(m-n)\theta} Y_{m,n}. \tag{C.31}$$

This implies that

$$U^{-1} \Big( \sum_{m+n=\ell} \Lambda_0^{m,n} X_{m,n} \Big) U = U^{-1} \Big( \sum_{m+n=\ell} \Xi_0^{m,n} Y_{m,n} \Big) U$$

$$= \sum_{m+n=\ell} \Xi_0^{m,n} e^{\frac{i}{2}(m-n)\theta} Y_{m,n} \tag{C.32}$$

$$= \sum_{m+n=\ell} \Xi_0^{m,n} e^{\frac{i}{2}(m-n)\theta} i^{m-n} (q_2)^\ell + \text{'more'}.$$

We thus see that the coefficient of the lowest-weight basis vector $X_{0,\ell}$ in $U^{-1} \big( \Sigma \Lambda_0^{m,n} X_{m,n} \big) U$ contains all the information on the exact symmetry $\Lambda^{\text{AdS}}$: its Fourier coefficients give directly the coefficients $\Xi_0^{m,n}$ or equivalently, through the change of basis (C.28), the coefficients $\Lambda_0^{m,n}$ that characterize $\Lambda^{\text{AdS}}$.

For the spin $s$ representation, there are $(2s+1)$ Fourier exponentials in the expansion of the above left-hand side along the $X_{(0,\ell)}$, namely, $e^{-is\theta}$, $e^{i(-s+1)\theta}$, ..., $e^{i(s-1)\theta}$, and $e^{is\theta}$. This precisely matches the number of coefficients $\Lambda_0^{m,n}$ ($m+n = \ell = 2s$), as it should from what we have just seen. Also note that half-integer spins have Fourier exponentials with half-integer frequencies ($e^{\frac{i}{2}\theta}$, $e^{\frac{3i}{2}\theta}$, etc.) and thus correspond to anti-periodic functions, obeying Neveu–Schwarz-like boundary conditions.

Recapitulating our analysis, we found that the lowest-weight components $\Lambda^{\text{AdS}\,0,\ell}(\varphi)$ of the $\text{sl}(2|\mathbb{R})$ spin-$s$ generators of the exact symmetries of the AdS connection contain Fourier components with frequencies $-s$, $-s+1$, ..., $s-1$, $s$. From the knowledge of these lowest-weight components, we can reconstruct the complete symmetry, either by applying the route inverse to the one explained above — viz. read the $\Lambda_0^{m,n}$ from the Fourier coefficients — or, alternatively and equivalently, by following a method close to the analysis of asymptotic symmetries given in the text. This method proceeds as follows. One solves the symmetry equation (C.25). In this case, it amounts to solving

$$(\Lambda^{\text{AdS}})' - \tfrac{1}{2}[X_{11} + X_{22}, \Lambda^{\text{AdS}}] = 0, \qquad (C.33)$$

starting from their lowest-weight components. The lowest-weight components of the equation give the coefficient $\Lambda^{\text{AdS}\,1,\ell-1}(\theta)$ of the symmetry generators along the basis vectors $X_{1,\ell-1}$ in terms of the coefficients $\Lambda^{\text{AdS}\,0,\ell}(\theta)$, then the next equations give $\Lambda^{\text{AdS}\,2,\ell-1}(\theta)$, etc. (see Subsection 2.2.2). The last, highest-weight component equations, which give the variation of the highest-weight component of the connection, are identically fulfilled because we are considering an exact symmetry.

### C.2.2 Wedge Subalgebra

The above way of describing the symmetries of the AdS connection shows explicitly how the symmetries are embedded in the algebra of asymptotic symmetries, which are constructed from the lowest-weight components in exactly the same manner. A generic asymptotic symmetry is characterized, for each spin representation, by an arbitrary periodic (integer spin) or anti-periodic (half-integer spin) function $\Lambda^{0,\ell}(\theta)$. Only the frequencies $-s \leq k \leq s$ correspond to the AdS symmetries, and the higher Fourier components correspond to asymptotic symmetries that are not background symmetries. Thus, for instance, in the case of the bosonic higher-spin algebra, the following Fourier components of the bosonic $\mathcal{W}_\infty$ algebra of [1] generate the hs(1, 1) algebra:

$$L_{-1}, L_0, L_1, \tag{C.34a}$$

$$M_{-2}, M_{-1}, M_0, M_1, M_2, \tag{C.34b}$$

$$N_{-3}, N_{-2}, N_{-1}, N_0, N_1, N_2, N_3, \tag{C.34c}$$

$$\ldots,$$

where we have also renamed $M^{(3)} \equiv M$ and $M^{(4)} \equiv N$, as in the main text. The higher Fourier components (e.g. $L_2$ or $M_3$, or $N_{-4}$, etc.) generate asymptotic symmetries which are beyond hs(1, 1). In the case of the $N = 1$ Supersymmetry, one must add $Q_{-1/2}, Q_{1/2}, R_{-3/2}, R_{-1/2}, R_{1/2}, R_{3/2}$, etc. in order to get the superalgebra shs(1, 1) from the corresponding super-$\mathcal{W}_\infty$ (again following the conventions given in the text we rename $M^{(3/2)}$ as R). For $N \geq 2$ extended Supersymmetry, there is an additional color index as well as the zero modes $B_0^A$ of the affine currents.[1]

There is one important point that should be stressed, however. Even when restricted to these generators, the (super)-$\mathcal{W}_\infty$ algebras differ from the original bulk superalgebra shs$(N, 2|\mathbb{R})$ because of nonlinear terms. Nevertheless, as we shall show below, the linear terms reproduce exactly the superalgebras shs$(N, 2|\mathbb{R})$. Furthermore, the central charges vanish when restricted to this sector, provided the generators – determined up to a constant — are adjusted to be equal to zero on the AdS$_3$ connection. It is in that sense that the algebras shs$(N, 2|\mathbb{R})$ are embedded in the super-$\mathcal{W}_\infty$ algebras. The algebra formed by the generators $\{M_n^{(i/2+1)}\}$ with $|n| \leq \frac{i}{2}$ (without the nonlinear terms) is called the *wedge algebra* [121, 144]. Hence, one can say that, up to the nonlinear terms, shs$(N, 2|\mathbb{R})$ is embedded in s$\mathcal{W}_\infty$ as its wedge subalgebra. We also comment on these points in Chapter 3.

The emergence of nonlinear terms is easy to understand. Although the lowest-weight components of the asymptotic symmetries are the same for all connections asymptotic to the anti-de Sitter connection, the higher-weight ones depend on the configuration. This is because the solution to the recursive equations determining them depends on the connection (see Subsection 2.2.2). Thus ,even if we start with a lowest-weight component that corresponds to an exact symmetry of the AdS$_3$ connection, the solution involves the deviations from the AdS$_3$ background as one works one's way up. For that reason, the residual transformations of shs($N, 2|\mathbb{R}$) depend on the configuration and the algebra of their generators are nonlinearly deformed.

---

[1] Only the zero modes appear because the $B^A$ are in the spin-zero representation.

It is instructive to check explicitly the above embedding by direct computation. We shall partly do so here by computing some Poisson bracket relations, chosen for their simplicity. Translating the asymptotic relations of Subsection 2.2.3 to Fourier modes and performing the redefinition

$$\mathrm{L}_{n+m} \to \mathrm{L}_{n+m} - \tfrac{1}{4} k \delta_{n+m,0} \tag{C.35}$$

we obtain, in particular, the following relations

$$[\mathrm{Q}_n, \mathrm{Q}_m] = 2k\, \delta_{n+m,0} \left(n^2 - \tfrac{1}{4}\right) + 2\mathrm{L}_{n+m}, \tag{C.36a}$$

$$[\mathrm{L}_n, \mathrm{L}_m] = \tfrac{1}{2} k \delta_{n+m,0}\, n(n^2 - 1) + (n-m)\mathrm{L}_{n+m}, \tag{C.36b}$$

$$[\mathrm{R}_n, \mathrm{R}_m] = \frac{1}{18} k \delta_{n+m,0} \left(n^2 - \tfrac{1}{4}\right)\left(n^2 - \tfrac{9}{4}\right) - 20 \mathrm{N}_{n+m}$$
$$\qquad + \tfrac{1}{36}\left(6(n^2 + m^2) - 8nm - 9\right) \mathrm{L}_{n+m} + \text{`Q} \times \text{Q'}, \tag{C.36c}$$

$$[\mathrm{M}_n, \mathrm{M}_m] = \tfrac{1}{288} k\, \delta_{n+m,0}\, n(n^2-1)(n^2-4) - 5(n-m)\mathrm{N}_{n+m}$$
$$\qquad + \tfrac{1}{144}\left(2(n^3 - m^3) - 3nm(n-m) - 8(n-m)\right) \mathrm{L}_{n+m}$$
$$\qquad + \text{`Q} \times \text{Q'} + \text{`L} \times \text{L'}, \tag{C.36d}$$

$$[\mathrm{L}_n, \mathrm{M}_m^{(\ell/2+1)}] = \left(\tfrac{1}{2}\ell n - m\right) \mathrm{M}_{n+m}^{(\ell/2+1)} \quad (\ell \neq 2), \tag{C.36e}$$

$$[\mathrm{Q}_n, \mathrm{M}_m^{(\ell/2+1)}] = (\ell+1)\mathrm{M}_{n+m}^{((\ell+3)/2)} \quad (\ell \text{ odd} > 1), \tag{C.36f}$$

$$[\mathrm{Q}_n, \mathrm{M}_m^{(\ell/2+1)}] = \left(n - \tfrac{m}{\ell}\right) \mathrm{M}_{n+m}^{((\ell+1)/2)} \quad (\ell \text{ even}). \tag{C.36g}$$

Here, the '( )' in the right-hand side refers to quadratics in the generators projected onto the mode $(m + n)$. We observe that, as expected, when we restrict the Fourier modes to the subsector $\{\mathrm{M}_n^{(i/2+1)}\}$ with $|n| \leq i/2$, the central charges all vanish and the linear terms on the right-hand side of the above relations only contain modes belonging to that subsector. Up to nonlinear terms, this subsector thus forms a subalgebra.

Furthermore, one can also verify that these Poisson bracket relations are identical to those of the bulk gauge algebra shs($N, 2|\mathbb{R}$) once the appropriate redefinitions are made. Indeed, the analog of the above relations for

shs($N, 2|\mathbb{R}$) are, in the basis of the $X$'s, given by

$$[X_n^{(3/2)}, X_m^{(3/2)}] = -iX_{n+m}^{(2)}, \tag{C.37a}$$

$$[X_n^{(2)}, X_m^{(2)}] = 2(n-m)X_{n+m}^{(2)}, \tag{C.37b}$$

$$[X_n^{(5/2)}, X_m^{(5/2)}] = -iX_{n+m}^{(4)} + \tfrac{i}{2}\left(6(n^2+m^2) - 8nm - 9\right)X_{n+m}^{(2)}, \tag{C.37c}$$

$$[X_n^{(3)}, X_m^{(3)}] = 4(n-m)X_{n+m}^{(4)} \tag{C.37d}$$
$$- 2\left(2(n^3-m^3) - 3nm(n-m) - 8(n-m)\right)X_{n+m}^{(2)},$$

$$[X_n^{(2)}, X_m^{(\ell/2+1)}] = 2\left(\tfrac{1}{2}\ell n - m\right)X_{n+m}^{(\ell/2+1)}, \tag{C.37e}$$

$$[X_n^{(3/2)}, X_m^{(\ell/2+1)}] = -iX_{n+m}^{((\ell+3)/2)} \qquad (\ell \text{ odd}), \tag{C.37f}$$

$$[X_n^{(3/2)}, X_m^{(\ell/2+1)}] = (\ell n - m)X_{n+m}^{((\ell+1)/2)} \qquad (\ell \text{ even}), \tag{C.37g}$$

where we have used the definition

$$X_n^{(\ell/2+1)} \equiv \tfrac{1}{2i}(q_1)^{2n}(q_2)^{\ell-2n} \quad \left(|n| \leq \tfrac{1}{2}\ell\right). \tag{C.38}$$

Making the redefinition

$$X_n^{(\ell/2+1)} \mapsto \gamma^\ell X_n^{(\ell/2+1)} \tag{C.39}$$

with

$$\gamma^1 = \sqrt{i},\ \gamma^2 = \frac{1}{2},\ \gamma^3 = \pm\frac{\sqrt{-i}}{6},\ \gamma^4 = \frac{\sqrt{-i}}{4}\gamma^3,\ \gamma^5 = \frac{\sqrt{i}}{120},\ \gamma^6 = \frac{1}{720}, \tag{C.40}$$

one then finds the relations (C.37g) match precisely (C.36g). We believe this illustration to be convincing enough, but the zealous reader shall find it straightforward to work out higher-order checks.

At this stage, one might wonder whether nonlinear deformations are constrained by generic arguments. The steps given below show that this is not the case: quadratic and higher-order nonlinearities are in not constrained in general, and can in principle appear.

### C.2.3 Possible Deformations

Algebraically, the situation we are facing is the following: we have a set of asymptotic symmetries generated by the generators $G_A$ ($A = \alpha, i$), which close in the Poisson bracket according to

$$[G_A, G_B] = K_{AB} + C^C{}_{AB}G_C + \tfrac{1}{2}D^{CD}{}_{AB}G_CG_D + \cdots, \tag{C.41}$$

where $K_{AB}$, $C^C{}_{AB}$, $D^{CD}{}_{AB}$, etc. are constants. Among these asymptotic symmetries, a subset, generated by the $G_\alpha$, leaves a background (here the AdS$_3$ connection) invariant, while the others, denoted by $G_i$, do not. Let the background be such that the charges-generators $G_A$ evaluated on it are zero, that is,

$$G_A|_{\text{BACKGROUND}} = 0. \tag{C.42}$$

In our AdS$_3$ situation, the $G_\alpha$ are the generators associated with the lowest Fourier modes, as described above, while the $G_i$ correspond to the higher Fourier modes.

Now, the transformations $\delta_\alpha F$ generated by the $G_\alpha$:

$$\delta_\alpha F = [G_\alpha, F], \tag{C.43}$$

where $F$ is an arbitrary function of the fields, have the following properties:

1. When evaluated on the background, the variation $\delta_\alpha F$ vanishes,

$$\delta_\alpha F|_{\text{BKGD}} = 0, \tag{C.44}$$

   since the background is strictly invariant under the transformation generated by $G_\alpha$.

2. Likewise,

$$[\delta_\alpha, \delta_\beta] F|_{\text{BKGD}} = f^\gamma{}_{\alpha\beta} \delta_\gamma F|_{\text{BKGD}}, \tag{C.45}$$

   where $f^\gamma{}_{\alpha\beta}$ are the structure constants of the background symmetry algebra.

It then follows from these two properties that

$$K_{A\alpha} = -K_{\alpha A} = 0, \tag{C.46}$$

since $\delta_\alpha G_A|_{\text{BKGD}} = K_{A\alpha}$, and that

$$C^\gamma{}_{\alpha\beta} = f^\gamma{}_{\alpha\beta}, \qquad C^i{}_{\alpha\beta} = 0, \tag{C.47}$$

because $[\delta_\alpha, \delta_\beta] F = [[G_\alpha, G_\beta], F]$ — an expression that reduces to $C^C{}_{\alpha\beta}[G_C, F]$ on the background. Note that the above argument says nothing about $D^{CD}{}_{\alpha\beta}$ or the higher order terms. Indeed, these can be different from zero and are actually found to be so in the present context.

## C.3 Extended Asymptotic Symmetry Algebra

In this section we reproduce the supersymmetric extended version of the asymptotic analysis of Section 2.2, where the simpler case of shs(1, 1) is treated. The steps being the same as in the non-extended case, we confine ourselves to point out the features which differ in the generic case.

### C.3.1 Boundary Conditions and Residual Gauge

We wish to define some boundary conditions at asymptotic infinity for our higher-spin gauge superconnection $\Gamma_\mu \in \text{shs}(N, 2|\mathbb{R})$ — and similarly for the other chiral copy. As mentioned in the text, nothing tells us what these boundary conditions should be, so that we need to try something that makes sense. However, we already have quite some amount of information at our disposal from which we could guess a plausible Ansatz. Indeed, on the one hand we already know, in some gauge, what the 'correct' asymptotic behavior is for the low-spin sector $\Gamma_\mu \in \text{osp}(N, 2|\mathbb{R}) \subset \text{shs}(N, 2|\mathbb{R})$ [60, 135] (see Subsection 2.1.3); and on the other hand we also know that in the bosonic case the 'highest-weight' boundary conditions proposed in [1] for $\Gamma_\mu \in \text{hs}(1,1) \subset \text{shs}(N, 2|\mathbb{R})$ do make sense, and they are given in Subsection 2.2.1. Thus, we are naturally led to propose the following asymptotic behavior for our full superconnection, in the same gauge:

$$\Gamma(x) \to [-X_{22} + \sum \Delta^{p, i_1 \cdots i_N}(x^\pm) X_{p, 0; i_1 \cdots i_N}] \mathrm{d}x^+, \qquad \text{(C.48a)}$$

$$\tilde{\Gamma}(x) \to [+X_{11} + \sum \tilde{\Delta}^{q, i_1 \cdots i_N}(x^\pm) X_{0, q; i_1 \cdots i_N}] \mathrm{d}x^-, \qquad \text{(C.48b)}$$

where we sum on repeated indices over all their possible values, and we use the basis written down in Appendix B. Note in particular that the values $p = 0$ and $q = 0$ occur when the degree $K = i_1 + i_2 + \cdots + i_N$ does not vanish. We also point out that the internal part (with only $\psi$-oscillators) is left completely arbitrary, that is, we do not impose any 'highest-weight' condition thereon, just as in the supergravity case. These are the boundary conditions we shall work with in the sequel. Even though there is no asymptotic restriction on the weights of the representations of the internal algebra, we continue to call the boundary conditions (C.48a) and (C.48b) the 'highest-weight' (resp. 'lowest-weight') gauge boundary conditions, in analogy with the non-extended cases ($N = 0$ or $N = 1$).

Given the AdS$_3$ boundary conditions (C.48a) and (C.48b), the next step is to look for the residual gauge transformations that act non-trivially at asymptotic infinity while leaving the boundary conditions intact. With gauge parameter $\Lambda(x)$, the infinitesimal gauge transformation of $\Gamma$ reads

$$\Gamma \to \Gamma' = \Gamma + \delta\Gamma, \qquad \text{where} \qquad \delta\Gamma = d\Lambda + [\Gamma, \Lambda]. \tag{C.49}$$

We see that, in order for $\Gamma'$ to retain the given asymptotics, $\Lambda$ cannot possibly depend on $r$ or $x^-$ to leading order at infinity. Moreover, the gauge transformations should not generate any other components than the highest-weight ones already present. A similar argument goes for the other chiral copy, where we see that in order for $\tilde{\Gamma}'$ to retain the boundary behavior (C.48b), $\tilde{\Lambda}$ cannot possibly depend on $r$ or $x^+$. Furthermore, the gauge transformations should not generate any other components than the lowest-weight ones already present in (C.48b). Summarizing, we find that the gauge transformations $\Lambda(x^+)$ and $\tilde{\Lambda}(x^-)$ must be chiral, respectively, antichiral at the least. These functions must be subject to further conditions in order to retain the boundary conditions, and this is the task we will undertake next, treating explicitly for definiteness the positive chirality sector (the negative chirality sector can be treated similarly).

To proceed further, we find it convenient to decompose the gauge transformations in stacks of successively higher $sl(2, \mathbb{R})$-spin layers. This is because, for each spin, the highest-weight components are the only ones that appear in the boundary conditions for the gauge connection. We thus write

$$\Lambda(x^+) = \sum \Lambda^{m,n;i_1,\cdots,i_N}(x^+) X_{m,n;i_1,\cdots,i_N} \equiv \Lambda^{\text{LW}} + \lambda, \tag{C.50}$$

with

$$\Lambda^{\text{LW}} = \sum_{i_1+\cdots+i_N \geq 2} \Lambda^{0,0;i_1,\cdots,i_N} X_{0,0;i_1,\cdots,i_N} + \sum_{i_1+\cdots+i_N \geq 1} \Lambda^{0,1;i_1,\cdots,i_N} X_{0,1;i_1,\cdots,i_N}$$
$$+ \sum_{\ell=2}^{\infty} \sum_{i_1,\cdots,i_N} \Lambda^{0,\ell;i_1,\cdots,i_N} X_{0,\ell;i_1,\cdots,i_N} \tag{C.51}$$

and

$$\lambda = \sum_{i_1+\cdots+i_N \geq 1} \Lambda^{1,0;i_1,\cdots,i_N} X_{1,0;i_1,\cdots,i_N} + \sum_{\ell=2}^{\infty} \sum_{i_1,\cdots,i_N} \Lambda^{1,\ell-1;i_1,\cdots,i_N} X_{1,\ell-1;i_1,\cdots,i_N}$$
$$+ \sum_{\ell=2}^{\infty} \sum_{i_1,\cdots,i_N} \Lambda^{2,\ell-2;i_1,\cdots,i_N} X_{2,\ell-2;i_1,\cdots,i_N} \tag{C.52}$$
$$+ \cdots + \sum_{\ell \geq s} \sum_{i_1,\cdots,i_N} \Lambda^{s,\ell-s;i_1,\cdots,i_N} X_{s,\ell-s;i_1,\cdots,i_N} + \cdots.$$

In plain words, we collected all the lowest-weight states, which are the states involving $X_{0,s;i_1,\cdots,i_N}$ in $\Lambda^{\text{LW}}$ and, at the same time, all higher weight states, involving $X_{m,n;i_1,\cdots,i_N}$ with $m > 0$, are packaged together in $\lambda$. We should also stress that, although this is not written explicitly, the sums in the above expressions are always restricted to total even degree. So, for instance, $i_1 + \cdots + i_N$ must be even in the first term in the right-hand side of the expression for $\Lambda^{\text{LW}}$, while it must be odd in the second term. Such a convention will always be adopted in the sequel.

The reason for proceeding in this manner is that the requirement that the asymptotic boundary conditions be preserved will be seen to determine $\lambda$ in terms of $\Lambda^{\text{LW}}$, much like $\Lambda^+$ and $\Lambda^3$ are determine in terms of $\Lambda^-$ in (2.13). Indeed, let us compute $\delta\Gamma = d\Lambda + [\Gamma, \Lambda]$. Structurally,

$$\delta\Gamma = \sum_{m,n;i_1,\cdots,i_N} \gamma^{m,n;i_1,\cdots,i_N}(x^+) X_{m,n;i_1,\cdots,i_N}, \tag{C.53}$$

where

$$\gamma^{m,n;i_1,\cdots,i_N} = \partial_+ \Lambda^{m,n;i_1,\cdots,i_N} + [\Gamma, \Lambda]^{m,n;i_1,\cdots,i_N}. \tag{C.54}$$

Since the only non-vanishing components of $\Gamma$ at infinity are $\gamma^{m,0;i_1,\cdots,i_N}$ (apart from $\gamma^{0,2;0,\cdots,0}$, which is fixed to be equal to $-1$), the requirement that these global gauge transformations do not alter the boundary conditions is that

$$\gamma^{m,1;i_1,\cdots,i_N} = \gamma^{m,2;i_1,\cdots,i_N} = \cdots = 0 \qquad \text{for} \qquad m = 0, 1, 2, \ldots \tag{C.55}$$

or, equivalently,

$$\gamma^{s,\ell-s;i_1,\cdots,i_N} = 0 \qquad \text{for} \qquad \ell \geq s+1, \quad s = 0, 1, 2, \ldots. \tag{C.56}$$

The highest-weight terms $\gamma^{m,0;i_1,\cdots,i_N}$ are not constrained to be zero and are equal to $\Delta^{m,i_1\cdots i_N}$, according to (C.48a).

Now, since

$$[X_{22}, X_{m,n;i_1,\cdots,i_N}] \sim X_{m-1,n+1;i_1,\cdots,i_N} \quad (m \geq 1), \tag{C.57}$$

one may solve recursively the conditions for the higher-weight coefficients $\Lambda^{1,n;i_1,\cdots,i_N}$, $\Lambda^{2,n;i_1,\cdots,i_N}$, ..., given the lowest-weight ones $\Lambda^{0,k;i_1,\cdots,i_N}$, along exactly the same lines as developed in [1]. One starts from the lowest-weight conditions $\gamma^{0,\ell;i_1,\cdots,i_N} = 0$ ($\ell \geq 1$) to determine the level-one coefficients $\Lambda^{1,\ell-1;i_1,\cdots,i_N}$. Then one proceeds to solving the level-one conditions $\gamma^{1,\ell-1;i_1,\cdots,i_N} = 0$ ($\ell \geq 2$)) to determine the level-two coefficients $\Lambda^{2,\ell-2;i_1,\cdots,i_N}$. One walks one's way up step by step in this fashion. The last set of conditions $\gamma^{\ell-1,1;i_1,\cdots,i_N} = 0$ ($\ell \geq 1$) determine the highest-weight coefficients $\Lambda^{\ell,0;i_1,\cdots,i_N}$. It should be stressed that, during the process, the higher-weight coefficients depend not only on the lowest-weight coefficients but also on their derivatives. The solutions depend also on the (non-zero) coefficients of the connection and their derivatives.

Collecting the results of the above structure analysis, we conclude that the gauge transformations that leave the boundary conditions intact are completely specified by the lowest-weight components $\Lambda^{0,k;i_1,\cdots,i_N}$ of the gauge function, while all higher-weight components are determined functionally in terms of these lowest-weight components of the gauge function and the highest-weight components of the original gauge connection. Notice that, in exactly the same manner as in the higher-spin bosonic case as well as in the extended supergravity models, the solution for the higher-weight components of the gauge function $\Lambda$ in terms of the lowest-weight ones (and in terms of the free gauge potential components $\Delta^{m,i_1\cdots i_N}$ and their derivatives) is nonlinear. It is this feature that will render the resulting asymptotic algebra also nonlinear.

### C.3.2 Asymptotic Symmetry Superalgebra

The functional of gauge transformation $G[\Lambda]$ is given by

$$G[\Lambda] = \int_{\Sigma_2} \sum \Lambda^{m,n;i_1,\cdots,i_N} \mathcal{G}_{m,n;i_1,\cdots,i_N} + S_\infty, \qquad \text{(C.58)}$$

where $\mathcal{G}_{m,n;i_1,\cdots,i_N}$ are the Gauss law constraints for our theory and $S_\infty$ is the usual boundary term. Again, after a straightforward integration by part one finds (to leading order)

$$S_\infty = \oint \sum_{s,i_1,\cdots,i_N} \Lambda^{0,s;i_1,\cdots,i_N} \Delta^s{}_{i_1,\cdots,i_N}, \qquad \text{(C.59)}$$

where we have redefined the $\Delta$'s through the absorption of the factors that appear in front of the integral, which we denote by $\alpha_{s,0;i_1,\cdots,i_N}$,

$$\Delta^{s;i_1,\cdots,i_N} \equiv \Gamma^{s,0;i_1,\cdots,i_N} \alpha_{s,0;i_1,\cdots,i_N}. \qquad \text{(C.60)}$$

We thus see that (up to those factors) the generators of the asymptotic symmetries are indeed nothing but the leading terms in the asymptotic expansion of the highest-weight components $\Gamma^{s,0;i_1,\cdots,i_N}$ of the gauge connection.

The algebra of the asymptotic symmetry generators $\Delta^{s;i_1,\cdots,i_N}$ can be read off by equating their variations under an arbitrary asymptotic symmetry transformation, computed in two different ways. First, $\delta\Delta^{s;i_1,\cdots,i_N}$ can be derived from the gauge variation formula,

$$\begin{aligned}\delta\Delta^{s;i_1,\cdots,i_N}(\theta) &= \delta\Gamma^{s,0;i_1,\cdots,i_N}(\theta)\,\alpha_{s,0;i_1,\cdots,i_N} \\ &= \left(\partial\Lambda^{s,0;i_1,\cdots,i_N} + [\Gamma,\Lambda]^{s,0;i_1,\cdots,i_N}\right)\alpha_{s,0;i_1,\cdots,i_N}\,,\end{aligned} \quad (C.61)$$

with the $\Lambda^{m,n;i_1,\cdots,i_N}$ determined from the lowest-weight $\Lambda^{0,s;i_1,\cdots,i_N}$ along the lines explained above. Second, $\delta\Delta^{s;i_1,\cdots,i_N}$ can also be obtained via

$$\delta\Delta^{s;i_1,\cdots,i_N}(\theta) = \{\Delta^{s;i_1,\cdots,i_N}(\theta), \oint \sum_{s',j_1,\cdots,j_N} \Lambda^{0,s';j_1,\cdots,j_N}\Delta^{s'}{}_{j_1,\cdots,j_N}(\theta)\}_{\text{PB}}\,. \quad (C.62)$$

Here, $\theta$ denotes the angular coordinate of the asymptotic infinity. Equating these two ways of computing $\delta\Delta^{s;i_1,\cdots,i_N}$ yields the Poisson brackets

$$\{\Delta^{s;i_1,\cdots,i_N}(\theta), \Delta^{s';j_1,\cdots,j_N}(\theta')\}_{\text{PB}} \qquad \text{for} \qquad s,s' \in \mathbb{N}\,. \quad (C.63)$$

It is evident that this algebra is closed, since the variations $\delta\Gamma^{s,0;i_1,\cdots,i_N} = \gamma^{s,0;i_1,\cdots,i_N}$, determined through the recursive procedure explained above, are functionals of $\Gamma^{s,0;i_1,\cdots,i_N} \sim \Delta^{s;i_1,\cdots,i_N}$ only (in addition to depending linearly on the independent gauge parameters $\Lambda^{0,s;i_1,\cdots,i_N}$). However, the functional dependence of $\gamma^{s,0;i_1,\cdots,i_N}$ on $\Delta^{s;i_1,\cdots,i_N}$ is nonlinear, which implies that the algebra of the $\Delta$'s is nonlinear. The terms independent of $\Delta$ and linear in the gauge parameters corresponds to the central charges. Although nonlinear, the algebra obeys of course the Jacobi identity since the Poisson bracket does.

The actual computation of the algebra of the $\Delta^{s;i_1,\cdots,i_N}$'s is rather cumbersome but it can be identified to be of the $s\mathcal{W}_\infty$ type by following a general argument similar to the one given in the text for the non-extended case. For $N \geq 2$ extended supersymmetry, the derivation proceeds essentially in the same way. The salient new features that arise are have been pointed out in the main text, at the very end of Chapter 2.

# Details in Dimension D

In this appendix we detail some of the computations of the main text. We start with Appendix D.1, were the most important $\gamma$-matrix identities which we make use of are recapitulated (without proof). Then, in Appendix D.2 we detail the matching of our electromagnetic vertices with those found by Sagnotti and Taronna in [25], and in Appendix D.3 we explicitate some computational steps related to the finding of our non-abelian $2-\frac{5}{2}-\frac{5}{2}$ vertices. Finally, Appendix D.4 is concerned with reproducing the reasoning of Section 5.4 in the gravitational case, again arriving at the conclusion that abelian vertices also preserve the original gauge symmetries in our context.

## D.1 Identities for Dirac Matrices

Let us summarize, without proof, the various $\gamma$-matrix identities that we use in the bulk of this work. We point out once again that, in the derivation of some of them we have used the Mathematica package 'GAMMA', discussed in [194]. Note that the conventions for our matrices are given in Appendix A.2.

One of our most-used identities is the following:

$$\eta^{\mu\nu|\alpha\beta} + \tfrac{1}{2}\gamma^{\mu\nu\alpha\beta} = -\tfrac{1}{2}\eta^{\mu\nu}\gamma^{\alpha\beta} + \tfrac{1}{2}\gamma^{\mu}\gamma^{\nu}\gamma^{\alpha\beta} - 2\gamma^{[\mu}\eta^{\nu][\alpha}\gamma^{\beta]} \qquad \text{(D.1)}$$
$$= 2\left(\eta^{\mu\nu|\alpha\beta} - \tfrac{1}{4}\eta^{\mu\nu}\gamma^{\alpha\beta}\right) + \tfrac{1}{4}\left(\gamma^{\mu}\gamma^{\nu}\gamma^{\alpha\beta} + \gamma^{\alpha\beta}\gamma^{\mu}\gamma^{\nu}\right),$$

the two right-hand sides of which are related by

$$2\gamma^{[\mu}\eta^{\nu][\alpha}\gamma^{\beta]} = \tfrac{1}{4}\left(\gamma^{\mu}\gamma^{\nu}\gamma^{\alpha\beta} - \gamma^{\alpha\beta}\gamma^{\mu}\gamma^{\nu}\right) - 2\eta^{\mu\nu|\alpha\beta}. \qquad \text{(D.2)}$$

Another form of the above identity can be obtained by making use of

$$\left[\gamma^\mu, \gamma^{\alpha\beta}\right] = 4\eta^{\mu[\alpha}\gamma^{\beta]}, \tag{D.3}$$

with which we can rewrite (D.1) as

$$\gamma^{\alpha\nu\rho\sigma} = -2\eta^{\alpha\nu|\rho\sigma} + \tfrac{1}{2}\left(\gamma^\alpha \gamma^{\rho\sigma}\gamma^\nu - \gamma^\nu\gamma^{\rho\sigma}\gamma^\alpha\right), \tag{D.4}$$

or even as

$$\gamma^{\lambda\nu\rho\sigma} = 2\eta^{\lambda\nu|\rho\sigma} + \tfrac{1}{2}\left(\gamma^{\lambda\nu}\gamma^{\rho\sigma} + \gamma^{\rho\sigma}\gamma^{\lambda\nu}\right), \tag{D.5}$$

which is our favorite form. We also make intensive use of the generic decomposition

$$\gamma^{\mu_1\cdots\mu_n} = \tfrac{1}{2}\left(\gamma^{\mu_1}\gamma^{\mu_2\cdots\mu_n} - (-)^n \gamma^{\mu_2\cdots\mu_n}\gamma^{\mu_1}\right), \tag{D.6}$$

which we most use for $n = 4, 5$. In particular, for $n = 3$ we also have the handy identity

$$\gamma^{\sigma\tau\lambda} = \tfrac{1}{2}\left(\gamma^\sigma\gamma^\tau\gamma^\lambda - \gamma^\lambda\gamma^\tau\gamma^\sigma\right). \tag{D.7}$$

More involved relations are also used, and one of them is

$$\gamma^{\mu\nu\lambda\alpha\beta}{}_{,\tau\rho\sigma} = -60\,\delta^{[\mu\nu\lambda}_{\tau\rho\sigma}\gamma^{\alpha\beta]} + 15\,\delta^{[\mu}_{[\tau}\gamma_{\rho\sigma]}{}^{\nu\lambda\alpha\beta]}. \tag{D.8}$$

## D.2 Comparative Study of Electromagnetic Vertices

Let us compare our obtained vertices with the corresponding expressions given in [25] by Sagnotti and Taronna. We will denote a *p*-derivative off-shell vertex of ours as $V^{(p)}$, and its 'Sagnotti–Taronna' (ST) counterpart as $V^{(p)}_{\text{ST}}$. The corresponding vertex in the transverse-traceless gauge will be denoted as $V^{(p)}_{\text{TT}}$ [1].

### D.2.1  1−3/2−3/2 Vertices Compared

**1-Derivative 1−3/2−3/2 Vertex**

Our 1-derivative off-shell $1-\tfrac{3}{2}-\tfrac{3}{2}$ vertex is given by

$$V^{(1)} = \bar\psi_\mu F^{+\mu\nu}\psi_\nu = \bar\psi_\mu\bigl(\eta^{\mu\nu|\alpha\beta} + \tfrac{1}{2}\gamma^{\mu\nu\alpha\beta}\bigr)F_{\alpha\beta}\psi_\nu. \tag{D.9}$$

---

[1] Both the spins $m + \tfrac{1}{2}$ and $m + \tfrac{3}{2}$ will have one vertex with $2m + 1$ derivatives. Our notation should not cause any confusion, as we will be considering one spin at a time.

To see what it reduces to in the TT gauge, let us use Identity (D.1), yielding

$$V^{(1)} = 2\left(\bar\psi_\mu F^{\mu\nu}\psi_\nu - \tfrac{1}{4}\bar\psi_\mu \slashed{F}\psi^\mu\right) + \tfrac{1}{4}\left(\bar{\slashed\psi}\gamma^\mu \slashed{F}\psi_\mu + \bar\psi_\mu \slashed{F}\gamma^\mu \slashed\psi\right). \quad \text{(D.10)}$$

On the other hand, the 1-derivative ST vertex reads [25]

$$\begin{aligned}V^{(1)}_{\text{ST}} =& -(\partial_\mu\bar\psi_\nu)\psi^\nu A^\mu + \bar\psi_\mu\psi_\nu(\partial^\mu A^\nu) + \bar\psi^\mu(\partial_\nu\psi_\mu)A^\nu \\ & -\bar\psi_\mu(\partial^\mu\psi^\nu)A_\nu + (\partial_\mu\bar\psi_\nu)\psi^\mu A^\nu - \bar\psi_\mu\psi_\nu(\partial^\nu A^\mu).\end{aligned} \quad \text{(D.11)}$$

Integrating by parts the 1rst, 4th and 5th terms on the right-hand side, we obtain

$$V^{(1)}_{\text{ST}} \doteq 2\bar\psi_\mu F^{\mu\nu}\psi_\nu + 2\bar\psi_\mu A\cdot\partial\psi^\mu + \bar\psi_\mu(\partial\cdot A)\psi^\mu + (\partial\cdot\bar\psi)A\cdot\psi - \bar\psi\cdot A(\partial\cdot\psi). \quad \text{(D.12)}$$

Let us take the 2nd term on the right-hand side and replace $\eta^{\alpha\beta} = \gamma^{(\alpha}\gamma^{\beta)}$ in the operator $(A\cdot\partial)$. Also in the 3rd term we replace $\eta^{\alpha\beta} = \gamma^\alpha\gamma^\beta - \gamma^{\alpha\beta}$ in $(\partial\cdot A)$. The result is

$$2\bar\psi_\mu A\cdot\partial\psi^\mu + \bar\psi_\mu(\partial\cdot A)\psi^\mu \doteq -\tfrac{1}{2}\bar\psi_\mu \slashed{F}\psi^\mu + \bar\psi_\mu \slashed{A}(\slashed\partial\psi^\mu) - (\slashed\partial\bar\psi_\mu)\slashed{A}\psi^\mu, \quad \text{(D.13)}$$

which, when plugged into the vertex (D.12) gives

$$V^{(1)}_{\text{ST}} \doteq 2\left(\bar\psi_\mu F^{\mu\nu}\psi_\nu - \tfrac{1}{4}\bar\psi_\mu \slashed{F}\psi^\mu\right) + \left[\bar\psi_\mu \slashed{A}(\slashed\partial\psi^\mu) - \bar\psi\cdot A(\partial\cdot\psi) + \text{h.c.}\right]. \quad \text{(D.14)}$$

It is obvious that both the off-shell vertices (D.10) and (D.14) reduce in the TT gauge to

$$V^{(1)}_{\text{TT}} = 2\left(\bar\psi_\mu F^{\mu\nu}\psi_\nu - \tfrac{1}{4}\bar\psi_\mu \slashed{F}\psi^\mu\right). \quad \text{(D.15)}$$

This is precisely the on-shell 1-derivative vertex reported by Metsaev [62]. To see explicitly that the off-shell vertices are also equivalent, we subtract (D.14) from (D.10) to get

$$V^{(1)} - V^{(1)}_{\text{ST}} \doteq \tfrac{1}{4}\left(\bar{\slashed\psi}\gamma^\mu \slashed{F}\psi_\mu + \bar\psi_\mu \slashed{F}\gamma^\mu \slashed\psi\right) - \left(\bar\psi_\mu \slashed{A}(\slashed\partial\psi^\mu) - \bar\psi\cdot A(\partial\cdot\psi) + \text{h.c.}\right). \quad \text{(D.16)}$$

Now we make use of the identity

$$\left[\gamma^\mu, \gamma^{\alpha\beta}\right] = 4\eta^{\mu[\alpha}\gamma^{\beta]}, \quad \text{(D.17)}$$

in order to be able to pass $\gamma^\mu$ past $\slashed{F}$ in both the terms in the parentheses on the right-hand side of Eq. (D.16). As a result, we will obtain, among others, the term $\tfrac{1}{2}\bar{\slashed\psi}\slashed{F}\slashed\psi$, in which we replace $\slashed{F} = \slashed\partial\slashed{A} - \partial\cdot A$. Now in all the resulting terms we perform integrations by parts such that no derivative acts on the photon field. The final result is

$$V^{(1)} - V^{(1)}_{\text{ST}} \doteq \left(2\bar{\psi}^{[\mu}A^{\nu]}\gamma_\mu \left(\slashed{\partial}\psi_\nu - \partial_\nu\slashed{\psi}\right) - \bar{\psi}^\mu A^\nu \gamma_{\mu\nu}\left(\partial \cdot \psi - \slashed{\partial}\slashed{\psi}\right) + \text{h.c.}\right). \tag{D.18}$$

This is manifestly $\Delta$-exact modulo d, which proves the equivalence of the off-shell vertices:

$$V^{(1)} \approx V^{(1)}_{\text{ST}}. \tag{D.19}$$

**2-Derivatives 1−3/2−3/2 Vertex**

Next, we consider the 2-derivative vertex,

$$V^{(2)} = \left(\bar{\Psi}_{\mu\nu}\,\gamma^{\mu\nu\alpha\beta\lambda}\,\Psi_{\alpha\beta}\right)A_\lambda \approx -2\left(\bar{\Psi}_{\mu\nu}\,\gamma^\rho\,\Psi^{\mu\nu}\right)A_\rho. \tag{D.20}$$

One can use the definition $\Psi_{\mu\nu} = 2\partial_{[\mu}\psi_{\nu]}$ to rewrite it as

$$V^{(2)} \approx -4\bar{\psi}_\alpha \overleftarrow{\partial}_\mu \slashed{A}\,\partial^\mu\psi^\alpha + 2\left(\bar{\psi}_\alpha \overleftarrow{\partial}_\mu \slashed{A}\,\partial^\alpha\psi^\mu + \text{h.c.}\right). \tag{D.21}$$

In the 1st term, we can use the 3-box rule, already given above,

$$2\partial_\mu X \partial^\mu Y = \Box(XY) - X(\Box Y) - (\Box X)Y, \tag{D.22}$$

and perform a double integration by parts in order to have a $\Box$ acting on the photon field. In the 2nd term on the right hand side of (D.20) one can integrate by parts w.r.t. any of the derivatives. When the derivative acts on the photon field, one can use $\partial_\mu A_\nu = F_{\mu\nu} + \partial_\nu A_\mu$ to rewrite it in terms of the field strength. The result is

$$\begin{aligned}V^{(2)} \approx &-2\left[\left(\bar{\psi}_\alpha \gamma^\mu \partial^\alpha \psi^\nu\right)\partial_\mu A_\nu + \text{h.c.}\right] + \left(\bar{\psi}_\alpha \gamma^\mu \partial^\alpha \psi^\nu - \bar{\psi}^\mu \overleftarrow{\partial}{}^\alpha \gamma^\nu \psi_\alpha\right)F_{\mu\nu} \\ &- 2\bar{\psi}_\alpha \Box \slashed{A}\,\psi^\alpha + 2\left[\bar{\psi}_\alpha \slashed{A}\left(\Box\psi^\alpha - \partial^\alpha \partial \cdot \psi\right) + \text{h.c.}\right].\end{aligned} \tag{D.23}$$

Now, in the last term of the first line we perform integration by parts so that no derivative acts on the photon field. On the other hand, the last term in the second line is $\Delta$-exact, and therefore can be dropped. Thus we are left with

$$\begin{aligned}V^{(2)} \approx\ & 2\left(\bar{\psi}_\alpha \gamma^\mu \partial^\alpha \psi^\nu - \bar{\psi}^\mu \overleftarrow{\partial}{}^\alpha \gamma^\nu \psi_\alpha\right)F_{\mu\nu} \\ &+ 2\left[\left(\bar{\psi}_\alpha \partial^\alpha \slashed{\partial}\psi^\nu + \bar{\psi}_\alpha \overleftarrow{\slashed{\partial}}\,\partial^\alpha \psi^\nu\right)A_\nu + \text{h.c.}\right] - 2\bar{\psi}_\alpha \Box \slashed{A}\,\psi^\alpha.\end{aligned} \tag{D.24}$$

As one reads off the 2-derivative ST vertex, it gives

$$\begin{aligned}V^{(2)}_{\text{ST}} =\ &-(\partial \cdot \bar{\psi})\slashed{A}\,\partial \cdot \psi - \bar{\slashed{\psi}}\,\psi^\nu \partial_\nu \partial \cdot A + \bar{\slashed{\psi}}\,\overleftarrow{\partial}_\nu \psi^\nu \partial \cdot A + \bar{\psi}^\mu(\partial_\mu \slashed{\psi})\partial \cdot A \\ &- \bar{\psi}^\mu \slashed{\psi}\,\partial_\mu \partial \cdot A - \bar{\slashed{\psi}}\,(\partial \cdot \psi)\slashed{\psi}\,\partial \cdot A - (\partial \cdot \bar{\psi})\slashed{\psi}\,\partial \cdot A - \bar{\psi}^\mu \overleftarrow{\partial}{}^\nu \gamma^\alpha \psi_\nu \partial_\mu A_\alpha \\ &- \bar{\psi}^\mu \gamma^\alpha(\partial_\mu \psi_\nu)\partial^\nu A_\alpha + \bar{\psi}^\mu \gamma^\alpha \psi^\nu \partial_\mu \partial_\nu A_\alpha + \bar{\psi}^\mu \overleftarrow{\partial}{}^\nu \gamma^\alpha (\partial_\mu \psi_\nu) A_\alpha,\end{aligned} \tag{D.25}$$

As we mentioned already, in this form it is not evident at all that this vertex vanishes for $D = 4$. Let us integrate by parts the 2nd and 3rd terms in the third line above w.r.t. $\partial_\mu$. The 2nd term in the first line and the 1st term in the second line contain the gradient of $\partial \cdot A$; we integrate by parts the gradient in both these terms. Thus we have

$$V_{\text{ST}}^{(2)} \doteq - 2\bar{\psi}_\mu \gamma^\alpha (\partial^\mu \psi^\nu) \partial_\nu A_\alpha - (\partial \cdot \bar{\psi})(\partial_\nu \slashed{A})\psi^\nu - 2\bar{\psi}^\mu \overleftarrow{\partial}^\nu \gamma^\alpha \psi_\nu \partial_\mu A_\alpha \\ - (\partial \cdot \bar{\psi})\slashed{A}\, \partial \cdot \psi + 2(\partial \cdot A)\big(\bar{\psi}^\mu \partial_\mu \slashed{\psi} + \bar{\slashed{\psi}}\overleftarrow{\partial}^\mu \psi_\mu\big) - (\partial \cdot \bar{\psi})\overleftarrow{\partial}_\nu \slashed{A}\psi^\nu. \tag{D.26}$$

Notice that the 2nd, 4th and 6th terms combine into a total derivative. One can rewrite the 1st and 3rd terms in terms of the photon field strength by using $\partial_\mu A_\nu = F_{\mu\nu} + \partial_\nu A_\mu$. Also, one can extract a $\Delta$-exact piece, by using EoMs: $\slashed{\partial}\psi_\mu - \partial_\mu \slashed{\psi} = 0$, in the term containing $(\partial \cdot A)$. This leaves us with

$$V_{\text{ST}}^{(2)} \approx - 2\left[\big(\bar{\psi}_\alpha \gamma^\mu \partial^\alpha \psi^\nu\big)\partial_\mu A_\nu + \text{h.c.}\right] + 2\big(\bar{\psi}_\alpha \gamma^\mu \partial^\alpha \psi^\nu - \bar{\psi}^\mu \overleftarrow{\partial}^\alpha \gamma^\nu \psi_\alpha\big) F_{\mu\nu} \\ + 2(\partial \cdot A)\big(\bar{\psi}^\mu \slashed{\partial}\psi_\mu + \bar{\psi}^\mu \overleftarrow{\slashed{\partial}} \psi_\mu\big). \tag{D.27}$$

Again, we integrate by parts the first term of the first line, so that no derivatives act on the photon field. In the second line as well we perform integration by parts to have 2 derivatives acting on the photon field. This finally gives

$$V_{\text{ST}}^{(2)} \approx 2\big(\bar{\psi}_\alpha \gamma^\mu \partial^\alpha \psi^\nu - \bar{\psi}^\mu \overleftarrow{\partial}^\alpha \gamma^\nu \psi_\alpha\big) F_{\mu\nu} \tag{D.28}$$
$$+ 2\left[\big(\bar{\psi}_\alpha \partial^\alpha \slashed{\partial}\psi^\nu + \bar{\psi}_\alpha \overleftarrow{\slashed{\partial}}\partial^\alpha \psi^\nu\big) A_\nu + \text{h.c.}\right] - 2\bar{\psi}_\alpha \big(\slashed{\partial}\partial \cdot A\big)\psi^\alpha.$$

It is clear that, in the TT gauge, both the off-shell vertices (D.24) and (D.28) reduce to

$$V_{\text{TT}}^{(2)} = 2\big(\bar{\psi}_\alpha \gamma^\mu \partial^\alpha \psi^\nu - \bar{\psi}^\mu \overleftarrow{\partial}^\alpha \gamma^\nu \psi_\alpha\big) F_{\mu\nu}, \tag{D.29}$$

which is nothing but the 2-derivative on-shell vertex given in [62]. The equivalence of the two off-shell vertices is also evident as, upon subtracting (D.28) from (D.24), we have

$$V^{(2)} - V_{\text{ST}}^{(2)} = 2\bar{\psi}_\alpha \gamma^\mu \big(\partial^\nu F_{\mu\nu}\big) \psi^\alpha = \Delta\text{-exact}. \tag{D.30}$$

**3-Derivatives $1-3/2-3/2$ Vertex**

Finally, we consider the vertex with 3 derivatives, which reads

$$V^{(3)} = \bar{\Psi}_{\mu\alpha} \Psi^{\alpha}{}_{\nu} F^{\mu\nu} = \big(\partial_\mu \bar{\psi}^\alpha - \partial^\alpha \bar{\psi}_\mu\big)\big(\partial_\alpha \psi_\nu - \partial_\nu \psi_\alpha\big) F^{\mu\nu}. \tag{D.31}$$

Integration by parts w.r.t. $\partial_\mu$, appearing in the 1st term inside the first parentheses, gives

$$V^{(3)} \approx - \bar\psi^\alpha \left(\partial_\mu\partial_\alpha\psi_\nu - \partial_\mu\partial_\nu\psi_\alpha\right) F^{\mu\nu} - \bar\psi^\alpha \Psi_{\alpha\nu}\partial_\mu F^{\mu\nu} \\ - \bar\psi_\mu \overleftarrow{\partial}{}^\alpha \partial_\alpha\psi_\nu F^{\mu\nu} + \bar\psi_\mu \overleftarrow{\partial}{}^\alpha \partial_\nu\psi_\alpha F^{\mu\nu}. \tag{D.32}$$

Here the 2nd term inside the parentheses on the right side is identically zero, while the term containing $\partial_\mu F^{\mu\nu}$ is $\Delta$-exact. We use the 3-box rule (D.22) in the penultimate term. Also we integrate by parts w.r.t. $\partial_\nu$ in the last term, and it produces a $\Delta$-exact piece, containing $\partial_\nu F^{\mu\nu}$, that we discard. The result is

$$V^{(3)} \approx - \left(\bar\psi^\alpha\partial_\alpha\partial_\mu\psi_\nu + \bar\psi_\mu\overleftarrow{\partial}_\nu\overleftarrow{\partial}_\alpha\psi^\alpha\right)F^{\mu\nu} \\ + \tfrac{1}{2}\left(\bar\psi_\mu \Box \psi_\nu + \bar\psi_\mu \overleftarrow\Box \psi_\nu - \Box\left(\bar\psi_\mu\psi_\nu\right)\right)F^{\mu\nu}. \tag{D.33}$$

Now, one can perform a double integration by parts in the last term in the brackets in order to have $\Box F^{\mu\nu}$, which gives a $\Delta$-exact piece, so that we finally have

$$V^{(3)} \approx \tfrac{1}{2}\left(\bar\psi_\mu \Box \psi_\nu + \bar\psi_\mu \overleftarrow\Box \psi_\nu\right)F^{\mu\nu} - \left(\bar\psi^\alpha\partial_\alpha\partial_\mu\psi_\nu + \bar\psi_\mu\overleftarrow{\partial}_\nu\overleftarrow{\partial}_\alpha\psi^\alpha\right)F^{\mu\nu} \tag{D.34}$$

On the other hand, the 3-derivative off-shell ST vertex contains as many as 14 terms:

$$\begin{aligned} V^{(3)}_{\rm ST} = &- \bar\psi^\mu(\partial_\alpha\partial_\mu\psi^\nu)\partial_\nu A^\alpha + (\partial_\alpha\bar\psi^\mu)(\partial_\mu\psi^\nu)\partial_\nu A^\alpha - (\partial_\alpha\partial_\nu\bar\psi^\mu)(\partial_\mu\psi^\nu)A^\alpha \\ &- (\partial_\alpha\bar\psi^\mu)\psi^\nu\partial_\mu\partial_\nu A^\alpha + \bar\psi^\mu(\partial_\alpha\psi^\nu)\partial_\mu\partial_\nu A^\alpha + (\partial_\alpha\partial_\nu\bar\psi^\mu)\psi^\nu\partial_\mu A^\alpha \\ &+ (\partial_\mu\partial\cdot\bar\psi)(\partial\cdot\psi)A^\mu - (\partial\cdot\bar\psi)(\partial_\mu\partial\cdot\psi)A^\mu + \bar\psi^\mu(\partial_\mu\partial_\alpha\psi^\alpha)\partial\cdot A \\ &- \bar\psi^\mu(\partial\cdot\psi)\partial_\mu\partial_\alpha A^\alpha - (\partial_\mu\partial_\alpha\bar\psi^\alpha)\psi^\mu\partial\cdot A + (\partial\cdot\bar\psi)\psi^\mu\partial_\mu\partial_\alpha A^\alpha. \\ &+ (\partial_\nu\bar\psi^\mu)(\partial_\mu\partial_\alpha\psi^\nu)A^\alpha - (\partial_\nu\bar\psi^\mu)(\partial_\alpha\psi^\nu)\partial_\mu A^\alpha. \end{aligned} \tag{D.35}$$

Here we will perform a number of integrations by parts. In the first line, we integrate by parts the 2nd term w.r.t. $\partial_\alpha$, the 3rd w.r.t. $\partial_\mu$, and the 4th w.r.t. $\partial_\nu$. In the second line, the 1st, 2nd and 4th terms are integrated by parts respectively w.r.t. $\partial_\nu$, $\partial_\mu$ and $\partial_\alpha$. In the third line, this is done only on the 3rd term w.r.t. $\partial_\alpha$. Finally, in the fourth line, the 1st and 3rd terms are integrated by parts w.r.t. both $\partial_\mu$ and $\partial_\alpha$, while the 2nd one only w.r.t. $\partial_\alpha$. Dropping total derivatives, the result is

$$\begin{aligned} V^{(3)}_{\rm ST} \approx &+ 4\bar\psi_\mu\overleftarrow\partial_\nu\overleftarrow\partial_\alpha\psi^\alpha\partial^\mu A^\nu + 2\left(\bar\psi^\mu\overleftarrow\partial_\alpha\psi^\alpha - \bar\psi^\alpha\partial_\alpha\psi^\mu\right)\partial_\mu\partial\cdot A \\ &- 4\bar\psi^\alpha(\partial_\alpha\partial_\mu\psi_\nu)\partial^\nu A^\mu + \left(\bar\psi\cdot\overleftarrow\partial\overleftarrow\partial_\mu\partial_\alpha\psi^\mu - \bar\psi^\mu\overleftarrow\partial_\alpha\partial_\mu\partial\cdot\psi\right)A^\alpha \\ &+ 2\left(\bar\psi\cdot\overleftarrow\partial\overleftarrow\partial_\mu\overleftarrow\partial_\alpha\psi^\alpha - \bar\psi^\alpha\partial_\alpha\partial_\mu\partial\cdot\psi\right)A^\mu + \left[(\partial_\alpha\partial\cdot\bar\psi)\partial\cdot\psi + {\rm h.c.}\right]A^\alpha \\ &+ \left[(\partial_\alpha\bar\psi_\mu)\partial\cdot\psi + {\rm h.c.}\right]\partial^\mu A^\alpha. \end{aligned} \tag{D.36}$$

Let us rewrite the first term in the first and second line in terms of the photon field strength by using $\partial_\mu A_\nu = F_{\mu\nu} + \partial_\nu A_\mu$, and use the 3-box rule (D.22) in the additional terms. Also, we notice that the last two terms together reduce exactly to the 2nd term on the second line, up to a total derivative. Then, the vertex reads

$$V_{\text{ST}}^{(3)} \approx + 4\big(\bar{\psi}^\alpha \partial_\alpha \partial_\mu \psi_\nu + \bar{\psi}_\mu \overleftarrow{\partial}_\nu \overleftarrow{\partial}_\alpha \psi^\alpha\big) F^{\mu\nu} + 2\big(\bar{\psi}_\mu \overleftarrow{\partial}_\nu \Box \psi^\nu - \bar{\psi}^\nu \overleftarrow{\Box} \partial_\nu \psi_\mu\big) A^\mu$$
$$+ 2\big[\bar{\psi}^\alpha \partial_\alpha \big(\Box \psi_\mu - \partial_\mu \partial \cdot \psi\big) - \big(\bar{\psi}_\mu \overleftarrow{\Box} - \bar{\psi} \cdot \overleftarrow{\partial} \overleftarrow{\partial}_\mu\big) \overleftarrow{\partial}_\alpha \psi^\alpha\big] A^\mu$$
$$- 2\big(\bar{\psi}^\mu \overleftarrow{\partial}_\alpha \psi^\alpha - \bar{\psi}^\alpha \partial_\alpha \psi^\mu\big) \big(\Box A_\mu - \partial_\mu \partial \cdot A\big) \quad \text{(D.37)}$$
$$+ 2\big(\bar{\psi} \cdot \overleftarrow{\partial} \overleftarrow{\partial}_\mu \partial_\alpha \psi^\mu - \bar{\psi}^\mu \overleftarrow{\partial}_\alpha \partial_\mu \partial \cdot \psi\big) A^\alpha.$$

Clearly, the second and third lines are $\Delta$-exact, while, modulo $\Delta$-exact pieces, the 2nd term in the first line can have $\Box \psi^\nu$ replaced by $\partial^\nu \partial \cdot \psi$. The latter result can be combined with the last line to give

$$V_{\text{ST}}^{(3)} \approx 4\big(\bar{\psi}^\alpha \partial_\alpha \partial_\mu \psi_\nu + \bar{\psi}_\mu \overleftarrow{\partial}_\nu \overleftarrow{\partial}_\alpha \psi^\alpha\big) F^{\mu\nu} - 2\big(\bar{\psi} \cdot \overleftarrow{\partial} \overleftarrow{\partial}_\mu \Psi^{\mu\nu} - \bar{\Psi}^{\mu\nu} \partial_\mu \partial \cdot \psi\big) A_\nu$$
$$\approx 4\big(\bar{\psi}^\alpha \partial_\alpha \partial_\mu \psi_\nu + \bar{\psi}_\mu \overleftarrow{\partial}_\nu \overleftarrow{\partial}_\alpha \psi^\alpha\big) F^{\mu\nu} + \big(\bar{\psi} \cdot \overleftarrow{\partial} \Psi_{\mu\nu} - \bar{\Psi}_{\mu\nu} \partial \cdot \psi\big) F^{\mu\nu},$$
(D.38)

where we have reached the second step by performing integration by parts w.r.t. $\partial_\mu$ in the 2nd term of the first step, and dropping $\Delta$-exact terms containing $\partial_\mu \Psi^{\mu\nu}$. In the 2nd term of the second step, one can write $\Psi_{\mu\nu} = 2\partial_{[\mu} \psi_{\nu]}$, and integrate by parts to obtain, among others, $\Delta$-exact terms containing $\partial_\mu F^{\mu\nu}$, which can be dropped. The result is

$$V_{\text{ST}}^{(3)} \approx \tfrac{1}{2}\big(\bar{\psi}_\mu \partial_\nu \partial \cdot \psi + \bar{\psi} \cdot \overleftarrow{\partial} \overleftarrow{\partial}_\mu \psi_\nu\big) F^{\mu\nu} - \big(\bar{\psi}^\alpha \partial_\alpha \partial_\mu \psi_\nu + \bar{\psi}_\mu \overleftarrow{\partial}_\nu \overleftarrow{\partial}_\alpha \psi^\alpha\big) F^{\mu\nu},$$
(D.39)

where we have performed the rescaling $A_\mu \to -\tfrac{1}{4} A_\mu$, for convenience of comparison with our vertex $V^{(3)}$. One finds that both the vertices reduce in the TT gauge to

$$V_{\text{TT}}^{(3)} = -\big(\bar{\psi}^\alpha \partial_\alpha \partial_\mu \psi_\nu + \bar{\psi}_\mu \overleftarrow{\partial}_\nu \overleftarrow{\partial}_\alpha \psi^\alpha\big) F^{\mu\nu}, \quad \text{(D.40)}$$

which indeed is the 3-derivatives on-shell vertex reported in [62]. In view of Eq. (D.34) and (D.39), one also finds that the two vertices differ by $\Delta$-exact terms:

$$V^{(3)} - V_{\text{ST}}^{(3)} \approx \tfrac{1}{2}\big(\bar{\psi}_\mu \big(\Box \psi_\nu - \partial_\nu \partial \cdot \psi\big) + \big(\bar{\psi}_\mu \overleftarrow{\Box} - \bar{\psi} \cdot \overleftarrow{\partial} \overleftarrow{\partial}_\mu\big) \psi_\nu\big] F^{\mu\nu} = \Delta(...).$$
(D.41)

This shows the equivalence of the full off-shell vertices.

### D.2.2  1−5/2−5/2 Vertices Compared

For the sake of simplicity, from now on we restrict our attention to the on-shell equivalence of vertices. As we already mentioned, if two vertices match in some gauge, say the TT one, they should also be off-shell equivalent. With this end in view, we read off the ST vertices [25], which would generally contain a lot of terms to begin with, even in the TT gauge. However, one can perform integrations by parts to see that actually the on-shell vertices are extremely simple, containing no more than a few non-trivial terms.

**3-Derivatives 1−5/2−5/2 Vertex**

For example, one can take the 3-derivative $1-\frac{5}{2}-\frac{5}{2}$ ST vertex in the TT gauge, and integrate by parts in order to have one derivative on each field. The result is simply

$$V^{(3)}_{\text{ST}} \sim \bar{\psi}_{\mu\alpha}\overleftarrow{\partial}_\beta F^{\mu\nu}\partial^\alpha\psi^\beta{}_\nu + \bar{\psi}_{\mu\alpha}\overleftarrow{\partial}_\rho \left(\partial_\beta A^\rho\right)\partial^\alpha\psi^{\beta\mu}, \tag{D.42}$$

where $\sim$ means equivalence in the TT gauge up to an overall factor. In the 2nd term we integrate by parts to avoid derivatives on the photon field. We get

$$V^{(3)}_{\text{ST}} \sim \bar{\psi}_{\mu\alpha}\overleftarrow{\partial}_\beta F^{\mu\nu}\partial^\alpha\psi^\beta{}_\nu - \bar{\psi}_{\mu\alpha}\overleftarrow{\partial}_\beta \left(\overleftarrow{\partial}\cdot A\right)\partial^\alpha\psi^{\beta\mu}. \tag{D.43}$$

One can now make use of $\overleftarrow{\partial}\cdot A = \frac{1}{2}\overleftarrow{\partial}_\rho A_\sigma\left(\gamma^\rho\gamma^\sigma + \gamma^\sigma\gamma^\rho\right)$, in the 2nd term on the right-hand side of the above equation, and then integrate by parts w.r.t. this derivative. Dropping some $\Delta$-exact terms in the TT gauge, we get

$$V^{(3)}_{\text{ST}} \sim \bar{\psi}_{\mu\alpha}\overleftarrow{\partial}_\beta F^{\mu\nu}\partial^\alpha\psi^\beta{}_\nu + \tfrac{1}{2}\bar{\psi}_{\mu\alpha}\overleftarrow{\partial}_\beta \left(\partial_\rho A_\sigma\right)\gamma^\sigma\gamma^\rho\partial^\alpha\psi^{\beta\mu}. \tag{D.44}$$

Because $\partial\cdot A = 0$ in our gauge choice, we can write $(\partial_\rho A_\sigma)\gamma^\sigma\gamma^\rho = -\frac{1}{2}\not{F}$, by making use of the identity $\gamma^\sigma\gamma^\rho = \eta^{\sigma\rho} - \gamma^{\rho\sigma}$. Therefore, we are left with

$$V^{(3)}_{\text{ST}} \sim \bar{\psi}_{\mu\alpha}\overleftarrow{\partial}_\beta \left(F^{\mu\nu} - \tfrac{1}{4}\eta^{\mu\nu}\not{F}\right)\partial^\alpha\psi^\beta{}_\nu. \tag{D.45}$$

We would like to see how this compares with our 3-derivatives $1-\frac{5}{2}-\frac{5}{2}$ vertex,

$$V^{(3)} = \bar{\psi}^{(1)}_{\alpha\beta\|\,\mu}F^{+\mu\nu}\psi^{(1)\alpha\beta\|}{}_\nu. \tag{D.46}$$

The same steps as took us from Eq. (D.9) to Eq. (D.10) for the spin-$\frac{3}{2}$ case lead to

$$V^{(3)} \sim 2\,\bar{\psi}^{(1)}_{\alpha\beta\|\,\mu}\left(F^{\mu\nu} - \tfrac{1}{4}\eta^{\mu\nu}\not{F}\right)\psi^{(1)\alpha\beta\|}{}_\nu. \tag{D.47}$$

Now, one can rewrite the fermionic 1-curl in terms of the original field. There will be terms that have at least one pair of mutually contracted derivatives: one acting on $\bar\psi_\mu$ and the other on $\psi_\mu$. For such terms one can make use of the 3-box rule (D.22) to see that they are trivial in the TT gauge. Up to a trivial factor, one then has

$$V^{(3)} \sim \bar\psi_{\mu\alpha}\overleftarrow\partial_\beta \left(F^{\mu\nu} - \tfrac{1}{4}\eta^{\mu\nu}\slashed{F}\right)\partial^\alpha\psi^\beta{}_\nu. \tag{D.48}$$

From Eq. (D.45) and (D.48), we see that the two vertices are indeed on-shell equivalent.

### 4-Derivatives $1-5/2-5/2$ Vertex

Let us move on to the 4-derivative $1-\tfrac{5}{2}-\tfrac{5}{2}$ vertex. The ST one is found to be

$$V^{(4)}_{\text{ST}} \sim \bar\psi_{\mu\nu}\overleftarrow\partial_\rho\overleftarrow\partial_\sigma \slashed{A}\,\partial^\mu\partial^\nu\psi^{\rho\sigma}, \tag{D.49}$$

whereas our one is given by

$$V^{(4)} = \left(\bar\Psi_{\mu\nu|\rho\sigma}\,\gamma^{\mu\nu\alpha\beta\lambda}\,\Psi_{\alpha\beta|}{}^{\rho\sigma}\right)A_\lambda \approx -2\left(\bar\Psi_{\mu\nu|\rho\sigma}\,\gamma^\lambda\,\Psi^{\mu\nu|\rho\sigma}\right)A_\lambda. \tag{D.50}$$

We rewrite the curvature in terms of the spin-$\tfrac{5}{2}$ field. Among the resulting terms those with contracted pair(s) of derivatives are, again, trivial in the TT gauge, thanks to the 3-box rule (D.22). The other terms clearly add up to reproduce the expression (D.49). Therefore,

$$V^{(4)} \approx V^{(4)}_{\text{ST}} \sim \bar\psi_{\mu\nu}\overleftarrow\partial_\rho\overleftarrow\partial_\sigma\slashed{A}\,\partial^\mu\partial^\nu\psi^{\rho\sigma}. \tag{D.51}$$

### 5-Derivatives $1-5/2-5/2$ Vertex

For spin $\tfrac{5}{2}$, the only other vertex is the 5-derivative one. The ST one reads

$$V^{(5)}_{\text{ST}} \sim \left(\bar\psi_{\mu\nu}\overleftarrow\partial_\rho\overleftarrow\partial_\sigma\overleftarrow\partial{}^\lambda\,\partial^\mu\partial^\nu\psi^{\rho\sigma}\right)A_\lambda. \tag{D.52}$$

On the other hand, we have the 5-derivative Born–Infeld type vertex:

$$V^{(5)} = \bar\Psi_{\alpha\beta|\,\mu\rho}\Psi^{\alpha\beta|\,\rho}{}_\nu F^{\mu\nu} \approx \tfrac{1}{2}\left(\bar\Psi_{\mu\nu|\,\rho\sigma}\overleftarrow\partial{}^\lambda\,\Psi^{\mu\nu|\,\rho\sigma}\right)A_\lambda. \tag{D.53}$$

The off-shell equivalence can be understood in view of Eq. (5.29)–(5.31), which pertain to spin $\tfrac{3}{2}$. In the equivalent vertex, again, we rewrite the fermionic curvature in terms of the spin-$\tfrac{5}{2}$ field, and massage the resulting terms the same way as was done for $V^{(4)}$. Thus, up to overall factors, we reproduce on-shell (D.52), so that

$$V^{(5)} \approx V^{(5)}_{\text{ST}} \sim \left(\bar\psi_{\mu\nu}\overleftarrow\partial_\rho\overleftarrow\partial_\sigma\overleftarrow\partial{}^\lambda\,\partial^\mu\partial^\nu\psi^{\rho\sigma}\right)A_\lambda. \tag{D.54}$$

### D.2.3  $1-s-s$ Vertices Compared

For arbitrary spin, $s = n + \frac{1}{2}$, the story is very similar, and there are no further complications other than cluttering of indices. One can write down the ST vertices in the TT gauge from Eq. (A.16) of [25]. They turn out to be

$$V_{\text{ST}}^{(2n-1)} \sim \bar{\psi}_{\mu\,\alpha_1...\alpha_{n-1}} \overleftarrow{\partial}_{\beta_1}...\overleftarrow{\partial}_{\beta_{n-1}} F^{\mu\nu} \partial^{\alpha_1}...\partial^{\alpha_{n-1}} \psi^{\beta_1...\beta_{n-1}}{}_\nu$$
$$- \bar{\psi}_{\mu\,\alpha_1...\alpha_{n-1}} \overleftarrow{\partial}_{\beta_1}...\overleftarrow{\partial}_{\beta_{n-1}} (\overleftarrow{\partial} \cdot A) \partial^{\alpha_1}...\partial^{\alpha_{n-1}} \psi^{\beta_1...\beta_{n-1}\,\mu}, \quad \text{(D.55a)}$$

$$V_{\text{ST}}^{(2n)} \sim \bar{\psi}_{\mu_1...\mu_n} \overleftarrow{\partial}_{\nu_1}...\overleftarrow{\partial}_{\nu_n} \slashed{A}\, \partial^{\mu_1}...\partial^{\mu_n} \psi^{\nu_1...\nu_n}, \quad \text{(D.55b)}$$

$$V_{\text{ST}}^{(2n+1)} \sim (\bar{\psi}_{\mu_1...\mu_n} \overleftarrow{\partial}_{\nu_1}...\overleftarrow{\partial}_{\nu_n} \overleftrightarrow{\partial}^\lambda \partial^{\mu_1}...\partial^{\mu_n} \psi^{\nu_1...\nu_n}) A_\lambda. \quad \text{(D.55c)}$$

Their similarity with the lower-spin counterparts is obvious. Indeed, setting $n = 2$ produces exactly the respective $1-\frac{5}{2}-\frac{5}{2}$ vertices given in Eq. (D.43), (D.49) and (D.52). One can massage the $(2n-1)$-derivative vertex, in particular, the same way as its spin-$\frac{5}{2}$ counterpart to obtain an arbitrary-spin generalization of Eq. (D.45), namely

$$V_{\text{ST}}^{(2n-1)} \sim \bar{\psi}_{\mu\,\alpha_1...\alpha_{n-1}} \overleftarrow{\partial}_{\beta_1}...\overleftarrow{\partial}_{\beta_{n-1}} \left( F^{\mu\nu} - \tfrac{1}{4}\eta^{\mu\nu} \slashed{F} \right) \partial^{\alpha_1}...\partial^{\alpha_{n-1}} \psi^{\beta_1...\beta_{n-1}}{}_\nu. \quad \text{(D.56)}$$

Our arbitrary-spin vertices are also straightforward generalizations of their lower-spin examples. In view of the spin-$\frac{5}{2}$ counterparts, Eq. (D.46), (D.50) and (D.53), one can write

$$V^{(2n-1)} \approx \bar{\psi}^{(n-1)}_{\alpha_1\beta_1|...|\alpha_{n-1}\beta_{n-1}\|\,\mu} F^{+\mu\nu} \psi^{(n-1)\alpha_1\beta_1|...|\alpha_{n-1}\beta_{n-1}\|}{}_\nu, \quad \text{(D.57a)}$$

$$V^{(2n)} \approx (\bar{\Psi}_{\mu_1\nu_1|...|\mu_n\nu_n} \gamma^\lambda\, \Psi^{\mu_1\nu_1|...|\mu_n\nu_n}) A_\lambda, \quad \text{(D.57b)}$$

$$V^{(2n+1)} \approx (\bar{\Psi}_{\mu_1\nu_1|...|\mu_n\nu_n} \overleftrightarrow{\partial}^\lambda\, \Psi^{\mu_1\nu_1|...|\mu_n\nu_n}) A_\lambda. \quad \text{(D.57c)}$$

Again, one can use the 2nd identity in (D.1) to rewrite the $F^{+\mu\nu}$ in the first vertex, and express the fermionic $(n-1)$- and $n$-curls in all the vertices (D.57a)–(D.57c) in terms of the original field. The terms with contracted pair(s) of derivatives are, as usual, subject to the 3-box rule (D.22), and hence trivial in the TT gauge. One finds that our vertices indeed reduce on shell respectively to (D.56), (D.55b) and (D.55c). This proves the on-shell (and therefore off-shell) equivalence of the $1-s-s$ vertices:

$$V^{(p)} \sim V_{\text{ST}}^{(p)}, \qquad p = 2n-1,\, 2n,\, 2n+1. \quad \text{(D.58)}$$

## D.3 Details of Spin-5/2 Gravitational Couplings

Throughout the bulk of the paper, we have omitted the proof of some cumbersome technical steps for the sake of readability. The detailed proof of those steps appears in this Appendix.

### D.3.1 2-Derivatives 1−5/2−5/2 Vertex

In Eq. (6.19), the part of $a_2$ that contains the fermionic antighost is given by
$$a_{2\tilde{g}} = -\tilde{g}\,\bar{\xi}^*_\rho \gamma^{\alpha\beta\rho\mu\nu} \xi_{\mu\nu} \mathfrak{C}_{\alpha\beta} + \text{h.c.}, \tag{D.59}$$

which comes with five $\gamma$-matrices. But it can be cast into an equivalent form that contains just one, like that appearing in Eq. (6.10). To see this, let us first use the $\gamma$-matrix identity:
$$\gamma^{\alpha\beta\rho\mu\nu} = \tfrac{1}{2}\left(\gamma^\alpha \gamma^{\beta\rho\mu\nu} + \gamma^{\beta\rho\mu\nu}\gamma^\alpha\right), \tag{D.60}$$

and then another one for the antisymmetric product of four $\gamma$-matrices, namely
$$\gamma^{\beta\rho\mu\nu} = -2\eta^{\beta\rho|\mu\nu} + \gamma^{\beta\rho}\gamma^{\mu\nu} - 4\gamma^{[\beta}\eta^{\rho][\mu}\gamma^{\nu]}. \tag{D.61}$$

The result is
$$\begin{aligned}a_{2\tilde{g}} = &- \tilde{g}\,\bar{\xi}^*_\rho \gamma^\alpha \big(-\eta^{\beta\rho|\mu\nu} + \tfrac{1}{2}\gamma^{\beta\rho}\gamma^{\mu\nu} - 2\gamma^{[\beta}\eta^{\rho]\mu}\gamma^{\nu]}\big)\xi_{\mu\nu}\mathfrak{C}_{\alpha\beta} \\ &- \tilde{g}\,\bar{\xi}^*_\rho\big(-\eta^{\beta\rho|\mu\nu} + \tfrac{1}{2}\gamma^{\beta\rho}\gamma^{\mu\nu} - 2\gamma^{[\beta}\eta^{\rho]\mu}\gamma^{\nu]}\big)\gamma^\alpha \xi_{\mu\nu}\mathfrak{C}_{\alpha\beta} + \text{h.c.}\end{aligned} \tag{D.62}$$

In both the first and the second lines on the right-hand side, the first term is of the desired form with a single $\gamma$-matrix, while the second and third terms give rise to the $\Gamma$-exact pieces $\gamma^\nu \xi_{\mu\nu}$ and $\gamma^{\mu\nu}\xi_{\mu\nu}$, either directly or through the relations: $\gamma^{\mu\nu}\gamma^\alpha = \gamma^\alpha\gamma^{\mu\nu} - 4\eta^{\alpha[\mu}\gamma^{\nu]}$ and $\gamma^\nu \gamma^\alpha = -\gamma^\alpha\gamma^\nu + 2\eta^{\nu\alpha}$. Finally, on account of the $\gamma$-tracelessness of $\bar{\xi}^*_\rho$, one obtains

$$a_{2\tilde{g}} = -\tilde{g}\,\bar{\xi}^*_\rho\big(-2\eta^{\beta\rho|\mu\nu}\gamma^\alpha - 2\gamma^\beta \eta^{\rho\mu}\eta^{\nu\alpha}\big)\xi_{\mu\nu}\mathfrak{C}_{\alpha\beta} + \text{h.c.} + \Gamma\text{-exact}, \tag{D.63}$$

which is indeed equivalent to the $p=2$ piece presented in Eq. (6.10), since more explicitly,
$$a_{2\tilde{g}} = 4\tilde{g}\,\bar{\xi}^{*\mu}\gamma^\alpha \xi^\beta{}_\mu \mathfrak{C}_{\alpha\beta} + \text{h.c.} + \Gamma\text{-exact}. \tag{D.64}$$

<div style="text-align:center">♮ ♮ ♮</div>

Now we will fill up the gaps between Eqs. (6.24) and (6.25). First we use the definition (E.25) of $\chi^{*\mu\nu}$, and Eqs. (E.14) to write

$$\Delta\chi^*_{\rho\sigma} = \mathcal{S}_{\rho\sigma} - \tfrac{1}{2}\gamma_\sigma \mathcal{\slashed{S}}_\rho - \tfrac{1}{2}\eta_{\rho\sigma}\mathcal{S}'. \tag{D.65}$$

One can take a curl of the above equation, and relate the 1-curl of the Fronsdal tensor to the $\gamma$-trace of the curvature through Eq. (E.19), which yields

$$2\Delta\partial_{[\nu}\chi^*_{\rho]\sigma} = \tfrac{i}{2}\gamma_{\sigma\tau\lambda}\Psi_{\nu\rho}{}^{\tau\lambda} + \eta_{\sigma[\nu}\partial_{\rho]}\mathcal{S}'. \tag{D.66}$$

When this expression is used in Eq. (6.24), the $\mathcal{S}'$-terms vanish because of the Bianchi identity, $\bar{\psi}_{[\alpha\beta\|\nu]}$, imposed by the antisymmetric 5-$\gamma$. The result is

$$\beta^\mu_C = i\tilde{g}\,\bar{\Psi}_{\nu\rho|\tau\lambda}\gamma^{\sigma\tau\lambda}\gamma^{\mu\nu\rho\alpha\beta}\psi_{\alpha\beta\|\sigma} - i\tilde{g}^*\bar{\psi}_{\alpha\beta\|\sigma}\gamma^{\mu\nu\rho\alpha\beta}\gamma^{\sigma\tau\lambda}\Psi_{\nu\rho|\tau\lambda}. \tag{D.67}$$

Now one can take $\partial_\tau$ out of the curvature and use Leibniz rule to find a total derivative plus some terms that can be identified as $-\beta^\mu_C$ only if $\tilde{g}$ is real. With this, one obtains

$$\beta^\mu_C = 2i\tilde{g}\,\partial_\nu\left(\bar{\psi}_{\rho\sigma\|\lambda}\,\gamma^{\mu\rho\sigma\alpha\beta,\,\nu\lambda\gamma}\,\psi_{\alpha\beta\|\gamma}\right). \tag{D.68}$$

Note that the quantity inside the parentheses can be made symmetric under $\mu \leftrightarrow \nu$ for free, thanks to the Bianchi identities playing role when $\partial_\nu$ hits the 1-curls. This leads us to Eq. (6.25) under the stated condition: $\tilde{g}$ is real.

♮ ♮ ♮

Next, we will derive Eq. (6.27) from Eq. (6.26). We will simply show (dropping the quite similar proof for hermitian conjugates) that

$$-i\bar{\xi}_\lambda R_{\mu\nu\rho\sigma}\gamma^{\mu\nu\lambda\alpha\beta,\,\tau\rho\sigma}\psi_{\alpha\beta\|\tau} \doteq 2ih_{\mu\nu}\,\Gamma\,\bar{\psi}_{\rho\sigma\|\lambda}\,\gamma^{\mu\rho\sigma\alpha\beta,\,\nu\lambda\gamma}\,\psi_{\alpha\beta\|\gamma} \\ - 2\mathfrak{h}_{\mu\nu\|}{}^\sigma\bar{\xi}_{\alpha\beta}\gamma^{\mu\nu\rho\alpha\beta}\Delta\chi^*_{\rho\sigma}. \tag{D.69}$$

Let us rewrite the right-hand side of Eq. (D.69) as

$$\text{R.H.S.} = -2ih_{\mu\nu}\partial_\lambda\bar{\xi}_{\rho\sigma}\,\gamma^{\mu\rho\sigma\alpha\beta,\,\nu\lambda\gamma}\,\psi_{\alpha\beta\|\gamma} - 2\mathfrak{h}_{\mu\nu\|}{}^\sigma\bar{\xi}_{\alpha\beta}\gamma^{\mu\nu\rho\alpha\beta}\Delta\chi^*_{\rho\sigma}. \tag{D.70}$$

In the first term on the right-hand side, let us pull out the derivative $\partial_\sigma$ off the ghost-curl and integrate by parts. This is followed by another integration by parts w.r.t. $\partial_\lambda$. In the second term, on the other hand, we pull out the derivative $\partial_\beta$ off the ghost-curl to integrate by parts. In these steps, we exploit the antisymmetry of the products of $\gamma$-matrices, which kills some terms by enforcing the Bianchi identities given in Appendix A.2. Then, we are left with

$$\text{R.H.S.} \doteq i\bar{\xi}_\rho \left( \mathfrak{h}_{\sigma\nu\|\mu}\, \gamma^{\sigma\nu\rho\alpha\beta,\,\mu\lambda\gamma}\, \Psi_{\alpha\beta|\lambda\gamma} + R_{\sigma\nu\lambda\mu}\, \gamma^{\sigma\nu\rho\alpha\beta,\,\mu\lambda\gamma}\, \psi_{\alpha\beta\|\gamma} \right) \\ - 4\bar{\xi}_\alpha \mathfrak{h}_{\mu\nu\|}{}^\sigma \gamma^{\mu\nu\rho\alpha\beta} \Delta \partial_{[\beta} \chi^*_{\rho]\sigma}. \tag{D.71}$$

In the last term on the right-hand side above, one can again use Eq. (D.66) and then drop the $\mathcal{S}'$-terms on account of the Bianchi identities. The result is

$$\text{R.H.S.} \doteq i\bar{\xi}_\rho \left( \mathfrak{h}_{\sigma\nu\|\mu}\, \gamma^{\sigma\nu\rho\alpha\beta,\,\mu\lambda\gamma}\, \Psi_{\alpha\beta|\lambda\gamma} + R_{\sigma\nu\lambda\mu}\, \gamma^{\sigma\nu\rho\alpha\beta,\,\mu\lambda\gamma}\, \psi_{\alpha\beta\|\gamma} \right) \\ - i\bar{\xi}_\alpha \mathfrak{h}_{\mu\nu\|\sigma} \gamma^{\mu\nu\rho\alpha\beta} \gamma^{\sigma\tau\lambda} \Psi_{\beta\rho|\tau\lambda}. \tag{D.72}$$

Now, in the last term on the above right-hand side, the matrices $\gamma^{\mu\nu\rho\alpha\beta}$ and $\gamma^{\sigma\tau\lambda}$ actually commute. This can be seen by first writing $\gamma^{\sigma\tau\lambda} = \frac{1}{2}\left(\gamma^\sigma\gamma^\tau\gamma^\lambda - \gamma^\lambda\gamma^\tau\gamma^\sigma\right)$, and then noticing that any of these $\gamma$-matrices commutes past $\gamma^{\mu\nu\rho\alpha\beta}$, on account of the identity:

$$\gamma^{\mu\nu\rho\alpha\beta}\gamma^\sigma = \gamma^\sigma\gamma^{\mu\nu\rho\alpha\beta} - 2\gamma^{\sigma\mu\nu\rho\alpha\beta}, \tag{D.73}$$

and similar ones for $\gamma^\tau$ and $\gamma^\lambda$, and the fact that the antisymmetric products of six $\gamma$-matrices are always eliminated by the Bianchi identities. This enables us to rewrite the last term on the right-hand side of Eq. (D.72) as the first one, but with an opposite sign, so that these terms actually cancel each other. Therefore, we are left only with

$$\text{L.H.S.} \doteq -i\bar{\xi}_\lambda R_{\mu\nu\rho\sigma} \gamma^{\mu\nu\lambda\alpha\beta,\,\tau\rho\sigma}\, \psi_{\alpha\beta\|\tau}. \tag{D.74}$$

This is precisely the left-hand side of Eq. (D.69), which, therefore, is proved.

♮ ♮ ♮

Now we will show how Eq. (6.28) follows from Eq. (6.27). In other words, we will prove

$$-i\tilde{g}\,\bar{\xi}_\lambda R_{\mu\nu\rho\sigma} \gamma^{\mu\nu\lambda\alpha\beta,\,\tau\rho\sigma}\, \psi_{\alpha\beta\|\tau} + \text{h.c.} \doteq -8i\tilde{g}\,\Gamma\left(\bar{\psi}_{\mu\alpha} R^{+\mu\nu\alpha\beta}\psi_{\nu\beta} \right. \\ \left. + \tfrac{1}{2}\,\bar{\psi}_\mu R^{\mu\nu} \psi_\nu\right) - \Delta\mathfrak{a}, \tag{D.75}$$

for some $\Delta\mathfrak{a}$ to be determined. First, let us write down a $\gamma$-matrix identity:

$$\gamma^{\mu\nu\lambda\alpha\beta,}{}_{\tau\rho\sigma} = -60\,\delta^{[\mu\nu\lambda}_{\tau\rho\sigma}\,\gamma^{\alpha\beta]} + 15\,\delta^{[\mu}_{[\tau}\,\gamma_{\rho\sigma]}{}^{\nu\lambda\alpha\beta]}. \tag{D.76}$$

Using this identity, one can rewrite the left-hand side of Eq. (D.75) as

$$\text{L.H.S.} = 60i\tilde{g}\,\bar{\xi}_\lambda\,\delta^{[\mu\nu\lambda]}_{\rho\sigma\tau}\,\gamma^{\alpha\beta]}R_{\mu\nu}{}^{\rho\sigma}\,\psi_{\alpha\beta\|}{}^\tau + \text{h.c.}, \qquad (\text{D.77})$$

where the potential terms with six $\gamma$-matrices have all been eliminated by the Bianchi identities. Whenever the index $\lambda$ is on a $\gamma$-matrix, we will get rid of $\gamma$-matrices altogether by using the identity: $\gamma^{\lambda\alpha} = \gamma^\lambda\gamma^\alpha - \eta^{\lambda\alpha}$, and the $\gamma$-tracelessness of the fermionic ghost. Otherwise, if just one of the indices $\alpha$ and $\beta$ appears on a $\gamma$-matrix, we will use the same identity to obtain a single $\gamma$-trace of $\psi_{\alpha\beta\|}{}^\tau$. These steps leave us with our L.H.S. given by

$$\begin{aligned}&+ 12i\tilde{g}\,\bar{\xi}_\lambda R_{\mu\nu}{}^{\rho\sigma}\left(\delta^{\alpha\beta\mu}_{\rho\sigma\tau}\,\eta^{\nu\lambda} + \delta^{\mu\nu\alpha}_{\rho\sigma\tau}\,\eta^{\beta\lambda} + 2\delta^{\lambda\mu\alpha}_{\rho\sigma\tau}\,\eta^{\nu\beta} + \tfrac{1}{2}\delta^{\lambda\mu\nu}_{\rho\sigma\tau}\,\gamma^{\alpha\beta}\right)\psi_{\alpha\beta\|}{}^\tau + \text{h.c.}\\ &- 24i\tilde{g}\,\bar{\xi}_\lambda R_{\mu\nu}{}^{\rho\sigma}\,\delta^{\lambda\mu\alpha}_{\rho\sigma\tau}\,\gamma^\nu\gamma^\beta\,\psi_{\alpha\beta\|}{}^\tau + 6i\tilde{g}\,\bar{\xi}_\lambda\,\delta^{\lambda\alpha\beta}_{\rho\sigma\tau}\,\slashed{R}^{\rho\sigma}\psi_{\alpha\beta\|}{}^\tau + \text{h.c.}.\end{aligned} \quad (\text{D.78})$$

It is rather easy to see that the entire first line reduces, up to total derivatives, to a $\Gamma$-exact piece modulo $\Delta$-exact terms. Although more difficult to see, the same is true for the second line as well. Let us call the first and the second lines on the right-hand side of Eq. (D.78) respectively as 1st Line and 2nd Line. In 1st Line we can use the relation (E.18), and carry out an explicit computation to write down

$$\begin{aligned}\text{1st Line} = &+ 6i\tilde{g}\,\bar{\xi}_\lambda\left[iR_{\mu\nu}{}^{[\lambda\mu}\slashed{S}^{\nu]} - \partial^{[\lambda}\left(R_{\mu\nu}{}^{\mu\nu]}\psi'\right)\right] + \text{h.c.}\\ &- 8i\tilde{g}\,\bar{\xi}_\lambda\Big[\left(R^{\lambda\nu\alpha\beta} + R^{\alpha\nu\lambda\beta}\right)\partial_\alpha\psi_{\nu\beta} + 2R^{\lambda\beta}\psi_{\alpha\beta\|}{}^\alpha \qquad (\text{D.79})\\ &\qquad - R^{\alpha\beta}\psi^\lambda{}_{\alpha\|\beta} + \tfrac{1}{2}R\psi^{\lambda\alpha\|}{}_\alpha\Big].\end{aligned}$$

One can integrate by parts w.r.t. $\partial_\alpha$ in the terms containing the Riemann tensor, and thereby extract a $\Gamma$-exact piece. The result is

$$\begin{aligned}\text{1st Line} \doteq &- 8i\tilde{g}\,\Gamma\left(\bar{\psi}_{\mu\alpha}R^{\mu\nu\alpha\beta}\psi_{\nu\beta}\right)\\ &+ 2i\tilde{g}\left[3i\bar{\xi}_\lambda R_{\mu\nu}{}^{[\lambda\mu}\slashed{S}^{\nu]} - \bar{\xi}_\lambda\partial^\lambda\left(R\psi'\right) + 2\bar{\xi}_\lambda\partial_\mu\left(R^{\mu\lambda}\psi'\right) - \text{h.c.}\right]\\ &+ 16i\tilde{g}\Big[\bar{\xi}_\lambda\left(\partial_\alpha R^{\lambda\nu\alpha\beta}\psi_{\nu\beta} - R^{\lambda\beta}\psi_{\alpha\beta\|}{}^\alpha\right. \qquad (\text{D.80})\\ &\qquad\qquad\left. + \tfrac{1}{2}R^{\alpha\beta}\psi^\lambda{}_{\alpha\|\beta} - \tfrac{1}{4}R\psi^{\lambda\alpha\|}{}_\alpha\right) - \text{h.c.}\Big],\end{aligned}$$

which is manifestly of the form $\Gamma$-exact plus $\Delta$-exact.

Similarly, in the first term of the 2nd Line, one can use Eq. (E.17) to rewrite that line as

$$24i\tilde{g}\,\bar{\xi}_\lambda\gamma^\mu\left[iR_{\mu\nu}{}^{[\alpha\lambda}\mathcal{S}_\alpha{}^{\nu]} - \partial^{[\lambda}\left(R_{\mu\nu}{}^{\nu\alpha]}\slashed{\psi}_\alpha\right)\right] + 12i\tilde{g}\,\bar{\xi}_\lambda\slashed{R}^{[\alpha\beta}\partial_\alpha\psi_\beta{}^{\lambda]} + \text{h.c.}. \quad (\text{D.81})$$

The second term in the brackets contains manifestly $\Delta$-exact pieces, which we separate:

$$\text{2nd Line} = 8i\tilde{g}\,\bar{\xi}_\lambda \gamma^\mu \big[3iR_{\mu\nu}{}^{[\alpha\lambda}\mathcal{S}_\alpha{}^{\nu]} + \partial^\lambda\left(R_{\mu\alpha}\psi^\alpha\right) - \partial^\alpha\left(R_\mu{}^\lambda\psi_\alpha\right)\big] + \text{h.c.}$$
$$+ 4i\tilde{g}\,\bar{\xi}_\lambda\big[3\slashed{R}^{[\alpha\beta}\partial_\alpha\psi_\beta{}^{\lambda]} - 2\gamma^\mu\partial^\nu\left(R_{\mu\nu}{}^{\alpha\lambda}\psi_\alpha\right)\big] + \text{h.c.} \qquad (D.82)$$

The first line on the right-hand side is now manifestly $\Delta$-exact, whereas the second line can be written as $\Gamma$-exact plus $\Delta$-exact modulo total derivatives, which we will now show.

To this end, we will first compute the $\Gamma$ variation of the following quantity:
$$Z \equiv -4i\tilde{g}\left(\bar{\psi}_{\lambda\alpha}\gamma^{\lambda\nu\rho\sigma}R_{\rho\sigma}{}^{\alpha\beta}\psi_{\nu\beta} + \bar{\psi}_\mu \slashed{R}^{\mu\nu}\psi_\nu\right). \qquad (D.83)$$

The $\Gamma$ variation gives derivatives of the ghost, but integrations by parts will yield

$$\Gamma Z \doteq -2i\tilde{g}\,\bar{\xi}_\lambda\left[\gamma^{\lambda\nu\rho\sigma}R_{\rho\sigma}{}^{\alpha\beta}\psi_{\alpha\beta\|\nu} + \gamma^{\alpha\nu\rho\sigma}R_{\rho\sigma}{}^{\lambda\beta}\psi_{\alpha\nu\|\beta}\right]$$
$$- 4i\tilde{g}\,\bar{\xi}_\lambda\gamma^{\lambda\nu\rho\sigma}\partial_\alpha R_{\rho\sigma}{}^{\alpha\beta}\psi_{\nu\beta} - 4i\tilde{g}\bar{\xi}_\lambda\slashed{\partial}\left(\slashed{R}^{\lambda\alpha}\psi_\alpha\right) + \text{h.c.} \qquad (D.84)$$

Let us use the $\gamma$-matrix identity: $\gamma^{\lambda\nu\rho\sigma} = 2\eta^{\lambda\nu|\rho\sigma} + \frac{1}{2}\left(\gamma^{\lambda\nu}\gamma^{\rho\sigma} + \gamma^{\rho\sigma}\gamma^{\lambda\nu}\right)$ for the first term in the brackets, and $\gamma^{\alpha\nu\rho\sigma} = -2\eta^{\alpha\nu|\rho\sigma} + \frac{1}{2}\left(\gamma^\alpha\gamma^{\rho\sigma}\gamma^\nu - \gamma^\nu\gamma^{\rho\sigma}\gamma^\alpha\right)$ for the second one. Furthermore, we break $\gamma^{\lambda\nu}$ to obtain the $\gamma$-trace of either the ghost (which is zero) or the fermion 1-curl (for which we use Eq. (E.17)). The result is

$$4i\tilde{g}\,\bar{\xi}_\lambda\slashed{R}^{\alpha\beta}\partial_\alpha\psi_\beta{}^\lambda + 2i\tilde{g}\,\bar{\xi}_\lambda\left[\left(\gamma^\beta\slashed{R}^{\lambda\alpha} - \slashed{R}^{\alpha\beta}\gamma^\lambda\right)\partial_\alpha\psi_\beta - 2\partial_\beta\left(\gamma^\beta\slashed{R}^{\lambda\alpha}\psi_\alpha\right)\right]$$
$$- 2i\tilde{g}\,\bar{\xi}_\lambda\left(i\gamma^\alpha\slashed{R}^{\lambda\beta}\mathcal{S}_{\alpha\beta} + 2\gamma^{\lambda\nu\rho\sigma}\partial_\alpha R_{\rho\sigma}{}^{\alpha\beta}\psi_{\nu\beta}\right) + \text{h.c.} \qquad (D.85)$$

In the first line above, for all three quantities inside the brackets, we commute the $\gamma$-matrix past the double $\gamma$-trace of the Riemann tensor. This leaves us with

$$\Gamma Z \doteq -4i\tilde{g}\,\bar{\xi}_\lambda\left(\gamma^{\lambda\nu\rho\sigma}\partial_\alpha R_{\rho\sigma}{}^{\alpha\beta}\psi_{\nu\beta} + 2\gamma_\sigma\partial_\alpha R^{\lambda\beta\alpha\sigma}\psi_\beta + \slashed{R}^{\lambda\alpha}\overleftarrow{\slashed{\partial}}\psi_\alpha\right) + \text{h.c.}$$
$$+ 4i\tilde{g}\,\bar{\xi}_\lambda\left(3\slashed{R}^{[\alpha\beta}\partial_\alpha\psi_\beta{}^{\lambda]} - 2\gamma^\mu\partial^\nu\left(R_{\mu\nu}{}^{\alpha\lambda}\psi_\alpha\right)\right) \qquad (D.86)$$
$$+ 2\tilde{g}\,\bar{\xi}_\lambda\left(\gamma^\alpha\slashed{R}^{\lambda\beta}\mathcal{S}_{\alpha\beta} + \slashed{R}^{\lambda\alpha}\slashed{\mathcal{S}}_\alpha\right).$$

Combining all the results, i.e., Eqs. (D.78), (D.80), (D.82) and (D.86), we finally arrive at Eq. (D.75), where $\Delta\mathfrak{a}$ is given, up to its hermitian

conjugate, by

$$\begin{aligned}\Delta\mathfrak{a} = &- 16i\tilde{g}\,\bar{\xi}_\lambda\big[\partial_\alpha R^{\lambda\nu\alpha\beta}\psi_{\nu\beta} - R^{\lambda\beta}\psi_{\alpha\beta\|}{}^\alpha + \tfrac{1}{2}R^{\alpha\beta}\psi^\lambda{}_{\alpha\|\beta} - \tfrac{1}{4}R\psi^{\lambda\alpha\|}{}_\alpha\big] \\ &- 8i\tilde{g}\,\bar{\xi}_\lambda\big[\partial^\lambda\left(\gamma^\mu R_{\mu\alpha}\slashed{\psi}^\alpha\right) - \partial^\alpha\left(\gamma^\mu R_\mu{}^\lambda\slashed{\psi}_\alpha\right) + \tfrac{1}{2}\partial_\mu\left(R^{\mu\lambda}\psi'\right) - \tfrac{1}{4}\partial^\lambda\left(R\psi'\right)\big] \\ &- 4i\tilde{g}\,\bar{\xi}_\lambda\big[\gamma^{\lambda\nu\rho\sigma}\partial_\alpha R_{\rho\sigma}{}^{\alpha\beta}\psi_{\nu\beta} + 2\gamma_\sigma\partial_\alpha R^{\lambda\beta\alpha\sigma}\slashed{\psi}_\beta + \slashed{R}^{\lambda\alpha}\overleftarrow{\slashed{\partial}}\psi_\alpha\big] \\ &+ 2\tilde{g}\,\bar{\xi}_\lambda\big[3R_{\mu\nu}{}^{[\lambda\mu}\slashed{\mathcal{S}}^{\nu]} + 12\gamma^\mu R_{\mu\nu}{}^{[\alpha\lambda}\mathcal{S}_\alpha{}^{\nu]} + \gamma^\alpha\slashed{R}^{\lambda\beta}\mathcal{S}_{\alpha\beta} + \slashed{R}^{\lambda\alpha}\slashed{\mathcal{S}}_\alpha\big]. \end{aligned} \quad (D.87)$$

This completes our proof.

<div align="center">♮ ♮ ♮</div>

Having found $\Delta\mathfrak{a}$, we will now see how this quantity may be related to $\Delta a_{1g}$, given by Eq. (6.23). This will lead us to the desired relation (6.29). Note from Eq. (6.23) that the graviton EoMs in $\Delta a_{1g}$ appear only through the Einstein tensor $G^{\mu\nu}$. Therefore, we will rewrite all the $\Delta$-exact terms in the first, second and third lines on the right-hand side of Eq. (D.87) in terms of the Einstein tensor, by making use of the relations (E.3)–(E.7). For he antisymmetric 4-$\gamma$, we use the identity (D.61) in order to kill some terms that give the $\gamma$-trace of $\bar{\xi}_\lambda$. We find that all the terms proportional to the trace of the Einstein tensor (Ricci scalar) combine into $\Gamma$-exact pieces. After some simplifications, the result is

$$\begin{aligned}\Delta\mathfrak{a} \doteq &+ 8i\tilde{g}\,\bar{\xi}_\lambda\left[2G^{\mu\nu}\partial^\lambda\psi_{\mu\nu} - 3\partial_\mu\left(G^{\mu\nu}\psi_\nu{}^\lambda\right) + \partial^\nu\left(G^{\lambda\mu}\psi_{\mu\nu}\right)\right] \\ &+ 4i\tilde{g}\,\Gamma\left[\slashed{\bar{\psi}}_\mu\slashed{G}_\nu\psi^{\mu\nu} + \bar{\psi}_{\mu\nu}\slashed{G}^\mu\slashed{\psi}^\nu - \slashed{\bar{\psi}}_\mu G^{\mu\nu}\slashed{\psi}_\nu + 2\bar{\psi}_{\mu\alpha}G^{\mu\nu}\psi_\nu{}^\alpha\right] \\ &- 6i\tilde{g}\,\Gamma\left[\bar{\psi}'G^{\mu\nu}\psi_{\mu\nu} + \bar{\psi}_{\mu\nu}G^{\mu\nu}\psi' - \tfrac{1}{3}\bar{\psi}_{\mu\nu}R\psi^{\mu\nu} + \tfrac{1}{2}\bar{\psi}'R\psi'\right] \quad (D.88) \\ &+ 4\tilde{g}\,\bar{\xi}_\lambda\big[\tfrac{3}{2}R_{\mu\nu}{}^{[\lambda\mu}\slashed{\mathcal{S}}^{\nu]} + 6\gamma^\mu R_{\mu\nu}{}^{[\alpha\lambda}\mathcal{S}_\alpha{}^{\nu]} + \gamma^{\alpha\rho\sigma}R_{\rho\sigma}{}^{\lambda\beta}\mathcal{S}_{\alpha\beta} \\ &\qquad\qquad + 2\slashed{G}_\alpha\mathcal{S}^{\alpha\lambda} - G^{\beta\lambda}\slashed{\mathcal{S}}_\beta\big] + \text{h.c.} \end{aligned}$$

The entire first line on the right-hand side plus its hermitian conjugate is easily identified, up to an overall factor, as $\Delta a_{1g}$. All the remaining terms, on the other hand, are $\Delta$ variations of $\Gamma$-closed quantities, and can be identified as $\Delta\tilde{a}_1$. Explicitly,

$$\begin{aligned}\tilde{a}_1 = &+ 8i\tilde{g}\,\Gamma\left(\slashed{\bar{\psi}}_\mu h^{*\mu\nu}_\nu\psi^{\mu\nu} - \tfrac{1}{2}\slashed{\bar{\psi}}_\mu h^{*\mu\nu}\slashed{\psi}_\nu + \bar{\psi}_{\mu\alpha}h^{*\mu\nu}\psi_\nu{}^\alpha - \tfrac{3}{2}\bar{\psi}'h^{*\mu\nu}\psi_{\mu\nu}\right) \\ &- \left(\tfrac{4}{D-2}\right)i\tilde{g}\,\Gamma\left(\bar{\psi}_{\mu\nu}h^{*\prime}\psi^{\mu\nu} - \tfrac{3}{2}\bar{\psi}'h^{*\prime}\psi'\right) + 4\tilde{g}\,\bar{\xi}_\lambda\big(\tfrac{3}{2}R_{\mu\nu}{}^{[\lambda\mu}\slashed{\varphi}^{*\nu]} \\ &+ 6\gamma^\mu R_{\mu\nu}{}^{[\alpha\lambda}\varphi^{*\,\nu]}_\alpha + \gamma^{\alpha\rho\sigma}R_{\rho\sigma}{}^{\lambda\beta}\varphi^*_{\alpha\beta} + 2\slashed{G}_\alpha\varphi^{*\alpha\lambda} - G^{\beta\lambda}\slashed{\varphi}^*_\beta\big) + \text{h.c.} \end{aligned} \quad (D.89)$$

Thus we have proved the relation (6.29), where the ambiguity is given by the above expression.

### D.3.2  3-Derivatives $1-5/2-5/2$ Vertex

First, we will show that the $a_2$ presented in Eq. (6.33) is equivalent to that appearing in the third line of Eq. (6.9). Given the identities (D.60) and (D.61), we rewrite Eq. (6.33) as

$$\begin{aligned}a_2 = &- ig\, C^*_\lambda \bar\xi_{\mu\nu}\gamma^\lambda\big( -\eta^{\mu\nu|\alpha\beta} + \tfrac{1}{2}\gamma^{\mu\nu}\gamma^{\alpha\beta} - 2\gamma^\mu\eta^{\nu\alpha}\gamma^\beta\big)\xi_{\alpha\beta} \\ &- ig\, C^*_\lambda \bar\xi_{\mu\nu}\big( -\eta^{\mu\nu|\alpha\beta} + \tfrac{1}{2}\gamma^{\mu\nu}\gamma^{\alpha\beta} - 2\gamma^\mu\eta^{\nu\alpha}\gamma^\beta\big)\gamma^\lambda\xi_{\alpha\beta}.\end{aligned} \quad \text{(D.90)}$$

It is clear that only the first terms in both the lines on the right-hand side are nontrivial, since the $\gamma$-trace of the ghost-curl $\xi_{\alpha\beta}$ is $\Gamma$-exact. This leaves us with

$$a_2 = 2ig\, C^*_\lambda\, \bar\xi_{\mu\nu}\gamma^\lambda\xi^{\mu\nu} + \Gamma\text{-exact}, \quad \text{(D.91)}$$

thereby proving the claimed equivalence.

♮ ♮ ♮

Now we will prove the statements that follow Eq. (6.37). Let us take the first term on the right-hand side of Eq. (6.37),

$$igR_{\mu\nu\rho\sigma}\bar\xi_\lambda\,\gamma^{\lambda\mu\nu\alpha\beta}\,\Psi_{\alpha\beta|}{}^{\rho\sigma}, \quad \text{(D.92)}$$

and use the identity (D.61) to rewrite it as

$$igR_{\mu\nu\rho\sigma}\bar\xi_\lambda\,\gamma^{\lambda\mu\nu\alpha\beta}\,\big( -\tfrac{1}{2}\gamma^{\rho\sigma\gamma\text{d}} + \tfrac{1}{2}\gamma^{\rho\sigma}\gamma^{\gamma\text{d}} - 2\gamma^{[\rho}\eta^{\sigma][\gamma}\gamma^{\delta]}\big)\Psi_{\alpha\beta|\gamma\text{d}}. \quad \text{(D.93)}$$

The first term above (when expanding the brackets) plus its hermitian conjugate is $\Gamma$-exact modulo d, while the remaining terms are $\Delta$-exact. To see this, let us massage these terms. We have

$$\text{1rst T.+h.c.} = -\tfrac{i}{2}gR_{\mu\nu\rho\sigma}\big(\bar\xi_\lambda\gamma^{\lambda\mu\nu\alpha\beta,\,\rho\sigma\gamma\text{d}}\,\Psi_{\alpha\beta|\gamma\text{d}} - \bar\Psi_{\alpha\beta|\gamma\text{d}}\,\gamma^{\lambda\mu\nu\alpha\beta,\,\rho\sigma\gamma\text{d}}\xi_\lambda\big), \quad \text{(D.94)}$$

by virtue of the fact that the antisymmetric products of 5-$\gamma$ and 4-$\gamma$ commute for exactly the same reason as presented in between Eqs. (D.72) and (D.74). Now we can pull $\partial_\mu$ off the Riemann tensor to integrate by parts. Because of the Bianchi identities, we get

$$\text{1rst+h.c.} \doteq -\tfrac{i}{2}g\,\mathfrak{h}_{\rho\sigma\|\lambda}\big(\bar\xi_{\mu\nu}\gamma^{\lambda\mu\nu\alpha\beta,\,\rho\sigma\gamma\text{d}}\,\Psi_{\alpha\beta|\gamma\text{d}} - \bar\Psi_{\alpha\beta|\gamma\text{d}}\,\gamma^{\lambda\mu\nu\alpha\beta,\,\rho\sigma\gamma\text{d}}\xi_{\mu\nu}\big), \quad \text{(D.95)}$$

Finally, we pull $\partial_\gamma$ off the spin-$\tfrac{5}{2}$ curvature and integrate by parts to obtain a derivative $\partial_\gamma\xi_{\mu\nu}$ of the ghost-curl, which is $\Gamma$-exact. Thus we end up having

$$\text{First Term} + \text{h.c.} \doteq -ig\,\Gamma\big(\mathfrak{h}^{\rho\sigma\|\lambda}\bar\psi_{\mu\nu\|\gamma}\,\gamma^{\lambda\mu\nu\alpha\beta,\,\rho\sigma\gamma\text{d}}\,\psi_{\alpha\beta\|\text{d}}\big). \quad \text{(D.96)}$$

On the other hand, it is manifest that the second and third terms appearing in Eq. (6.37) are $\Delta$-exact quantities. Moreover, they are $\Delta$ variations of some $\Gamma$-closed objects. The following choice of the ambiguity will eliminate these terms:

$$\Delta \tilde{a}_1 = -ig R_{\mu\nu\rho\sigma} \bar{\xi}_\lambda \, \gamma^{\lambda\mu\nu\alpha\beta} \left( \tfrac{1}{2} \gamma^{\rho\sigma} \gamma^{\gamma d} - 2\gamma^{[\rho} \eta^{\sigma][\gamma} \gamma^{\delta]} \right) \Psi_{\alpha\beta|\gamma\delta} + \text{h.c.} \,. \quad \text{(D.97)}$$

This choice is tantamount to

$$\tilde{a}_1 = -g R_{\mu\nu\rho\sigma} \, \bar{\xi}_\lambda \big( 4\gamma^{\lambda\mu\nu\alpha\beta,\, \rho} \, \partial_{[\alpha} \varphi^*_{\beta]}{}^\sigma + \tfrac{1}{D} \, \gamma^{\lambda\mu\nu\alpha\beta,\, \rho\sigma} \, \psi^*_{\alpha\beta} \big) + \text{h.c.,} \quad \text{(D.98)}$$

and with this we arrive at Eq. (6.38).

## D.4 Gauge Algebra-Preserving Vertices

To prove this, first we note that it is always possible to rewrite a cubic vertex as

$$a_0 = T^{\mu\nu} h_{\mu\nu}, \quad \text{(D.99)}$$

i.e., the graviton field $h_{\mu\nu}$ contracted with a symmetric fermion-bilinear current $T^{\mu\nu}$. If the vertex is abelian, we will see that the latter can be chosen to satisfy

$$\Gamma T^{\mu\nu} = 0, \qquad \partial_\nu T^{\mu\nu} = \Delta M^\mu \quad \text{with} \quad \Gamma M^\mu = 0. \quad \text{(D.100)}$$

For $s = n + \tfrac{1}{2}$, let us write the most general form of the $a_1$ corresponding to (D.99):

$$a_1 = 2M^\mu C_\mu + \big( \bar{P}_{\mu_1\ldots\mu_{n-1}} \xi^{\mu_1\ldots\mu_{n-1}} - \bar{\xi}_{\mu_1\ldots\mu_{n-1}} P^{\mu_1\ldots\mu_{n-1}} \big) + a_1', \quad \text{(D.101)}$$

where $M^\mu$ and $P_{\mu_1\ldots\mu_{n-1}}$ belong to H($\Gamma$) and have $pgh = 0$, $\text{agh}\# = 1$, and $a_1'$ stands for expansion terms in the ghost-curls. The consistency condition (6.78) now reads

$$0 \doteq \Delta a_1' + 2\Delta M^\mu C_\mu + \big( \Delta \bar{P}_{\mu_1\ldots\mu_{n-1}} \xi^{\mu_1\ldots\mu_{n-1}} - \bar{\xi}_{\mu_1\ldots\mu_{n-1}} \Delta P^{\mu_1\ldots\mu_{n-1}} \big)$$
$$+ \Gamma \left( T^{\mu\nu} h_{\mu\nu} \right). \quad \text{(D.102)}$$

It is clear from the properties of $P_{\mu_1\ldots\mu_{n-1}}$ that it may consist of two kinds of terms: one contains the antifield $h^{*\mu\nu}$ and its derivatives, and the other contains the antifield $\psi^{*\nu_1\ldots\nu_n}$ and its derivatives. The former kind also contains (derivatives of) the Fronsdal tensor $\mathcal{S}_{\nu_1\ldots\nu_n}$ or (derivatives of) the curvature $\Psi_{\mu_1\nu_1|\ldots|\mu_n\nu_n}$, while the latter one contains (derivatives of) the linearized Riemann tensor $R_{\mu\nu\rho\sigma}$. By using the Leibniz rule, however, one can choose to get rid of derivatives on $h^{*\mu\nu}$ and $R_{\mu\nu\rho\sigma}$. Thus one can write

$$P_{\mu_1\ldots\mu_{n-1}} = h^{*\mu\nu} \left[ \vec{P}^{(\mathcal{S})\nu_1\ldots\nu_n}_{\mu\nu,\,\mu_1\ldots\mu_{n-1}} \mathcal{S}_{\nu_1\ldots\nu_n} + \vec{P}^{(\Psi)\nu_1\rho_1|\ldots|\nu_n\rho_n}_{\mu\nu,\,\mu_1\ldots\mu_{n-1}} \Psi_{\nu_1\rho_1|\ldots|\nu_n\rho_n} \right]$$
$$+ R^{\mu\nu\rho\sigma} \vec{P}^{(\psi^*)\nu_1\ldots\nu_n}_{\mu\nu\rho\sigma,\,\mu_1\ldots\mu_{n-1}} \psi^*_{\nu_1\ldots\nu_n} + \partial^{\mu_n} p_{\mu_1\ldots\mu_n}, \tag{D.103}$$

where $\Gamma p_{\mu_1\ldots\mu_n} = 0$, and the $\vec{P}$'s are differential operators acting to the right. Notice that in the above expression both terms in the brackets are not only $\Gamma$-closed but also $\Delta$-exact.[2] Now, taking the $\Delta$ variation of $P_{\mu_1\ldots\mu_{n-1}}$ one finds from Eq. (D.103) that

$$\Delta P_{\mu_1\ldots\mu_{n-1}} = \tfrac{1}{4} R^{\mu\nu\rho\sigma} \Delta Q_{\mu\nu\rho\sigma,\,\mu_1\ldots\mu_{n-1}} + \partial^{\mu_n} \Delta q_{\mu_1\ldots\mu_n}, \tag{D.104}$$

where the quantity $Q_{\mu\nu\rho\sigma,\,\mu_1\ldots\mu_{n-1}}$ is $\Gamma$-closed and enjoys the same symmetries in its first four indices as the Riemann tensor, and $\Gamma q_{\mu_1\ldots\mu_n} = 0$. Therefore, one finds that

$$\bar{\xi}^{\mu_1\ldots\mu_{n-1}} \Delta P_{\mu_1\ldots\mu_{n-1}} \doteq h^{\mu\nu} \Delta \left[ \partial^\alpha \partial^\beta \left( \bar{\xi}_{\mu_1\ldots\mu_{n-1}} Q_{\mu\alpha\nu\beta,}{}^{\mu_1\ldots\mu_{n-1}} \right) \right]$$
$$- \bar{\xi}_{\mu_1\ldots\mu_{n-1}} \overleftarrow{\partial}_{\mu_n} \Delta q^{\mu_1\ldots\mu_n}. \tag{D.105}$$

The last term on the right-hand side above is $\Gamma$-closed, and can be broken into a $\Gamma$-exact piece plus terms involving the fermionic ghost-curls. The latter can always be canceled in the cocycle condition (D.102) by appropriately choosing $a'_1$. One is thus left with

$$\Gamma \left[ T^{\mu\nu} h_{\mu\nu} + \Delta \left( \tfrac{1}{n} \bar{\psi}_{\mu_1\ldots\mu_n} q^{\mu_1\ldots\mu_n} + \text{h.c.} \right) \right] + 2 \Delta M^\mu C_\mu \tag{D.106}$$
$$- h^{\mu\nu} \Delta \left[ \partial^\alpha \partial^\beta \left( \bar{\xi}_{\mu_1\ldots\mu_{n-1}} Q_{\mu\alpha\nu\beta,}{}^{\mu_1\ldots\mu_{n-1}} + \text{h.c.} \right) \right] \doteq 0.$$

Now, one can drop the $\Delta$-exact terms added to the original vertex $T^{\mu\nu} h_{\mu\nu}$ to write

$$h^{\mu\nu} \left[ \Gamma T_{\mu\nu} - \partial^\alpha \partial^\beta \left( \bar{\xi}_{\mu_1\ldots\mu_{n-1}} \Delta Q^{\mu_1\ldots\mu_{n-1}}_{\mu\alpha\nu\beta,} + \text{h.c.} \right) \right] + 2 \left( \Delta M^\mu - \partial_\nu T^{\mu\nu} \right) C_\mu \doteq 0. \tag{D.107}$$

Taking a functional derivative w.r.t. $C_\mu$ then yields the second condition in Eq. (D.100):

$$\partial_\nu T^{\mu\nu} = \Delta M^\mu, \tag{D.108}$$

---

[2] While the $\Delta$-exactness of the first term therein is manifest, the second term contains the spin-$s$ curvature, which admits only $\Delta$-exact terms like its own ($\gamma$-)traces and divergences (see Appendix E), thanks to the way the indices are contracted.

with $\Gamma M^\mu = 0$ by assumption. On the other hand, a functional derivative w.r.t. $h_{\mu\nu}$ gives

$$\Gamma T_{\mu\nu} = \partial^\alpha \partial^\beta \left( \bar{\xi}_{\mu_1...\mu_{n-1}} \Delta Q_{\mu\alpha\nu\beta,}{}^{\mu_1...\mu_{n-1}} \right) + \text{h.c.}, \tag{D.109}$$

which means, in particular, that the quantity on the right-hand side must be $\Gamma$-exact. This is possible if $\partial^\alpha \partial^\beta Q_{\mu\alpha\nu\beta,\,\mu_1...\mu_{n-1}}$ is $\Delta$-closed, and the indices of $Q$ have the interchange symmetry $\alpha \leftrightarrow \mu_i$ and $\beta \leftrightarrow \mu_i$ with $i = 1, 2, \ldots, n-1$. This enables one to conclude

$$\begin{aligned} T_{\mu\nu} = \tilde{T}_{\mu\nu} + \tfrac{1}{n} \, \Delta \big[ & 2\bar{\psi}_{\mu_1...\mu_n} \partial_\alpha Q_{(\mu}{}^\alpha{}_{\nu)}{}^{\mu_1,\,\mu_2...\mu_n} \\ & + \partial_\alpha \bar{\psi}_{\mu_1...\mu_n} Q_{(\mu}{}^\alpha{}_{\nu)}{}^{\mu_1,\,\mu_2...\mu_n} + \text{h.c.} \big], \end{aligned} \tag{D.110}$$

where $\Gamma \tilde{T}_{\mu\nu} = 0$. Therefore, one can render the current gauge invariant by field redefinitions without affecting the form (D.108) of its divergence. This completes the proof of Eq. (D.100). Then the $a_1$ following from Eq. (6.78) reads

$$a_1 = 2M^\mu C_\mu. \tag{D.111}$$

We will now prove a sufficient condition for the triviality of $a_1$, given by (D.111), and hence of the deformation of the gauge transformations. It is

$$\Delta M^\mu = \partial_\nu \mathcal{X}^{(\mu\nu)} + \partial_\rho \partial_\sigma \mathcal{Y}^{\mu\rho\sigma}, \qquad \text{with } \mathcal{X}^{(\mu\nu)}, \mathcal{Y}^{\mu\rho\sigma} \; \Delta\text{-exact and } \Gamma\text{-closed.} \tag{D.112}$$

If Eq. (D.112) is true, then from Eq. (D.111) we can write $\Delta a_1$ as

$$\Delta a_1 = 2\big(\partial_\nu \mathcal{X}^{(\mu\nu)} + \partial_\nu \partial_\rho \mathcal{Y}^{\mu\nu\rho}\big) C_\mu \doteq -2\mathcal{X}^{(\mu\nu)} \partial_{(\mu} C_{\nu)} + 2\mathcal{Y}^{\mu\nu\rho} \partial_\nu \partial_\rho C_\mu. \tag{D.113}$$

But the derivatives of the bosonic ghost are $\Gamma$-exact: $2\partial_{(\mu} C_{\nu)} = \Gamma h_{\mu\nu}$ and $2\partial_\nu \partial_\rho C_\mu = \partial_\rho \Gamma h_{\mu\nu} - \Gamma \mathfrak{h}_{\mu\nu\|\rho}$. Because $\mathcal{X}^{(\mu\nu)}$ and $\mathcal{Y}^{\mu\nu\rho}$ are $\Gamma$-closed, one can write

$$\Delta a_1 \doteq -\Gamma \big[ \big( \mathcal{X}^{(\mu\nu)} + \partial_\rho \mathcal{Y}^{\mu\nu\rho} \big) h_{\mu\nu} + \mathcal{Y}^{\mu\nu\rho} \mathfrak{h}_{\mu\nu\|\rho} \big]. \tag{D.114}$$

In view of the the cocycle condition $\Delta a_1 \doteq -\Gamma a_0$, one can therefore write

$$\Gamma \big[ a_0 - \big( \mathcal{X}^{(\mu\nu)} + \partial_\rho \mathcal{Y}^{\mu\nu\rho} \big) h_{\mu\nu} - \mathcal{Y}^{\mu\nu\rho} \mathfrak{h}_{\mu\nu\|\rho} \big] \doteq 0. \tag{D.115}$$

Because the quantities added to $a_0$ on the left-hand side are $\Delta$-exact by assumption, one can render the vertex gauge-invariant only up to a total derivative, by field redefinitions. This proves the triviality of $a_1$ if Eq. (D.112) holds.

## D.5 Beyond Cubic Order: Gravitational Case

Let us recall from Chapter 4 that consistent second-order deformations require
$$(S_1, S_1) = -2sS_2 = -2\Gamma S_2 - 2\Delta S_2, \tag{D.116}$$
and that this antibracket is zero for the abelian vertices, which go unobstructed beyond the cubic level. The non-abelian vertices, on the other hand, have nontrivial $a_1$ and $a_2$ and may not fulfill this requirement. Here we will prove by contradiction that indeed they do not. The line of reasoning is close to that of Section 5.4, where we perform a similar analysis (with same conclusions) for the photon-coupled massless fermions.

Notice that $S_2$ is at most linear in the antifields $\Phi_A^*$, on which $\Gamma$ does not act. On the other hand, only the $\Delta$ variation of an antighost can produce an antifield. Therefore, the general form of the antibracket evaluated at zero antifields is
$$(S_1, S_1)|_{\Phi_A^*=0} = \Gamma N + \Delta M, \quad N \equiv -2\left[S_2\right]_{\Phi_A^*=0}, \; M \equiv -2\left[S_2\right]_{\mathcal{C}_\alpha^*=0}. \tag{D.117}$$
Let us also note that in the antibracket of $S_1 = \int (a_2 + a_1 + a_0)$ with itself, among all the possibilities, only the antibracket between $\int a_0$ and $\int a_1$ survives when the antifields are set to zero. Thus one is left with
$$(S_1, S_1)|_{\Phi_A^*=0} = 2\left(\int a_0, \int a_1\right) \equiv \int b. \tag{D.118}$$

It is relatively easier to compute the quantity $b$, which must satisfy the following requirement in view of Eqs. (D.117) and (D.118):
$$b \doteq \Gamma\text{-exact} + \Delta\text{-exact}. \tag{D.119}$$

For simplicity, we shall again stick to the simplest non-trivial case, which this time is that of the spin-$\frac{5}{2}$ field. As we now have two different non-abelian vertices, below we discuss them separately — recall that abelian vertices always go unobstructed beyond the cubic order.

**The 2-Derivatives Vertex**

For the 2-derivatives $2-\frac{5}{2}-\frac{5}{2}$ vertex, let us write down the deformations $a_0$ and $a_1$. First, from Eq. (6.32), one can rewrite the vertex as $a_0 \doteq T^{\mu\nu} h_{\mu\nu}$. The result is
$$\begin{aligned}T^\mu{}_\nu = 4ig\big[&\partial^\rho\partial_\sigma\{\bar{\psi}_{\nu\lambda}\big(\eta^{\mu\sigma|\lambda\tau} + \tfrac{1}{2}\gamma^{\mu\sigma\lambda\tau}\big)\psi_{\rho\tau} + \tfrac{1}{2}\bar{\slashed{\psi}}_{[\nu}\gamma^{\mu\sigma}\slashed{\psi}_{\rho]}\} \\ &+ \tfrac{1}{16}\bar{\psi}_{\rho\sigma\|}{}^\lambda\gamma^{\mu\rho\sigma\alpha\beta,}{}_{\nu\lambda\gamma}\psi_{\alpha\beta\|}{}^\gamma\big].\end{aligned} \tag{D.120}$$

On the other hand, from Eqs. (6.21)–(6.22) and Eq. (D.89) we can write

$$a_1 = igh^{*\mu}{}_\nu \left( \bar{\xi}_{\mu\lambda}\psi^{\nu\lambda} + \bar{\psi}^{\nu\lambda}\xi_{\mu\lambda} - 2\bar{\xi}^\lambda \psi_{\mu\lambda\|}{}^\nu - 2\bar{\psi}_{\mu\lambda\|}{}^\nu \xi^\lambda + \Gamma j_\mu{}^\nu \right) + \cdots, \qquad (D.121)$$

where the ellipses stand for terms containing the antifield $\psi^{*\mu\nu}$, and $j_\mu{}^\nu$ is some spin-$\frac{5}{2}$ bilinear. Then the quantity $b$ will contain 4-fermions terms plus fermion bilinears:[3]

$$b = 2ig\, T^\mu{}_\nu \left( \bar{\xi}_{\mu\lambda}\psi^{\nu\lambda} + \bar{\psi}^{\nu\lambda}\xi_{\mu\lambda} - 2\bar{\xi}^\lambda \psi_{\mu\lambda\|}{}^\nu - 2\bar{\psi}_{\mu\lambda\|}{}^\nu \xi^\lambda + \Gamma j_\mu{}^\nu \right) + \cdots. \qquad (D.122)$$

Note that the two kinds of terms are completely different and we can treat them separately. If Eq. (D.119) is fulfilled, a functional derivative thereof w.r.t. $\bar{\xi}_\mu$ has to be $\Delta$-exact up to the divergence of a symmetric tensor. This functional derivative reads:

$$\frac{\delta b}{\delta \bar{\xi}_\mu} = 4ig \left[ \partial_\nu \left( T_\rho^{[\mu} \psi^{\nu]\rho} \right) + T_{\rho\sigma} \psi^{\mu\rho\|\sigma} \right] + \cdots. \qquad (D.123)$$

and because the vertex is nontrivial, $T^\mu{}_\nu$ cannot be $\Delta$-exact. Now, the right-hand side of Eq. (D.123) is trilinear in the spin-$\frac{5}{2}$ field. Given that possible Fierz rearrangements cannot redistribute the derivatives among the fields, let us consider, among others, the terms in which three derivatives act on a single fermion. By inspection, it is clear that these terms cannot be written as $\Delta$-exact quantities up to the divergence of a symmetric tensor. Therefore, it is not possible to satisfy Eq. (D.119). Then, in a local theory, the non-abelian $2-\frac{5}{2}-\frac{5}{2}$ vertex with two derivatives is obstructed beyond the cubic order.

**The 3-Derivatives Vertex**

The proof for the 3-derivatives case is in the same spirit as the previous example. Let us rewrite, from (6.39), the vertex as $a_0 \doteq T^{\mu\nu} h_{\mu\nu}$, with the current given by

$$T_\mu{}^\nu = 2ig \partial^\lambda \left( \bar{\psi}_{\rho\sigma\|\mu}\, \gamma^{\nu\rho\sigma\alpha\beta}\, \psi_{\alpha\beta\|\lambda} - \bar{\psi}_{\rho\sigma\|\lambda}\, \gamma^{\nu\rho\sigma\alpha\beta}\, \psi_{\alpha\beta\|\mu} \right.\\ \left. + \bar{\psi}_{\rho\sigma\|}{}^\gamma\, \gamma^{\nu\rho\sigma\alpha\beta,}{}_{\mu\lambda\gamma\delta}\, \psi_{\alpha\beta\|}{}^\delta \right). \qquad (D.124)$$

Now $a_1$ is given by Eqs. (6.34) and (D.98), and it has the form:

$$a_1 = -2ig h^{*\,\nu}_\mu \left( \bar{\xi}_{\rho\sigma}\gamma^{\mu\rho\sigma\alpha\beta}\psi_{\alpha\beta\|\nu} - \bar{\psi}_{\rho\sigma\|\nu}\gamma^{\mu\rho\sigma\alpha\beta}\xi_{\alpha\beta} \right) + \cdots. \qquad (D.125)$$

---

[3] The latter terms, which we do not make explicit, come from the ellipses in (D.121).

Again, the quantity $b$ will contain 4-fermions terms and fermion bilinears:

$$b = -4ig\, T_\mu{}^\nu \left( \bar{\xi}_{\rho\sigma} \gamma^{\mu\rho\sigma\alpha\beta} \psi_{\alpha\beta\|\nu} - \bar{\psi}_{\rho\sigma\|\nu} \gamma^{\mu\rho\sigma\alpha\beta} \xi_{\alpha\beta} \right) + \cdots. \qquad (D.126)$$

Let us consider, in the functional derivative of $b$ w.r.t. $\bar{\xi}_\mu$, the terms with three derivatives acting on a single fermion to find that they cannot be written as $\Delta$-exact objects modulo the divergence of a symmetric tensor. Therefore, Eq. (D.119) will not be satisfied, and so in a local theory the 3-derivatives $2-\tfrac{5}{2}-\tfrac{5}{2}$ vertex is also inconsistent at the quartic order. Finally, upon inspection one easily deduces that, even if linearly combined, the two vertices above would suffer from the same obstruction.

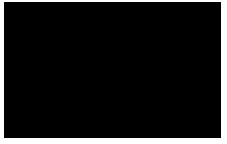

APPENDIX E

# Cohomologies

The basic cohomological relations that we use in the main part of the text are summarized here. For our higher-spin fields, we only present the fermionic case which is of interest to us, and refer the interested reader to [248] for an analogous treatment involving higher-spin bosons.

## E.1 The Cohomology of $\Delta$: Equations of Motion

### The Photon

The original[1] equations of motion for the photon are given by

$$\partial^\mu F_{\mu\nu} = \Box A_\nu - \partial_\nu(\partial \cdot A) = \Delta A^*_\nu, \qquad (E.1)$$

which, upon taking a 1-curl thereof yields

$$\Box F_{\mu\nu} = 2\Delta(\partial_{[\mu} A^*_{\nu]}), \qquad (E.2)$$

showing us a less well-known form of the spin-1 equations of motion.

### The Graviton

The original equations of motion for the graviton are expressed in terms of the linearized Einstein tensor,

$$G_{\mu\nu} \equiv R_{\mu\nu} - \tfrac{1}{2}\eta_{\mu\nu} R = \Delta h^*_{\mu\nu}. \qquad (E.3)$$

---

[1] By 'original' we mean those which follow from equating to zero the variation of the action principle given in the main text.

Taking a trace, it follows immediately that

$$R_{\mu\nu} = \Delta\left(h^*_{\mu\nu} - \tfrac{1}{D-2}\eta_{\mu\nu}h^{*\prime}\right), \qquad R = -\left(\tfrac{2}{D-2}\right)\Delta h^*, \qquad \text{(E.4)}$$

which one can use to show the contracted Bianchi identity, which says that the divergence of the Riemann tensor is $\Delta$-exact:

$$\partial^\rho R_{\mu\nu\rho\sigma} = 2\partial_{[\mu}R_{\nu]\sigma} = 2\partial_{[\mu}G_{\nu]\sigma} - \eta_{\sigma[\mu}\partial_{\nu]}R = \Delta\text{-exact.} \qquad \text{(E.5)}$$

From the above it of course follows that

$$\partial^\rho \slashed{R}_{\rho\sigma} = (\gamma^\mu\gamma^\nu - \eta^{\mu\nu})\left(2\partial_\mu G_{\nu\sigma} - \eta_{\sigma\mu}\partial_\nu R\right) = 2\slashed{\partial}\slashed{G}_\sigma - \gamma_\sigma\slashed{\partial}R + \slashed{\partial}_\sigma R = \Delta(\ldots), \qquad \text{(E.6)}$$

from which one can then derive, making use of $\gamma^\alpha\gamma^{\rho\sigma} = 2\gamma^{\alpha\rho\sigma} - \gamma^{\rho\sigma}\gamma^\alpha$ and the Bianchi identity $\partial_{[\rho}R_{\mu\nu]\alpha\beta} = 0$, that

$$\slashed{\partial}\,\slashed{R}_{\mu\nu} = -\slashed{R}_{\mu\nu}\overleftarrow{\slashed{\partial}} = 4\gamma^\rho\partial_{[\mu}R_{\nu]\rho} = \Delta\text{-exact.} \qquad \text{(E.7)}$$

Another consequence of (E.5) is of course the divergenceless property of the Einstein tensor, namely $\partial^\mu G_{\mu\nu} = 0$, which is most well-known. Other forms of the equations of motion, that we do not use, include $\gamma^\mu\slashed{R}_{\mu\nu}$ and $\Box R_{\mu\nu\rho\sigma}$.

**The Rarita–Schwinger**

For spin $\tfrac{3}{2}$, the original equations of motion read

$$\mathcal{R}^\mu = \gamma^{\mu\alpha\beta}\Psi_{\alpha\beta} = -2i\Delta\psi^{*\mu}, \qquad \text{(E.8a)}$$

$$\bar{\mathcal{R}}^\mu = \bar{\Psi}_{\alpha\beta}\gamma^{\alpha\beta\mu} = 2i\Delta\bar{\psi}^{*\mu}. \qquad \text{(E.8b)}$$

One can take the $\gamma$-trace of (E.8a), and use $\gamma_\mu\gamma^{\mu\alpha\beta} = (D-2)\gamma^{\alpha\beta}$ to obtain

$$\gamma^{\mu\nu}\Psi_{\mu\nu} = 2\left(\slashed{\partial}\slashed{\psi} - \partial\cdot\psi\right) = -2i\Delta\left(\tfrac{1}{D-2}\slashed{\psi}^*\right). \qquad \text{(E.9)}$$

Alternatively, one can start from the same (E.8a) but instead use the identity $\gamma^{\mu\alpha\beta} = \gamma^\mu\gamma^{\alpha\beta} - 2\eta^{\mu[\alpha}\gamma^{\beta]}$ and then the above equation, which leaves one with the very useful form

$$\gamma^\mu\Psi_{\mu\nu} = -iS_\nu = \slashed{\partial}\psi_\nu - \partial_\nu\slashed{\psi} = -i\Delta\left(\psi^*_\nu - \tfrac{1}{D-2}\gamma_\nu\slashed{\psi}^*\right). \qquad \text{(E.10)}$$

Further taking a curl of the above relation and using (A.13) one finds

$$\slashed{\partial}\,\Psi_{\mu\nu} = -2i\Delta\left(\partial_{[\mu}\psi^*_{\nu]} - \tfrac{1}{D-2}\gamma_{[\nu}\partial_{\mu]}\slashed{\psi}^*\right). \qquad \text{(E.11)}$$

Another useful form can be obtained by applying the Dirac operator on (E.10) and then getting rid of $\not{\partial}\psi$ in the resulting expression by using (E.9). The result is

$$\partial^\mu \Psi_{\mu\nu} = \Box \psi_\nu - \partial_\nu (\partial \cdot \psi) = -i\Delta \big[\not{\partial} \psi^*_\nu + \tfrac{1}{D-2}\gamma_{\nu\rho}\partial^\rho \not{\psi}^*\big]. \quad (\text{E.12})$$

Similarly, one could have started with (E.8b) to derive the following:

$$\bar{\Psi}_{\mu\nu}\gamma^{\mu\nu} = 2\big(\bar{\psi}\cdot\overleftarrow{\partial} - \bar{\not{\psi}}\overleftarrow{\not{\partial}}\big) = 2i\Delta\big(\tfrac{1}{D-2}\bar{\not{\psi}}^*\big), \quad (\text{E.13a})$$

$$\bar{\Psi}_{\mu\nu}\gamma^\nu = \bar{\not{\psi}}\overleftarrow{\partial}_\mu - \bar{\psi}_\mu \overleftarrow{\not{\partial}} = i\Delta\big(\bar{\psi}^*_\mu - \tfrac{1}{D-2}\bar{\not{\psi}}^*\gamma_\mu\big), \quad (\text{E.13b})$$

$$\bar{\Psi}_{\mu\nu}\overleftarrow{\not{\partial}} = 2i\Delta\big(\bar{\psi}^*_{[\mu}\overleftarrow{\partial}_{\nu]} - \tfrac{1}{D-2}\bar{\not{\psi}}^*\gamma_{[\mu}\overleftarrow{\partial}_{\nu]}\big), \quad (\text{E.13c})$$

$$\bar{\Psi}_{\mu\nu}\overleftarrow{\partial}^\nu = \big(\bar{\psi}\cdot\overleftarrow{\partial}\big)\overleftarrow{\partial}_\mu - \bar{\psi}_\mu\overleftarrow{\Box} = i\Delta\big[\bar{\psi}^*_\mu \overleftarrow{\not{\partial}} + \tfrac{1}{D-2}\bar{\not{\psi}}^*\overleftarrow{\partial}^\rho\gamma_{\rho\mu}\big]. \quad (\text{E.13d})$$

**Spin 5/2**

For spin $\tfrac{5}{2}$, let us recall that the original EoMs are given by

$$\mathcal{R}_{\mu\nu} = \mathcal{S}_{\mu\nu} - \gamma_{(\mu}\not{\mathcal{S}}_{\nu)} - \tfrac{1}{2}\eta_{\mu\nu}\mathcal{S}' = \Delta\psi^*_{\mu\nu}, \quad (\text{E.14a})$$

$$\bar{\mathcal{R}}_{\mu\nu} = \bar{\mathcal{S}}_{\mu\nu} - \bar{\not{\mathcal{S}}}_{(\mu}\gamma_{\nu)} - \tfrac{1}{2}\eta_{\mu\nu}\bar{\mathcal{S}}' = \Delta\bar{\psi}^*_{\mu\nu}. \quad (\text{E.14b})$$

One can easily rewrite these in terms of the Fronsdal tensor,

$$\mathcal{S}_{\nu_1\nu_2} \equiv i\big[\not{\partial}\psi_{\nu_1\nu_2} - 2\partial_{(\nu_1}\not{\psi}_{\nu_2)}\big] = \Delta\varphi^*_{\nu_1\nu_2}, \quad (\text{E.15})$$

and similarly $\bar{\mathcal{S}}_{\nu_1\nu_2} = \Delta\bar{\varphi}^*_{\nu_1\nu_2}$ for its Dirac conjugate, where

$$\varphi^*_{\mu\nu} \equiv \psi^*_{\mu\nu} - \tfrac{2}{D}\gamma_{(\mu}\not{\psi}^*_{\nu)} - \tfrac{1}{D}\eta_{\mu\nu}\psi^{*\prime}. \quad (\text{E.16})$$

From the definition of the Fronsdal tensor, one easily finds that

$$\gamma^\sigma \psi_{\rho\sigma\|\alpha} = i\mathcal{S}_{\rho\alpha} - \partial_\alpha \not{\psi}_\rho, \quad (\text{E.17})$$

whose $\gamma$-trace, in turn, gives:

$$\gamma^{\rho\sigma}\psi_{\rho\sigma\|\alpha} = i\not{\mathcal{S}}_\alpha - \partial_\alpha \psi', \qquad \psi' = \psi^\mu_\mu. \quad (\text{E.18})$$

Now we see that the quantity $\gamma^{\mu_1}\Psi_{\mu_1\nu_1|\mu_2\nu_2}$ is given by the 1-curl of Eq. (E.17), and that it is $\Delta$-exact:

$$\gamma^{\mu_1}\Psi_{\mu_1\nu_1|\mu_2\nu_2} = -2i\partial_{[\mu_2}\mathcal{S}_{\nu_2]\nu_1} = \Delta\text{-exact}. \quad (\text{E.19})$$

Similarly, from a 1-curl of Eq. (E.18), we obtain another useful form:

$$\gamma^{\mu_1\nu_1}\Psi_{\mu_1\nu_1|\mu_2\nu_2} = 2i\gamma^{\nu_1}\partial_{[\mu_2}\mathcal{S}_{\nu_2]\nu_1} = \Delta\text{-exact}. \quad (\text{E.20})$$

Taking a curl of (E.19), one finds yet another form,

$$\partial\!\!\!/\, \Psi^{\mu_1\nu_1}|_{\mu_2\nu_2} = -4i\partial^{[\mu_1}\partial_{[\mu_2}\mathcal{S}_{\nu_2]}{}^{\nu_1]} = \Delta\text{-exact}. \tag{E.21}$$

In fact, the relations (E.19), (E.20) and (E.21) also mean that $\Psi^{\mu}{}_{\nu|\mu\sigma} = \Delta$-exact and $\Box \Psi_{\mu\nu|\rho\sigma} = \Delta$-exact. Finally, by using the identity $\partial^{\mu_1} = \frac{1}{2}\left(\partial\!\!\!/\,\gamma^{\mu_1} + \gamma^{\mu_1}\partial\!\!\!/\,\right)$, we derive from (E.19) and (E.21) that

$$\partial^{\mu_1}\Psi_{\mu_1\nu_1|\mu_2\nu_2} = -2i\partial\!\!\!/\,\partial_{[\mu_2}\mathcal{S}_{\nu_2]\nu_1} + i\partial_{\nu_1}\gamma^{\rho}\partial_{[\mu_2}\mathcal{S}_{\nu_2]\rho} = \Delta\text{-exact}. \tag{E.22}$$

Similarly, one can find the various forms of the EoMs for the Dirac conjugate spinor.

Now from the definition of the Fronsdal tensor, one can find the identity

$$\partial \cdot \mathcal{S}_\mu = \tfrac{1}{2}\partial\!\!\!/\,\mathcal{S}\!\!\!/\,_\mu + \tfrac{1}{2}\partial_\mu \mathcal{S}'. \tag{E.23}$$

Taking a divergence of (E.14a), and then using the above identity, one can then write

$$\partial_\nu \mathcal{R}^{\mu\nu} = -\tfrac{1}{2}\gamma^\mu\,\partial\cdot\mathcal{S}\!\!\!/\,, \tag{E.24}$$

which can be rewritten, by using (E.14a), (E.15) and (E.16), as

$$\Delta\left(\partial_\nu \chi^{*\mu\nu}\right) = 0, \qquad \chi^{*\mu\nu} \equiv \psi^{*\mu\nu} - \tfrac{1}{D}\gamma^\mu \psi\!\!\!/\,^{*\nu}. \tag{E.25}$$

**Arbitrary Spin**

For spin $s = n + \tfrac{1}{2}$ the original EoMs are given by

$$\mathcal{R}_{\mu_1\ldots\mu_n} = \mathcal{S}_{\mu_1\ldots\mu_n} - \tfrac{1}{2}n\,\gamma_{(\mu_1}\mathcal{S}\!\!\!/\,_{\mu_2\ldots\mu_n)} - \tfrac{1}{4}n(n-1)\,\eta_{(\mu_1\mu_2}\mathcal{S}'_{\mu_3\ldots\mu_n)}$$
$$= \Delta\psi^*_{\mu_1\ldots\mu_n}, \tag{E.26a}$$
$$\bar{\mathcal{R}}_{\mu_1\ldots\mu_n} = \bar{\mathcal{S}}_{\mu_1\ldots\mu_n} - \tfrac{1}{2}n\,\bar{\mathcal{S}}\!\!\!/\,_{(\mu_1\ldots\mu_{n-1}}\gamma_{\mu_n)} - \tfrac{1}{4}n(n-1)\,\eta_{(\mu_1\mu_2}\bar{\mathcal{S}}'_{\mu_3\ldots\mu_n)}$$
$$= \Delta\bar{\psi}^*_{\mu_1\ldots\mu_n}. \tag{E.26b}$$

One can reexpress the EoMs in terms of the Fronsdal tensor as follows:

$$\mathcal{S}_{\nu_1\ldots\nu_n} \equiv i\bigl(\partial\!\!\!/\,\psi_{\nu_1\ldots\nu_n} - n\partial_{(\nu_1}\psi\!\!\!/\,_{\nu_2\ldots\nu_n)}\bigr) = \Delta\varphi^*_{\nu_1\ldots\nu_n}, \tag{E.27}$$

and similarly $\bar{\mathcal{S}}_{\nu_1\ldots\nu_n} = \Delta\bar{\varphi}^*_{\nu_1\ldots\nu_n}$ for its Dirac conjugate, where

$$\varphi^*_{\nu_1\ldots\nu_n} \equiv \psi^*_{\nu_1\ldots\nu_n} - \tfrac{n}{2n+D-4}\,\gamma_{(\nu_1}\psi\!\!\!/\,^*_{\nu_2\ldots\nu_n)} - \tfrac{n(n-1)}{2(2n+D-4)}\,\eta_{(\nu_1\nu_2}\psi^{*\prime}_{\nu_3\ldots\nu_n)}. \tag{E.28}$$

Taking an $(n-2)$-curl of the the Fronsdal tensor (E.27), one finds the relation

$$\gamma^{\nu_{n-1}} \psi^{(n-1)}_{\mu_1\nu_1|\ldots|\mu_{n-1}\nu_{n-1}\|\nu_n} = i\mathcal{S}^{(n-2)}_{\mu_1\nu_1|\ldots|\mu_{n-2}\nu_{n-2}\|\nu_{n-1}\nu_n} \\ - \partial_{\nu_n} \slashed{\psi}^{(n-2)}_{\mu_1\nu_1|\ldots|\mu_{n-2}\nu_{n-2}\|\nu_{n-1}}, \quad \text{(E.29)}$$

whose $\gamma$-trace, in turn, gives:

$$\gamma^{\mu_{n-1}\nu_{n-1}} \psi^{(n-1)}_{\mu_1\nu_1|\ldots|\mu_{n-1}\nu_{n-1}\|\nu_n} = i\slashed{\mathcal{S}}^{(n-2)}_{\mu_1\nu_1|\ldots|\mu_{n-2}\nu_{n-2}\|\nu_n} \\ - \partial_{\nu_n} \psi'^{(n-2)}_{\mu_1\nu_1|\ldots|\mu_{n-2}\nu_{n-2}}. \quad \text{(E.30)}$$

The arbitrary spin generalization of the relations (E.19)–(E.22) is rather straightforward, and can be derived in the same way. The corresponding expressions respectively read

$$\gamma^{\mu_1} \Psi_{\mu_1\nu_1|\ldots|\mu_n\nu_n} = -i\mathcal{S}^{(n-1)}_{\mu_2\nu_2|\ldots|\mu_n\nu_n\|\nu_1} = \Delta(\ldots), \quad \text{(E.31a)}$$

$$\gamma^{\mu_1\nu_1} \Psi_{\mu_1\nu_1|\ldots|\mu_n\nu_n} = i\slashed{\mathcal{S}}^{(n-1)}_{\mu_2\nu_2|\ldots|\mu_n\nu_n} = \Delta(\ldots), \quad \text{(E.31b)}$$

$$\slashed{\partial} \Psi_{\mu_1\nu_1|\ldots|\mu_n\nu_n} = -i\mathcal{S}^{(n)}_{\mu_1\nu_1|\ldots|\mu_n\nu_n} = \Delta(\ldots), \quad \text{(E.31c)}$$

$$\partial^{\mu_1} \Psi_{\mu_1\nu_1|\ldots|\mu_n\nu_n} = -i\slashed{\partial}\, \mathcal{S}^{(n-1)}_{\mu_2\nu_2|\ldots|\mu_n\nu_n\|\nu_1} \\ + \tfrac{i}{2}\partial_{\nu_1} \slashed{\mathcal{S}}^{(n-1)}_{\mu_2\nu_2|\ldots|\mu_n\nu_n} = \Delta(\ldots). \quad \text{(E.31d)}$$

Obvious consequences of the above equations include the $\Delta$-exactness of $\eta^{\mu_1\mu_2} \Psi_{\mu_1\nu_1|\ldots|\mu_n\nu_n}$ and $\Box \Psi_{\mu_1\nu_1|\ldots|\mu_n\nu_n}$. Similar forms of the EoMs can be written for the Dirac conjugate spinor.

Finally, we have the following generalization of the identity (E.23):

$$\partial \cdot \mathcal{S}_{\mu_1\ldots\mu_{n-1}} = \tfrac{1}{2}\, \slashed{\partial}\, \slashed{\mathcal{S}}_{\mu_1\ldots\mu_{n-1}} + \tfrac{n-1}{2}\, \partial_{(\mu_1} \mathcal{S}'_{\mu_2\ldots\mu_{n-1})}, \quad \text{(E.32)}$$

which, when used in the divergence of (E.26a) gives

$$\partial \cdot \mathcal{R}_{\mu_1\ldots\mu_{n-1}} = -\tfrac{n-1}{2}\, \gamma_{(\mu_1} \partial \cdot \slashed{\mathcal{S}}_{\mu_2\ldots\mu_{n-1})} - \tfrac{(n-1)(n-2)}{4}\, \eta_{(\mu_1\mu_2} \partial \cdot \mathcal{S}'_{\mu_3\ldots\mu_{n-1})}. \quad \text{(E.33)}$$

Given the equations (E.26)–(E.28), this can then be rewritten as

$$\Delta\left(\partial^{\mu_n} \chi^*_{\mu_1\ldots\mu_n}\right) = 0, \quad \text{(E.34)}$$

where

$$\chi^*_{\mu_1\ldots\mu_n} \equiv \psi^*_{\mu_1\ldots\mu_n} - \tfrac{n-1}{2n+D-4}\, \gamma_{(\mu_1} \slashed{\psi}^*_{\mu_2\ldots\mu_{n-1})\mu_n} \\ - \tfrac{(n-1)(n-2)}{2(2n+D-4)}\, \eta_{(\mu_1\mu_2} \psi'^*_{\mu_3\ldots\mu_{n-1})\mu_n}. \quad \text{(E.35)}$$

## E.2 The Cohomology of $\Gamma$: Gauge Invariance

This Appendix is devoted to clarifying and providing proofs of the statements about the cohomology of $\Gamma$ appearing in the main text. We recall that the action of $\Gamma$ on the various fields is defined by

$$\Gamma A_\mu = \partial_\mu C, \tag{E.36a}$$

$$\Gamma h_{\mu\nu} = 2\partial_{(\mu} C_{\nu)}, \tag{E.36b}$$

$$\Gamma \psi_{\nu_1...\nu_n} = n\partial_{(\nu_1} \xi_{\nu_2...\nu_n)}, \qquad \Gamma \bar\psi_{\nu_1...\nu_n} = -n\partial_{(\nu_1} \bar\xi_{\nu_2...\nu_n)}. \tag{E.36c}$$

The nontrivial elements in the cohomology of $\Gamma$ are nothing but gauge-invariant objects that themselves are not gauge variations of something else. In the following we consider one by one all such elements, and also prove some useful relations involving $\Gamma$-exact terms.

**The Curvatures**

The curvatures $\{F_{\mu\nu}, R_{\mu\nu\alpha\beta}, \Psi_{\mu_1\nu_1|...|\mu_n\nu_n}\}$ and their derivatives belong to the cohomology of $\Gamma$. Seeing that the curvatures are $\Gamma$-closed is straightforward. For the photon it follows directly from the commutativity of partial derivatives as one takes a curl of the above $\Gamma$-variation of $A_\mu$,

$$\Gamma F_{\mu\nu} = \Gamma \left(2\partial_{[\mu} A_{\nu]}\right) = 2\partial_{[\mu}\partial_{\nu]} C = 0, \tag{E.37}$$

and for the graviton the same reasoning yields

$$\Gamma R_{\mu\nu}{}^{\rho\sigma} = \Gamma\left(4\partial^{[\rho}\partial_{[\mu} h_{\nu]}{}^{\sigma]}\right) = 4\partial^{[\rho}\partial_{[\mu}\partial_{\nu]}C^{\sigma]} + 4\partial^{[\rho}\partial_{[\mu}\partial^{\sigma]}C_{\nu]} = 0. \tag{E.38}$$

One can also take a 1-curl of the above fermion gauge-variation to obtain

$$\Gamma \psi^{(1)\mu_1\nu_1\|}{}_{\nu_2...\nu_n} = (n-1)\partial_{(\nu_2} \xi^{(1)\mu_1\nu_1\|}{}_{\nu_3...\nu_n)}, \tag{E.39}$$

and similarly for the Dirac conjugate. Likewise, an $m$-curl of Eq. (E.36c) gives, for $m \leq n$,

$$\Gamma \psi^{(m)\mu_1\nu_1|...|\mu_m\nu_m\|}{}_{\nu_{m+1}...\nu_n} = (n-m)\partial_{(\nu_{m+1}} \xi^{(m)\mu_1\nu_1|...|\mu_m\nu_m\|}{}_{\nu_{m+2}...\nu_n)}. \tag{E.40}$$

In particular, when $m = n$, we have the $\Gamma$-variation of the curvature; it vanishes:

$$\Gamma \Psi^{\mu_1\nu_1|...|\mu_n\nu_n} = 0. \tag{E.41}$$

Note that the $\Gamma$-closure of the curvature holds without requiring any constraints on the fermionic ghost. To see that the curvatures are not Γ-exact, we simply notice that these are pgh-0 objects, whereas any Γ-exact piece must have $pgh > 0$. Therefore, the curvatures are nontrivial elements in the cohomology of Γ, and so are their derivatives.

As we have already seen, only the highest curl ($n$-curl) of the field $\psi_{\nu_1\ldots\nu_n}$ is Γ-closed, while no lower curls are. It is the commutativity of partial derivatives that plays a crucial role. Clearly, an arbitrary derivative of the field will not be Γ-closed in general. Yet, some particular linear combination of such objects (or their $\gamma$-traces) can be Γ-closed under the constrained ghost. The latter possibility is exhausted precisely by the Fronsdal tensor and its derivatives, which will be discussed later.

### The Antifields

The antifields $\{A^{*\mu}, C^*, h^{*\mu\nu}, C^{*\mu}, \bar\psi^{*\mu_1\ldots\mu_n}, \bar\xi^{*\mu_1\ldots\mu_{n-1}}\}$ and their derivatives belong to the cohomology of Γ as well. These objects are Γ-closed simply because Γ does not act on the antifields. On the other hand, having $pgh = 0$, they cannot be Γ-exact.

### The Ghosts & their Curls

The *undifferentiated* ghosts $\{C, C_\mu, \xi_{\mu_1\ldots\mu_{n-1}}\}$ are Γ-closed objects simply because Γ does not act on them. Also they cannot be Γ-exact, because any Γ-exact piece must contain at least one derivative of any of the ghosts (which is obvious from the gauge variations).

Any derivatives of the ghosts will also be Γ-closed. Some derivatives, however, will be Γ-exact, and therefore trivial in the cohomology of Γ. The bosonic case is easy: one can immediately dismiss as trivial *any* derivative of the bosonic ghost $C$, because of definition $\partial_\mu C = \Gamma A_\mu$. Also, any *symmetrized* derivatives of the bosonic ghost is trivial: $\partial_{(\mu} C_{\nu)} = \frac{1}{2}\Gamma h_{\mu\nu}$, but its 1-curl is not, and we have

$$\partial_\mu C_\nu = \partial_{(\mu} C_{\nu)} + \partial_{[\mu} C_{\nu]} = \tfrac{1}{2}\Gamma h_{\mu\nu} + \tfrac{1}{2}\mathfrak{C}_{\mu\nu}. \tag{E.42}$$

By taking a curl of the above equation, one however finds that any derivative of $\mathfrak{C}_{\mu\nu}$ is Γ-exact:

$$\partial_\rho \mathfrak{C}_{\mu\nu} = \Gamma \mathfrak{h}_{\mu\nu\|\rho}. \tag{E.43}$$

Derivatives of the fermionic ghost are more interesting. For the Rarita–Schwinger, the corresponding ghost $\xi$ has no spacetime indices, so that the situation is trivial in the sense that it is the same as that of the photon (the spinor index plays no role). In the simplest non-trivial case of a spin-$\frac{5}{2}$ field, with $n = 2$, we see that

$$\partial_\mu \xi_\nu = \partial_{(\mu} \xi_{\nu)} + \partial_{[\mu} \xi_{\nu]} = \tfrac{1}{2}\Gamma \psi_{\mu\nu} + \tfrac{1}{2}\xi_{\mu\nu}. \tag{E.44}$$

The 1-curl $\xi_{\mu\nu}$ is a nontrivial element in the cohomology of $\Gamma$, but its $\gamma$-trace is not:
$$\gamma^\alpha \xi_{\alpha\beta} = \slashed{\partial} \xi_\alpha = \Gamma \slashed{\psi}_\alpha, \tag{E.45}$$

thanks to the $\gamma$-tracelessness of the ghost. Again, a derivative of the 1-curl is trivial:
$$\partial_\rho \xi_{\mu\nu} = \Gamma \psi_{\mu\nu\|\rho}, \qquad \partial_\rho \bar{\xi}_{\mu\nu} = -\Gamma \bar{\psi}_{\mu\nu\|\rho}, \tag{E.46}$$

which is obtained directly from Eq. (E.39) by setting $n = 2$. We thus see that the $n = 2$ case is akin to that of the graviton, but one has the extra freedom of taking $\gamma$-traces, which complicates matters.

The $n = 3$ counterpart of Eq. (E.44) reads
$$\partial_\mu \xi_{\nu\rho} = \partial_{(\mu} \xi_{\nu\rho)} + \tfrac{4}{3} \partial_{[\mu} \xi_{\nu]\rho} + \tfrac{2}{3} \partial_{[\nu} \xi_{\rho]\mu} = \tfrac{1}{3} \Gamma \psi_{\mu\nu\rho} + \tfrac{2}{3} \xi^{(1)}_{\mu\nu\|\rho} + \tfrac{1}{3} \xi^{(1)}_{\nu\rho\|\mu}. \tag{E.47}$$

The generalization to arbitrary spin is straightforward. One obtains
$$\begin{aligned}
\partial_\rho \xi_{\nu_1 \ldots \nu_{n-1}} &= \partial_{(\rho} \xi_{\nu_1 \ldots \nu_{n-1})} + 2\left(1 - \tfrac{1}{n}\right) \partial_{[\rho} \xi_{\nu_1] \nu_2 \ldots \nu_{n-1}} \\
&\quad + 2 \sum_{m=1}^{n-2} \left(1 - \tfrac{m+1}{n}\right) \partial_{[\nu_m} \xi_{\nu_{m+1}]\rho \, \nu_1 \ldots \nu_{m-1} \nu_{m+2} \ldots \nu_{n-1}} \\
&= \tfrac{1}{n} \Gamma \psi_{\rho \, \nu_1 \ldots \mu_{n-1}} + \left(1 - \tfrac{1}{n}\right) \xi^{(1)}_{\rho\nu_1 \| \nu_2 \ldots \nu_{n-1}} \\
&\quad + \sum_{m=1}^{n-2} \left(1 - \tfrac{m+1}{n}\right) \xi^{(1)}_{\nu_m \nu_{m+1} \| \rho \, \nu_1 \ldots \nu_{m-1} \nu_{m+2} \ldots \nu_{n-1}}.
\end{aligned} \tag{E.48}$$

We conclude that any first derivative of the fermionic ghost is a linear combination of 1-curls, up to $\Gamma$-exact terms. Therefore, it suffices to consider only 1-curls of the ghost in the cohomology of $\Gamma$. More generally, for $m$ derivatives, with $m \leq n-1$, one can consider only the $m$-curls in the cohomology of $\Gamma$. To see this, we can first take a curl of Eq. (E.48) to convince ourselves that only 2-curls of the ghost are nontrivial. Similarly, we can continue step by step to show that for any $m$-derivative combination of the fermionic ghost, with $m \leq n-1$, it suffices to consider only $m$-curls thereof.

It is clear that the derivative of an $m$-curl, $\partial_{\nu_n} \xi^{(m)}_{\mu_1 \nu_1 | \ldots | \mu_m \nu_m \| \nu_{m+1} \ldots \nu_{n-1}}$, contains non-trivial $(m+1)$-curls. Only when symmetrized w.r.t. the indices $\{\nu_{m+1}, \ldots, \nu_n\}$, may this quantity be $\Gamma$-exact. This fact is nothing but a restatement of Eq. (E.40) for $0 \leq m \leq n-1$:
$$\partial_{(\nu_n} \xi^{(m)\mu_1\nu_1|\ldots|\mu_m\nu_m\|}{}_{\nu_{m+1}\ldots\nu_{n-1})} = \tfrac{1}{n-m} \Gamma \psi^{(m)\mu_1\nu_1|\ldots|\mu_m\nu_m\|}{}_{\nu_{m+1}\ldots\nu_n}. \tag{E.49}$$

Setting $m = n-1$, it follows immediately that a derivative of the highest ghost-curl is always $\Gamma$-exact:

$$\partial_{\nu_n}\xi^{(n-1)}_{\mu_1\nu_1|...|\mu_{n-1}\nu_{n-1}} = \Gamma\,\psi^{(n-1)}_{\mu_1\nu_1|...|\mu_{n-1}\nu_{n-1}\|\nu_n}, \qquad (E.50)$$

which generalizes Eq. (E.46) for arbitrary spin.

However, the $\gamma$-trace of any $m$-curl, $\xi^{(m)}_{\mu_1\nu_1|...|\mu_m\nu_m\|\nu_{m+1}...\nu_{n-1}}$, is always $\Gamma$-exact. If the $\gamma$-matrix carries one of the unpaired indices $\{\nu_{m+1},...,\nu_{n-1}\}$, this quantity vanishes since the ghost is $\gamma$-traceless. Otherwise, the same constraint gives rise to the following:

$$\gamma^{\mu_1}\xi^{(m)}_{\mu_1\nu_1|...|\mu_m\nu_m\|\nu_{m+1}...\nu_{n-1}} = \slashed{\partial}\,\xi^{(m-1)}_{\mu_2\nu_2|...|\mu_m\nu_m\|\nu_1\nu_{m+1}...\nu_{n-1}}. \qquad (E.51)$$

But one can take a $\gamma$-trace of Eq. (E.49) to see that the above quantity is actually $\Gamma$-exact. Thus one finds the arbitrary-spin generalization of Eq. (E.45):

$$\gamma^{\mu_1}\xi^{(m)}_{\mu_1\nu_1|...|\mu_m\nu_m\|\nu_{m+1}...\nu_{n-1}} = \tfrac{1}{n-m}\,\Gamma\slashed{\psi}^{(m-1)}_{\mu_1\nu_1|...|\mu_m\nu_m\|\nu_{m+1}...\nu_{n-1}}. \qquad (E.52)$$

So, one may exclude from the cohomology of $\Gamma$ the $\gamma$-traces of the fermionic ghost-curls.

**The Fronsdal Tensor**

The Fronsdal tensor $\mathcal{S}_{\mu_1...\mu_n}$ and derivatives thereof also belong to the cohomology of $\Gamma$. From the definition, one finds that its $\Gamma$ variation is given by

$$\begin{aligned}\Gamma\mathcal{S}_{\mu_1...\mu_n} &= i\big[\slashed{\partial}\,\Gamma\psi_{\mu_1...\mu_n} - n\partial_{(\mu_1}\Gamma\slashed{\psi}_{\mu_2...\mu_n)}\big]\\ &= in\left[\slashed{\partial}\,\partial_{(\mu_1}\xi_{\mu_2...\mu_n)} - n\gamma^\rho\partial_{(\mu_1}\partial_{(\rho}\xi_{\mu_2...\mu_n))}\right]\\ &= -in(n-1)\partial_{(\mu_1}\partial_{(\mu_2}\slashed{\xi}_{\mu_3...\mu_n))}.\end{aligned}$$

This quantity vanishes since the ghost is $\gamma$-traceless. $\mathcal{S}_{\mu_1...\mu_n}$, being a 0-pgh# object, is not $\Gamma$-exact either. Therefore, the Fronsdal tensor and its derivatives belong to H($\Gamma$).

In view of Eq. (E.31a) and (E.31c), however, we see that the two highest curls of the Fronsdal tensor boil down to objects already enlisted, and therefore do not need separate consideration. Consequently, for the spin-$\tfrac{5}{2}$ case, it suffices to consider only symmetrized derivatives of the Fronsdal tensor. Let us also note that the aforementioned equations are generalizations of the Damour–Deser relations [249–251] (see also [248]).

# References

[1] **M. Henneaux**, **S.-J. Rey**, "Nonlinear $W_{infinity}$ as Asymptotic Symmetry of Three-Dimensional Higher Spin Anti-de Sitter Gravity", *JHEP* 1012 (2010), p. 007, arXiv:1008.4579 (cit. on pp. 1, 5, 11, 69, 70, 73, 79, 80, 84, 85, 89, 217, 221, 223)

[2] **A. Campoleoni**, **S. Fredenhagen**, **S. Pfenninger**, **S. Theisen**, "Asymptotic symmetries of three-dimensional gravity coupled to higher-spin fields", *JHEP* 1011 (2010), p. 007, arXiv:1008.4744 (cit. on pp. 1, 33, 39, 43, 89)

[3] **C. Pope**, "Lectures on W algebras and W gravity", (1991), arXiv:hep-th/9112076 (cit. on p. 1)

[4] **P. Bouwknegt**, **K. Schoutens**, "W symmetry in conformal field theory", *Phys.Rept.* 223 (1993), pp. 183–276, arXiv:hep-th/9210010 (cit. on pp. 1, 10, 11, 68, 81, 83, 87, 89, 90)

[5] **M. R. Gaberdiel**, **R. Gopakumar**, "An $AdS_3$ Dual for Minimal Model CFTs", *Phys.Rev.* D83 (2011), p. 066007, arXiv:1011.2986 (cit. on p. 1)

[6] **M. R. Gaberdiel**, **R. Gopakumar**, "Minimal Model Holography", *J.Phys.* A46 (2013), p. 214002, arXiv:1207.6697 (cit. on pp. 1, 10, 53, 55, 88, 91)

[7] **I. Klebanov**, **A. Polyakov**, "AdS dual of the critical O(N) vector model", *Phys.Lett.* B550 (2002), pp. 213–219, arXiv:hep-th/0210114 (cit. on p. 1)

[8] **E. Sezgin**, **P. Sundell**, "Holography in 4D (super) higher spin theories and a test via cubic scalar couplings", *JHEP* 0507 (2005), p. 044, arXiv:hep-th/0305040 (cit. on p. 1)

[9] **S. Giombi**, **X. Yin**, "Higher Spin Gauge Theory and Holography: The Three-Point Functions", *JHEP* 1009 (2010), p. 115, arXiv:0912.3462 (cit. on p. 1)

[10] **S. Konstein**, **M. Vasiliev**, **V. Zaikin**, "Conformal higher spin currents in any dimension and AdS / CFT correspondence", *JHEP* 0012 (2000), p. 018, arXiv:hep-th/0010239 (cit. on p. 1)

[11] **B. Sundborg**, "Stringy gravity, interacting tensionless strings and massless higher spins", *Nucl.Phys.Proc.Suppl.* 102 (2001), pp. 113–119, arXiv:hep-th/0103247 (cit. on p. 1)

[12] **E. Witten**, "Spacetime Reconstruction", *Talk given at the John Schwarz 60th Birthday Symposium* (2001) (cit. on p. 1)

[13] **A. Mikhailov**, "Notes on higher spin symmetries", (2002), arXiv:hep-th/0201019 (cit. on p. 1)

[14] **E. Sezgin**, **P. Sundell**, "Massless higher spins and holography", *Nucl.Phys.* B644 (2002), pp. 303–370, arXiv:hep-th/0205131 (cit. on p. 1)

[15] **J. M. Maldacena**, "The Large N limit of superconformal field theories and supergravity", *Adv.Theor.Math.Phys.* 2 (1998), pp. 231–252, arXiv:hep-th/9711200 (cit. on p. 2)

[16] **S. Gubser**, **I. R. Klebanov**, **A. M. Polyakov**, "Gauge theory correlators from noncritical string theory", *Phys.Lett.* B428 (1998), pp. 105–114, arXiv:hep-th/9802109 (cit. on p. 2)

[17] **E. Witten**, "Anti-de Sitter space and holography", *Adv.Theor.Math.Phys.* 2 (1998), pp. 253–291, arXiv:hep-th/9802150 (cit. on p. 2)

[18] **S. Giombi**, **X. Yin**, "The Higher Spin/Vector Model Duality", *J.Phys.* A46 (2013), p. 214003, arXiv:1208.4036 (cit. on p. 2)

[19] **E. Fradkin**, **M. A. Vasiliev**, "Cubic Interaction in Extended Theories of Massless Higher Spin Fields", *Nucl.Phys.* B291 (1987), p. 141 (cit. on pp. 2, 4, 23, 176)

[20] **E. Fradkin**, **M. A. Vasiliev**, "On the Gravitational Interaction of Massless Higher Spin Fields", *Phys.Lett.* B189 (1987), pp. 89–95 (cit. on pp. 2, 176)

[21] **E. Joung**, **M. Taronna**, "Cubic interactions of massless higher spins in (A)dS: metric-like approach", *Nucl.Phys.* B861 (2012), pp. 145–174, arXiv:1110.5918 (cit. on pp. 2, 177)

[22] **X. Bekaert**, **N. Boulanger**, **S. Cnockaert**, "Spin three gauge theory revisited", *JHEP* 0601 (2006), p. 052, arXiv:hep-th/0508048 (cit. on p. 2)

[23] **I. Buchbinder**, **A. Fotopoulos**, **A. C. Petkou**, **M. Tsulaia**, "Constructing the cubic interaction vertex of higher spin gauge fields", *Phys.Rev.* D74 (2006), p. 105018, arXiv:hep-th/0609082 (cit. on p. 2)

[24] **N. Boulanger**, **S. Leclercq**, "Consistent couplings between spin-2 and spin-3 massless fields", *JHEP* 0611 (2006), p. 034, arXiv:hep-th/0609221 (cit. on pp. 2, 175)

[25] **A. Sagnotti**, **M. Taronna**, "String Lessons for Higher-Spin Interactions", *Nucl.Phys.* B842 (2011), pp. 299–361, arXiv:1006.5242 (cit. on pp. 2, 6, 94, 95, 172–174, 227–229, 234, 236)

[26] **M. Ammon**, **M. Gutperle**, **P. Kraus**, **E. Perlmutter**, "Black holes in three dimensional higher spin gravity: A review", *J.Phys.* A46 (2013), p. 214001, arXiv:1208.5182 (cit. on p. 2)

[27] **A. Perez**, **D. Tempo**, **R. Troncoso**, "Brief review on higher spin black holes", (2014), arXiv:1402.1465 (cit. on p. 2)

[28] **W. Rarita**, **J. Schwinger**, "On a theory of particles with half integral spin", *Phys.Rev.* 60 (1941), p. 61 (cit. on p. 3)

[29] **M. Fierz**, "Force-free particles with any spin", *Helv.Phys.Acta* 12 (1939), pp. 3–37 (cit. on p. 3)

[30] **M. Fierz**, **W. Pauli**, "On relativistic wave equations for particles of arbitrary spin in an electromagnetic field", *Proc.Roy.Soc.Lond.* A173 (1939), pp. 211–232 (cit. on p. 3)

[31] **E. Majorana**, "Relativistic theory of particles with arbitrary intrinsic momentum", *Nuovo Cim.* 9 (1932), pp. 335–344 (cit. on p. 3)

[32] **P. Dirac**, "Relativistic wave equations", *Proc.Roy.Soc.Lond.* 155A (1936), pp. 447–459 (cit. on p. 3)

[33] **C. Fronsdal**, "Massless Fields with Integer Spin", *Phys.Rev.* D18 (1978), p. 3624 (cit. on pp. 3, 34)

[34] **J. Fang**, **C. Fronsdal**, "Massless Fields with Half Integral Spin", *Phys.Rev.* D18 (1978), p. 3630 (cit. on pp. 3, 187)

[35] **L. Singh**, **C. Hagen**, "Lagrangian formulation for arbitrary spin. 1. The boson case", *Phys.Rev.* D9 (1974), pp. 898–909 (cit. on p. 3)

[36] **L. Singh**, **C. Hagen**, "Lagrangian formulation for arbitrary spin. 2. The fermion case", *Phys.Rev.* D9 (1974), pp. 910–920 (cit. on p. 3)

[37] **E. P. Wigner**, "On Unitary Representations of the Inhomogeneous Lorentz Group", *Annals Math.* 40 (1939), pp. 149–204 (cit. on p. 3)

[38] **V. Bargmann**, **E. P. Wigner**, "Group Theoretical Discussion of Relativistic Wave Equations", *Proc.Nat.Acad.Sci.* 34 (1948), p. 211 (cit. on p. 3)

[39] **A. K. Bengtsson**, **I. Bengtsson**, **L. Brink**, "Cubic Interaction Terms for Arbitrary Spin", *Nucl.Phys.* B227 (1983), p. 31 (cit. on pp. 3, 175)

[40] **F. A. Berends**, **G. Burgers**, **H. Van Dam**, "On Spin Three Self Interactions", *Z.Phys.* C24 (1984), pp. 247–254 (cit. on p. 3)

[41] **F. A. Berends**, **G. Burgers**, **H. van Dam**, "On the Theoretical Problems in Constructing Interactions Involving Higher Spin Massless Particles", *Nucl.Phys.* B260 (1985), p. 295 (cit. on p. 3)

[42] **C. Aragone**, **S. Deser**, "Consistency Problems of Hypergravity", *Phys.Lett.* B86 (1979), p. 161 (cit. on pp. 3, 32, 94, 95, 152)

[43] **C. Aragone**, **S. Deser**, "Higher Spin Vierbein Gauge Fermions and Hypergravities", *Nucl.Phys.* B170 (1980), p. 329 (cit. on pp. 3, 94, 95, 152)

[44] **S. Weinberg**, **E. Witten**, "Limits on Massless Particles", *Phys.Lett.* B96 (1980), p. 59 (cit. on pp. 3, 94, 95)

[45] **S. Weinberg**, "Photons and Gravitons in s Matrix Theory: Derivation of Charge Conservation and Equality of Gravitational and Inertial Mass", *Phys.Rev.* 135 (1964), B1049–B1056 (cit. on pp. 4, 94, 95)

[46] **M. A. Vasiliev**, "'Gauge' form of description of massless fields with arbitrary spin", *Sov. J. Nucl. Phys.* 32 (1980), p. 439 (cit. on pp. 4, 33, 37, 38)

[47] **M. A. Vasiliev**, "Free Massless Fields of Arbitrary Spin in the de Sitter Space and Initial Data for a Higher Spin Superalgebra", *Fortsch.Phys.* 35 (1987), pp. 741–770 (cit. on pp. 4, 23, 33, 38)

[48] **E. Fradkin**, **M. A. Vasiliev**, "Candidate to the Role of Higher Spin Symmetry", *Annals Phys.* 177 (1987), p. 63 (cit. on p. 4)

[49] **M. A. Vasiliev**, "Equations of Motion of Interacting Massless Fields of All Spins as a Free Differential Algebra", *Phys.Lett.* B209 (1988), pp. 491–497 (cit. on p. 4)

- [50] **M. A. Vasiliev**, "Consistent equation for interacting gauge fields of all spins in (3+1)-dimensions", *Phys.Lett.* B243 (1990), pp. 378–382 (cit. on pp. 4, 33, 177)
- [51] **X. Bekaert**, **S. Cnockaert**, **C. Iazeolla**, **M. Vasiliev**, "Nonlinear higher spin theories in various dimensions", (2005), arXiv:`hep-th/0503128` (cit. on p. 4)
- [52] **X. Bekaert**, **N. Boulanger**, **P. Sundell**, "How higher-spin gravity surpasses the spin two barrier: no-go theorems versus yes-go examples", *Rev.Mod.Phys.* 84 (2012), pp. 987–1009, arXiv:`1007.0435` (cit. on pp. 4, 32, 94)
- [53] **V. Didenko**, **E. Skvortsov**, "Elements of Vasiliev theory", (2014), arXiv:`1401.2975` (cit. on pp. 4, 34)
- [54] **E. Sezgin**, **P. Sundell**, "Doubletons and 5-D higher spin gauge theory", *JHEP* 0109 (2001), p. 036, arXiv:`hep-th/0105001` (cit. on p. 4)
- [55] **C.-M. Chang**, **S. Minwalla**, **T. Sharma**, **X. Yin**, "ABJ Triality: from Higher Spin Fields to Strings", *J.Phys.* A46 (2013), p. 214009, arXiv:`1207.4485` (cit. on p. 4)
- [56] **M. Blencowe**, "A Consistent Interacting Massless Higher Spin Field Theory in $D = (2+1)$", *Class.Quant.Grav.* 6 (1989), p. 443 (cit. on pp. 5, 10, 13, 33, 36, 39, 195)
- [57] **E. Witten**, "(2+1)-Dimensional Gravity as an Exactly Soluble System", *Nucl.Phys.* B311 (1988), p. 46 (cit. on pp. 5, 10, 13, 22, 23, 25, 26)
- [58] **A. Achucarro**, **P. Townsend**, "A Chern-Simons Action for Three-Dimensional anti-De Sitter Supergravity Theories", *Phys.Lett.* B180 (1986), p. 89 (cit. on pp. 5, 10, 22, 31, 32, 194, 197)
- [59] **G. Lucena Gómez**, "Higher-Spin Theory - Part II: enter dimension three", *Proceedings of Science* ModaveVIII (2012), p. 003, arXiv:`1307.3200` (cit. on p. 5)
- [60] **M. Henneaux**, **L. Maoz**, **A. Schwimmer**, "Asymptotic dynamics and asymptotic symmetries of three-dimensional extended AdS supergravity", *Annals Phys.* 282 (2000), pp. 31–66, arXiv:`hep-th/9910013` (cit. on pp. 5, 32, 66, 67, 69, 81, 82, 84, 195, 221)
- [61] **M. Henneaux**, **G. Lucena Gómez**, **J. Park**, **S.-J. Rey**, "Super-W(infinity) Asymptotic Symmetry of Higher-Spin $AdS_3$ Supergravity", *JHEP* 1206 (2012), p. 037, arXiv:`1203.5152` (cit. on pp. 5, 11, 90) [62] **R. Metsaev**, "Cubic interaction vertices for fermionic and bosonic arbitrary spin fields", *Nucl.Phys.* B859 (2012), pp. 13–69, arXiv:0712.3526 (cit. on pp. 6, 95, 132, 172, 173, 176, 229, 231, 233)

[63] **M. Henneaux**, **G. Lucena Gómez**, **R. Rahman**, "Higher-Spin Fermionic Gauge Fields and Their Electromagnetic Coupling", *JHEP* 1208 (2012), p. 093, arXiv:1206.1048 (cit. on pp. 6, 95)

[64] **M. Henneaux**, **G. Lucena Gómez**, **R. Rahman**, "Gravitational Interactions of Higher-Spin Fermions", *JHEP* 1401 (2014), p. 087, arXiv:1310.5152 (cit. on pp. 6, 95)

[65] **E. Abbott Abbott**, *Flatland: A Romance of Many Dimensions*, Seely & Co., 1884 (cit. on p. 9)

[66] **S. Carlip**, *Quantum Gravity in 2+1 Dimensions*, Cambridge University Press, 2003 (cit. on pp. 10, 94)

[67] **J. D. Brown**, **M. Henneaux**, "Central Charges in the Canonical Realization of Asymptotic Symmetries: An Example from Three-Dimensional Gravity", *Commun.Math.Phys.* 104 (1986), pp. 207–226 (cit. on pp. 10, 56–58, 65, 69)

[68] **M. A. Vasiliev**, "Higher spin gauge theories in four-dimensions, three-dimensions, and two-dimensions", *Int.J.Mod.Phys.* D5 (1996), pp. 763–797, arXiv:hep-th/9611024 (cit. on p. 10)

[69] **K. Papadodimas**, **S. Raju**, "Correlation Functions in Holographic Minimal Models", *Nucl.Phys.* B856 (2012), pp. 607–646, arXiv:1108.3077 (cit. on p. 10)

[70] **M. Ammon**, **P. Kraus**, **E. Perlmutter**, "Scalar fields and three-point functions in D=3 higher spin gravity", *JHEP* 1207 (2012), p. 113, arXiv:1111.3926 (cit. on p. 10)

[71] **T. Creutzig**, **Y. Hikida**, **P. B. Ronne**, "Three point functions in higher spin $AdS_3$ supergravity", *JHEP* 1301 (2013), p. 171, arXiv:1211.2237 (cit. on p. 10)

[72] **H. Moradi**, **K. Zoubos**, "Three-Point Functions in $N = 2$ Higher-Spin Holography", *JHEP* 1304 (2013), p. 018, arXiv:1211.2239 (cit. on p. 10)

[73] **M. R. Gaberdiel**, **T. Hartman**, **K. Jin**, "Higher Spin Black Holes from CFT", *JHEP* 1204 (2012), p. 103, arXiv:1203.0015 (cit. on p. 10)

[74] **A. Perez**, **D. Tempo**, **R. Troncoso**, "Higher spin gravity in 3D: Black holes, global charges and thermodynamics", *Phys.Lett.* B726 (2013), pp. 444–449, arXiv:1207.2844 (cit. on p. 10)

[75] **A. Perez**, **D. Tempo**, **R. Troncoso**, "Higher spin black hole entropy in three dimensions", *JHEP* 1304 (2013), p. 143, arXiv:1301.0847 (cit. on p. 10)

[76] **M. R. Gaberdiel**, **R. Gopakumar**, **T. Hartman**, **S. Raju**, "Partition Functions of Holographic Minimal Models", *JHEP* 1108 (2011), p. 077, arXiv:1106.1897 (cit. on p. 10)

[77] **P. Kraus**, **E. Perlmutter**, "Partition functions of higher spin black holes and their CFT duals", *JHEP* 1111 (2011), p. 061, arXiv:1108.2567 (cit. on p. 10)

[78] **M. Beccaria**, **G. Macorini**, "On the partition functions of higher spin black holes", *JHEP* 1312 (2013), p. 027, arXiv:1310.4410 (cit. on p. 10)

[79] **M. R. Gaberdiel**, **R. Gopakumar**, **A. Saha**, "Quantum $W$-symmetry in $AdS_3$", *JHEP* 1102 (2011), p. 004, arXiv:1009.6087 (cit. on p. 10)

[80] **M. R. Gaberdiel**, **R. Gopakumar**, "Triality in Minimal Model Holography", *JHEP* 1207 (2012), p. 127, arXiv:1205.2472 (cit. on pp. 10, 90)

[81] **T. Creutzig**, **Y. Hikida**, **P. B. Ronne**, "Higher spin $AdS_3$ supergravity and its dual CFT", *JHEP* 1202 (2012), p. 109, arXiv:1111.2139 (cit. on pp. 11, 88)

[82] **C. Candu**, **M. R. Gaberdiel**, "Supersymmetric holography on $AdS_3$", *JHEP* 1309 (2013), p. 071, arXiv:1203.1939 (cit. on p. 11)

[83] **C. Candu**, **M. R. Gaberdiel**, "Duality in $N = 2$ Minimal Model Holography", *JHEP* 1302 (2013), p. 070, arXiv:1207.6646 (cit. on pp. 11, 88)

[84] **S. Fredenhagen**, **C. Restuccia**, "The geometry of the limit of $N = 2$ minimal models", *J.Phys.* A46 (2013), p. 045402, arXiv:1208.6136 (cit. on pp. 11, 88)

[85] **T. Creutzig**, **Y. Hikida**, **P. B. Ronne**, "$N = 1$ supersymmetric higher spin holography on $AdS_3$", *JHEP* 1302 (2013), p. 019, arXiv:1209.5404 (cit. on pp. 11, 88)

[86] **C. Peng**, "Dualities from higher-spin supergravity", *JHEP* 1303 (2013), p. 054, arXiv:1211.6748 (cit. on pp. 11, 88)

[87] **C. Candu**, **C. Vollenweider**, "The $\mathcal{N}=1$ algebra $\mathcal{W}_\infty[\mu]$ and its truncations", *JHEP* 1311 (2013), p. 032, arXiv:1305.0013 (cit. on p. 11)

[88] **M. Beccaria**, **C. Candu**, **M. R. Gaberdiel**, **M. Groher**, "$\mathcal{N}=1$ extension of minimal model holography", (2013), arXiv:1305.1048 (cit. on p. 11)

[89] **M. R. Gaberdiel**, **R. Gopakumar**, "Large $N=4$ Holography", *JHEP* 1309 (2013), p. 036, arXiv:1305.4181 (cit. on pp. 11, 88)

[90] **T. Creutzig**, **Y. Hikida**, **P. B. Ronne**, "Extended higher spin holography and Grassmannian models", *JHEP* 1311 (2013), p. 038, arXiv:1306.0466 (cit. on pp. 11, 88)

[91] **M. R. Gaberdiel**, **C. Peng**, "The symmetry of large $N=4$ holography", (2014), arXiv:1403.2396 (cit. on p. 11)

[92] **A. Campoleoni**, **S. Fredenhagen**, **S. Pfenninger**, **S. Theisen**, "Towards metric-like higher-spin gauge theories in three dimensions", *J.Phys.* A46 (2013), p. 214017, arXiv:1208.1851 (cit. on pp. 14, 33)

[93] **D. Z. Freedman**, **A. Van Proyen**, *Supergravity*, Cambridge University Press, 2012 (cit. on pp. 14, 16, 17, 19, 28, 31)

[94] **M. Nakahara**, *Geometry, Topology, and Physics*, Institute of Physics Publishing, 2003 (cit. on pp. 14, 16)

[95] **L. Castellani**, **R. D'Auria**, **P. Fré**, *Supergravity and Superstrings, Vol. 1*, World Scientific, 1991 (cit. on pp. 14, 28)

[96] **A. Einstein**, "The Foundation of the General Theory of Relativity", *Annalen Phys.* 49 (1916), pp. 769–822 (cit. on p. 14)

[97] **C. Misner**, **K. Thorne**, **J. Wheeler**, *Gravitation*, W. Freeman, 1973 (cit. on p. 18)

[98] **S. MacDowell**, **F. Mansouri**, "Unified Geometric Theory of Gravity and Supergravity", *Phys.Rev.Lett.* 38 (1977), p. 739 (cit. on p. 23)

[99] **K. Stelle**, **P. C. West**, "Spontaneously Broken De Sitter Symmetry and the Gravitational Holonomy Group", *Phys.Rev.* D21 (1980), p. 1466 (cit. on p. 23)

[100] **A. H. Chamseddine**, **P. C. West**, "Supergravity as a Gauge Theory of Supersymmetry", *Nucl.Phys.* B129 (1977), p. 39 (cit. on p. 23)

[101] **V. Lopatin**, **M. A. Vasiliev**, "Free Massless Bosonic Fields of Arbitrary Spin in $d$-dimensional De Sitter Space", *Mod.Phys.Lett.* A3 (1988), p. 257 (cit. on pp. 23, 39)

[102] **J. Fuchs**, **C. Schweigert**, *Symmetries, Lie Algebras and Representations*, Cambridge University Press, 1997 (cit. on pp. 24, 44, 45)

[103] **M. Henneaux**, **C. Teitelboim**, *Quantization of Gauge Systems*, Princeton University Press, 1992 (cit. on pp. 30, 63, 97, 104, 107, 173, 198, 212)

[104] **M. Vasiliev**, "Nonlinear equations for symmetric massless higher spin fields in (A)dS(d)", *Phys.Lett.* B567 (2003), pp. 139–151, arXiv:hep-th/0304049 (cit. on pp. 33, 177)

[105] **M. A. Vasiliev**, "Properties of equations of motion of interacting gauge fields of all spins in (3+1)-dimensions", *Class.Quant.Grav.* 8 (1991), pp. 1387–1417 (cit. on pp. 33, 177)

[106] **Y. Zinoviev**, "Spin 3 cubic vertices in a frame-like formalism", *JHEP* 1008 (2010), p. 084, arXiv:1007.0158 (cit. on p. 33)

[107] **A. Campoleoni**, "Higher Spins in D = 2+1", (2011), arXiv:1110.5841 (cit. on p. 33)

[108] **I. Buchbinder**, **T. Snegirev**, **Y. Zinoviev**, "On gravitational interactions for massive higher spins in $AdS_3$", *J.Phys.* A46 (2013), p. 214015, arXiv:1208.0183 (cit. on p. 33)

[109] **R. Rahman**, "Higher Spin Theory - Part I", *PoS* ModaveVIII (2012), p. 004, arXiv:1307.3199 (cit. on p. 35)

[110] **N. Boulanger**, **P. Sundell**, "An action principle for Vasiliev's four-dimensional higher-spin gravity", *J.Phys.* A44 (2011), p. 495402, arXiv:1102.2219 (cit. on p. 35)

[111] **B. Binegar**, "RELATIVISTIC FIELD THEORIES IN THREE-DIMENSIONS", *J.Math.Phys.* 23 (1982), p. 1511 (cit. on p. 36)

[112] **M. A. Vasiliev**, "Free Massless Fermionic Fields of Arbitrary Spin in $d$-dimensional De Sitter Space", *Nucl.Phys.* B301 (1988), p. 26 (cit. on p. 39)

[113] **C. Pope**, **P. Townsend**, "Conformal Higher Spin in (2+1)-dimensions", *Phys.Lett.* B225 (1989), p. 245 (cit. on p. 42)

[114] **A. Castro**, **E. Hijano**, **A. Lepage-Jutier**, "Unitarity Bounds in $AdS_3$ Higher Spin Gravity", *JHEP* 1206 (2012), p. 001, arXiv:1202.4467 (cit. on pp. 44, 45)

[115] **P. Grozman**, **D. Leites**, *Defining relations associated with the principal sl(2)-subalgebras of simple Lie algebras*, tech. rep. math-ph/0510013, Oct. 2005 (cit. on p. 45)

[116] **M. Gary**, **D. Grumiller**, **R. Rashkov**, "Towards non-AdS holography in 3-dimensional higher spin gravity", *JHEP* 1203 (2012), p. 022, arXiv:1201.0013 (cit. on p. 45)

[117] **X. Bekaert**, *Universal enveloping algebras and some applications in physics*, tech. rep. IHES-P-2005-26, Bures-sur-Yvette: Inst. Hautes Etud. Sci., Oct. 2005 (cit. on p. 46)

[118] **B. L. Feigin**, "The Lie algebras $\mathfrak{gl}(\lambda)$ and cohomologies of Lie algebras of differential operators", *Russian Mathematical Surveys* 43.2 (1988), p. 169 (cit. on p. 46)

[119] **E. Bergshoeff**, **M. Blencowe**, **K. Stelle**, "Area Preserving Diffeomorphisms and Higher Spin Algebra", *Commun.Math.Phys.* 128 (1990), p. 213 (cit. on pp. 46, 50, 195)

[120] **M. Bordemann**, **J. Hoppe**, **P. Schaller**, "INFINITE DIMENSIONAL MATRIX ALGEBRAS", *Phys.Lett.* B232 (1989), p. 199 (cit. on pp. 46, 50, 195)

[121] **C. Pope**, **L. Romans**, **X. Shen**, "$W$(infinity) and the Racah-Wigner Algebra", *Nucl.Phys.* B339 (1990), pp. 191–221 (cit. on pp. 46, 86, 87, 217)

[122] **M. A. Vasiliev**, "Higher Spin Algebras and Quantization on the Sphere and Hyperboloid", *Int.J.Mod.Phys.* A6 (1991), pp. 1115–1135 (cit. on pp. 46, 48, 50, 51)

[123] **E. Fradkin**, **V. Y. Linetsky**, "Infinite dimensional generalizations of simple Lie algebras", *Mod.Phys.Lett.* A5 (1990), pp. 1967–1977 (cit. on p. 46)

[124] **E. Fradkin**, **V. Y. Linetsky**, "Infinite dimensional generalizations of finite dimensional symmetries", *J.Math.Phys.* 32 (1991), pp. 1218–1226 (cit. on p. 46)

[125] **S. Prokushkin**, **M. A. Vasiliev**, "3-d higher spin gauge theories with matter", (1998), arXiv:hep-th/9812242 (cit. on pp. 47, 53)

[126] **S. Prokushkin**, **M. A. Vasiliev**, "Higher spin gauge interactions for massive matter fields in 3-D AdS space-time", *Nucl.Phys.* B545 (1999), p. 385, arXiv:hep-th/9806236 (cit. on pp. 47, 53)

[127] **N. Boulanger**, **E. Skvortsov**, "Higher-spin algebras and cubic interactions for simple mixed-symmetry fields in AdS spacetime", *JHEP* 1109 (2011), p. 063, arXiv:1107.5028 (cit. on pp. 47, 53)

[128] **E. Joung**, **K. Mkrtchyan**, "Notes on higher-spin algebras: minimal representations and structure constants", (2014), arXiv:1401.7977 (cit. on pp. 47, 51)

[129] **N. Boulanger**, **D. Ponomarev**, **E. Skvortsov**, **M. Taronna**, "On the uniqueness of higher-spin symmetries in AdS and CFT", *Int.J.Mod.Phys.* A28 (2013), p. 1350162, arXiv:1305.5180 (cit. on p. 47)

[130] **M. A. Vasiliev**, "Extended Higher Spin Superalgebras and Their Realizations in Terms of Quantum Operators", *Fortsch.Phys.* 36 (1988), pp. 33–62 (cit. on pp. 48, 195, 201)

[131] **M. A. Vasiliev**, "Higher spin gauge theories: Star product and AdS space", (1999), arXiv:hep-th/9910096 (cit. on p. 48)

[132] **M. Henneaux**, **C. Teitelboim**, "Asymptotically anti-De Sitter Spaces", *Commun.Math.Phys.* 98 (1985), pp. 391–424 (cit. on pp. 57, 58, 69)

[133] **O. Coussaert**, **M. Henneaux**, **P. van Driel**, "The Asymptotic dynamics of three-dimensional Einstein gravity with a negative cosmological constant", *Class.Quant.Grav.* 12 (1995), pp. 2961–2966, arXiv:gr-qc/9506019 (cit. on p. 59)

[134] **T. Regge**, **C. Teitelboim**, "Role of Surface Integrals in the Hamiltonian Formulation of General Relativity", *Annals Phys.* 88 (1974), p. 286 (cit. on pp. 62, 76)

[135] **M. Banados** et al., "Anti-de Sitter / CFT correspondence in three-dimensional supergravity", *Phys.Rev.* D58 (1998), p. 085020, arXiv:hep-th/9805165 (cit. on pp. 66, 221)

[136] **J. M. Figueroa-O'Farrill**, **J. Mas**, **E. Ramos**, "The Topography of W(infinity) type algebras", *Phys.Lett.* B299 (1993), pp. 41–48, arXiv:hep-th/9208077 (cit. on pp. 83, 89, 90)

[137] **A. Campoleoni**, **S. Fredenhagen**, **S. Pfenninger**, "Asymptotic W-symmetries in three-dimensional higher-spin gauge theories", *JHEP* 1109 (2011), p. 113, arXiv:1107.0290 (cit. on pp. 85, 88)

[138] **A. Zamolodchikov**, "Infinite Additional Symmetries in Two-Dimensional Conformal Quantum Field Theory", *Theor.Math.Phys.* 65 (1985), pp. 1205–1213 (cit. on pp. 85, 89)

[139] **H. Tan**, "Exploring Three-dimensional Higher-Spin Supergravity based on sl(N |N - 1) Chern-Simons theories", *JHEP* 1211 (2012), p. 063, arXiv:1208.2277 (cit. on p. 85)

[140] **C. Ahn**, **K. Schoutens**, **A. Sevrin**, "The full structure of the super $W_3$ algebra", *Int. J. Mod. Phys. A* 6.ITP-SB-90-66 (Sept. 1990), 3467–3488. 28 p (cit. on p. 85)

[141] **J. M. Figueroa-O'Farrill**, **S. Schrans**, "The Conformal bootstrap and super W algebras", *Int.J.Mod.Phys.* A7 (1992), pp. 591–618 (cit. on p. 85)

[142] **G. Compère**, **W. Song**, **A. Strominger**, "Chiral Liouville Gravity", *JHEP* 1305 (2013), p. 154, arXiv:1303.2660 (cit. on p. 86)

[143] **J. D. Brown**, **M. Henneaux**, "On the Poisson Brackets of Differentiable Generators in Classical Field Theory", *J.Math.Phys.* 27 (1986), pp. 489–491 (cit. on p. 86)

[144] **C. Pope**, **L. Romans**, **X. Shen**, "Ideals of Kac-Moody Algebras and Realizations of $W(\text{infinity})$", *Phys.Lett.* B245 (1990), pp. 72–78 (cit. on pp. 86, 217)

[145] **E. Bergshoeff**, **B. de Wit**, **M. A. Vasiliev**, "The Structure of the superW(infinity) (lambda) algebra", *Nucl.Phys.* B366 (1991), pp. 315–346 (cit. on p. 87)

[146] **P. Bowcock**, **G. Watts**, "On the classification of quantum W algebras", *Nucl.Phys.* B379 (1992), pp. 63–95, arXiv:hep-th/9111062 (cit. on p. 87)

[147] **M. R. Gaberdiel**, **T. Hartman**, "Symmetries of Holographic Minimal Models", *JHEP* 1105 (2011), p. 031, arXiv:1101.2910 (cit. on pp. 87, 88)

[148] **C. Pope**, **L. Romans**, **X. Shen**, "The Complete Structure of W(Infinity)", *Phys.Lett.* B236 (1990), p. 173 (cit. on p. 87)

[149] **C. Pope**, **L. Romans**, **X. Shen**, "A New Higher Spin Algebra and the Lone Star Product", *Phys.Lett.* B242 (1990), pp. 401–406 (cit. on p. 87)

[150] **K. Hanaki**, **C. Peng**, "Symmetries of Holographic Super-Minimal Models", *JHEP* 1308 (2013), p. 030, arXiv:1203.5768 (cit. on p. 88)

[151] **M. Gunaydin**, **G. Sierra**, **P. Townsend**, "The Unitary Supermultiplets of $d = 3$ Anti-de Sitter and $d = 2$ Conformal Superalgebras", *Nucl.Phys.* B274 (1986), p. 429 (cit. on pp. 88, 194, 205)

[152] **A. Della Selva**, **A. Sciarrino**, "Realization of exceptional superalgebras in terms of fermion - boson creation - annihilation operators", *J.Math.Phys.* 33 (1992), pp. 1538–1545 (cit. on p. 88)

[153] **H. Afshar** et al., "Higher spin theory in 3-dimensional flat space", *Phys.Rev.Lett.* 111 (2013), p. 121603, arXiv:1307.4768 (cit. on p. 88)

[154] **H. A. Gonzalez**, **J. Matulich**, **M. Pino**, **R. Troncoso**, "Asymptotically flat spacetimes in three-dimensional higher spin gravity", *JHEP* 1309 (2013), p. 016, arXiv:1307.5651 (cit. on p. 88)

[155] **J. M. Figueroa-O'Farrill**, **J. Mas**, **E. Ramos**, "Bihamiltonian structure of the KP hierarchy and the W(KP) algebra", *Phys.Lett.* B266 (1991), pp. 298–302 (cit. on p. 89)

[156] **A. K. Das**, **W.-J. Huang**, **S. Panda**, "The Hamiltonian structures of the KP hierarchy", *Phys.Lett.* B271 (1991), pp. 109–115 (cit. on p. 89)

[157] **I. Bakas**, **E. Kiritsis**, "Beyond the large N limit: Nonlinear W(infinity) as symmetry of the SL(2,R) / U(1) coset model", *Int.J.Mod.Phys.* A7S1A (1992), pp. 55–81, arXiv:hep-th/9109029 (cit. on p. 89)

[158] **F. Yu**, **Y.-S. Wu**, "Nonlinear W(hat)(infinity) current algebra in the SL(2,R) / U(1) coset model", *Phys.Rev.Lett.* 68 (1992), pp. 2996–2999, arXiv:hep-th/9112009 (cit. on p. 89)

[159] **F. Yu**, **Y.-S. Wu**, "Nonlinearly deformed W(infinity) algebra and second Hamiltonian structure of KP hierarchy", *Nucl.Phys.* B373 (1992), pp. 713–734 (cit. on p. 89)

[160] **F. Yu**, **Y.-S. Wu**, "On the KP hierarchy, W(infinity) algebra, and conformal SL(2,R) / U(1) model. 1. The Classical case", *J.Math.Phys.* 34 (1993), pp. 5851–5871, arXiv:hep-th/9210117 (cit. on p. 89)

[161] **F. Yu**, **Y.-S. Wu**, "On the KP hierarchy, W(infinity) algebra, and conformal SL(2,R) / U(1) model: The Classical and quantum cases", (1992), arXiv:hep-th/9210162 (cit. on p. 89)

[162] **F. Yu**, **Y.-S. Wu**, "On the KP hierarchy, W(infinity) algebra, and conformal SL(2,R) / U(1) model. 2. The Quantum case", *J.Math.Phys.* 34 (1993), pp. 5872–5896, arXiv:hep-th/9210118 (cit. on p. 89)

[163] **J. M. Figueroa-O'Farrill**, **J. Mas**, **E. Ramos**, "A One parameter family of Hamiltonian structures for the KP hierarchy and a continuous deformation of the nonlinear W(KP) algebra", *Commun.Math.Phys.* 158 (1993), pp. 17–44, arXiv:hep-th/9207092 (cit. on p. 89)

[164] **C. Castro**, "A Universal W(infinity) algebra and quantization of integrable deformations of selfdual gravity", (1994) (cit. on p. 89)

[165] **C. Castro**, "Nonlinear W(infinity) algebras from nonlinear integrable deformations of selfdual gravity", *Phys.Lett.* B353 (1995), pp. 201–208 (cit. on p. 89)

[166] **S. Ghosh**, **S. K. Paul**, "N=2 superW(infinity) algebra and its nonlinear realization through superKP formulation", *Phys.Lett.* B341 (1995), pp. 293–301, arXiv:hep-th/9406124 (cit. on pp. 89, 90)

[167] **V. Drinfeld**, **V. Sokolov**, "Lie algebras and equations of Korteweg-de Vries type", *J.Sov.Math.* 30 (1984), pp. 1975–2036 (cit. on p. 90)

[168] **M. Porrati**, "Universal Limits on Massless High-Spin Particles", *Phys.Rev.* D78 (2008), p. 065016, arXiv:0804.4672 (cit. on pp. 94, 95)

[169] **A. Fotopoulos**, **M. Tsulaia**, "Interacting higher spins and the high energy limit of the bosonic string", *Phys.Rev.* D76 (2007), p. 025014, arXiv:0705.2939 (cit. on p. 94)

[170] **A. Fotopoulos**, **M. Tsulaia**, "On the Tensionless Limit of String theory, Off - Shell Higher Spin Interaction Vertices and BCFW Recursion Relations", *JHEP* 1011 (2010), p. 086, arXiv:1009.0727 (cit. on p. 94)

[171] **M. Taronna**, "Higher-Spin Interactions: four-point functions and beyond", *JHEP* 1204 (2012), p. 029, arXiv:1107.5843 (cit. on pp. 95, 175)

[172] **C. Becchi**, **A. Rouet**, **R. Stora**, "Renormalization of the Abelian Higgs-Kibble Model", *Commun.Math.Phys.* 42 (1975), pp. 127–162 (cit. on pp. 97, 104)

[173] **C. Becchi**, **A. Rouet**, **R. Stora**, "Renormalization of Gauge Theories", *Annals Phys.* 98 (1976), pp. 287–321 (cit. on pp. 97, 104)

[174] **I. Tyutin**, "Gauge Invariance in Field Theory and Statistical Physics in Operator Formalism", (1975), arXiv:0812.0580 (cit. on pp. 97, 104)

[175] **I. Batalin**, **G. Vilkovisky**, "Gauge Algebra and Quantization", *Phys.Lett.* B102 (1981), pp. 27–31 (cit. on p. 97)

[176] **G. Barnich**, **M. Henneaux**, "Consistent couplings between fields with a gauge freedom and deformations of the master equation", *Phys.Lett.* B311 (1993), pp. 123–129, arXiv:`hep-th/9304057` (cit. on pp. 97, 175)

[177] **G. Barnich**, **F. Brandt**, **M. Henneaux**, "Local BRST cohomology in the antifield formalism. 1. General theorems", *Commun.Math.Phys.* 174 (1995), pp. 57–92, arXiv:`hep-th/9405109` (cit. on pp. 97, 130, 141, 144, 165)

[178] **G. Barnich**, **F. Brandt**, **M. Henneaux**, "Local BRST cohomology in the antifield formalism. II. Application to Yang-Mills theory", *Commun.Math.Phys.* 174 (1995), pp. 93–116, arXiv:`hep-th/9405194` (cit. on pp. 97, 111, 112, 130, 141, 144, 165)

[179] **G. Barnich**, **F. Brandt**, **M. Henneaux**, "Conserved currents and gauge invariance in Yang-Mills theory", *Phys.Lett.* B346 (1995), pp. 81–86, arXiv:`hep-th/9411202` (cit. on pp. 97, 113, 144)

[180] **M. Henneaux**, "Consistent interactions between gauge fields: The Cohomological approach", *Contemp.Math.* 219 (1998), p. 93, arXiv:`hep-th/9712226` (cit. on p. 97)

[181] **M. Henneaux**, "Lectures on the Antifield-BRST Formalism for Gauge Theories", *Nucl.Phys.Proc.Suppl.* 18A (1990), pp. 47–106 (cit. on pp. 97, 100, 106, 107, 116)

[182] **G. Barnich**, **F. Brandt**, **M. Henneaux**, "Local BRST cohomology in gauge theories", *Phys.Rept.* 338 (2000), pp. 439–569, arXiv:`hep-th/0002245` (cit. on p. 97)

[183] **M. Henneaux**, "Space-time Locality of the BRST Formalism", *Commun.Math.Phys.* 140 (1991), pp. 1–14 (cit. on p. 109)

[184] **N. Boulanger**, **S. Leclercq**, **P. Sundell**, "On The Uniqueness of Minimal Coupling in Higher-Spin Gauge Theory", *JHEP* 0808 (2008), p. 056, arXiv:`0805.2764` (cit. on pp. 111, 125, 172, 175, 176)

[185] **G. Barnich**, **M. Henneaux**, "Isomorphisms between the Batalin-Vilkovisky anti-bracket and the Poisson bracket", *J.Math.Phys.* 37 (1996), pp. 5273–5296, arXiv:`hep-th/9601124` (cit. on p. 116)

[186] **M. Gerstenhaber**, "On the Deformation of Rings and Algebras", *Annals Math.* 79 (1963), pp. 59–103 (cit. on p. 118)

[187] **N. Boulanger**, **T. Damour**, **L. Gualtieri**, **M. Henneaux**, "Inconsistency of interacting, multigraviton theories", *Nucl.Phys.* B597 (2001), pp. 127–171, arXiv:`hep-th/0007220` (cit. on p. 118)

[188] **N. Boulanger**, **M. Esole**, "A Note on the uniqueness of D = 4, N=1 supergravity", *Class.Quant.Grav.* 19 (2002), pp. 2107–2124, arXiv:`gr-qc/0110072` (cit. on p. 118)

[189] **G. Barnich**, **M. Henneaux**, **R. Tatar**, "Consistent interactions between gauge fields and the local BRST cohomology: The Example of Yang-Mills models", *Int.J.Mod.Phys.* D3 (1994), pp. 139–144, arXiv:`hep-th/9307155` (cit. on p. 120)

[190] **N. Bouatta**, **G. Compere**, **A. Sagnotti**, "An Introduction to free higher-spin fields", (2004), arXiv:`hep-th/0409068` (cit. on p. 126)

[191] **E. Joung**, **M. Taronna**, "Cubic-interaction-induced deformations of higher-spin symmetries", (2013), arXiv:`1311.0242` (cit. on p. 141)

[192] **G. Barnich**, **R. Constantinescu**, **P. Gregoire**, "BRST - anti-BRST antifield formalism: The Example of the Freedman-Townsend model", *Phys.Lett.* B293 (1992), pp. 353–360, arXiv:`hep-th/9209007` (cit. on p. 141)

[193] **T. Damour**, **S. Deser**, "Higher Derivative Interactions of Higher Spin Gauge Fields", *Class.Quant.Grav.* 4 (1987), p. L95 (cit. on p. 168)

[194] **U. Gran**, "GAMMA: A Mathematica package for performing gamma matrix algebra and Fierz transformations in arbitrary dimensions", (2001), arXiv:`hep-th/0105086` (cit. on pp. 171, 227)

[195] **A. Sagnotti**, **M. Tsulaia**, "On higher spins and the tensionless limit of string theory", *Nucl.Phys.* B682 (2004), pp. 83–116, arXiv:`hep-th/0311257` (cit. on p. 174)

[196] **D. Polyakov**, "Interactions of Massless Higher Spin Fields From String Theory", *Phys.Rev.* D82 (2010), p. 066005, arXiv:`0910.5338` (cit. on p. 174)

[197] **D. Polyakov**, "Gravitational Couplings of Higher Spins from String Theory", *Int.J.Mod.Phys.* A25 (2010), pp. 4623–4640, arXiv:`1005.5512` (cit. on p. 174)

[198] **A. Sagnotti**, "Notes on Strings and Higher Spins", *J.Phys.* A46 (2013), p. 214006, arXiv:`1112.4285` (cit. on p. 174)

[199] **S. Ferrara**, **P. van Nieuwenhuizen**, "Consistent Supergravity with Complex Spin 3/2 Gauge Fields", *Phys.Rev.Lett.* 37 (1976), p. 1669 (cit. on p. 174)

[200] **D. Z. Freedman**, **A. K. Das**, "Gauge Internal Symmetry in Extended Supergravity", *Nucl.Phys.* B120 (1977), p. 221 (cit. on p. 174)

[201] **X. Bekaert**, **N. Boulanger**, **S. Leclercq**, "Strong obstruction of the Berends-Burgers-van Dam spin-3 vertex", *J.Phys.* A43 (2010), p. 185401, arXiv:1002.0289 (cit. on p. 174)

[202] **D. Francia**, **A. Sagnotti**, "Free geometric equations for higher spins", *Phys.Lett.* B543 (2002), pp. 303–310, arXiv:hep-th/0207002 (cit. on p. 175)

[203] **D. Francia**, **A. Sagnotti**, "On the geometry of higher spin gauge fields", *Class.Quant.Grav.* 20 (2003), S473–S486, arXiv:hep-th/0212185 (cit. on p. 175)

[204] **D. Francia**, **A. Sagnotti**, "Higher-spin geometry and string theory", *J.Phys.Conf.Ser.* 33 (2006), p. 57, arXiv:hep-th/0601199 (cit. on p. 175)

[205] **D. Francia**, **J. Mourad**, **A. Sagnotti**, "Current Exchanges and Unconstrained Higher Spins", *Nucl.Phys.* B773 (2007), pp. 203–237, arXiv:hep-th/0701163 (cit. on p. 175)

[206] **A. Sagnotti**, "Higher Spins and Current Exchanges", *PoS* CORFU2011 (2011), p. 106, arXiv:1002.3388 (cit. on p. 175)

[207] **P. Benincasa**, **F. Cachazo**, "Consistency Conditions on the S-Matrix of Massless Particles", (2007), arXiv:0705.4305 (cit. on p. 175)

[208] **P. Benincasa**, **E. Conde**, "Exploring the S-Matrix of Massless Particles", *Phys.Rev.* D86 (2012), p. 025007, arXiv:1108.3078 (cit. on p. 175)

[209] **P. Benincasa**, "Exploration of the Tree-Level S-Matrix of Massless Particles", *Fortsch.Phys.* 60 (2012), pp. 889–895, arXiv:1201.3191 (cit. on p. 175)

[210] **D. Polyakov**, "Higher Spins and Open Strings: Quartic Interactions", *Phys.Rev.* D83 (2011), p. 046005, arXiv:1011.0353 (cit. on p. 175)

[211] **P. Dempster**, **M. Tsulaia**, "On the Structure of Quartic Vertices for Massless Higher Spin Fields on Minkowski Background", *Nucl.Phys.* B865 (2012), pp. 353–375, arXiv:1203.5597 (cit. on p. 175)

[212] **A. K. Bengtsson**, **I. Bengtsson**, **N. Linden**, "Interacting Higher Spin Gauge Fields on the Light Front", *Class.Quant.Grav.* 4 (1987), p. 1333 (cit. on p. 175)

[213] **R. Manvelyan**, **K. Mkrtchyan**, **W. Ruhl**, "General trilinear interaction for arbitrary even higher spin gauge fields", *Nucl.Phys.* B836 (2010), pp. 204–221, arXiv:1003.2877 (cit. on p. 175)

[214] **R. Manvelyan**, **K. Mkrtchyan**, **W. Ruehl**, "A Generating function for the cubic interactions of higher spin fields", *Phys.Lett.* B696 (2011), pp. 410–415, arXiv:1009.1054 (cit. on p. 175)

[215] **R. Metsaev**, "Cubic interaction vertices of massive and massless higher spin fields", *Nucl.Phys.* B759 (2006), pp. 147–201, arXiv:hep-th/0512342 (cit. on p. 176)

[216] **R. Metsaev**, "Gravitational and higher-derivative interactions of massive spin 5/2 field in (A)dS space", *Phys.Rev.* D77 (2008), p. 025032, arXiv:hep-th/0612279 (cit. on p. 176)

[217] **R. Metsaev**, "Massless mixed symmetry bosonic free fields in d-dimensional anti-de Sitter space-time", *Phys.Lett.* B354 (1995), pp. 78–84 (cit. on p. 177)

[218] **R. Metsaev**, "Fermionic fields in the d-dimensional anti-de Sitter space-time", *Phys.Lett.* B419 (1998), pp. 49–56, arXiv:hep-th/9802097 (cit. on p. 177)

[219] **A. Fotopoulos**, **K. L. Panigrahi**, **M. Tsulaia**, "Lagrangian formulation of higher spin theories on AdS space", *Phys.Rev.* D74 (2006), p. 085029, arXiv:hep-th/0607248 (cit. on p. 177)

[220] **M. Vasiliev**, "Cubic Vertices for Symmetric Higher-Spin Gauge Fields in $(A)dS_d$", *Nucl.Phys.* B862 (2012), pp. 341–408, arXiv:1108.5921 (cit. on p. 177)

[221] **N. Boulanger**, **D. Ponomarev**, **E. Skvortsov**, "Non-abelian cubic vertices for higher-spin fields in anti-de Sitter space", *JHEP* 1305 (2013), p. 008, arXiv:1211.6979 (cit. on p. 177)

[222] **M. Porrati**, "Massive spin 5/2 fields coupled to gravity: Tree level unitarity versus the equivalence principle", *Phys.Lett.* B304 (1993), pp. 77–80, arXiv:gr-qc/9301012 (cit. on p. 177)

[223] **S. Ferrara**, **M. Porrati**, **V. L. Telegdi**, "g = 2 as the natural value of the tree level gyromagnetic ratio of elementary particles", *Phys.Rev.* D46 (1992), pp. 3529–3537 (cit. on p. 177)

[224] **J. Scherk**, **J. H. Schwarz**, "How to Get Masses from Extra Dimensions", *Nucl.Phys.* B153 (1979), pp. 61–88 (cit. on p. 177)

- [225] **J. Scherk**, **J. H. Schwarz**, "Spontaneous Breaking of Supersymmetry Through Dimensional Reduction", *Phys.Lett.* B82 (1979), p. 60 (cit. on p. 177)
- [226] **B. de Wit**, **P. Lauwers**, **A. Van Proeyen**, "Lagrangians of N=2 Supergravity - Matter Systems", *Nucl.Phys.* B255 (1985), p. 569 (cit. on p. 177)
- [227] **S. Deser**, **B. Zumino**, "Broken Supersymmetry and Supergravity", *Phys.Rev.Lett.* 38 (1977), p. 1433 (cit. on p. 177)
- [228] **M. Porrati**, **R. Rahman**, "Electromagnetically Interacting Massive Spin-2 Field: Intrinsic Cutoff and Pathologies in External Fields", (2008), arXiv:0809.2807 (cit. on p. 177)
- [229] **M. Porrati**, **R. Rahman**, "Intrinsic Cutoff and Acausality for Massive Spin 2 Fields Coupled to Electromagnetism", *Nucl.Phys.* B801 (2008), pp. 174–186, arXiv:0801.2581 (cit. on p. 177)
- [230] **M. Porrati**, **R. Rahman**, "A Model Independent Ultraviolet Cutoff for Theories with Charged Massive Higher Spin Fields", *Nucl.Phys.* B814 (2009), pp. 370–404, arXiv:0812.4254 (cit. on p. 177)
- [231] **M. Porrati**, **R. Rahman**, "Causal Propagation of a Charged Spin 3/2 Field in an External Electromagnetic Background", *Phys.Rev.* D80 (2009), p. 025009, arXiv:0906.1432 (cit. on p. 177)
- [232] **M. Porrati**, **R. Rahman**, **A. Sagnotti**, "String Theory and The Velo-Zwanziger Problem", *Nucl.Phys.* B846 (2011), pp. 250–282, arXiv:1011.6411 (cit. on p. 177)
- [233] **M. Porrati**, **R. Rahman**, "Notes on a Cure for Higher-Spin Acausality", *Phys.Rev.* D84 (2011), p. 045013, arXiv:1103.6027 (cit. on p. 177)
- [234] **R. Rahman**, "Helicity-1/2 Mode as a Probe of Interactions of Massive Rarita-Schwinger Field", *Phys.Rev.* D87 (2013), p. 065030, arXiv:1111.3366 (cit. on p. 177)
- [235] **Y. Zinoviev**, "On massive spin 2 interactions", *Nucl.Phys.* B770 (2007), pp. 83–106, arXiv:hep-th/0609170 (cit. on p. 177)
- [236] **Y. Zinoviev**, "On massive spin 2 electromagnetic interactions", *Nucl.Phys.* B821 (2009), pp. 431–451, arXiv:0901.3462 (cit. on p. 177)
- [237] **Y. Zinoviev**, "On spin 2 electromagnetic interactions", *Mod.Phys.Lett.* A24 (2009), pp. 17–23, arXiv:0806.4030 (cit. on p. 177)

[238] **I. Buchbinder**, **T. Snegirev**, **Y. Zinoviev**, "Cubic interaction vertex of higher-spin fields with external electromagnetic field", *Nucl.Phys.* B864 (2012), pp. 694–721, arXiv:`1204.2341` (cit. on p. 177)

[239] **C. Lorce**, "Electromagnetic properties for arbitrary spin particles: Natural electromagnetic moments from light-cone arguments", *Phys.Rev.* D79 (2009), p. 113011, arXiv:`0901.4200` (cit. on p. 177)

[240] **M. Banados**, **C. Teitelboim**, **J. Zanelli**, "The Black hole in three-dimensional space-time", *Phys.Rev.Lett.* 69 (1992), pp. 1849–1851, arXiv:`hep-th/9204099` (cit. on p. 183)

[241] **M. Banados**, **M. Henneaux**, **C. Teitelboim**, **J. Zanelli**, "Geometry of the (2+1) black hole", *Phys.Rev.* D48 (1993), pp. 1506–1525, arXiv:`gr-qc/9302012` (cit. on p. 183)

[242] **S. Weinberg**, "Photons and gravitons in perturbation theory: Derivation of Maxwell's and Einstein's equations", *Phys.Rev.* 138 (1965), B988–B1002 (cit. on p. 186)

[243] **B. de Wit**, **D. Z. Freedman**, "Systematics of Higher Spin Gauge Fields", *Phys.Rev.* D21 (1980), p. 358 (cit. on p. 186)

[244] **M. Dubois-Violette**, **M. Henneaux**, "Generalized cohomology for irreducible tensor fields of mixed Young symmetry type", *Lett.Math.Phys.* 49 (1999), pp. 245–252, arXiv:`math/9907135` (cit. on p. 186)

[245] **V. Kac**, "A Sketch of Lie Superalgebra Theory", *Commun.Math.Phys.* 53 (1977), pp. 31–64 (cit. on p. 194)

[246] **W. Nahm**, "Supersymmetries and their Representations", *Nucl.Phys.* B135 (1978), p. 149 (cit. on p. 194)

[247] **V. Kac**, "Lie Superalgebras", *Adv.Math.* 26 (1977), pp. 8–96 (cit. on p. 194)

[248] **X. Bekaert**, **N. Boulanger**, "Gauge invariants and Killing tensors in higher-spin gauge theories", *Nucl.Phys.* B722 (2005), pp. 225–248, arXiv:`hep-th/0505068` (cit. on pp. 251, 259)

[249] **T. Damour**, **S. Deser**, "'Geometry' of Spin 3 Gauge Theories", *Annales Poincare Phys.Theor.* 47 (1987), p. 277 (cit. on p. 259)

[250] **X. Bekaert**, **N. Boulanger**, "On geometric equations and duality for free higher spins", *Phys.Lett.* B561 (2003), pp. 183–190, arXiv:`hep-th/0301243` (cit. on p. 259)

[251] **X. Bekaert**, **N. Boulanger**, "Tensor gauge fields in arbitrary representations of GL(D,R). II. Quadratic actions", *Commun.Math.Phys.* 271 (2007), pp. 723–773, arXiv:hep-th/0606198 (cit. on p. 259)